\documentclass[a4paper,11pt]{book}

\usepackage{ando2}
\usepackage{amsmath}
\usepackage{feynmf}
\usepackage{latexsym}
\usepackage{amssymb}
\usepackage[dvips]{graphicx}
\usepackage{caption}
\usepackage{rotating}
\usepackage{latexsym}
\usepackage{verbatim}
\usepackage{hyperref}

\def\be{\begin{equation}}
\def\ee{\end{equation}}
\def\bea{\begin{eqnarray}}
\def\eea{\end{eqnarray}}
\def\bei{\begin{itemize}}
\def\eei{\end{itemize}}
\def\bs{\begin{slide}}
\def\es{\end{slide}}
\def\nn{\nonumber}
\def\half{\frac{1}{2}}
\def\pd{\partial}

\def\L{\mathcal{L}}
\def\M{\mathcal{M}}
\def\a{\alpha}
\def\ad{{\dot\alpha}}
\def\b{\beta}
\def\bd{{\dot\beta}}
\def\g{\gamma}
\def\G{\Gamma}
\def\d{\delta}
\def\D{\Delta}
\def\m{\mu}
\def\n{\nu}
\def\t{\tau}
\def\r{\rho}
\def\th{\theta}
\def\w{\omega}

\def\l{\lambda}
\def\lb{{\bar\l}}

\def\s{\sigma}

\def\sb{\bar{\s}}
\def\psib{{\bar{\psi}}}
\def\e{\epsilon}
\def\f{\phi}

\def\MZp{M_{Z'}}
\def\MZO{M_{Z_0}}
\def\MW{M_W}

\def\({\left(}
\def\){\right)}
\def\[{\left[}
\def\]{\right]}

\def\gt{\tilde{g}}
\def\qt{\tilde{q}}
\def\ut{\tilde{u}}
\def\dt{\tilde{d}}
\def\ct{\tilde{c}}
\def\st{\tilde{s}}
\def\topt{\tilde{t}}
\def\bt{\tilde{b}}
\def\hti{\tilde{h}}
\def\utcs{{{\tilde{u}^{c \dag}}}}
\def\dtcs{{{\tilde{d}^{c \dag}}}}
\def\lt{\tilde{l}}
\def\et{\tilde{e}}
\def\mt{\tilde{\mu}}
\def\taut{\tilde{\tau}}
\def\nt{\tilde{\nu}}
\def\Nt{\tilde{N}}
\def\Ct{\tilde{C}}
\def\etcs{{{\tilde{e}^{c \dag}}}}

\def\vf{\varphi}
\def\la{\langle}
\def\ra{\rangle}
\def\Tr{\textnormal{Tr}}
\def\STr{\textnormal{STr}}

\def\Eps{\epsilon^{\mu\nu\rho\sigma}}
\def\emn{\eta_{\m\n}}

\def\thth{\theta^2 \bar\theta^2}

\def\QQ{{Q_Q}}
\def\QU{{Q_{U^c}}}
\def\QD{{Q_{D^c}}}
\def\QL{{Q_L}}
\def\QE{{Q_{E^c}}}
\def\QHu{{Q_{H_u}}}
\def\QHd{{Q_{H_d}}}
\def\cA{\mathcal{A}}

\newcommand{\bth}{{\bf 3}}
\newcommand{\btw}{{\bf 2}}
\newcommand{\bon}{{\bf 1}}
\def\slashed{\ds}
\def\as{\slashed a}
\def\bs{\slashed b}
\def\cs{\slashed c}
\def\conj{{{\rm h.c.}}}
\def\MPlanck{M_{\rm P}}
\def\gu{\underline{g}}
\def\Baryon{{\rm B}}
\def\Lepton{{\rm L}}
\def\Stuckelberg{{St\"uckelberg }}
\def\axion{\text{axion}}
\def\chiral{\text{chiral}}
\def\GS{\text{GS}}
\def\GCS{\text{GCS}}
\def\NG{\text{NG}}
\def\GeV{\text{GeV}}
\def\BR{\text{BR}}
\def\numeq{n^\text{eq}}

\def\hri#1#2{\href{http://arxiv.org/abs/#1}{arXiv:#1 #2}}
\def\hre#1#2{\href{http://arxiv.org/abs/#1/#2}{arXiv:#1/#2}}

\def\ds#1{#1\kern-1ex\hbox{/}}
\def\sla{\raise.15ex\hbox{$/$}\kern-.57em}

\begin{document}

\thispagestyle{empty}

\vspace*{-1.5truecm}

\begin{center}
 \vspace{.7cm}
\makebox[\textwidth]{\raisebox{0.68cm}{
\begin{minipage}[h]{13truecm}
        \hspace*{0.2cm}
        {\sc \LARGE UNIVERSIT\`A DEGLI STUDI DI ROMA\\
        \hspace*{3.36cm}
        ``TOR \rule{0pt}{24pt}VERGATA''}\\
    \end{minipage}}}
\end{center}

\vspace{0.2cm}

\begin{center}
    \includegraphics[height=1.45truecm]{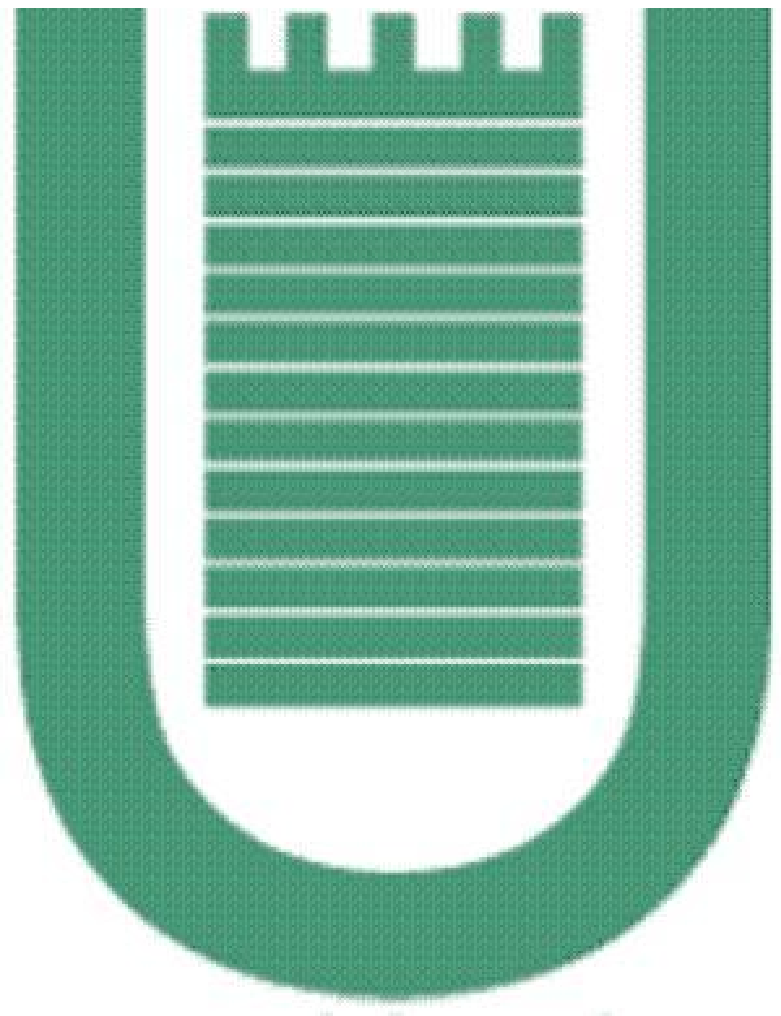}
\end{center}

\begin{center}
FACOLT\`A DI SCIENZE MATEMATICHE, FISICHE E NATURALI \\
Dipartimento di Fisica \\
\end{center}

\vspace{2cm}

\begin{center}
{\LARGE {\bf Anomalies, $U(1)'$ and the MSSM}}
\end{center}

\vspace{1.5cm}

\begin{center}
{\large Tesi di Dottorato di Ricerca in Fisica} \\\vspace{.3cm}
\rule{0pt}{22pt}{\large {\sl Antonio Racioppi} \rule{0pt}{22pt}}
\end{center}

\vspace{1.5cm}  \noindent
{\large
 {\centering
  \mbox{Relatore
  \phantom{xxxxxxxxxxxxxxxxxxxxxxxxxxxxxxxxx}
  Coordinatore del dottorato}\\
 \mbox{Prof. {\it Francesco Fucito} \phantom{xxxxxxxxxxxxxxxxxxxxxxx}
 Prof. {\it Piergiorgio Picozza}}\\}
\quad\\

\vspace{1.0cm}
\begin{center}
\underline{Ciclo XXI}\\
\end{center}

\newpage\thispagestyle{empty}
\quad\\
\newpage\thispagestyle{empty}

\quad\\
\vskip 2cm
\begin{center}
 {\Large \bf Abstract}
\end{center}
This Thesis reviews an extension of the MSSM by the addition of an anomalous abelian vector multiplet and contains
some original results concerning the phenomenology of an anomalous $Z'$.

The review part covers an introduction of the MSSM focusing on its main features, a discussion on the chiral
anomalies and how to cancel them in the Standard Model and by the Green-Schwarz mechanism.

Then, the original results are presented. We build the Lagrangian for the Minimal Anomalous $U(1)'$ Extension of the MSSM where
the anomalies are cancelled by the Green-Schwarz mechanism and the addition of Chern-Simons terms,
stressing the main differences between our model and the MSSM.
The advantage of this choice over the standard one is that it allows for arbitrary values of the quantum numbers
of the extra $U(1)$. As a first step towards the study of hadron annihilations producing four leptons in the final
state (a clean signal which might be studied at LHC) we then compute the decays $Z'\to Z_0 \g$ and $Z'\to Z_0 Z_0$.
We find that the largest values of the decay rate are $\sim 10^{-4}$ GeV, while the expected number of
events per year at LHC is at most of the order of 10.
Then we compute the relic density predicted by our model with a new dark matter candidate, the axino,
which is the LSP of the theory.
We find agreement with experimental data admitting a bino-higgsino NLSP or a wino-like NLSP, almost
degenerate in mass to the axino.

\thispagestyle{empty}
\newpage\thispagestyle{empty}
\quad\\
\newpage\thispagestyle{empty}

\chapter*{Acknowledgements}
\begin{figure}[t!]
``There are many path to the top of the mountain.\\
\phantom{``}But the view is always the same.''\\
\phantom{``But the view is always the same.''}\emph{Asian proverb}
\end{figure}

\thispagestyle{empty}

This Thesis work includes most of the research I have done during my PhD course at the University of Rome ``Tor
Vergata''. I wish to thank the Department of Physics and the Coordinator of the PhD program Prof. Piergiorgio
Picozza for supporting my work. I spent several months of my PhD absent from Tor Vergata, at DAMTP in Cambridge,
both of which I would like to thank for the hospitality and the resources they made available. I am also grateful to
INFN for travelling support and to Marie Curie Research Training Network ``Superstring Theory'' for financial support.

First of all, my gratitude goes to my advisor Prof. Francesco Fucito, who first introduced and then continued to
support me in a fascinating research field. Many thanks also to Prof. Massimo Bianchi, Dr. Andrea Lionetto, Prof.
Yassen Stanev, Prof. Gianfranco Pradisi and Dr. Pascal Anastasopoulos for their kind assistance and support in Tor
Vergata, to Prof. Michael Green for his hospitality in DAMTP and to Prof. Claudio Corian\`o for useful
collaborations and discussions.

I am very grateful to all the friends I have been in touch with during these years.
I would like to thank for their friendship: Andrea and Matteo (all started at the ``campetto" and all will finish there).
Valentina for her support and help in my troubles.
Alberto, Ilaria and Maria Vittoria for their sustain and their visits.
Elisa, Piero, Francesco and Laura for their support.
``La Dottorandi" and in particular Simona, Francesco G., Giulia, Marine, Flavio, Andrea S., Maurizio and Andrea M.;
a special thanks goes to Francesco S. for several intense Friday afternoons and his friendship.
``The Lost Legacy": Enrico, Regula and Alessandra. Gianluca and the YouTube searches. Paola and her verve.
I also wish to thank Roberta, Michele and Sven for useful discussions and their friendship.
I am grateful to UmbroTeam, Avis Frassati and Darwin College Team for giving me the possibility to play
the best sport ever: basketball.
Thanks to all of them, everywhere I was, I felt at home.

Last but not least. My deepest gratitude and love go to my parents and my brother for their support,
encouragement and care throughout my studies and travels. Without them, nothing of this would be possible.

\thispagestyle{empty}

\newpage\thispagestyle{empty}

\newpage\thispagestyle{empty}

\pagenumbering{roman} \setcounter{page}{1}
\tableofcontents
\newpage\thispagestyle{empty}

\chapter{Introduction} \label{chap:Introduction}
\pagenumbering{arabic} \setcounter{page}{1}

\fancyhead[RO]{\bfseries\leftmark}

The Standard Model (SM) of particle physics has been confirmed to a great accuracy in many experiments. Despite the fact that the
Higgs particle remains experimentally elusive, few scientists doubt that there will be major surprises in this direction.
The whole scientific community, however, knows that the SM needs to be improved.
Apart from the experimental discover of a tiny neutrino mass, there are also several
theoretical issues that make physicists believe that the SM is only an effective manifestation of a more Fundamental Theory.

We know that a new framework must appear at the Planck scale ($\sim 10^{19}$ GeV) where a theory of quantum gravity reveals.
Since the ratio between this scale and the electroweak (EW) scale ($\sim 100$ GeV) is huge, all quantum corrections will turn out
to be many orders of magnitude greater then tree values. This is the so called hierarchy problem, related to fine-tuning and naturalness.
Quantum corrections are power-law (usually quadratic) divergent which means that the highest energy physics is most important.
More technically, the question is why the Higgs boson is so much lighter than the Planck scale. Indeed one would expect that the large
quadratically divergent quantum contributions to the square of the Higgs boson mass would be inevitably make the mass huge, unless
there is a fine-tuned cancellation between the quadratic radiative corrections and the bare mass.

Besides the hierarchy problem, there some unexplained features in cosmological observations.
Measurements constraint the quantity of baryon matter, dark matter and dark energy in the
universe revealing that the known particles make up only a small fraction of the total energy density
of the universe. So particle physics must provide candidates for dark matter (DM) and dark energy and experiments to check the inherited models.

During the last years many ideas were used to overcome these problems. The most appealing theory is supersymmetry (SUSY), even if its historical
motivation was to provide fermions in string theory. In 1981 it was proposed the Minimal Supersymmetric Standard Model (MSSM) i.e. the minimal
extension of the SM which provides supersymmetry.
SUSY pairs bosons with fermions, so each SM particle has a corresponding partner (unfortunately not discovered yet).
SUSY removes the quadratic divergences of the radiative corrections to the Higgs mass. However, there is no understanding of why the Higgs
mass is so small from the beginning ($\m$ problem). One of the most intriguing feature of the MSSM (and its extensions) is that it provides
DM candidates with relic density compatible with experimental data. In fact, if $R$-parity is conserved, the lightest supersymmetric
particle of the MSSM is stable and makes it a good cold dark matter (CDM) particle.

As cited before, string theory was the framework where SUSY was born. Moreover it is also the most appealing candidate for a quantum theory
of gravity. One of the prediction of string theory is the existence of anomalous U(1)'s. In standard theories the chiral anomalies are cancelled
satisfying that the trace of the product of the gauge generators is zero.
Instead in string theory this is not always possible, so the U(1)'s are anomalous,
but the theory is still consistent and anomaly free, thanks to the Green-Schwarz (GS) mechanism.
These anomalous gauge fields must be massive, otherwise we already discovered them, and the mass is acquired with the \Stuckelberg mechanism which was the
first attempt to give a vector mass preserving gauge invariance, but phenomenologically ruled out in the SM.
This procedure involves an axion-like scalar field, which plays also an important role in the GS mechanism for the anomaly cancellation.
D-brane models contain several anomalous abelian factors living on each stack of branes. In the presence of these anomalous $U(1)$'s,
the GS couplings with the axions cancels mixed anomalies\footnote{Irreducible anomalies are cancelled by the tadpole cancellation.}, and
the \Stuckelberg mixing renders the ``anomalous" gauge fields massive. The masses depend non-trivially on the internal volumes and on other
moduli, allowing the physical masses of the anomalous $U(1)$ gauge bosons to be much smaller than the string scale
(even at a few TeV range). An important role is played by the so-called Generalized Chern-Simons terms (GCS) which are local
gauge non-invariant terms. Indeed, these trilinear gauge bosons anomalous couplings
are responsible for the cancellation of mixed anomalies between anomalous $U(1)'$s and non anomalous factors ensuring the
consistency of the theory.

In this Thesis, we are interested in {\it anomaly related} $Z'$ bosons and we present the results obtained in \cite{mypaper} and \cite{mypaper2}.
The theory is non-renormalizable since it has an energy cut-off related to the $Z'$ mass.
More precisely, we study an extension of the MSSM by the addition of an abelian vector multiplet
$V^{(0)}$ and we assume that generically all MSSM particles are charged with respect to the new $U(1)$.
The anomalies are cancelled using the GS mechanism and the GCS terms, so the anomalous abelian boson becomes massive and behaves like a $Z'$.

In approximately one year, the Large Hadron Collider (LHC) at CERN will start to operate at energies of order of 14 TeV in the center of
mass (CM). Apart from the search for the Higgs boson, it will probably give us some answers about the parameter space of the
physics beyond the SM, the eventual existence of new particles, such as the $Z'$, and their nature.

This Thesis is organized as follows.
In Chapter~\ref{chap:OverviewMSSM} we introduce the MSSM. We explain its main characteristics,
focusing on the Renormalization Group Equations (RGEs) and the mass spectrum of the theory.
In Chapter~\ref{chap:Anomalies} we discuss the problem of the 1 loop chiral anomalies in field theory.
We show how to cancel them with the standard procedure and with the GS mechanism, both in a standard and in a
supersymmetric framework.
In Chapter~\ref{chap:MiAUMSSM. Model building} we present our model. We discuss the Lagrangian and the differences between
our theory and the MSSM.
In Chapter~\ref{chap:MiAUMSSM. Phenomenology} we analyze some phenomenological consequences of our model such as the $Z'$ anomalous decays
and the prediction of the relic density with a new dark matter candidate.
In Chapter~\ref{chap:Conclusions} we summarize the conclusions and the results of our work.
Finally in the Appendix we give further details of the computations omitted in the main body of the Thesis.

\newpage\thispagestyle{empty}

\fancyhead[RO]{\bfseries\rightmark}

\chapter{Overview of the MSSM} \label{chap:OverviewMSSM}
Supersymmetry is the only possible non trivial unification of internal and space-time symmetries compatible with
quantum field theory. It emerges naturally in string theory but it can be introduced apart of it.
A SUSY transformation turns a bosonic state into a fermionic state increasing its spin by $1/2$, and viceversa.
The single particle states of a supersymmetric theory fall into irreducible representations of the SUSY algebra,
called \emph{supermultiplets} which, in turn can be organized into \emph{superfield}.
Each supermultiplet contains both fermion and boson states that are \emph{superpartners}
of each other. One may show that no particle of the SM can be the superpartner of another, so SUSY
predicts a list of new particles, none of which has been discovered yet. In reason of this, superpartners cannot be mass
degenerate and so SUSY must be a broken symmetry. There are several ways to introduce a breaking of SUSY,
however the simplest way is to explicit break it adding \emph{ad hoc} mass terms.
This is done in the MSSM.
The following treatment of the MSSM is based on the works of Martin~\cite{Martin:1997ns} and Derendinger~\cite{Derendinger}.
\newpage\section{The superpotential} \label{sect:superpot}
In a renormalizable supersymmetric field theory, the interactions and masses of all particles are determined just
by their gauge transformation properties and by the superpotential $W$. $W$ is an analytic function of the complex
chiral superfield $\Phi$. The gauge quantum numbers and mass dimensions of a chiral superfield are given by these
of its scalar component. In the superfield formulation we write the superpotential as
\be
W = \frac{1}{2} M^{ij} \Phi_i \Phi_j + \frac{1}{6} y^{ijk} \Phi_i \Phi_j \Phi_k  \label{generalW}
\ee
Given the supermultiplet content of the theory, the form of the superpotential is guided by the requirement
of gauge invariance, so only a subset of the parameters $M^{ij}$ and $y^{ijk}$ are allowed to be different from zero.
The entries of the mass matrix $M^{ij}$ can be non zero only for $i$ and $j$ such that the superfield $\Phi_i$ and $\Phi_j$
have gauge transformation conjugates of each other. Similarly for the Yukawa couplings $y^{ijk}$.
The interactions implied by the superpotential were shown in Fig.~\ref{fig:dim0coupl}-\ref{fig:dimcoupl}.
\begin{figure}[b]
\vskip 1.5cm
\centering
\raisebox{-4ex}[0cm][0cm]{\unitlength=0.4mm
\begin{fmffile}{Ayukawa1}
\begin{fmfgraph*}(60,40)
\fmfpen{thin} \fmftop{i1} \fmfbottom{i2,i3} \fmf{scalar}{i1,c1} \fmf{fermion}{i2,c1} \fmf{fermion}{i3,c1}
\fmflabel{$i$}{i1} \fmflabel{$j$}{i2} \fmflabel{$k$}{i3}
\end{fmfgraph*}
\end{fmffile}}
~~~~~~~~
\raisebox{-4ex}[0cm][0cm]{\unitlength=0.4mm
\begin{fmffile}{Ayukawa2}
\begin{fmfgraph*}(60,40)
\fmfpen{thin} \fmftop{i1} \fmfbottom{i2,i3} \fmf{scalar}{c1,i1} \fmf{fermion}{c1,i2} \fmf{fermion}{c1,i3}
\fmflabel{$i$}{i1} \fmflabel{$j$}{i2} \fmflabel{$k$}{i3}
\end{fmfgraph*}
\end{fmffile}}
~~~~~~~~
\raisebox{-4ex}[0cm][0cm]{\unitlength=0.4mm
\begin{fmffile}{Aquarticyukawa}
\begin{fmfgraph*}(60,40)
\fmfpen{thin} \fmftop{o1,o2} \fmfbottom{i1,i2} \fmf{scalar}{c1,o1} \fmf{scalar}{c1,o2} \fmf{scalar}{i1,c1}
\fmf{scalar}{i2,c1} \fmflabel{$i$}{i1} \fmflabel{$j$}{i2} \fmflabel{$k$}{o1} \fmflabel{$l$}{o2}
\end{fmfgraph*}
\end{fmffile}}
\vskip 0.5cm (a) ~~~~~~~~~~~~~~~~~~~~~~~~~~~~ (b) ~~~~~~~~~~~~~~~~~~~~~~~~~~~~ (c)
\caption{The dimensionless non-gauge interaction vertices in a supersymmetric theory: (a) scalar-fermion-fermion
Yukawa interaction $y^{ijk}$, (b) the complex conjugate interaction $y_{ijk}$, and (c) quartic scalar interaction
$y^{ijn}y^*_{kln}$.} \label{fig:dim0coupl}
\end{figure}
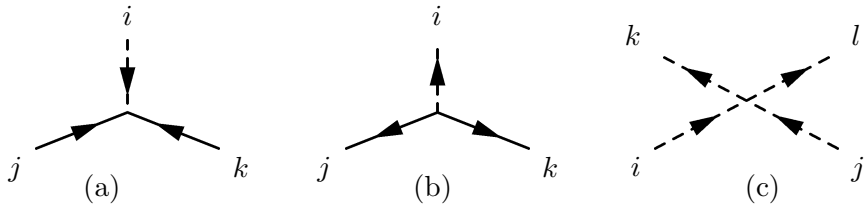
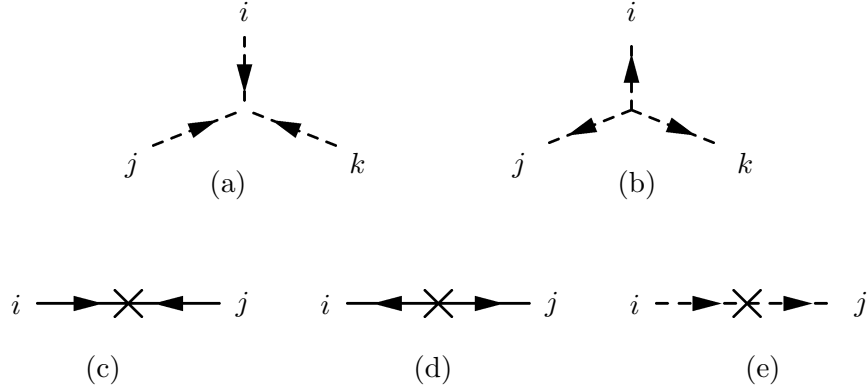
\begin{figure}[p]
\vskip 1.5cm
\centering
\raisebox{-4ex}[0cm][0cm]{\unitlength=0.4mm
\begin{fmffile}{A3scalar1}
\begin{fmfgraph*}(60,40)
\fmfpen{thin} \fmftop{i1} \fmfbottom{i2,i3} \fmf{scalar}{i1,c1} \fmf{scalar}{i2,c1} \fmf{scalar}{i3,c1}
\fmflabel{$i$}{i1} \fmflabel{$j$}{i2} \fmflabel{$k$}{i3}
\end{fmfgraph*}
\end{fmffile}}
~~~~~~~~~~~~~~~~
\raisebox{-4ex}[0cm][0cm]{\unitlength=0.4mm
\begin{fmffile}{A3scalar2}
\begin{fmfgraph*}(60,40)
\fmfpen{thin} \fmftop{i1} \fmfbottom{i2,i3} \fmf{scalar}{c1,i1} \fmf{scalar}{c1,i2} \fmf{scalar}{c1,i3}
\fmflabel{$i$}{i1} \fmflabel{$j$}{i2} \fmflabel{$k$}{i3}
\end{fmfgraph*}
\end{fmffile}}
\vskip 0.5cm (a) ~~~~~~~~~~~~~~~~~~~~~~~~~~~~~~~~~~~~ (b) \vskip 1cm
\raisebox{-4ex}[0cm][0cm]{\unitlength=0.4mm
\begin{fmffile}{Afermionmass1}
\begin{fmfgraph*}(60,40)
\fmfpen{thin} \fmfleft{i1} \fmfright{i2} \fmf{fermion}{i1,c1} \fmf{fermion}{i2,c1} \fmfv{decor.shape=cross}{c1}
\fmflabel{$i$}{i1} \fmflabel{$j$}{i2}
\end{fmfgraph*}
\end{fmffile}}
~~~~~~~~
\raisebox{-4ex}[0cm][0cm]{\unitlength=0.4mm
\begin{fmffile}{Afermionmass2}
\begin{fmfgraph*}(60,40)
\fmfpen{thin} \fmfleft{i1} \fmfright{i2} \fmf{fermion}{c1,i1} \fmf{fermion}{c1,i2} \fmfv{decor.shape=cross}{c1}
\fmflabel{$i$}{i1} \fmflabel{$j$}{i2}
\end{fmfgraph*}
\end{fmffile}}
~~~~~~~~
\raisebox{-4ex}[0cm][0cm]{\unitlength=0.4mm
\begin{fmffile}{Ascalarmass}
\begin{fmfgraph*}(60,40)
\fmfpen{thin} \fmfleft{i1} \fmfright{i2} \fmf{scalar}{i1,c1} \fmf{scalar}{c1,i2} \fmfv{decor.shape=cross}{c1}
\fmflabel{$i$}{i1} \fmflabel{$j$}{i2}
\end{fmfgraph*}
\end{fmffile}}
\vskip 0.2cm (c) ~~~~~~~~~~~~~~~~~~~~~~~~~~~~ (d) ~~~~~~~~~~~~~~~~~~~~~~~~~~~~ (e)
\caption{Supersymmetric dimensionful couplings: (a) (scalar)$^3$ interaction vertex $M^*_{in} y^{jkn}$ and (b) the
conjugate interaction $M^{in} y^*_{jkn}$, (c) fermion mass term $M^{ij}$ and (d) conjugate fermion mass term
$M^*_{ij}$, and (e) scalar squared-mass term $M^*_{ik}M^{kj}$.} \label{fig:dimcoupl}
\end{figure}
%
%
%
%
%
\begin{table}[p]
\centering
\begin{tabular}[h]{|c|c|c|c|c|c|}
 \hline \multicolumn{3}{|c|}{Supermultiplet} & \multicolumn{3}{|c|}{Gauge Representation} \\
 \hline
 \hline Superfield  &           Spin $0$           &        Spin $1/2$           & $SU(3)_c$    & $SU(2)_L$ & $U(1)_Y$  \\
 \hline $Q_i$       &   $(\ut_{i L} \ \dt_{i L})$  &  $(u_{i L} \ d_{i L})$      & $\bth$       &  $\btw$   & $1/6$     \\
 \hline $U^c_i$     &   $\ut^c_i=\ut_{i R}^\dag $  &  $u^c_i=u^\dag_{i R}$       & $\bar \bth$  &  $\bon$   & $-2/3$    \\
 \hline $D^c_i$     &   $\dt^c_i=\dt_{i R}^\dag$   &  $d^c_i=d^\dag_{i R}$       & $\bar \bth$  &  $\bon$   & $1/3$     \\
 \hline $L_i$       &   $(\nt_{i L} \ \et_{i L})$  &  $(\n_{i L} \ e_{i L})$     & $\bon$       &  $\btw$   & $-1/2$    \\
 \hline $E^c_i$     &   $\et^c_i=\et_{i R}^\dag$   &  $e^c_i=e^\dag_{i R}$       & $\bon$       &  $\bon$   &  $1$      \\
 \hline $H_u$       &   $(h_u^+ \ h_u^0)$          &  $(\hti_u^+ \ \hti_u^0)$    & $\bon$       &  $\btw$   & $1/2$     \\
 \hline $H_d$       &   $(h_d^0 \ h_d^-)$          &  $(\hti_d^0 \ \hti_d^-)$    & $\bon$       &  $\btw$   & $-1/2$    \\
 \hline
\end{tabular}
\caption{Chiral supermultiplets in the Minimal Supersymmetric Standard Model. The spin-$0$ fields are complex
scalars, and the spin-$1/2$ fields are left-handed two-component Weyl fermions. The index $i$ runs over the three
families.}\label{tab:chiral}
\end{table}
%
%
\begin{table}[p]
\centering
\begin{tabular}[h]{|c|c|c|c|c|c|}
 \hline \multicolumn{3}{|c|}{Supermultiplet} & \multicolumn{3}{|c|}{Gauge Representation} \\
 \hline
 \hline Superfield  &   Spin $1/2$                             &   Spin $1$      & $SU(3)_c$ & $SU(2)_L$ & $U(1)_Y$  \\
 \hline $V_3$       &   $\tilde G$ or $\l^{(3)}$                   &  $G$            & $\bf 8$   &  $\bf 1$  & 0         \\
 \hline $V_2$       &   $\tilde W^\pm \  \tilde W^0$ or $\l^{(2)}_i$ &  $W^\pm \ W^0$  & $\bf 1$   &  $\bf 3$  & 0         \\
 \hline $V_1$       &   $\tilde B$                   or $\l^{(1)}$ &   $B$           & $\bf 1$   &  $\bf 1$  & 0         \\
 \hline
\end{tabular}
\caption{Gauge supermultiplets in the Minimal Supersymmetric Standard Model. The $\bf 1$ means that the field is a
singlet i.e. uncharged respect to $SU(n)$ gauge groups.\label{tab:gauge}}
\end{table}

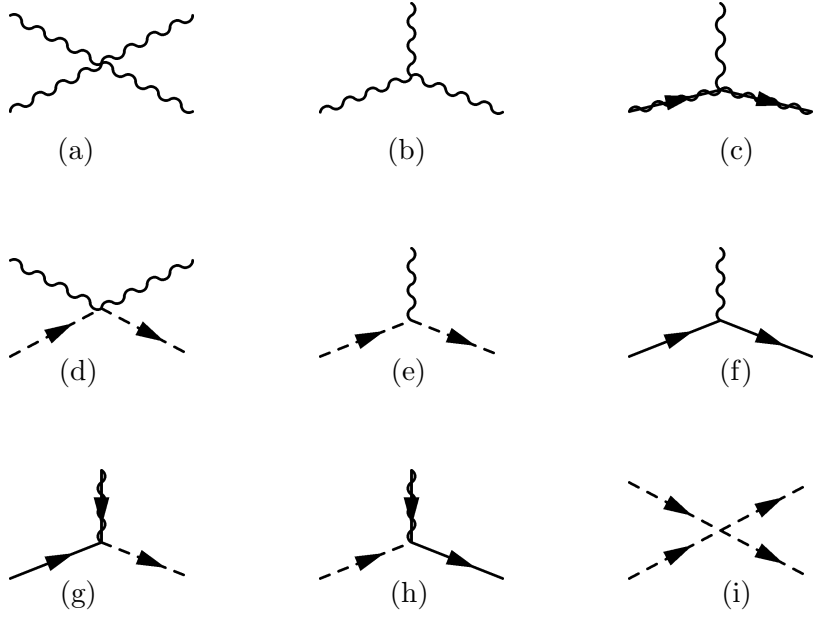
\begin{figure}[h]
\vskip 1.5cm \centering
\raisebox{-4ex}[0cm][0cm]{\unitlength=0.4mm
\begin{fmffile}{A4vectors}
\begin{fmfgraph*}(60,40)
\fmfpen{thin} \fmftop{o1,o2} \fmfbottom{i1,i2} \fmf{boson}{c1,o1} \fmf{boson}{c1,o2} \fmf{boson}{i1,c1}
\fmf{boson}{i2,c1}
\end{fmfgraph*}
\end{fmffile}}
~~~~~~~~
\raisebox{-4ex}[0cm][0cm]{\unitlength=0.4mm
\begin{fmffile}{A3vectors}
\begin{fmfgraph*}(60,40)
\fmfpen{thin} \fmftop{i1} \fmfbottom{i2,i3} \fmf{boson}{c1,i1} \fmf{boson}{c1,i2} \fmf{boson}{c1,i3}
\end{fmfgraph*}
\end{fmffile}}
~~~~~~~~
\raisebox{-4ex}[0cm][0cm]{\unitlength=0.4mm
\begin{fmffile}{A1vectors2gauginos}
\begin{fmfgraph*}(60,40)
\fmfpen{thin} \fmftop{i1} \fmfbottom{i2,i3} \fmf{boson}{c1,i1} \fmf{boson}{c1,i2} \fmf{boson}{c1,i3}
\fmf{fermion}{c1,i3} \fmf{fermion}{i2,c1}
\end{fmfgraph*}
\end{fmffile}}
\vskip 0.5cm (a) ~~~~~~~~~~~~~~~~~~~~~~~~~~~~ (b) ~~~~~~~~~~~~~~~~~~~~~~~~~~~~ (c) \vskip 1.5cm
\raisebox{-4ex}[0cm][0cm]{\unitlength=0.4mm
\begin{fmffile}{A2vectors2scalars}
\begin{fmfgraph*}(60,40)
\fmfpen{thin} \fmftop{o1,o2} \fmfbottom{i1,i2} \fmf{boson}{c1,o1} \fmf{boson}{c1,o2} \fmf{scalar}{i1,c1}
\fmf{scalar}{c1,i2}
\end{fmfgraph*}
\end{fmffile}}
~~~~~~~~
\raisebox{-4ex}[0cm][0cm]{\unitlength=0.4mm
\begin{fmffile}{A1vectors2scalars}
\begin{fmfgraph*}(60,40)
\fmfpen{thin} \fmftop{o1} \fmfbottom{i1,i2} \fmf{boson}{c1,o1} \fmf{scalar}{i1,c1}  \fmf{scalar}{c1,i2}
\end{fmfgraph*}
\end{fmffile}}
~~~~~~~~
\raisebox{-4ex}[0cm][0cm]{\unitlength=0.4mm
\begin{fmffile}{A1vectors2fermions}
\begin{fmfgraph*}(60,40)
\fmfpen{thin} \fmftop{o1} \fmfbottom{i1,i2} \fmf{boson}{c1,o1} \fmf{fermion}{i1,c1}  \fmf{fermion}{c1,i2}
\end{fmfgraph*}
\end{fmffile}}
\vskip 0.2cm (d) ~~~~~~~~~~~~~~~~~~~~~~~~~~~~ (e) ~~~~~~~~~~~~~~~~~~~~~~~~~~~~ (f)
\vskip 1.5cm
\raisebox{-4ex}[0cm][0cm]{\unitlength=0.4mm
\begin{fmffile}{A1gaugino1fermion1scalarv01}
\begin{fmfgraph*}(60,40)
\fmfpen{thin} \fmftop{o1} \fmfbottom{i1,i2} \fmf{boson}{o1,c1}  \fmf{fermion}{i1,c1}  \fmf{scalar}{c1,i2} \fmffreeze
\fmf{fermion}{o1,c1}
\end{fmfgraph*}
\end{fmffile}}
~~~~~~~~
\raisebox{-4ex}[0cm][0cm]{\unitlength=0.4mm
\begin{fmffile}{A1gaugino1fermion1scalarv02}
\begin{fmfgraph*}(60,40)
\fmfpen{thin} \fmftop{o1} \fmfbottom{i1,i2} \fmf{boson}{o1,c1}  \fmf{scalar}{i1,c1}  \fmf{fermion}{c1,i2} \fmffreeze
\fmf{fermion}{o1,c1}
\end{fmfgraph*}
\end{fmffile}}
~~~~~~~~
\raisebox{-4ex}[0cm][0cm]{\unitlength=0.4mm
\begin{fmffile}{A4scalars}
\begin{fmfgraph*}(60,40)
\fmfpen{thin} \fmftop{o1,o2} \fmfbottom{i1,i2} \fmf{scalar}{o1,c1} \fmf{scalar}{c1,o2} \fmf{scalar}{i1,c1}
\fmf{scalar}{c1,i2}
\end{fmfgraph*}
\end{fmffile}}
\vskip 0.2cm (g) ~~~~~~~~~~~~~~~~~~~~~~~~~~~~ (h) ~~~~~~~~~~~~~~~~~~~~~~~~~~~~ (i)
\caption{Supersymmetric gauge interaction vertices.} \label{fig:gaugecoupl}
\end{figure}

The particle content of the MSSM is given in Table~\ref{tab:chiral}-\ref{tab:gauge}.
The superpotential for the MSSM is
\be
W_\text{MSSM} = y_u^{i j} Q_i U^c_j H_u - y_d^{i j} Q_i D^c_j H_d - y_e^{i j} L_i E^c_j H_d + \m H_u H_d
\label{MSSMsuperpot}
\ee
where the indices $i, j$ run over the three families. For the complete Langrangian of the MSSM see Appendix~\ref{app:MSSM Lagrangian}.

All of the gauge indices in eq.~(\ref{MSSMsuperpot}) are suppressed.
The ``$\mu$ term", as it is usually called, can be written out as $\m (H_u)_\a (H_d)_\b \e^{\a \b}$,
where $\e^{\a \b}$ is used to tie together $SU(2)_L$ weak isospin indices $\a, \b=1,2$ in a gauge-invariant way.
Similarly for the terms $Q_i H_u$, $Q_i H_d$ and $L_i H_d$.

The $\mu$ term in eq.~(\ref{MSSMsuperpot}) is the supersymmetric version of the Higgs boson mass in the SM.
It is unique, since terms $H_u^\dag H_u$ or $H_d^\dag H_d$ are forbidden in the superpotential, because it must
be analytic in the chiral superfields. We can also see from the form of eq.~(\ref{MSSMsuperpot}) why both
$H_u$ and $H_d$ are needed in order to give Yukawa couplings, and thus masses, to all of the quarks and leptons.
Since the superpotential must be analytic, for instance the $Q U^c H_u$ Yukawa terms cannot be replaced by
something like $Q U^c H_d^\dag$. So we need both $H_u$ and $H_d$, even without invoking the argument based
on anomaly cancellation (as we will see in Section~\ref{sect:Anomaly cancellation. SM and MSSM}).

Since the top quark, bottom quark and tau lepton are the heaviest fermions in SM, it is common to use the approximation
that only the third (3,3) component of $y_u$, $y_d$ and $y_e$ are important
\be
 y_u^{ij} \approx \( \begin{array}{ccc}
                      0 & 0 & 0 \\
                      0 & 0 & 0 \\
                      0 & 0 & y_t \\
                     \end{array} \) \qquad
 y_d^{ij} \approx \( \begin{array}{ccc}
                      0 & 0 & 0 \\
                      0 & 0 & 0 \\
                      0 & 0 & y_b \\
                     \end{array} \) \qquad
 y_\t^{ij} \approx \( \begin{array}{ccc}
                      0 & 0 & 0 \\
                      0 & 0 & 0 \\
                      0 & 0 & y_\t \\
                     \end{array} \) \qquad
\label{3familyapprox}
\ee
In this limit the MSSM superpotential reads in $SU(2)_L$ components
\bea
 W_{\rm MSSM} &\approx&  y_t  (t^c t h_u^0 - t^c b h_u^+) - y_b (b^c t h_d^- - b^c b h_d^0) -
                         y_\t (\t^c \n_\t h_d^- - \t^c \t h_d^0) \nn\\
                && + \m (h_u^+ h_d^- - h_u^0 h_d^0)
\label{Wthird}
\eea
where $Q_3=(t \ b)$, $L_3=(\n_\t \ \t)$, $h_u=(h_u^+ \ h_u^0)$, $h_d=(h_d^0 \ h_d^-)$, $u_3^c=t^c$, $d_3^c=b^c$
and $e_3^c=\t^c$. The minus sign inside the parentheses appear because of the use of $\e^{\a \b}$ to tie up internal
$SU(2)_L$ indices. These minus signs were chosen so that terms proportional to $y_t$, $y_b$ and $y_\t$ have
positive signs when they become the top, bottom and tau masses.

Since the Yukawa interactions $y^{ijk}$ in a general supersymmetric theory
must be completely symmetric under interchange of $i,j,k$, we know that
$y_u$, $y_d$ and $y_e$ imply not only Higgs-quark-quark
and Higgs-lepton-lepton couplings as in the SM, but also
squark-higgsino-quark and slepton-higgsino-lepton interactions.

There are also scalar quartic interactions with strength proportional to the square of Yukawa couplings.
The existence of all the quark and lepton Yukawa couplings in the superpotential~(\ref{MSSMsuperpot}) leads also to
the following scalar quartic couplings: (squark)$^4$, (slepton)$^4$, (squark)$^2$(slepton)$^2$, (squark)$^2$(Higgs)$^2$ and
(slepton)$^2$(Higgs)$^2$.
However, the dimensionless interactions determined by the superpotential are usually not the most important ones
of direct interest for phenomenology. This is because the Yukawa couplings are already known to
be very small, except for those of the third family (top, bottom, tau). Instead, production and decay processes
for superpartners in the MSSM are typically dominated by the supersymmetric interactions of gauge-coupling
strength. The couplings of the Standard Model gauge bosons ($\g$, $W^\pm$, $Z^0$ and gluons) to the MSSM
particles are determined completely by the gauge invariance of the kinetic terms in the Lagrangian.
The gauginos also couple to (squark, quark) and (slepton, lepton) and (Higgs, higgsino) pairs as illustrated in
the general case in Fig.~\ref{fig:gaugecoupl}.
For instance, each of the squark-quark-gluino couplings is given by
$\sqrt{2} g_3 (\tilde q \, T^{a} q \tilde g +\conj)$ where $T^a = \lambda^a/2 \quad (a=1\ldots 8)$
are the matrix generators for $SU(3)_C$.

The dimensionful couplings in the supersymmetric part of the MSSM Lagrangian are all dependent on $\mu$.
From eq.~(\ref{MSSMsuperpot}), we get the higgsino fermion mass terms
\be
 -\L_{\text{higgsino mass}}=  \m (\hti_u^+ \hti_d^- - \hti_u^0 \hti_d^0)+ \conj
\label{htmass}
\ee
as well as Higgs squared-mass terms in the scalar potential
\be
-\L_{\text{susy Higgs mass}} \,=\, |\mu|^2 \( |h_u^0|^2 + |h_u^+|^2 + |h_d^0|^2 + |h_d^-|^2 \)
\label{hmass}
\ee
Since eq.~(\ref{hmass}) positive definite, we cannot understand electroweak symmetry breaking (EWSB) without
including a negative supersymmetry-breaking squared-mass soft term for the Higgs scalars (see Section~\ref{subsect:EWSB}).
Moreover we expect that $\mu$ should be roughly of order $10^2$ or $10^3$ GeV,
not so far from the electroweak scale. But why $|\mu|^2$ is so small compared to $\MPlanck^2$?
Why should it be roughly of the same order as $m^2_{\rm soft}$?
The scalar potential of the MSSM seems to depend on two distinct dimensionful parameters, namely the
supersymmetry-preserving mass $\mu$ and the supersymmetry-breaking soft mass terms.
The observed value for the electroweak breaking scale suggests, both of these should be within an order of magnitude
or so of 100 GeV. This puzzle is called ``the $\mu$ problem". Several different
solutions to the $\mu$ problem have been proposed, but they all work in roughly the
same way\footnote{Some other attractive solutions are proposed in \cite{muproblemW}-\cite{muproblemGMSB}.}; the
$\mu$ term is absent at tree-level before symmetry breaking, and then it arises from the VEV(s) of some new
field(s). These VEVs are determined by minimizing a potential that depends on soft terms.
Thus, the value of the effective parameter $\mu$ is related to supersymmetry breaking.
From the point of view of the MSSM, however, we can just treat $\mu$ as an independent parameter.

Finally, the $\mu$-term and the Yukawa couplings in the superpotential eq.~(\ref{MSSMsuperpot}) combine to yield
(scalar)$^3$ couplings of the form
\bea \L_{\text{susy (scalar)$^3$}} &=& \m^* (
  \ut^c_i \, y_u^{ij} \, \ut_j \, h_d^{0\dag} + \dt^c_i \, y_d^{ij} \, \dt_j \, h_u^{0\dag}
+ \et^c_i \, y_e^{ij} \, \et_j \, h_u^{0\dag} +\nn\\
&&
  \ut^c_i \, y_u^{ij} \, \dt_j \, h_d^{-\dag} + \dt^c_i \, y_d^{ij} \, \ut_j \, h_u^{+\dag}
+ \et^c_i \, y_e^{ij} \, \nt_j \, h_u^{+\dag}
)+ \conj
\label{scalar3terms}
\eea
\section{Soft supersymmetry breaking} \label{sect:Softsusybreak}
A realistic phenomenological model must contain SUSY breaking. From a theoretical point of view we expect that SUSY
is an exact symmetry which is spontaneously broken. Many models have been proposed but we do not discuss them
because is beyond our scope. From a practical perspective it is sufficient to parametrize our ignorance introducing
extra terms that break explicitly SUSY. However this breaking should be soft because we do not want to lose the main
properties of a SUSY theory (see Chapter~\ref{chap:Introduction} and \cite{Martin:1997ns}).
As illustrated in \cite{Derendinger} the one loop divergent contributions
to the scalar potential are given by
\be
 \d V=\frac{\Lambda^2}{32 \pi^2} \STr \M^2 + \frac{1}{64 \pi^2} \STr \M^4 \ln \( \frac{\M^2}{\Lambda^2} \)
 \label{Vrunning}
\ee
where $\Lambda$ is the energy cut-off and $\M$ is the mass matrix of the theory. As it is well known a SUSY theory
is free of quadratic divergences, in fact $\STr \M^2=0$, so the soft breaking terms must be such that this property
still holds.
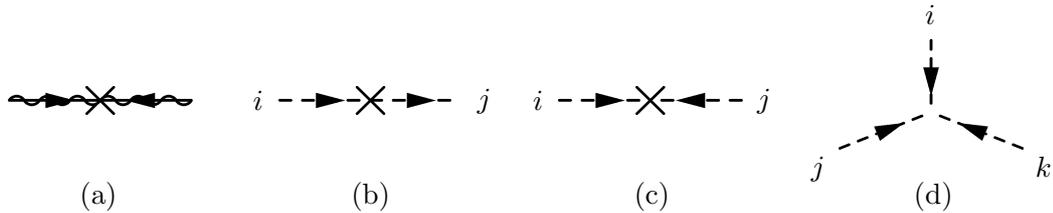
\begin{figure}[t]
\vskip 1.5cm \centering
\raisebox{-4ex}[0cm][0cm]{\unitlength=0.4mm
\begin{fmffile}{Agauginomass1}
\begin{fmfgraph*}(60,40)
\fmfpen{thin} \fmfleft{i1} \fmfright{i2} \fmf{boson}{i1,c1} \fmf{boson}{i2,c1} \fmffreeze
\fmfv{decor.shape=cross}{c1} \fmf{fermion}{i1,c1} \fmf{fermion}{i2,c1}
\end{fmfgraph*}
\end{fmffile}}
~~~~
\raisebox{-4ex}[0cm][0cm]{\unitlength=0.4mm
\begin{fmffile}{Ascalarmass}
\begin{fmfgraph*}(60,40)
\fmfpen{thin} \fmfleft{i1} \fmfright{i2} \fmf{scalar}{i1,c1} \fmf{scalar}{c1,i2} \fmfv{decor.shape=cross}{c1}
\fmflabel{$i$}{i1} \fmflabel{$j$}{i2}
\end{fmfgraph*}
\end{fmffile}}
~~~~~
\raisebox{-4ex}[0cm][0cm]{\unitlength=0.4mm
\begin{fmffile}{Ascalarmass2}
\begin{fmfgraph*}(60,40)
\fmfpen{thin} \fmfleft{i1} \fmfright{i2} \fmf{scalar}{i1,c1} \fmf{scalar}{i2,c1} \fmfv{decor.shape=cross}{c1}
\fmflabel{$i$}{i1} \fmflabel{$j$}{i2}
\end{fmfgraph*}
\end{fmffile}}
~~~~~
\raisebox{-4ex}[0cm][0cm]{\unitlength=0.4mm
\begin{fmffile}{A3scalar1}
\begin{fmfgraph*}(60,40)
\fmfpen{thin} \fmftop{i1} \fmfbottom{i2,i3} \fmf{scalar}{i1,c1} \fmf{scalar}{i2,c1} \fmf{scalar}{i3,c1}
\fmflabel{$i$}{i1} \fmflabel{$j$}{i2} \fmflabel{$k$}{i3}
\end{fmfgraph*}
\end{fmffile}}
~
\vskip 0.6cm
\begin{flushleft}
~~~~~~~~~~~(a)~~~~~~~~~~~~~~~~~~~~~~~~(b)~~~~~~~~~~~~~~~~~~~~~~~~~(c)~~~~~~~~~~~~~~~~~~~~~~~~~(d)
\end{flushleft}
\caption{Soft supersymmetry-breaking terms: (a) Gaugino mass $M_a$; (b) non-analytic scalar squared mass
$(m^2)_j^i$; (c) analytic scalar squared mass $b^{ij}$; and (d) scalar cubic coupling $a^{ijk}$. For each of the
interactions in a,c,d there is another with all arrows reversed, corresponding to the complex conjugate term in the
Lagrangian.}  \label{fig:soft}
\end{figure}
The possible soft supersymmetry breaking of a general renormalizable theory are
\be
 \L_{\text{soft}}= - \( \frac{1}{2} M_a \l^a \l^a + \frac{1}{6} a^{ijk} \f_i \f_j \f_k + \frac{1}{2} b^{ij} \f_i \f_j \) + \conj
                   - \frac{1}{2} \( m^2 \)^i_j \f^{j\dag} \f_i
 \label{generalLsoft}
\ee
They are gaugino masses $M_a$ for each gauge group, scalar squared mass terms $b^{ij}$ and $\( m^2 \)^i_j$ and
trilinear scalar coupling $a^{ijk}$. There are no mass terms for the chiral supermultiplet fermions, like $ -
\frac{1}{2} m^{ij} \psi_i \psi_j + \conj$, because in general they are not soft\footnote{In Section~\ref{subsec: softterms} we will
see a case in which is possible to have such a term.}. The Lagrangian $\L_{\text{soft}}$ breaks SUSY since it
involves only scalars and gauginos and not their respective superpartners. The gaugino masses $M_a$ are always
allowed by gauge symmetry. The $\( m^2 \)^i_j$ terms are allowed for $i,j$ such that $\f_i, \f^{j\dag}$ transform in
complex conjugate representations under all gauge groups. This is obviously true when $i=j$, so every scalar can
acquire mass in this way. The remaining soft terms are restricted by the symmetries. Because the $a^{ijk}$ and
$b^{ij}$ have the same form as $y^{ijk}$ and $M^{ij}$, they will be allowed if and only if there is a corresponding
superpotential term in the Lagrangian. The corresponding Feynman diagrams are given in Fig.~\ref{fig:soft}.

Applying what we have shown to the MSSM, we get that
\bea
 \L_{soft}^{MSSM} &=& - {1\over2}  \sum_{a=1}^3 \(M_a \l^{(a)} \l^{(a)} + \conj \) -
              \( m^2_{q_{ij}} \qt_i \qt_j^\dag + m^2_{u_{ij}} \ut^c_i \utcs_j + m^2_{d_{ij}} \dt^c_i \dtcs_j  \right. \nn\\
                 &&\left.+m^2_{l_{ij}} \lt_i \lt_j^\dag + m^2_{e_{ij}} \et^c_i \etcs_j + m^2_{h_u} |h_u|^2 + m^2_{h_d} |h_d|^2 \) \nn\\
                 &&-\( a_u^{ij} \qt_i \qt_j h_u - a_d^{ij} \qt_i \dt_j h_d - a_e^{ij} \lt_i \et_j h_d + b h_u h_d + \conj \)
\label{LsoftMSSM}
\eea
We see soft mass terms for the gluinos $\l^{(3)}$, the winos $\l^{(2)}$ and the bino $\l^{(1)}$. Then we have the (scalar)$^3$
couplings, which are in one to one correspondence with the Yukawa couplings of the superpotential and the sleptons
and squarks masses. As usual we suppressed the gauge indices. Finally we have the soft terms that contributes to the
Higgs potential, $m_{h_u}$, $m_{h_d}$ and $b$.

The above $\L_{soft}^{MSSM}$ introduces many new parameters which are absent in the ordinary SM.
A careful count \cite{dimsut} shows that there are 105 masses, phases and mixing angles in the MSSM Lagrangian that cannot
be rotated away by some redefinition of the basis of quark and lepton supermultiplets. However there are experimental
flavor mixing and CP violation data which limit this number \cite{flavorreview}. All of these potentially dangerous
effects can be avoided assuming that supersymmetry breaking is ``universal''. In particular we can suppose that
the squark and slepton masses are flavor-blind, then
\bea
 && m^2_{q_{ij}} \sim m^2_q \, \d_{ij} \qquad m^2_{u_{ij}} \sim m^2_u \, \d_{ij} \qquad
    m^2_{d_{ij}} \sim m^2_d \, \d_{ij}  \nn\\
 && m^2_{l_{ij}} \sim m^2_l \, \d_{ij} \qquad m^2_{e_{ij}} \sim m^2_e \, \d_{ij}  \label{scalarmassunification}
\eea
If so, then all sfermions mixing angles are trivial. Therefore MSSM contributions to FCNC processes will be very small
up to mixing induced by $a_u$, $a_d$ and $a_e$. Making the further assumption that these couplings are proportional to the
respective Yukawa coupling,
\be
 a_u = A_u y_u \qquad a_d = A_d y_d \qquad a_e = A_e y_e \label{softpropyukawa}
\ee
will ensure that only the sfermions of the third family can have large (scalar)$^3$ couplings. Finally, to avoid large
CP-violation effects it is usually assumed that soft terms do not introduce new complex phases. This is automatic
for the Higgses and the sfermions if eq.~(\ref{scalarmassunification}) is true. We can also fix the $\m$ and $b$ parameters
to be real, rotating the phase of $H_{u,d}$ supermultiplets. Assuming also that the $A_i$ in eq.~(\ref{softpropyukawa})
and the gaugino soft masses are real, then the only CP-violating phase in the theory will be the ordinary
CKM phase of the SM.

The conditions~(\ref{scalarmassunification}) and (\ref{softpropyukawa}) reflect the assumption of soft-breaking universality
invoked in the above. These must be taken as boundary conditions on the running of soft parameters at a
very high RG scale $Q_0$. Then we must run all the Lagrangian parameters down to the EW scale.
\section{RGE} \label{sect:RGE}
In this section we list the result for the RGEs in the MSSM.
For a method of computation see for instance \cite{Derendinger}.

The 1 loop RG equations for the gauge couplings are
\be
 \b \equiv \frac{d}{dt} \gu_a = \frac{1}{16 \pi^2} b_a \gu_a^3,
 \quad (b_1,b_2,b_3) = \left\{ \begin{array}{cc} (41/10, \ -19/6, \ -7) & \text{SM}\\
                                                (33/5, \ 1, \ -3) & \text{MSSM}
                               \end{array} \right.
 \label{grge}
\ee
where $t=\ln(Q/Q_0)$ with $Q$ the RG scale, $\gu_1 = \sqrt{5/3} \, g_1$, $\gu_2 = g_2$ and $\gu_3 = g_3$.
The normalization of $\gu_1$ is chosen to agree with Grand Unification Theories (GUTs) like $SU(5)$ or $SO(10)$.
The quantities $\a_a=\gu_a^2/4 \pi$ turn out to have their reciprocals that run linearly with the RG scale
\be
 \frac{d}{dt} \a_a^{-1} = -\frac{b_a}{2 \pi}
 \label{alpharge}
\ee
In Fig.~\ref{fig:gaugeunification} is plotted the RG evolution of the $\a_a^{-1}$. Unlike the SM, which has a very
rough unification, MSSM has a very ``precise'' unification at the scale $M_U \sim 2 \times 10^{16}$ GeV. This can be
an accident, but it may also be taken as a hint toward SUSY GUT or superstring models.
\begin{figure}[t]
 \centering
 \includegraphics[scale=0.4]{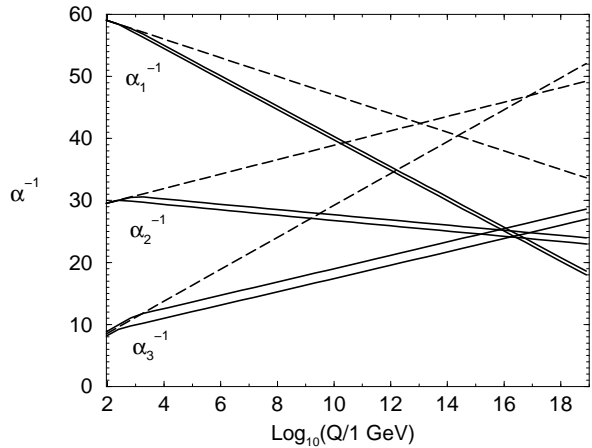}
 \caption{RG evolution of the inverse gauge couplings $\alpha_a^{-1}(Q)$ in the SM (dashed lines) and the MSSM (solid lines).
          In the MSSM case, the sparticle mass thresholds are varied between 250 GeV and 1 TeV,
          and $\alpha_3(\MZO)$ between $0.113$ and $0.123$. Two-loop effects are included.} \label{fig:gaugeunification}
\end{figure}

Next we consider the 1 loop RGEs of the parameters in the superpotential. Considering the third family
approximation~(\ref{3familyapprox}) we get the following equations for the Yukawa couplings
\bea
 \b_{y_t} \equiv {d\over dt} y_t &=&
    {y_t \over 16 \pi^2} \[ 6 y_t^* y_t + y_b^* y_b - {16\over 3} g_3^2 - 3 g_2^2 - {13\over 15} g_1^2 \]
 \label{eq:betayt}\\
 \b_{y_b} \equiv {d\over dt} y_b &=&
    {y_b \over 16 \pi^2} \[ 6 y_b^* y_b + y_t^* y_t + y_\tau^* y_\tau -{16\over 3} g_3^2 - 3 g_2^2 - {7\over 15} g_1^2 \]
 \label{eq:betayb}\\
 \b_{y_\t} \equiv {d\over dt} y_\t &=&
    {y_\t \over 16 \pi^2} \[ 4 y_\tau^* y_\tau + 3 y_b^* y_b - 3 g_2^2 - {9\over 5} g_1^2 \]
 \label{eq:betaytau}
\eea
while the RGE for the $\m$ term is
\be
 \b_\m \equiv {d\over dt} \m =
  {\m \over 16 \pi^2} \[ 3 y_t^*y_t + 3 y_b^* y_b + y_\t^* y_\t -3 g_2^2 - {3\over 5} g_1^2 \]
 \label{eq:betamu}
\ee

The one-loop RG equations for the three gaugino mass parameters in the MSSM are determined by the same quantities
$b_a^{\rm MSSM}$ that appear in the gauge coupling RG eqs.~(\ref{grge})
\be
\b_{M_a} \equiv {d\over dt} M_a \,=\, {1\over 8\pi^2} b_a g_a^2 M_a\qquad\>\>\> (b_a = 33/5, \>1,\>-3)
\label{gauginomassrge}
\ee
for $a=1,2,3$. It is easy to check that the three ratios $M_a/g_a^2$ are each constant (RG scale independent) at 1
loop order. Since the gauge couplings unify at $Q = M_U = 2 \times 10^{16}$ GeV, it is a natural assumption that the
gaugino masses also unify near that scale, with a value $m_{1/2}$. If so, then it follows that
\be
 {M_1 \over g_1^2} = {M_2 \over g_2^2} = {M_3 \over g_3^2} = {m_{1/2} \over g_U^2} \label{gauginomassunification}
\ee
at any RG scale, up to small two-loop effects. Here $g_U$ is the unified gauge coupling at $Q = M_U$.

Then we consider the 1-loop RG equations for the soft parameters connected to the superpotential.
Using the third family approximation~(\ref{3familyapprox}) we can write at any RG scale
\be
 a_u^{ij} \approx \( \begin{array}{ccc}
                      0 & 0 & 0 \\
                      0 & 0 & 0 \\
                      0 & 0 & a_t \\
                     \end{array} \) \qquad
 a_d^{ij} \approx \( \begin{array}{ccc}
                      0 & 0 & 0 \\
                      0 & 0 & 0 \\
                      0 & 0 & a_b \\
                     \end{array} \) \qquad
 a_\t^{ij} \approx \( \begin{array}{ccc}
                      0 & 0 & 0 \\
                      0 & 0 & 0 \\
                      0 & 0 & a_\t \\
                     \end{array} \)
\label{soft3familyapprox}
\ee
which defines the running parameters $a_t$, $a_b$ and $a_\t$.
In this approximation the RGEs for the trilinear soft couplings are
\bea
 16\pi^2 {d\over dt} a_t &=&
    a_t \[ 18 y_t^* y_t + y_b^* y_b - {16\over 3} g_3^2 - 3 g_2^2 - {13\over 15} g_1^2 \] + 2 a_b y_b^* y_t \nn\\
 &&+ y_t \[ {32\over 3} g_3^2 M_3 + 6 g_2^2 M_2 + {26\over 15} g_1^2 M_1 \] \label{atrge} \\
 16\pi^2 {d\over dt} a_b &=&
    a_b \[ 18 y_b^* y_b + y_t^* y_t + y_\tau^* y_\tau - {16\over 3} g_3^2 - 3 g_2^2 - {7\over 15} g_1^2 \]
            + 2 a_t y_t^* y_b   \nn\\
 &&+ 2 a_\tau y_\tau^* y_b
   + y_b \[ {32\over 3} g_3^2 M_3 + 6 g_2^2 M_2 + {14 \over 15} g_1^2 M_1 \] \qquad \label{abrge} \\
 16\pi^2 {d\over dt} a_\t &=&
    a_\t \[ 12 y_\t^* y_\t + 3 y_b^* y_b - 3 g_2^2 - {9\over 5} g_1^2 \] + 6 a_b y_b^* y_\t +\nn\\
 &&+y_\t \[ 6 g_2^2 M_2 + {18\over 5} g_1^2 M_1 \] \label{ataurge}
\eea
while the RGE for the $b$ parameter is
\bea
  16\pi^2{d\over dt} b &=&
         b \[ 3 y_t^* y_t + 3 y_b^* y_b + y_\t^* y_\t - 3 g_2^2 - {3\over 5} g_1^2 \]
  \phantom{xxxxxxxxxxxxxxxxxxxxx}
  \nn\\
  &&+ \mu \[ 6 a_t y_t^* + 6 a_b y_b^* + 2 a_\t y_\t^* + 6 g_2^2 M_2 + {6\over 5} g_1^2 M_1 \]
\label{brge}
\eea
The $\beta$-function for each of these soft parameters is not proportional to the parameter itself,
because couplings which break supersymmetry are not protected by supersymmetric non-renormalization theorems.

Next let us consider the RG equations for the scalar masses in the MSSM. In the approximation of
eqs.~(\ref{3familyapprox}) and (\ref{soft3familyapprox}), the scalar masses satisfy boundary conditions like
eq.~(\ref{scalarmassunification}) at an input RG scale, so when renormalized to any other RG scale, they will
stay almost diagonal,
\be
 m^2_{q_{ij}} \approx \( \begin{array}{ccc}
                           m^2_{q_1} & 0 & 0 \\
                           0 & m^2_{q_2} & 0 \\
                           0 & 0 & m^2_{q_3} \\
                          \end{array} \) \qquad
 m^2_{u_{ij}} \approx \( \begin{array}{ccc}
                           m^2_{u_1} & 0 & 0 \\
                           0 & m^2_{u_2} & 0 \\
                           0 & 0 & m^2_{u_3} \\
                          \end{array} \) \qquad
\dots \label{softscalarmassmatrix}
\ee
The first and second family sfermions remain very nearly degenerate
\be
 m^2_{q_2} \approx m^2_{q_1} \qquad m^2_{u_2} \approx m^2_{u_1} \qquad \dots \label{softscalarmass12approx}
\ee
but the third-family sfermions have larger Yukawa couplings and so their squared masses get renormalized
differently. The one-loop RGEs for the first and second family squark and slepton squared masses are
\be
 16 \pi^2 {d\over dt} m_{\f_i}^2 = -\sum_{a=1,2,3} 8 g_a^2 C_a (\f_i) |M_a|^2 + \frac{6}{5} Y_i g^{2}_1 S
 \label{rge12scalar}
\ee
for each scalar $\phi_i$, where $a$ runs over the three gauge groups $U(1)_Y$, $SU(2)_L$ and $SU(3)_C$,
$C_a(\f_i)$ are the quadratic Casimir invariants and $M_a$ are the corresponding running gaugino masses.
Also,
\be
 S \equiv \Tr[Y_j m^2_{\f_j}] = m_{H_u}^2 - m_{H_d}^2 +
   \Tr[ m^2_{q_{ij}} - m^2_{l_{ij}} - 2 m^2_{u_{ij}} + m^2_{d_{ij}} + m^2_{e_{ij}}]
\label{eq:defS}
\ee
The right-hand side of eq.~(\ref{rge12scalar}) is negative\footnote{The contributions proportional to $S$ is zero
in mSUGRA models.}, so the scalar squared-mass parameters grow as they are RG-evolved from the input scale down
to the EW scale. The scalars masses will obtain large positive contributions at the electroweak scale,
because of the effects of the gaugino masses.

The RG equations for the remaining scalars, Higgs and third family sfermions, also get contributions from
the large Yukawa ($y_{t,b,\tau}$) and soft ($a_{t,b,\tau}$) couplings. It is useful to define the following quantities
\bea
 X_t  &=& 2  |y_t|^2 (m_{h_u}^2 + m_{q_3}^2 + m_{u_3}^2) + 2  |a_t|^2  \label{Xt} \\
 X_b  &=& 2  |y_b|^2 (m_{h_d}^2 + m_{q_3}^2 + m_{d_3}^2) + 2  |a_b|^2  \label{Xb} \\
 X_\t &=& 2 |y_\t|^2 (m_{h_d}^2 + m_{l_3}^2 + m_{e_3}^2) + 2 |a_\t|^2  \label{Xtau}
\eea
In terms of these quantities, the RGEs for the soft Higgs squared-masses are
\bea
 16\pi^2 {d\over dt} m_{h_u}^2 &=&
         3 X_t - 6 g_2^2 |M_2|^2 - {6\over 5} g_1^2 |M_1|^2 + \frac{3}{5} g^{2}_1 S \label{mhurge}\\
 16\pi^2 {d\over dt} m_{h_d}^2 &=&
         3 X_b + X_\t - 6 g_2^2 |M_2|^2 - {6\over 5} g_1^2 |M_1|^2 - \frac{3}{5} g^{2}_1 S \label{mhdrge}
\eea
while the RG equations for the third family sfermions mass parameters are
\bea
 16\pi^2{d\over dt} m_{q_3}^2 &=&
  X_t + X_b-{32\over 3} g_3^2 |M_3|^2 -6 g_2^2 |M_2|^2 -{2\over 15} g_1^2 |M_1|^2 + \frac{1}{5} g^{2}_1 S
  \phantom{xxxxx}
   \label{mq3rge} \\
 16\pi^2 {d\over dt} m_{u_3}^2 &=&
  2 X_t - {32\over 3} g_3^2 |M_3|^2 - {32\over 15} g_1^2|M_1|^2 - \frac{4}{5} g^{2}_1 S
   \label{mt3rge} \\
 16\pi^2 {d\over dt} m_{d_3}^2 &=&
  2 X_b - {32\over 3} g_3^2 |M_3|^2 - {8\over 15} g_1^2|M_1|^2 + \frac{2}{5} g^{2}_1 S
   \label{md3rge} \\
 16\pi^2 {d\over dt} m_{l_3}^2 &=&
  X_\tau  - 6 g_2^2 |M_2|^2 - {6\over 5} g_1^2 |M_1|^2 - \frac{3}{5} g^{2}_1 S
   \label{mstaurge}\\
 16\pi^2 {d\over dt} m_{e_3}^2 &=& 2 X_\t - {24\over 5} g_1^2 |M_1|^2  + \frac{6}{5} g^{2}_1 S .
  \label{mstaucrge}
\eea

An examination of eq.~(\ref{atrge})-(\ref{brge}), (\ref{rge12scalar}), and (\ref{mhurge})-(\ref{mstaucrge}) reveals
that if the gaugino masses $M_1$, $M_2$, and $M_3$ are non-zero at the input scale, then all of the other soft
terms will be generated too. On the other hand, if the gaugino masses vanished at tree-level,
then they would not get any contributions to their masses at the one-loop order; in that case the gauginos would be
extremely light and the model would not be phenomenologically acceptable.
\section{Mass spectrum} \label{sect:Mass spectrum}
\subsection{Electroweak symmetry breaking} \label{subsect:EWSB}
In the MSSM, the description of electroweak symmetry breaking is slightly complicated by the fact that there
are two complex Higgs doublets $h_u =(h_u^+,\> h_u^0)$ and $h_d = (h_d^0,\> h_d^-)$ rather than just one in the
ordinary SM. The classical scalar potential for the Higgs sector in the MSSM is given by
\bea
 V &=& (|\m|^2 + m^2_{h_u}) (|h_u^0|^2 + |h_u^+|^2) + (|\m|^2 + m^2_{h_d}) (|h_d^0|^2 + |h_d^-|^2) \nn\\
  &&+\, [b\, (h_u^+ h_d^- - h_u^0 h_d^0) + \conj] \nn\\
  && + {1\over 8} (g_2^2 + g_1^2) (|h_u^0|^2 + |h_u^+|^2 - |h_d^0|^2 - |h_d^-|^2 )^2
 + \half g^2 |h_u^+ h_d^{0\dag} + h_u^0 h_d^{-\dag}|^2
 \phantom{xxxxx}
\label{VHiggscomplete}
\eea
The terms proportional to $|\mu |^2$ come from $F$-terms while the terms proportional to $g_1^2$ and $g_2^2$
are the $D$-term contributions. Finally, the terms proportional to $m_{h_u}^2$, $m_{h_d}^2$ and $b$ are coming
from the soft breaking Lagrangian~(\ref{LsoftMSSM}). The full scalar potential of the theory also includes many terms
involving the sparticle fields which we can ignore here, since they do not get VEVs because they have
large positive squared masses.

We now have to demand that the minimum of this potential should give a correct electroweak symmetry breaking
$SU(2)_L\times U(1)_Y \rightarrow U(1)_{\rm EM}$. It is easy to see that in the minimum we have $h_u^+=h_d^-=0$.
This is good, because it means that at the minimum of the potential electromagnetism remains unbroken,
since the charged components of the $H_{u,d}$ cannot get VEVs.
After setting $h_u^+=h_d^-=0$, we are left to consider the scalar potential
\bea
 V &=& (|\m|^2 + m^2_{h_u}) |h_u^0|^2 + (|\m|^2 + m^2_{h_d}) |h_d^0|^2 - (b\, h_u^0 h_d^0 + \conj) \nn\\
    && + {1\over 8} (g_2^2 + g_1^2) ( |h_u^0|^2 - |h_d^0|^2 )^2
 \label{Vhiggs}
\eea
The only term in this potential that depends on the phases of the fields is the $b$-term.
Therefore, redefining the phase of $h_u$ or $h_d$ we can absorb any phase in $b$,
so we take $b$ to be real and positive.
To have a bounded from below potential and to avoid the possibility to have $h_u^0=h_d^0=0$ in the minimum,
the parameters in the potential must obey the following constraints
\bea
 2 b &<& 2 |\m |^2 + m^2_{h_u} + m^2_{h_d}        \label{eq:boundedfrombelow} \\
 b^2 &>& (|\m|^2 + m^2_{h_u} )(|\m|^2 + m^2_{h_d})  \label{eq:destabilizeorigin}
\eea
Having established the conditions necessary for $h_u^0$ and $h_d^0$ to get
non-zero VEVs, we can now require that they give a correct EW phenomenology. Let us write
\be
 v_u/\sqrt2 = \langle h_u^0\rangle,
 \qquad\qquad
 v_d/\sqrt2 = \langle h_d^0\rangle.
 \label{defvuvd}
\ee
These VEVs are related to the known mass of the $Z_0$ boson and the gauge couplings (see eq.~(\ref{Z0mass}))
\be
 \frac{v_u^2 + v_d^2}{2} = \frac{v^2}{2} = 2 \MZO^2/(g_1^2 + g_2^2) \approx (174\>{\rm GeV})^2
\label{vuvdcon}
\ee
The ratio of the VEVs is usually written as
\be
 \tan\b \equiv v_u/v_d.
\label{deftanbeta}
\ee
The value of $\tan\b$ is not fixed by experiments, but it depends on the Lagrangian parameters of
the MSSM in a calculable way. Now we can write down the conditions for minimizing the potential
\bea
 &&m_{h_u}^2 + |\m |^2 -b \cot\b - (\MZO^2/2) \cos (2\b) \>=\> 0 \label{mubsub2} \\
 &&m_{h_d}^2 + |\m |^2 -b \tan\b + (\MZO^2/2) \cos (2\b) \>=\> 0
\label{condminVHiggs}
\eea
Using $|\mu|^2$, $b$, $m_{h_u}^2$ and $m_{h_d}^2$ as input parameters, we can solve these two equations to give the values
for $\b$ and $\MZO$
\bea
 \sin (2\b) &=& \frac{2 b}{m^2_{h_u} + m^2_{h_d} + 2|\m|^2} \label{eq:solvesintwobeta}\\
 \MZO^2 &=& \frac{|m^2_{h_d} - m^2_{h_u}|}{\sqrt{1 - \sin^2(2\b)}}- m^2_{h_u} - m^2_{h_d} -2|\m|^2 \label{eq:solvemzsq}
\eea
In the following subsections we will treat the mass spectrum for each particle sector of the theory at tree level.ù
\subsection{Higgs Bosons} \label{subsect:Higgs Bosons}
The Higgs scalar fields in the MSSM consist of two complex $SU(2)_L$-doublet, so eight real, scalar degrees of freedom.
When the electroweak symmetry is broken, three of them are the would-be Nambu-Goldstone (NG) bosons $G^0$, $G^\pm$,
which become the longitudinal modes of the $Z^0$ and $W^\pm$ massive vector bosons. The remaining five Higgs scalar mass
eigenstates consist of two CP-even neutral scalars $h^0$ and $H^0$, one CP-odd neutral scalar $A^0$, and a charge $+1$
scalar $H^+$ and its conjugate charge $-1$ scalar $H^-$.\footnote{Here we define $G^- = G^{+*}$ and $H^- = H^{+*}$.
Also, by convention, $h^0$ is lighter than $H^0$.} The gauge-eigenstate fields can be expressed in terms of the mass
eigenstate fields as
\bea
 \( \begin{array}{c} h_u^0 \\
                     h_d^0 \end{array} \) &=& {1\over \sqrt{2}}
 \( \begin{array}{c} v_u \\
                     v_d \end{array}\) + {1\over \sqrt{2}} R_\a
 \( \begin{array}{c} h^0 \\
                     H^0 \end{array}\) + {i\over \sqrt{2}} R_{\b}
 \( \begin{array}{c} G^0 \\
                     A^0 \end{array}\)
 \label{NeutralHiggs}\\
 \( \begin{array}{c} h_u^+ \\
                     h_d^{-*} \end{array} \) &=&  R_{\b}
 \( \begin{array}{c} G^+ \\
                     H^+ \end{array}\)
\eea
where the orthogonal rotation matrices
\be
R_{\a} =  \( \begin{array}{cc}  \cos\a & \sin\a \\
                                   -\sin\a & \cos\a \end{array} \) \qquad\quad
R_{\b} = \( \begin{array}{cc}  \sin\b & \cos\b \\
                                   -\cos\b & \sin\b \end{array} \) \label{Ralpha and Rbeta}
\ee
are chosen to diagonalize quadratic part of the potential
\bea
 V_{\text{Higgs mass}} &=& \half m_{h^0}^2 (h^{0})^2 + \half m_{H^0}^2 (H^{0})^2 + \half m_{G^0}^2 (G^{0})^2
       + \half m_{A^0}^2 (A^{0})^2\nn\\
    && + m_{G^\pm}^2 |G^+|^2 +  m_{H^\pm}^2 |H^+|^2
\eea
The mass eigenvalues are given by
\bea
 m_{A^0}^2 &=& 2|\m|^2 + m^2_{h_u} + m^2_{h_d} \label{eq:m2A0}\\
 m^2_{h^0, H^0} &=& \frac{1}{2} \[ m^2_{A^0} + \MZO^2 \mp
                    \sqrt{\( m_{A^0}^2 - \MZO^2\)^2 + 4 \MZO^2 m_{A^0}^2  \sin^2 (2\b)} \]\>\>\>\>\>{} \label{eq:m2hH}\\
 m^2_{H^\pm} &=& m^2_{A^0} + \MW^2 =  m^2_{A^0} +  g_2^2 \frac{v^2}{4} \label{eq:m2Hpm}\\
 m^2_{G^0, G^\pm} &=& 0
\eea
The mixing angle $\a$ is determined by
\be
 {\sin 2\a \over \sin 2\b} = -{ m_{H^0}^2 + m_{h^0}^2 \over m_{H^0}^2 - m_{h^0}^2 }  \qquad\qquad
 {\tan 2\a \over \tan 2\b} =  { m_{A^0}^2 + \MZO^2 \over  m_{A^0}^2 - \MZO^2 } \label{alphaangle}
\ee
and is traditionally chosen to be negative, so $-\pi/2 < \a <0$ (provided $m_{A_0}>\MZO$).

The masses of $A^0$, $H^0$ and $H^\pm$ can in principle be arbitrarily large since they all grow with $b/\sin (2\b)$.
In contrast, the mass of $h^0$ is bounded above.
From eq.~(\ref{eq:m2hH}), one finds at tree-level \cite{treelevelhiggsbound}
\be
 m_{h^0} \><\> \MZO  |\cos (2\beta) |
\label{eq:higgsineq}
\ee
If this inequality were robust, we would have discovered the lightest Higgs boson at LEP2. In fact the tree level
formula~(\ref{eq:m2hH}) is  subject to drastic quantum corrections and it is possible to show that  $m^2_{h^0}$ can
exceed the LEP bounds. Assuming that none of the sfermions masses exceed 1 TeV and that all the couplings in the model
remain perturbative up the GUT scale, we get \cite{KKW}
\be
 m_{h^0} \lesssim 150\>{\rm GeV}.
\label{generalhiggsbound}
\ee
However the previous bound is weakened if we relax one of the previous assumptions.
\subsection{Vector Bosons} \label{subsect:Vector Bosons}
The gauge bosons in the MSSM get masses in the same way of the SM with the identification $v^2=v_u^2+v_d^2$.
We recall very briefly the Lagrangian
\bea
\!\!\!\!\!\!\!\!
 \L_{\rm Vect.Mass}&=& g_2^2 \frac{v^2}{8} \(A_{1\m} A_1^\m+A_{2\m} A_2^\m\) + \nn\\
                     &&+\frac{1}{2} \(\ Y_\m \ A_{3\m} \)
                     \( \begin{array}{cc} g_1^2 \frac{v^2}{4}    & -g_1 g_2 \frac{v^2}{4}   \\
                                          -g_1 g_2 \frac{v^2}{4} & g_2^2 \frac{v^2}{4}     \\ \end{array} \)
                     \( \begin{array}{c} Y^\m\\ A_3^\m \end{array} \) \!\!\!\!\!\!\!\! \nn\\
                    &=&\MW^2 W^+_\m W^{-\m} + \MZO^2 Z_{0\m} Z_0^\m \!\!\!\!\!\!\!\!
\eea
where the eigenstates are (the Weinberg angle $\theta_W$ is defined by $\tan\theta_W=g_1/g_2$)
\bea
 W^\pm_\m &=& \frac{ A_{1\m} \mp i A_{2\m} }{\sqrt2}             \label{W}\\
 Z_{0\m}  &=& \frac{g_2 A_{3\m} - g_1 Y_\m}{\sqrt{g_1^2+g_2^2}}= \cos\theta_W A_{3\m} - \sin\theta_W Y_\m  \label{Z0}\\
 A_\m     &=& \frac{g_1 A_{3\m} + g_2 Y_\m}{\sqrt{g_1^2+g_2^2}}= \sin\theta_W A_{3\m} + \cos\theta_W Y_\m  \label{photon}
\eea
and their corresponding eigenvalues are
\bea
 \MW^2   &=&\frac{1}{4}       g_2^2 v^2     \label{Wmass}\\
 \MZO^2 &=&\frac{1}{4} \(g_1^2+g_2^2\) v^2 \label{Z0mass}\\
 M^2_{\g}  &=& 0
\eea
The corresponding coupling constants are
\bea
 g_W     &=& \frac{g_2}{\sqrt 2}        \label{gW} \\
 g_{Z_0} &=& \sqrt{g_1^2+g_2^2}         \label{gZ0} \\
       e &=& \frac{g_1 g_2}{\sqrt{g_1^2+g_2^2}}  \label{e}
\eea
Exactly the same results of SM.
\subsection{Neutralinos and Charginos} \label{subsect:Neutralinos and Charginos}
The higgsinos and electroweak gauginos mix with each other because of the effects of EWSB.
The neutral higgsinos ($\hti_u^0$ and $\hti_d^0$) and the neutral gauginos ($\l^{(1)}$, $\l^{(2)}_3$)
combine to form four mass eigenstates called {\it neutralinos} ($\Nt_i,\ i=1 \dots 4$).
The charged higgsinos ($\hti_u^+$ and $\hti_d^-$) and winos ($\tilde W^+$ and $\tilde W^-$) mix to form
two mass eigenstates with charge $\pm 1$ called {\it charginos} ($\Ct^\pm_i,\ i=1,2$).\footnote{
Other common notations use $\tilde \chi_i^0$ or $\tilde Z_i$ for neutralinos, and
$\tilde \chi^\pm_i$ or $\tilde W^\pm_i$ for charginos.} By convention, these are labelled in ascending order, so that
$m_{\Nt_1} < m_{\Nt_2} <m_{\Nt_3} <m_{\Nt_4}$ and $m_{\Ct_1} < m_{\Ct_2}$.

With the basis $\psi^0 = (\l^{(1)}, \l^{(2)}, \hti_d^0, \hti_u^0)$, the neutralino mass part of the Lagrangian is
\be
 \L_{\text{neutralino mass}} = -\frac{1}{2} (\psi^{0})^T {\bf M}_{\Nt} \psi^0 + \conj \label{neutrmassL}
\ee
where ${\bf M}_{\Nt}$ is the following matrix
\be
 {\bf M}_{\Nt}
  =   \(\begin{array}{cccc}
          M_1 &    0  & -g_1 v_d/2 & g_1 v_u/2 \\
        \dots &   M_2 & g_2 v_d/2 & -g_2 v_u/2 \\
        \dots & \dots &     0      & -\m  \\
        \dots & \dots &   \dots    & 0
      \end{array}\) \label{Neutralinomassmatrix}
\ee
where the lower dots denote the obvious terms under symmetrization.
The entries $M_1$ and $M_2$ are the soft gaugino masses in~(\ref{LsoftMSSM}), while the entries $-\mu$ are
the supersymmetric higgsino mass terms in~(\ref{MSSMsuperpot}). The terms proportional to $g_{1,2}$ are the
result of Higgs-higgsino-gaugino couplings (see Fig.~\ref{fig:gaugecoupl} g,h), with the Higgs replaced by their VEVs.

The parameters $M_1$, $M_2$, and $\m$ in the equations above can have arbitrary complex phases.
A redefinition of the phases of $\l^{(1)}$ and $\l^{(2)}$ permit us to choose $M_1$ and $M_2$ both real and positive.
The phase of $\m$ within that convention is then really a physical parameter and cannot be rotated away.
However, if $\m$ is not real, then there can be potentially disastrous CP-violating effects in low-energy physics,
so it is usual to assume that $\m$ is real in the same set of phase conventions that make
$M_1$, $M_2$, $b$, $v_u$ and $v_d$ real and positive.
The sign of $\m$ is still undetermined by this constraint.

There is a not-unlikely limit in which EWSB is a small perturbation inside ${\bf M}_{\Nt}$
\bea
 \MZO \,\ll\, |\mu \pm M_{1}|,\, |\mu \pm M_{2}|
\label{gauginolike}
\eea
In this case the neutralino eigenstates are very nearly a ``bino-like" $\Nt_1 \approx \l^{(1)}$, a ``wino-like"
$\Nt_2 \approx \l^{(2)}$ and ``higgsino-like" $\Nt_{3,4} \approx(\hti_u^0 \pm \hti_d^0)/\sqrt{2}$, with masses
\bea
 m_{\Nt_1} &=& M_1 - {\MZO^2 s^2_W (M_1 + \m \sin 2\b ) \over \m^2 - M_1^2 } + \dots\\
 \label{mNt1approx}
 m_{\Nt_2} &=& M_2 - {    \MW^2    (M_2 + \m \sin 2 \b) \over \m^2 - M_2^2 } + \dots\\
 \label{mNt2approx}
 m_{\Nt_3} &=& |\m|+ {\MZO^2 (I - \sin 2 \b) (\m + M_1 c^2_W + M_2 s^2_W) \over 2 (\m + M_1) (\m + M_2) } + \dots\\
 \label{mNt3approx}
 m_{\Nt_4} &=& |\m|+ {\MZO^2 (I + \sin 2 \b) (\m - M_1 c^2_W - M_2 s^2_W) \over 2 (\m - M_1) (\m - M_2) } + \dots
 \label{mNt4approx}
\eea
where $c_W$ and $s_W$ are respectively the cosine and sine of the Weinberg angle $\th_W$ and $I = \pm 1$
is the sign of the $\m$ parameter. The neutralino labels should be rearranged depending on the
numerical values of the parameters; in particular the above labelling is a common situation in mSUGRA models.

The chargino spectrum can be treated in a similar way. In the gauge-eigenstate basis
$\psi^\pm = (\tilde W^+,\, \hti_u^+,\,\tilde W^- ,\, \hti_d^- )$, the chargino mass terms are
\be
 \L_{\text{chargino mass}} = -\half (\psi^\pm)^T {\bf M}_{\Ct} \psi^\pm + \conj \label{chargmassL}
\ee
where, in $2\times 2$ block form,
\be
 {\bf M}_{\Ct} =  \(\begin{array}{cc}
                            0  & {\bf X}^T \\
                       {\bf X} &  0
                      \end{array}\)
\qquad \text{with} \quad
 {\bf X} =   \(\begin{array}{cc}
                       M_2     & g_2 v_u \\
                       g_2 v_d & \m
                 \end{array}\)  \label{charginomassmatrix}
\ee
Since these are only 2$\times$2 matrices, it is easy to find the mass eigenvalues
\bea
 m^2_{\Ct_{1,2}}&=&{1\over 2} \( |M_2|^2 + |\m|^2 + 2\MW^2 \)\nn\\
                   &&\mp {1\over 2} \sqrt{(|M_2|^2 + |\m |^2 + 2 \MW^2 )^2 - 4 | \m M_2 - \MW^2 \sin 2\b |^2 }
\label{charginomasses}
\eea
In the limit of eq.~(\ref{gauginolike}), the charginos consist of a wino-like $\Ct_1^\pm$ and a higgsino-like $\Ct_2^\pm$, with masses
\bea
 m_{\Ct_1} &=&  M_2 - {  \MW^2  (M_2 + \m \sin 2\b ) \over \m^2 - M_2^2 } + \label{Ct1mass} \dots\\
 m_{\Ct_2} &=& |\m| + { \MW^2 I (\m + M_2 \sin 2\b)  \over \m^2 - M^2_2 } + \label{Ct2mass} \dots
\eea
The labelling is given in the same way as the neutralino case. Amusingly, $\Ct_1$ is nearly degenerate with $\Nt_2$
in the approximation shown, but it is not an exact result.
Similarly the higgsino-like fermions $\Nt_3$, $\Nt_4$ and $\Ct_2$ have masses close to $|\m|$.
\subsection{Gluinos} \label{subsect:Gluinos}
The gluino is a color octet fermion, it cannot mix with any other particle in the MSSM. Because it is not involved
in EWSB, the only source of its mass is the soft parameter $M_3$. In mSUGRA models this soft mass is related
to the other soft gaugino masses, giving the rough prediction
\be
 M_3 : M_2 : M_1 \approx 6:2:1
\ee
near the TeV scale. Then it is reasonable to suppose that the gluino is much heavier than the lighter neutralinos
and charginos.
\subsection{Sfermions} \label{subsect:Sfermions}
A priori, any scalars with the same electric charge, $R$-parity\footnote{See Subsection~\ref{subsect:R parity}.},
and color quantum numbers can mix with each other. This means that in general, the mass eigenstates of the squarks and
sleptons of the MSSM should be obtained by diagonalizing three $6\times 6$ squared-mass matrices for up-type squarks
($\ut_L$, $\ct_L$, $\topt_L$, $\ut_R$, $\ct_R$, $\topt_R$),
down-type squarks ($\dt_L$, $\st_L$, $\bt_L$, $\dt_R$, $\st_R$, $\bt_R$),
and charged sleptons ($\et_L$, $\mt_L$, $\taut_L$, $\et_R$, $\mt_R$, $\taut_R$),
and one $3\times 3$ matrix for sneutrinos ($\nt_e$, $\nt_\m$, $\nt_\t$).
However with the assumptions~(\ref{3familyapprox}), (\ref{soft3familyapprox}) and the result (\ref{softscalarmass12approx})
 we have that the first and second families are almost eigenstates while we can have the previous mixings only for the
third family sfermions. So the masses for the squarks and sleptons of the first two families are
\bea
 m^2_{\ut_L} = m^2_{\ct_L}  &=& m^2_{q_1} + \( \frac{1}{3} g_1^2 - g_2^2 \) \frac{\D v^2}{8}  \label{mutL}\\
 m^2_{\ut_R} = m^2_{\ct_R}  &=& m^2_{u_1}     -g_1^2                        \frac{\D v^2}{6}  \label{mutR}\\
 m^2_{\dt_L} = m^2_{\st_L}  &=& m^2_{q_1} + \( \frac{1}{3} g_1^2 + g_2^2 \) \frac{\D v^2}{8}  \label{mdtL}\\
 m^2_{\dt_R} = m^2_{\st_R}  &=& m^2_{d_1} +    g_1^2                        \frac{\D v^2}{12} \label{mdtR}\\
 m^2_{\nt_e} = m^2_{\nt_\m} &=& m^2_{l_1} - \( g_1^2 + g_2^2 \) \frac{\D v^2}{8} \label{mnetL}\\
 m^2_{\et_L} = m^2_{\mt_L}  &=& m^2_{l_1} - \( g_1^2 - g_2^2 \) \frac{\D v^2}{8} \label{metL}\\
 m^2_{\et_R} = m^2_{\mt_R}  &=& m^2_{e_1} +    g_1^2            \frac{\D v^2}{4} \label{metR}
\eea
where $\D v^2=v_u^2-v_d^2=-v^2 \cos(2\b)$. The first and most important contribution are the soft masses given
in~(\ref{scalarmassunification}) while the second (and the third) one comes from $D$ terms.
We can see a degeneracy between the sfermions with the same charges.
Instead the mass matrix for the third family squarks is
\be
 \L_{\topt-\bt \ \text{mass}}=
  -\( {\tilde{t}}_L^\dag \ {\tilde{t}}_R^\dag \ {\tilde{b}}_L^\dag \ {\tilde{b}}_R^\dag\)
                     \( \begin{array}{cccc} m^2_{t_Lt_L} & m^2_{t_Lt_R} & 0 & 0 \\
                                            m^2_{t_Lt_R} & m^2_{t_Rt_R} & 0  & m^2_{t_Rb_R}  \\
                                            0 & 0 & m^2_{b_Lb_L} & m^2_{b_Lb_R}\\
                                            0 & m^2_{t_Rb_R} & m^2_{b_Lb_R} &  m^2_{b_Rb_R} \\\end{array} \)
                     \( \begin{array}{c} {\tilde{t}}_L \\
                                         {\tilde{t}}_R\\
                                         {\tilde{b}}_L \\
                                         {\tilde{b}}_R \end{array} \) \label{toptbtmassmatrix}
\ee
where
\bea
 &&m^2_{t_Lt_L} = m^2_{q_3} + \( \frac{1}{3} g_1^2 - g_2^2 \) \frac{\D v^2}{8} + y_t^2 \frac{v_u^2}{2}  \nn\\
 &&m^2_{t_Rt_R} = m^2_{u_3}     -g_1^2                        \frac{\D v^2}{6} + y_t^2 \frac{v_u^2}{2}  \nn\\
 &&m^2_{b_Lb_L} = m^2_{q_3} + \( \frac{1}{3} g_1^2 + g_2^2 \) \frac{\D v^2}{8} + y_b^2 \frac{v_d^2}{2}  \nn\\
 &&m^2_{b_Rb_R} = m^2_{d_3} +    g_1^2                        \frac{\D v^2}{12} + y_b^2 \frac{v_d^2}{2}  \nn\\
 &&m^2_{t_Lt_R} = \m y_t  \frac{v_d}{\sqrt2}+a_{t} \frac{v_u}{\sqrt2}  \nn\\
 &&m^2_{t_Rb_R} = - y_t \frac{v_u v_d}{2} \nn\\
 &&m^2_{b_Lb_R} = -\m y_b  \frac{v_u}{\sqrt2}+a_{b} \frac{v_d}{\sqrt2}  \label{toptbtmassmatrixelements}
\eea
We can see that the diagonal entries are similar contributions as the first two families plus the one
coming from Yukawa couplings which now are not negligible. Finally the mass terms for the tau sleptons
\be
 \L_{\taut \, \text{mass}}=
 -\( {\tilde{\t}}_L^\dag \ {\tilde{\t}}_R^\dag \)
  \( \begin{array}{cc} m^2_{l_3} - \frac{\( g_1^2 - g_2^2 \) \D v^2}{8}+y_\t^2 \frac{v_d^2}{2} & -\m y_\t \frac{v_u}{\sqrt2}+a_\t \frac{v_d}{\sqrt2} \\
                     -\m y_\t \frac{v_u}{\sqrt2}+a_\t \frac{v_d}{\sqrt2} & m^2_{e_3} + g_1^2 \frac{\D v^2}{4}+y_\t^2 \frac{v_d^2}{2} \\
     \end{array} \)
  \( \begin{array}{c} {\tilde{\t}}_L \\
                                          {\tilde{\t}}_R\\\end{array} \)
 \label{staumassmatrix}
\ee
and their corresponding sneutrino
\be
 m^2_{\nt_\t}=m^2_{l_3} -  \frac{\( g_1^2 + g_2^2 \) \D v^2}{8} \label{mntautL}
\ee
As in the previous case the diagonal entries for~(\ref{staumassmatrix}) have the similar first(second) family contribution
plus the Yukawa correction.
\section{$R$ parity} \label{subsect:R parity}
The superpotential~(\ref{MSSMsuperpot}) is minimal in the sense that it is sufficient to produce a phenomenologically
viable model. However we can write other terms (gauge-invariant, renormalizable and analytic in the chiral superfields)
which are not included in the MSSM since they do not respect either baryon number ($\Baryon$) or total lepton number ($\Lepton$).
These terms are
\bea
 W_{\D \Lepton =1} &=& {1\over 2} \l^{ijk} L_i L_j E^c_k + \l'^{ijk} L_i Q_j D^c_k + \m'^{i} L_i H_u
                       \label{WLviol} \\
 W_{\D \Baryon =1} &=& {1\over 2} \l''^{ijk} U^c_i D^c_j D^c_k
                       \label{WBviol}
\eea
where $i,j,k$ runs over the three families. The baryon and lepton number assignment are the usual ones: $\Baryon=+1/3$
for $Q_i$, $\Baryon=-1/3$ for $U^c_i, D^c_i$ and $\Baryon=0$ for all others; $\Lepton=+1$ for $L_i$,
$\Lepton=-1$ for $E^c_i$, and $\Lepton=0$ for all others. So the terms
in eq.~(\ref{WLviol}) violate total lepton number by 1 unit (as well as the individual lepton flavors) and those
in eq.~(\ref{WBviol}) violate baryon number by 1 unit.

The possible existence of such terms might seem rather disturbing, since corresponding $\Baryon$- and $\Lepton$-violating
processes (for example the proton decay) have not been seen experimentally. One could simply try to postulate
$\Baryon$ and $\Lepton$ conservation in the MSSM by adding a new symmetry, which has the effect of eliminating
the possibility of $\Baryon$ and $\Lepton$ violating terms in the renormalizable superpotential,
while allowing the good terms in eq.~(\ref{MSSMsuperpot}). This new symmetry is called ``$R$-parity" \cite{Rparity}
and it is defined as
\be
 P_R = (-1)^{3(\Baryon-\Lepton) + 2 s} \label{defRparity}
\ee
where $s$ is the spin of the particle. Therefore particles within the same supermultiplet do not have the same
$R$-parity\footnote{In general, symmetries with the property that fields within the same supermultiplet have different
transformations are called $R$ symmetries; they do not commute with supersymmetry.}.
The $R$-parity assignment is very useful for phenomenology because all of the SM particles and the Higgs bosons have
$P_R=+1$, while all of the squarks, sleptons, gauginos, and higgsinos have $P_R=-1$.
The $R$-parity odd particles are known as ``sparticles" (``supersymmetric particles"), and they are distinguished
by a tilde (see Tables~\ref{tab:chiral} and \ref{tab:gauge}).
$R$-parity conservation implies that there can be no mixing between the $P_R=-1$ and the $P_R=+1$ particles.
Furthermore, every interaction vertex in the theory contains an even number of sparticles.
This has three extremely important phenomenological consequences:
\bei
\item[i.] The lightest $P_R=-1$ particle, called the LSP (``lightest supersymmetric particle"), must be absolutely stable.
         If the LSP is electrically neutral, it interacts only weakly with ordinary matter, and it can be
         an attractive candidate \cite{neutralinodarkmatter} for the non-baryonic dark matter.
\item[ii.] Each sparticle other than the LSP must decay into a state that contains an odd number of sparticles.%
\item[iii.] In collider experiments, sparticles can only be produced in even numbers.
\eei

We {\it define} the MSSM to conserve $R$-parity. This decision seems to be well-motivated phenomenologically
by proton decay constraints, but it might appear somewhat artificial from a theoretical point of view.
One attractive solution could occur if B$-$L is a continuous gauge symmetry because it forbids the renormalizable terms
that violate B and L \cite{Rparityoriginone,Rparityorigintwo}.
However, the $U(1)_{\Baryon-\Lepton}$ must be spontaneously broken at some very high energy scale, because there is no
corresponding massless vector boson.
Moreover, if it is only broken by scalar VEVs (or other order parameters) that carry even integer values of $3($B$-$L$)$,
then $P_M=(-1)^{3(\Baryon-\Lepton)}$ will automatically survive as an exactly conserved discrete remnant subgroup
\cite{Rparityorigintwo}. $P_M$ is called ``matter-parity" and can be treated in a similar way as $P_R$.
A variety of extensions of the MSSM in which exact $R$-parity conservation is guaranteed
in just this way have been proposed (see for example \cite{Rparityorigintwo,Rparityoriginthree}).

\newpage\thispagestyle{empty}

\chapter{Anomalies} \label{chap:Anomalies}
A theory is anomalous if at the quantum (loop) level a classical symmetry $G$ of the Lagrangian is no longer conserved.
If $G$ is an anomalous global group the quantum theory remains consistent (for example $\pi^0 \to \g \, \g$).
Instead, if $G$ is an anomalous gauge group, the quantum theory becomes inconsistent, so the condition of
anomaly cancellation may be used as a constraint on the physical gauge theories. The first studies on anomalies emerged
in trying to understand the decay rate of the already cited neutral pion, in the form of an anomaly that violates some
global symmetry of the strong interactions. In 1969 the source of this anomaly was traced by Bell and
Jackiw~\cite{Bell:1969ts} to the violation of chiral symmetry by the regulator that is needed in order to derive
consequences of the conservation of the neutral axial vector current for one loop Feynman diagrams.
Their result was confirmed and generalized by Adler~\cite{Adler:1969gk}.
After 10 years, Fujikawa~\cite{Fujikawa:1979ay} showed that the chiral symmetry breaking anomaly
enters only in the measure used to define the path integral over fermion fields. The following treatment of anomalies
is based on Chapter 22 of the book by Weinberg~\cite{Weinberg2} and Chapter 6 of the book by Cheng and Li~\cite{Cheng:1985bj}.
We will discuss the anomalies only from a Feynman diagram point of view, leaving the Fujikawa argument to the reader
(for a review see also Section 22.2 of \cite{Weinberg2})
\newpage\section{Triangular Anomaly} \label{sect:Triangular Anomaly}
Consider a chiral gauge theory given by a certain number of massless left(right) Weyl fermions charged under two
$U(1)$ gauge fields\footnote{The extension to a larger number of gauge fields and to a
                                non abelian case can be treated in a similar way.}
$B_\m$ and $C_\m$
\be
 \L_{\chiral} =-\frac{1}{4} F_{\m \n}^B F^{B \m \n}-\frac{1}{4} F_{\m \n}^C F^{C \m \n}+
      i \sum_f \Big( f_L^\dag  \sb^\m D_\m f_L +  f_R^\dag  \s^\m D_\m f_R \Big)\label{chiralLWeyl}
\ee
where $f_{L(R)}$ are the left(right) Weyl fermions in the model, and
\be
 D_\m=\pd_\m + i Q_{f_{L(R)}}^B B_\m + i Q_{f_{L(R)}}^C C_\m \label{covdevBC}
\ee
is the covariant derivative.
We can express the Lagrangian in terms of the Dirac spinors $ \Psi_f = \( \begin{array}{c} f_L\\ f_R \end{array} \)$
obtaining
\bea
 \L_{\chiral} &=&-\frac{1}{4} F_{\m \n}^B F^{B \m \n}-\frac{1}{4} F_{\m \n}^C F^{C \m \n}+\nn\\
              &&i \sum_f \bar \Psi_f \g^\m \Big[ \pd_\m + \frac{i}{2} (v_f^B - a_f^B \g_5) B_\m
             + \frac{i}{2} (v_f^C - a_f^C \g_5) C_\m \Big] \Psi_f \label{chiralLDirac}
\eea
where the vectorial (V) and axial (A) couplings are given by
\bea
  v_f^k &=& Q^k_{f_L}+Q^k_{f_R}  \label{chiralvec}\\
  a_f^k &=& Q^k_{f_L}-Q^k_{f_R}  \label{chiralaxi}
\eea
with $k=B,C$. For the conventions on the sigma matrices and the properties of gamma matrices see Appendix~\ref{app:Gammology}.

We may be interested in computing the one loop amplitude and the corresponding Ward Identities (WIs) for an incoming
$C_\m$ and two outgoing $B_\m$. The Feynman diagrams
involved are depicted in Fig.~\ref{TriangleDiagramCBB}. The Feynman rules for the interacting vertices are given in Appendix \ref{sec:C-B model}.
We can split the amplitude in four contributions: AVV, VAV, VVA and AAA
\bea
 \D_{\r \m \n}^{CBB} &=&- \frac{1}{8} \sum_f \Big(a_f^C \, v_f^B \, v_f^B \, \G_{\r \m \n}^{AVV}(p,q;0) +
                                        v_f^C \, a_f^B \, v_f^B \, \G_{\r \m \n}^{VAV}(p,q;0)  + \nn\\
&& \phantom{- \frac{1}{8} \sum_f \Big(} v_f^C \, v_f^B \, a_f^B \, \G_{\r \m \n}^{VVA}(p,q;0) +
                                        a_f^C \, a_f^B \, a_f^B \, \G_{\r \m \n}^{AAA}(p,q;0) \Big)
\label{CBBchiraltrian}
\eea
where
\bea
\G_{\r \m \n}^{AVV}(p,q;0)&=&\int \frac{d^{4} \ell}{(2
\pi)^{4}} \, \Tr\(  \g_5  \g_\r  \frac{1}{\slashed \ell - \slashed
q} \g_\n \frac{1}{\slashed \ell} \g_\m
\frac{1}{\slashed \ell + \slashed p} \)    \nn\\
&&+ (p \leftrightarrow q, \m \leftrightarrow \n) \label{AVVtrian}\\
\G_{\r \m \n}^{VAV}(p,q;0)&=&\int \frac{d^{4} \ell}{(2 \pi)^{4}}
  \, \Tr\(  \g_\r  \frac{1}{\slashed \ell - \slashed q}
  \g_\n \frac{1}{\slashed \ell} \g_5 \g_\m
  \frac{1}{\slashed \ell + \slashed p} \)    \nn\\
  &&+ (q \leftrightarrow -(p+q), \n \leftrightarrow \r) \label{VAVtrian}\\
\G_{\r \m \n}^{VVA}(p,q;0)&=&\int \frac{d^{4} \ell}{(2 \pi)^{4}}
  \, \Tr\(   \g_\r  \frac{1}{\slashed \ell - \slashed q}
  \g_5 \g_\n  \frac{1}{\slashed \ell}  \g_\m
  \frac{1}{\slashed \ell + \slashed p} \)\nn\\
  &&+ (p \leftrightarrow -(p+q), \m \leftrightarrow \r)  \label{VVAtrian}\\
\G_{\r \m \n}^{AAA}(p,q;0)&=&\int \frac{d^{4} \ell}{(2 \pi)^{4}}
\, \Tr\(  \g_5  \g_\r  \frac{1}{\slashed \ell - \slashed q}
\g_5  \g_\n \frac{1}{\slashed \ell} \g_5 \g_\m
\frac{1}{\slashed \ell + \slashed p} \)\nn\\
&&+ (p \leftrightarrow q, \m \leftrightarrow \n)  \label{AAAtrian}
\eea
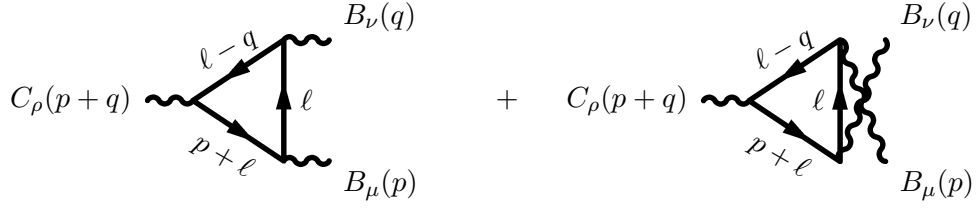
\begin{figure}[t]
 \vskip 1.5cm
      \centering
      \raisebox{-4.3ex}[0cm][0cm]{\unitlength=0.4mm
      \begin{fmffile}{loop3a}
      \begin{fmfgraph*}(60,40)
       \fmfpen{thick} \fmfleft{ii0,ii1,ii2} \fmfstraight \fmffreeze \fmftop{ii2,t1,t2,t3,oo2}
       \fmfbottom{ii0,b1,b2,b3,oo1} \fmf{phantom}{ii2,t1,t2} \fmf{phantom}{ii0,b1,b2} \fmf{phantom}{t1,v1,b1}
       \fmf{phantom}{t2,b2} \fmf{phantom}{t3,b3}
       \fmffreeze
       \fmf{photon}{ii1,v1}
       \fmf{fermion,label.side=right,label=\begin{rotate}{30}$\! \! \! \! \! \! \ell-q$\end{rotate}}{t3,v1}
       \fmf{fermion,label=$\ell$}{b3,t3}
       \fmf{fermion,label.dist=0.5cm,label=\begin{rotate}{-30} $\! \! \! \! \! \! p+\ell$\end{rotate}}{v1,b3}
       \fmf{photon}{b3,oo1} \fmf{photon}{t3,oo2}
       \fmflabel{$C_\r (p+q)$}{ii1} \fmflabel{$B_\m(p)$}{oo1} \fmflabel{$B_\n(q)$}{oo2}
      \end{fmfgraph*}
      \end{fmffile}}
~~~~~~~~~~~~~~~+~~~~~~~~~~~~~~~~
      \raisebox{-4.3ex}[0cm][0cm]{\unitlength=0.4mm
      \begin{fmffile}{loop3b}
      \begin{fmfgraph*}(60,40)
       \fmfpen{thick} \fmfleft{ii0,ii1,ii2} \fmfstraight \fmffreeze \fmftop{ii2,t1,t2,t3,oo2}
       \fmfbottom{ii0,b1,b2,b3,oo1} \fmf{phantom}{ii2,t1,t2} \fmf{phantom}{ii0,b1,b2} \fmf{phantom}{t1,v1,b1}
       \fmf{phantom}{t2,b2} \fmf{phantom}{t3,b3}
       \fmffreeze
       \fmf{photon}{ii1,v1}
       \fmf{fermion,label.side=right,label=\begin{rotate}{30}$\! \! \! \! \! \! \ell-q$\end{rotate}}{t3,v1}
       \fmf{fermion,label.side=left,label=$\ell$}{b3,t3}
       \fmf{fermion,label.dist=0.5cm,label=\begin{rotate}{-30} $\! \! \! \! \! \! p+\ell$\end{rotate}}{v1,b3}
       \fmf{photon}{b3,oo2} \fmf{photon}{t3,oo1}
       \fmflabel{$C_\r (p+q)$}{ii1} \fmflabel{$B_\m(p)$}{oo1} \fmflabel{$B_\n(q)$}{oo2}
      \end{fmfgraph*}
      \end{fmffile}}
    \vskip 1.5cm
    \caption{Fermionic triangle diagrams for the amplitude $C_\r(p+q)\to B_\m(p) \, B_\n(q)$.}
\label{TriangleDiagramCBB}
\end{figure}
The notation $\G_{\r \m \n}^{AVV}$ means that we have an axial vertex ($a_f^C$) on the $\r$ line,
a vectorial vertex ($v_f^B$) on the $\m$ line and
a vectorial vertex ($v_f^B$) on the $\n$ line. Analogously for the other $\G$'s.

Consider the first contribution. We have
\bea
 &&\!\!\!\!\!\!\!\!\!\!\!\!\!\!
 \left. \D_{\r \m \n}^{CBB} \right|_{AVV}=-\frac{1}{8} \sum_f a_f^C \, v_f^B \, v_f^B \, \G_{\r \m \n}^{AVV}(p,q;0)=
 \label{AVVtrianCBB}\\
 &&\qquad \ =-\frac{1}{8} \sum_f a_f^C \, v_f^B \, v_f^B \, \int \frac{d^{4} \ell}{(2\pi)^{4}} \,
 \Tr\( \g_5  \g_\r \frac{1}{\slashed \ell - \slashed q} \g_\n \frac{1}{\slashed \ell}
            \g_\m \frac{1}{\slashed \ell + \slashed p} \)
 + (p_\m \leftrightarrow q_\n) \nn
\eea
The integral $\G_{\r \m \n}^{AVV}(p,q;0)$ is superficially divergent so it is not uniquely defined
(see Appendix~\ref{app:Linearly divergent integrals}).
Eq.~\ref{AVVtrianCBB} implies a particular choice for the assignment of the internal momentum $\ell$: the fermion line between
the two $B$ lines carries momentum $\ell$. We could have chosen a different assignment of it so that this fermion line
carries $\ell+a$, where $a$ is some arbitrary combination of $p$ and $q$
\be
 a=\a p + (\a-\b) q \label{a}
\ee
The fact that the integral is linearly divergent implies that $\left. \G_{\r \m \n}^{CBB} \right|_{AVV}$ has an ambiguity
in its definition by an amount
\bea
 D_{\r \m \n}(a) &=& \G_{\r \m \n}^{AVV}(a) - \G_{\r \m \n}^{AVV} = \nn\\
 &=&  \left[ \int \frac{d^{4} \ell}{(2\pi)^{4}} \,
  \Tr\( \g_5  \g_\r \frac{1}{\slashed \ell - \slashed q + \slashed a}
  \g_\n \frac{1}{\slashed \ell + \slashed a}
  \g_\m \frac{1}{\slashed \ell + \slashed p + \slashed a} \) \right. \nn\\
 &&\left. - \int \frac{d^{4} \ell}{(2\pi)^{4}} \,
   \Tr\( \g_5  \g_\r \frac{1}{\slashed \ell - \slashed q} \g_\n \frac{1}{\slashed \ell}
   \g_\m \frac{1}{\slashed \ell + \slashed p} \) \right]
     + (p \leftrightarrow q, \m \leftrightarrow \n) \nn\\
 &=& D_{\r \m \n}^{(1)} + D_{\r \m \n}^{(2)} \label{Drmna}
\eea
Applying the result~(\ref{MinkIntdiv}), we have
\bea
 D_{\r \m \n}^{(1)} &=&  \int \frac{d^{4} \ell}{(2\pi)^{4}} \, a^\r \frac{\pd}{\pd \ell_\r}
   \Tr\( \g_5  \g_\r \frac{1}{\slashed \ell - \slashed q} \g_\n \frac{1}{\slashed \ell}
   \g_\m \frac{1}{\slashed \ell + \slashed p} \) \nn\\
 &=& \frac{2 i \pi^2}{(2 \pi)^4} a^\r \lim_{\ell \to \infty} \ell^2 \ell_\r
     \Tr[\g_5 \g_\r \slashed \ell \g_\n \slashed \ell \g_\m \slashed \ell] \frac{1}{\ell^6}
 =\frac{2 i \pi^2}{(2 \pi)^4} a_\r \lim_{\ell \to \infty} \frac{\ell^\r \ell^\d}{\ell^2} (-4i) \e_{\r\d\n\m}\nn\\
&=& \frac{1}{(8 \pi)^2} \e_{\d\m\n\r} a^\d \label{Drmn1}
\eea
where we used
\be
 \lim_{\ell \to \infty} \frac{\ell^\r \ell^\d}{\ell^2} = \frac{\eta^{\r \d}}{4} \label{etalim}
\ee
Since $D_{\r \m \n}^{(2)}$ is related to $D_{\r \m \n}^{(1)}$ by the exchanges $p \leftrightarrow q$ and $\m \leftrightarrow \n$,
we have from eq.~(\ref{a}), (\ref{Drmna}) and (\ref{Drmn1})
\be
 D_{\r \m \n}(a) = D_{\r \m \n}^{(1)} + D_{\r \m \n}^{(2)} = -\frac{\b}{8\pi^2}\e_{\r\m\n\d}(p-q)^\d \label{Drmnafinal}
\ee
Thus the definition of $\G_{\r \m \n}^{AVV}$ has an ambiguity parametrized by
\be
 \G_{\r \m \n}^{AVV}(p,q,\b;0) = \G_{\r \m \n}^{AVV}(p,q;0)-\frac{\b}{8\pi^2}\e_{\r\m\n\d}(p-q)^\d \label{AVVtrianbeta}
\ee

Now we are going to show what is the relation between the $\b$ ambiguity in the amplitude and the WIs.
In quantum field theory, the WIs are identities between correlation functions that follows from the global or
gauged symmetries of the theory, and which remains valid after renormalization.
The WIs are a quantum version of the classical Noether's theorem, and any symmetries in a quantum field theory
can lead to an equation of motion for correlation functions. We can express them using Feynman amplitude in momentum space
or by the path integral formulation. We will usually adopt the first formulation.
The WIs for massless gauge fields can be written in the diagrammatic form showed in Fig.~\ref{WI}.
\begin{figure}[t]
\vskip 0.5cm
\centering
$-i k_\m~\Bigg($~~~~~~~~
      \raisebox{-4.7ex}[0cm][0cm]{\unitlength=0.7mm
      \begin{fmffile}{vectorbubble5a}
      \begin{fmfgraph*}(40,25)
       \fmfpen{thick} \fmfleft{o1} \fmfright{i1,i2} \fmfstraight \fmftop{t1,t2,t3,t4,i2}\fmfbottom{b1,b2,b3,b4,i1}
       \fmf{boson}{o1,v1}   \fmf{boson}{v1,b4}       \fmf{boson}{v1,i2}
              \fmf{vanilla}{v1,t4}       \fmf{vanilla}{v1,i1}
              \fmffreeze     \fmf{boson}{v1,t3}
       \fmfv{decor.shape=circle,decor.filled=shaded, decor.size=.25w}{v1}
       \fmflabel{~~1PI}{v1} \fmflabel{$V^\m(k)$}{o1}
      \end{fmfgraph*}
      \end{fmffile}}~~~$\Bigg)= 0$
\vskip 0.5cm
\caption{Ward Identity for the massless gauge field $V^\m$. The blob denotes all the 1PI diagrams.} \label{WI}
\end{figure}
We start with the incoming momentum $(p+q)_\r$. Using eq.~(\ref{AVVtrianbeta}), we have
\be
 (p+q)^\r \G_{\r \m \n}^{AVV}(p,q,\b;0) = (p+q)^\r \G_{\r \m \n}^{AVV}(p,q;0)-
           (p+q)^\r \frac{\b}{8\pi^2} \e_{\r\m\n\d}(p-q)^\d \label{WIrhoAVVtrianbeta}
\ee
and we focus on the first contribution
\be
 (p+q)^\r \G_{\r \m \n}^{AVV}(p,q;0) = D_{\m \n}^{(A)} + D_{\m \n}^{(B)} \label{WIrhoAVVtrian}
\ee
where the first term is
\bea
 D_{\m \n}^{(A)}&=&(p+q)^\r \int \frac{d^{4} \ell}{(2\pi)^{4}} \,
   \Tr\[ \g_5  \g_\r \frac{1}{\slashed \ell - \slashed q} \g_\n \frac{1}{\slashed \ell}
   \g_\m \frac{1}{\slashed \ell + \slashed p} \] \nn\\
 &=&\int \frac{d^{4} \ell}{(2\pi)^{4}} \,
   \Tr\[ \g_5 \( \slashed p + \slashed q \) \frac{1}{\slashed \ell - \slashed q} \g_\n \frac{1}{\slashed \ell}
   \g_\m \frac{1}{\slashed \ell + \slashed p} \] \nn\\
 &=& - \int \frac{d^{4} \ell}{(2\pi)^{4}} \,
   \Tr \left\{ \[ \g_5 (\slashed \ell - \slashed q) + (\slashed \ell + \slashed p) \g_5 \]
     \[\frac{1}{\slashed \ell - \slashed q} \g_\n \frac{1}{\slashed \ell}
   \g_\m \frac{1}{\slashed \ell + \slashed p} \] \right\} \nn\\
&=& - \int \frac{d^{4} \ell}{(2\pi)^{4}} \,
   \Tr\[ \frac{\g_5 \g_\n \slashed\ell \g_\m (\slashed \ell + \slashed p)}{\ell^2 (l+p)^2} -
         \frac{(\slashed \ell - \slashed q) \g_5 \g_\n \slashed\ell \g_\m }{\ell^2 (l-q)^2}\]
\label{DmnA}
\eea
while the second one is $D_{\m \n}^{(A)}$ with the exchange $(p \leftrightarrow q, \m \leftrightarrow \n)$
\be
 D_{\m \n}^{(B)}=
 - \int \frac{d^{4} \ell}{(2\pi)^{4}} \,
   \Tr\[ \frac{\g_5 \g_\m \slashed\ell \g_\n (\slashed \ell + \slashed q)}{\ell^2 (l+q)^2} -
         \frac{(\slashed \ell - \slashed p) \g_5 \g_\m \slashed\ell \g_\n }{\ell^2 (l-p)^2}\]
\label{DmnB}
\ee
We rearrange the computation grouping the integrals with only a $p$ dependence or only a $q$ dependence. So
\be
 D_{\m \n}^{(A)} + D_{\m \n}^{(B)} = D_{\m \n}^{(1)} + D_{\m \n}^{(2)} \label{DABtoD12}
\ee
where
\bea
 D_{\m \n}^{(1)}&=&
 - \int \frac{d^{4} \ell}{(2\pi)^{4}} \,
   \Tr\[ \frac{\g_5 \g_\m \slashed\ell \g_\n (\slashed \ell + \slashed p)}{\ell^2 (l+p)^2} -
         \frac{(\slashed \ell - \slashed p) \g_5 \g_\m \slashed\ell \g_\n }{\ell^2 (l-p)^2}\] \label{Dmn1} \\
 D_{\m \n}^{(2)}&=&  \left. D_{\m \n}^{(1)} \right|_{(p \leftrightarrow q, \m \leftrightarrow \n)} \label{Dmn2}
\eea
Using the result~(\ref{MinkIntdiv}) we get
\bea
 D_{\m \n}^{(1)}&=& - \int \frac{d^{4} \ell}{(2\pi)^{4}} \, p^\r \frac{\pd}{\pd \ell_\r}
  \Tr \[ \frac{\g_5 \g_\n (\slashed\ell - \slashed p) \g_\m \slashed\ell}{\ell^2 (\ell -p)^2} \]\nn\\
 &=&- \frac{2 i \pi^2}{(2 \pi)^4} p^\r \lim_{\ell \to \infty} \ell^2 \ell_\r
     \Tr[\g_5 \g_\n (\slashed \ell-\slashed p) \g_\m \slashed \ell] \frac{1}{\ell^4}
 =-\frac{1}{8 \pi^2} \lim_{\ell \to \infty} \frac{\ell^\r \ell^\b}{\ell^2} \e_{\n\a\m\b} p_\r p^\a \nn\\
&=& - \frac{1}{(32 \pi)^2} \e_{\n\a\m\b} p^\b p^\a = 0 =D_{\m \n}^{(2)} \label{Dmn12final}
\eea
In the last line we used eq.~(\ref{etalim}) and the last identity follows simply by the definition~(\ref{Dmn2}).
Then using eq.~(\ref{WIrhoAVVtrianbeta}), (\ref{WIrhoAVVtrian}), (\ref{DABtoD12}) and (\ref{Dmn12final}) we find
\be
  (p+q)^\r \G_{\r \m \n}^{AVV}(p,q,\b;0) = \frac{\b}{4\pi^2} \e_{\m\n\a\b} p^\a q^\b \label{WIrhoAVVtrianbetafinal}
\ee
In a similar way we can compute the WIs for the $p^\m$ and $q^\n$ momenta, obtaining
\bea
 p^\m \G_{\r \m \n}^{AVV}(p,q,\b;0) &=& -\frac{2+\b}{8\pi^2} \e_{\n\r\a\b} q^\a p^\b \label{WImuAVVtrianbeta}\\
 q^\n \G_{\r \m \n}^{AVV}(p,q,\b;0) &=& -\frac{2+\b}{8\pi^2} \e_{\r\m\a\b} p^\a q^\b \label{WInuAVVtrianbeta}
\eea
A similar analysis can be done for the other VAV, VVA and AAA amplitudes in eq.~(\ref{VAVtrian})-(\ref{AAAtrian}).
So the amplitude $\D_{\r \m \n}^{CBB}$ has the following WIs
\bea
 (p+q)^\r \D_{\r \m \n}^{CBB}(\b) &=& - \sum_f t_f^{CBB} \,
                                 \frac{\b}{32\pi^2} \, \e_{\m\n\a\b} p^\a q^\b~~~~~~~~ \label{WIrhoCBBtrian}\\
    p^\m \D_{\r \m \n}^{CBB}(\b)  &=&\sum_f t_f^{CBB} \,
                                  \frac{2+\b}{64\pi^2} \, \e_{\n\r\a\b} q^\a p^\b \label{WImuCBBtrian}\\
    q^\n \D_{\r \m \n}^{CBB}(\b)  &=&\sum_f t_f^{CBB} \,
                                  \frac{2+\b}{64\pi^2} \, \e_{\r\m\a\b} p^\a q^\b \label{WInuCBBtrian}
\eea
with
\bea
 t_f^{CBB} &=& a_f^C \, v_f^B \, v_f^B + 2 v_f^C \, v_f^B \, a_f^B + a_f^C \, a_f^B \, a_f^B \nn\\
        &=&4 \[Q_{f_L}^C \, \( Q_{f_L}^B \)^2 - Q_{f_R}^C \, \( Q_{f_R}^B \)^2\] \label{tCBBchiral}
\eea
where we used eq.~(\ref{chiralvec}) and (\ref{chiralaxi}).
As we can see from eq.~(\ref{WIrhoCBBtrian}), (\ref{WImuCBBtrian}) and (\ref{WInuCBBtrian}), there is
no choice of $\b$ that cancels simultaneously all the WIs and preserves gauge invariance, recovering the result depicted in Fig.~\ref{WI}.
The same problem will arise in the computation of the $CCC$, $CCB$, and $BBB$
triangles. This is the ``triangular (or chiral) anomaly''. The classical gauge invariance is broken by the triangular
fermionic loop making the theory anomalous. Obviously, to have a physical gauge theory we must cancel this anomaly and the standard
(and simplest) way is to cancel the overall sum of charges in the WIs. In other words the theory will be anomaly free if
the charges satisfy the following constraints
\bea
 \cA^{CCC}&=&\sum_f \[ \( Q_{f_L}^C \)^3 -  \( Q_{f_R}^C \)^3 \] = 0 \label{CCCanofree}\\
 \cA^{CCB}&=&\sum_f \[ \( Q_{f_L}^C \)^2 \, Q_{f_L}^B - \( Q_{f_R}^C \)^2 \, Q_{f_R}^B \] = 0 \label{CCBanofree}\\
 \cA^{CBB}&=&\sum_f \[Q_{f_L}^C \, \( Q_{f_L}^B \)^2 - Q_{f_R}^C \, \( Q_{f_R}^B \)^2 \] = 0 \label{CBBanofree}\\
 \cA^{BBB}&=&\sum_f \[ \( Q_{f_L}^B \)^3 -  \( Q_{f_R}^B \)^3 \] = 0 \label{BBBanofree}
\eea

Depending on the symmetries of the problem, there are cases in which there is a preferential choice for the
$\b$ parameter.
The most common values are (see eq.~(\ref{WIrhoCBBtrian}), (\ref{WImuCBBtrian}) and (\ref{WInuCBBtrian}):
\begin{itemize}
 \item $\b=-2$, for which all the anomaly is concentrated in the $\rho$ vertex .
 \item $\b=-2/3$, for which the anomaly is equally distributed among the three vertices
      (to be used when we have three identical external vectors).
\end{itemize}

The extension to non abelian theories goes in the same way. The overall charge factor will become
\be
 \cA^{i j k} = \frac{1}{2} \Tr \[ \{ T_i^L,T_j^L\} T_k^L \] - \frac{1}{2} \Tr \[ \{ T_i^R,T_j^R\} T_k^R \]
\ee
where $T_{i,j,k}$ are the generators of the three external gauge fields.
In some theories the condition of anomaly cancellation will fix completely the charge content of the model.
This is the case for the SM (see for instance Section 22.4 of \cite{Weinberg2}).
\section{Anomaly cancellation. SM and MSSM} \label{sect:Anomaly cancellation. SM and MSSM}
The SM gauge group is $SU(3)_c \times SU(2)_L \times U(1)_Y$ and its fermionic content is displayed in Table~\ref{tab:SM}
%
%
\begin{table}[h]
\centering
\begin{tabular}[h]{|c|c|c|c|}
 \hline                    & $SU(3)_c$    & $SU(2)_L$ & $U(1)_Y$  \\
 \hline $(u_{i} \ d_{i})$  & $\bth$       &  $\btw$   & $1/6$     \\
 \hline $u^c_i$            & $\bar \bth$  &  $\bon$   & $-2/3$    \\
 \hline $d^c_i$            & $\bar \bth$  &  $\bon$   & $1/3$     \\
 \hline $(\n_{i} \ e_{i})$ & $\bon$       &  $\btw$   & $-1/2$    \\
 \hline $e^c_i$            & $\bon$       &  $\bon$   &  $1$      \\
 \hline
\end{tabular}
\caption{Fermionic charges in the Standard Model. The index $i$ runs over the three families.}\label{tab:SM}
\end{table}
%
%

Now we check that $\cA^{i j k}$ vanishes for all the generators of the SM gauge group. We need only to consider those
combinations of generators for which the product of $T_i$, $T_j$ and $T_k$ is neutral under
$SU(3)_c \times SU(2)_L \times U(1)_Y$, since $\cA^{i j k}$ obviously vanishes for all the others. We can make invariants
out of either zero, two or three $SU(3)_c$ or $SU(2)_L$ generators and any number of $U(1)_Y$ generators.
We will use the following notation. $Y_f$ is the hypercharge and
$T_{k_a}^{(a)}$, $a=2,3;\,\, k_a=1,\ldots,{\rm dim G}^{(a)}$ are the generators of the $G^{(2)}=SU(2)$ and $G^{(3)}=SU(3)$
algebras respectively. In our notation $\Tr[T_j^{(a)} T_k^{(a)}] = {1\over2}\d_{jk}$.
So we need to check the following cases
\begin{itemize}
\item $SU(3)_c-SU(3)_c-SU(3)_c$. Here $\cA^{i j k}$ vanishes because the left-handed fermions furnish a representation
$\bth + \bth + \bar\bth + \bar\bth + \bon + \bon +\bon$ of $SU(3)_c$ which is real. In other words, for each quark,
the $q$ contribution is cancelled exactly by the corresponding $q^c$ contribution.
\item $U(1)_Y-SU(3)_c-SU(3)_c$. Here the anomaly is
 \be
  \sum_{f \in \bth, \bar\bth} Y_f \Tr[T_{k_3}^{(3)} T_{k_3}^{(3)}] = \frac{1}{2} \sum_{f \in \bth, \bar\bth} Y_f =
   \frac{3}{2} \[2  \(\frac{1}{6}\)- \frac{2}{3}+\frac{1}{3}\]=0
 \ee
\item $SU(2)_L-SU(2)_L-SU(2)_L$. There is no anomaly here because $SU(2)_L$ only has real or pseudoreal representations.
\item $U(1)_Y-SU(2)_L-SU(2)_L$. There the anomaly is
 \be
  \sum_{f \in \btw} Y_f \Tr[T_{k_2}^{(2)} T_{k_2}^{(2)}] = \frac{1}{2} \sum_{f \in \btw} Y_f =
  \frac{3}{2} \[3  \(\frac{1}{6}\)- \frac{1}{2}\]=0
 \ee
\item $U(1)_Y-U(1)_Y-U(1)_Y$. There the anomaly is
 \be
  \sum_f \( Y_f \)^3 =   3 \[6 \(\frac{1}{6}\)^3+3 \(-\frac{2}{3}\)^3+3 \(\frac{1}{3}\)^3+2 \(-\frac{1}{2}\)^3+(1)^3\]=0
 \ee
\end{itemize}
There is one more anomaly that needs to be evaluated. All the fermions interact also with gravity. We can have
anomalies different from zero generated by the fermionic with an incoming gauge vector a two outgoing gravitons.
This anomaly is proportional to the trace of the gauge generator involved in the diagram.
For $SU(3)_c$ and $SU(2)_L$ there is no problem because the generator are traceless, so we need to check only
the hypercharge. We have
\be
 \sum_f Y_f = 3 \[6 \(\frac{1}{6}\)+3 \(-\frac{2}{3}\)+3 \(\frac{1}{3}\)+2 \(-\frac{1}{2}\)+1\]=0
\ee
so there are no gravitational anomalies in the SM.

The same happens for the MSSM. With respect to the SM now we have two new fermions that are involved in the anomalies,
the higgsinos. As we have just seen, the SM fermions contribution is zero, so the only relevant contribution comes
from the supersymmetric sector. However $\tilde h_u$ and $\tilde h_d$ have opposite charge assignment (see Table~\ref{tab:chiral})
so that the higgsino contribution is zero.
The final result is an anomaly free model.
\section{Anomaly cancellation. Green-Schwarz mechanism} \label{sect:Anomaly cancellation. Green-Schwarz mechanism}
\subsection{Introduction}
Before introducing the GS mechanism, we give a generalization of the chiral anomalies in even dimensions $d=2\w$.
In any even dimensions (and only in even dimensions) one can define a chirality matrix $\g_{2\w+1}$ that anticommutes with all
matrices $\g_\m$, $\m=0,\dots,2\w-1$ (compare with eq.~(\ref{ganticomm})). One can then have massless fermions that are chiral, i.e.
either left-handed or right-handed. As in Section \ref{sect:Triangular Anomaly}, the anomalies can again be traced to the presence of
the chirality matrix $\g_{2\w+1}$ (i.e. axial coupling) in the interaction vertices between fermions and vectors.
In $2\w$ dimensions the traces involving $\g_{2\w+1}$ will lead to an $\e^{\m_1 \dots \m_{2\w}}$,
and again the WIs on the amplitude with $k+1$ external vector, $\G^{\r \m_1 \dots \m_{k}} ( p_1, \dots, p_k )$,
must involve this $\e^{\m_1 \dots \m_{2\w}}$ tensor.
Since this $\e$-tensor is completely antisymmetric in all its $2\w$ indices, to get a non vanishing expression we need at least $\w$
indices $\m_i$ and $\w$ different independent momenta. Hence the minimal value for $k$ is $\w$.
It follows that in $2\w$ dimensions the anomaly first manifests itself in the loop amplitude $\G^{\r \m_1 \dots \m_{\w}}$ with
$(\w+1)$ external vectors.
In four dimensions this corresponds to the fermionic triangle loop. For further details and general properties of the anomalies,
we remand for instance to \cite{Bilal:2008qx}.
\begin{figure}[t]
 \centering
 \includegraphics[scale=1.5]{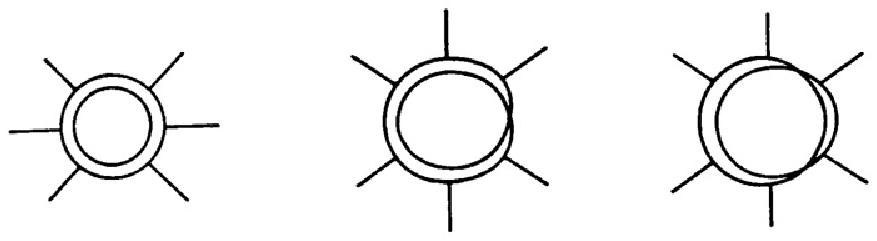}
 \begin{flushleft}~~~~~~~~~~~~~~~~(a)~~~~~~~~~~~~~~~~~~~~~~~~~~~~~~~~~~~~(b)~~~~~~~~~~~~~~~~~~~~~~~~~~~~~~~~~~~(c)
 \end{flushleft}
\caption{Hexagon string diagrams: planar (a), non orientable (b) and non planar (c).} \label{Fig:stringGSfig}
\end{figure}

In 1984, M. Green and J. H. Schwarz realized the cancellation of anomalies in type I superstring theory with the gauge group $SO(32)$.
Now we consider only the effective field theory point of view. For further details and the string formalism,
we remand to the original papers \cite{GSmechanism}.
In ten-dimensional theories gauge anomalies, mixed anomalies, and gravitational anomalies were expected to arise from a hexagon loop diagram
($\w=10/2=5$ then $k=6$).
The anomaly cancels because of an extra contribution from a 2-form field, $B_{\m \n}$, which is one of the new states predicted by string theory
(if we can think of a gauge vector $V_\m$ as column, we can think of $B_{\m \n}$ as a matrix).
For the special choice of the gauge group $SO(32)$ or $E_8 \times E_8$, however, the anomaly factorizes and may be cancelled by a tree diagram.
The tree diagram describes the exchange of a virtual $B$.
It is somewhat counterintuitive to see that a tree diagram cancels a one-loop diagram, but in reality,
both of these diagrams arise as one-loop diagrams in superstring theory, in which the anomaly cancellation is more transparent.
In Fig.~\ref{Fig:stringGSfig} and Fig.~\ref{Fig:GSstring} are depicted respectively the diagrams involved in the
string theory mechanism and in the effective field theory mechanism.
The diagram (a) and (b) of Fig.~\ref{Fig:stringGSfig} are generated by planar open string loops
and we can naively say that they produce the effective fermionic hexagon diagram, while the diagram (c) is a non planar open string loop.
However the latter can also be viewed as the tree level propagation of a closed string. One of the closed string states is the $B$ field, so
we can naively say that this diagram, apart from contributing to the diagram of Fig.~\ref{Fig:GSstring}(a),
produces the effective tree level diagram of Fig.~\ref{Fig:GSstring}(b).
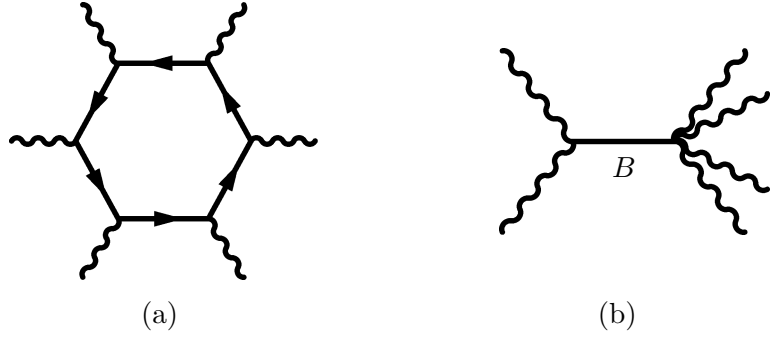
\begin{figure}[t]
\vskip 3cm
 \centering
 \raisebox{-4.3ex}[0cm][0cm]{\unitlength=0.4mm
 \begin{fmffile}{hexa}
 \begin{fmfgraph*}(100,100)
  \fmfpen{thick}
  \fmfsurroundn{e}{6}
  \begin{fmffor}{n}{1}{1}{6}
   \fmf{photon}{e[n],i[n]}
  \end{fmffor}
  \fmfcyclen{fermion,tension=6/8}{i}{6}
 \end{fmfgraph*}
 \end{fmffile}}
\qquad \qquad
 \raisebox{0.5ex}[0cm][0cm]{\unitlength=0.4mm
 \begin{fmffile}{anogaugeB_2}
 \begin{fmfgraph*}(100,60)
  \fmfpen{thick}
  \fmfleft{i1,i2,v0,i3,i4} \fmfright{o1,o2,v3,o3,o4}
  \fmf{phantom}{v0,v1,v2,v3}
  \fmffreeze
  \fmf{photon}{i1,v1}  \fmf{photon}{i4,v1}
  \fmf{plain,label=$B$}{v1,v2}  \fmf{photon}{i4,v1}
  \fmf{photon}{v2,o1} \fmf{photon}{v2,o2} \fmf{photon}{v2,o3} \fmf{photon}{v2,o4}
 \end{fmfgraph*}
 \end{fmffile}}
\vskip 0.5cm
(a) \qquad \qquad \qquad \qquad \qquad \qquad \qquad (b)
\vskip 0.5 cm
 \caption{Feynman diagrams with six external gauge fields, involved in the GS mechanism in 10-D theories: (a) the hexagon fermionic loop;
(b) the tree level diagram with the 2-form $B$ exchanged.} \label{Fig:GSstring}
\end{figure}

Let us see how the mechanism works for the $SO(32)$ gauge group. The low energy effective
field theory is $N=1$ $D=10$ supergravity coupled to supersymmetric Yang-Mills theory.
The bosonic action is given by
\bea
 S_0 &=& -\int d^{10}x \, e \, \Bigg\{ \frac{1}{2k^2} R  +\frac{1}{k^2} \vf^{-2} \pd_\m \vf \pd^\m \vf + \nn\\
     &&\phantom{-\int d^{10}x \, e \, \Bigg\{}
       \frac{1}{4g^2} \vf^{-1} F_{\m\n}^a F^{\m\n a} +\frac{3k^2}{2g^4} \vf^{-2} H_{\m\n\r} H^{\m\n\r}  \Bigg\}
 \label{StringLag}
\eea
where $g$ and $k$ are respectively the Yang-Mills and the gravitational coupling, $e$ is the determinant of the zehnbein $e_\m^m$
and $\vf$ is a scalar field. The Yang-Mills field strength is defined (in the language of forms) by
\be
 F=\frac{1}{2} F_{\m\n} dx^\m \wedge dx^\n = d A + A^2
\ee
where $A^2=A \wedge A$. The field strength $H$ associated with the two-form potential $B=B_{\m\n} dx^\m \wedge dx^\n$ of the
supergravity multiplet is defined by
\be
 H = dB - \w^0_{3Y} - \w^0_{3L}
\ee
where $\w^0_{3Y}$ is the Yang-Mills Chern-Simons three-form
\be
 \w^0_{3Y} = \Tr (A F - \frac{1}{3}A^3) \label{YMCS}
\ee
and $\w^0_{3L}$ is the Lorentz group Chern-Simons three-form
\be
 \w^0_{3L} = \Tr (\w R - \frac{1}{3}\w^3) \label{LCS}
\ee
Here $\w=\w_\m dx^\m$ is the local Lorentz connection and $R=d\w+\w^2$ is the Lorentz curvature two-form.
The Chern-Simons forms (\ref{YMCS}) and (\ref{LCS}) satisfy
\bea
 d \w^0_{3Y} = \Tr F^2 \\
 d \w^0_{3L} = \Tr R^2
\eea
Under infinitesimal local Yang-Mills transformation (with parameter $\Lambda$) and local Lorentz transformation
(with parameter $\Theta$), the fields transform as follows
\bea
 &&\d A = d \Lambda + [A,\Lambda] \qquad  \d F = [F,\Lambda]\\
 &&\d \w = d \Theta + [\w,\Theta] \qquad  \d R = [R,\Theta] \\
 &&\d B = \Tr (A \, d \Lambda) + \Tr (\w \, d \Theta) \qquad  \d H = 0 \label{B2trasf}
\eea
Consider the Yang-Mills hexagon anomaly. The analysis of \cite{Zumino:1983rz} relates the anomalies to a formal expression in 12 dimensions, namely
the gauge invariant two-form $\Omega_{12}$, where
\be
 \Omega_{12} = \Tr F^6 = 15 \Tr F^2 \Tr F^4
\ee
The last equality is true only for the $SO(32)$ gauge group. This can be expressed as
\be
\Tr F^2 \Tr F^4 = d (\w^0_{3Y} \Tr F^4) = d (\w^0_{7Y} \Tr F^2)
\ee
where $\w^0_{2n+1 Y}$ is defined by
\be
 d \w^0_{2n+1 Y} = \Tr F^{n+1}
\ee
Similarly, applying an infinitesimal gauge transformation, one defines $\w^1_{2n Y}$ by
\be
 \d \w^0_{2n+1 Y} = - d \w^1_{2n Y}
\ee
In the case of Yang-Mills hexagon, the consistent anomaly is
\be
 G = \cA \int \( \frac{1}{3} \w^1_{2 Y} \Tr F^4 + \frac{2}{3} \w^1_{6 Y} \Tr F^2 \)
\ee
where the anomaly factor $\cA$ can be deduced from \cite{Zumino:1983rz}. This anomaly can be cancelled by adding to the effective action
\be
 S_1 = \cA \int \( B \Tr F^4 + \frac{2}{3} \w^0_{3 Y} \w^0_{7 Y} \) \label{GSLagB}
\ee
The non trivial $\Lambda$ transformation of $B$ in eq.~(\ref{B2trasf}) is the key of the cancellation
\be
  G + \d_\Lambda S_1 = 0
\ee
The diagram (b) in Fig.~\ref{Fig:GSstring} is built by connecting the $B \Tr F^4$ vertex in eq.~(\ref{GSLagB}) to the
$\w^0_{3 Y} dB$  vertex (contained in the $H^2$ term) in eq.~(\ref{StringLag}) by a $B$ propagator.

This is the cancellation of the Yang-Mills anomaly for the gauge group $SO(32)$. The cancellation of the gravitational and mixed anomaly goes in a similar
way and we remand to the original paper \cite{GSmechanism} for details.

In the following section we will see that something similar happens in four-dimensional theories.
The anomaly, as shown in the previous section, is related to a triangle loop diagram and it can be cancelled by the contribution of an extra field,
the \emph{axion}\footnote{In four dimensions the axion is dual to a 2-form. See for instance \cite{Klein:1999im} for the GS-mechanism with this duality manifest.}.
\subsection{Triangular Anomaly}
Let us go back to the Lagrangian~(\ref{chiralLWeyl}). Let $B_\m$ be anomaly free i.e. eq.~(\ref{BBBanofree})
holds, but let $C_\m$ be anomalous i.e. $\cA^{CCC},\cA^{CCB},\cA^{CBB}\neq 0$.
We use the GS mechanism to restore gauge invariance. As stated before, we need to modify the Lagrangian inserting a new field, the axion $\f$.
So we get
\be
 \L =\L_{\chiral}+\L_{\axion}+\L_{\GCS} \label{chiralLWeylGS}
\ee
$\L_{\axion}$ contains the couplings of $\f$
\bea
 \L_{\axion}&=& {1\over2} \( \pd_\m \f +2 b_3 C_\m \)^2 \label{Laxionchiral}\\
  &&- {1\over4} b^{CCC}_2 \f \, \Eps  F_{\m \n}^{C} F_{\r \s}^{C}
    - {1\over4} b^{CCB}_2 \f \, \Eps  F_{\m \n}^{B} F_{\r \s}^{C}
    - {1\over4} b^{CBB}_2 \f \, \Eps  F_{\m \n}^{B} F_{\r \s}^{B} \nn
\eea

The first row is the \Stuckelberg Lagrangian \cite{StuckLag}. The axion is uncharged under $B_\m$ and $C_\m$ and it transforms as
\bea
 C_\m &\to& C_\m - \pd_\m \e \nn\\
 \f &\to& \f + 2 b_3 \e \label{axiontransfchiral}
\eea
so that the \Stuckelberg Lagrangian is gauge invariant. The reason to have these terms is to give a mass $2b_3$ to
the gauge vector $C_\m$ without using a proper scalar potential breaking some symmetries . The axion is the third
degree of freedom of the massive $C_\m$, so it is an unphysical particle, a Goldstone boson. The second row of terms
contains the GS couplings\footnote{We use the notation $\tilde F^{\m \n} = \Eps F_{\r \s}$.}
($\f F \tilde F$) which are not gauge invariant because of the $\f$ transformation.
These terms are directly involved in the anomaly cancellation procedure and their form is dictated by the form of the WIs
for the triangle amplitudes.

$\L_{\GCS}$ contains the so called Generalized Chern-Simons couplings
\be
 \L_{\GCS}=d^{CCB} ~\Eps C_\m B_\n F_{\r \s}^{C} - d^{CBB} ~\Eps C_\m B_\n F_{\r \s}^{B} \label{LGCSchiral}
\ee
which are antisymmetric trilinear interactions of gauge vectors. Also these terms are not gauge invariant and they will play some role
in the GS mechanism.

The cancellation of the anomalies will fix the GS and GCS couplings.
In the following we will show two different approaches to the GS mechanism: using WIs or using the effective Lagrangian.
Obviously we will get the same result.
\subsection{Ward Identities for massive gauge fields} \label{subsect:Ward Identities}
As stated before, the WIs on a Feynman amplitude reflect the gauge invariance of a quantum field theory.
They can be written in the diagrammatic form showed in Fig.~\ref{WI}.
However, it is well known that gauge invariance can be achieved only with massless vector fields.
On the other side we experience the existence of not only physical massless vectors, with two degrees of freedom, such as the photon,
but also of physical massive vectors, with three degrees of freedom, such as the $Z_0$.
The Proca Lagrangian describes the last one, but it is not gauge invariant since an explicit mass term for the vector breaks the gauge invariance.
Fortunately, it is well known that we can avoid such a problem with the Higgs or with the \Stuckelberg mechanism.
We add in a proper way (which depends on the mechanism used) a scalar field (Goldstone boson) and vector mass terms to the massless vector Lagrangian.
This scalar field will become the third degree of freedom of the massive vector. However this will become manifest after a unitary gauge fixing.
In this gauge the Goldstone bosons are set equal to zero, we turn back to the Proca Langrangian and the physical degrees of freedom are manifest.
In this way we can achieve a gauge theory with a massive vector and since we have the gauge invariance we expect that the WIs hold.
But now we have to take into account that we have a new degree of freedom coming from the Goldstone boson, which is directly related to the gauge sector.
This modifies the WIs, which get the diagrammatic form showed in Fig.~\ref{GoldWI} \cite{Chanowitz:1985hj}.
\begin{figure}[t]
\vskip 0.5cm
\centering
$-i k^\m~\Bigg($~~~~~~~~
      \raisebox{-4.7ex}[0cm][0cm]{\unitlength=0.7mm
      \begin{fmffile}{vectorbubble5a}
      \begin{fmfgraph*}(40,25)
       \fmfpen{thick} \fmfleft{o1} \fmfright{i1,i2} \fmfstraight \fmftop{t1,t2,t3,t4,i2}\fmfbottom{b1,b2,b3,b4,i1}
       \fmf{boson}{o1,v1}   \fmf{boson}{v1,b4}       \fmf{boson}{v1,i2}
              \fmf{vanilla}{v1,t4}       \fmf{vanilla}{v1,i1}
              \fmffreeze     \fmf{boson}{v1,t3}
       \fmfv{decor.shape=circle,decor.filled=shaded, decor.size=.25w}{v1}
       \fmflabel{~~1PI}{v1} \fmflabel{$V^\m(k)$}{o1}
      \end{fmfgraph*}
      \end{fmffile}}~~~$\Bigg)$
  $+ m_V$~
     \Bigg(~~~~~~~~
      \raisebox{-4.7ex}[0cm][0cm]{\unitlength=0.7mm
      \begin{fmffile}{goldbubble5a}
      \begin{fmfgraph*}(40,25)
       \fmfpen{thick} \fmfleft{o1} \fmfright{i1,i2} \fmfstraight \fmftop{t1,t2,t3,t4,i2}\fmfbottom{b1,b2,b3,b4,i1}
       \fmf{dashes}{o1,v1}    \fmf{boson}{v1,b4}       \fmf{boson}{v1,i2}
              \fmf{vanilla}{v1,t4}       \fmf{vanilla}{v1,i1}
              \fmffreeze    \fmf{boson}{v1,t3}
       \fmfv{decor.shape=circle,decor.filled=shaded,label=1PI , decor.size=.25w}{v1}
       \fmflabel{~~1PI}{v1} \fmflabel{$G_V(k)$}{o1}
      \end{fmfgraph*}
      \end{fmffile}}~~~$\Bigg)= 0$
\vskip 0.5cm
\caption{Ward Identity for the massive gauge field $V^\m$. $G_V$ is the
corresponding Goldstone boson and $m_V$ is the mass of $V_\m$. The blob denotes all the 1PI diagrams.} \label{GoldWI}
\end{figure}
As we can see, now we have a contribution coming from the Goldstone boson\footnote{For a brief demonstration see
Appendix~\ref{app:Ward Identities}.}.
To clarify the point we consider a specific example: the decay $Z_0 \to e^+ e^-$ in the SM.
\begin{figure}[t]
\vskip 1cm
$-i (p+q)^\m$~\Bigg(~~~~~~~~~~~~~~~
      \raisebox{-5.85ex}[0cm][0cm]{\unitlength=0.7mm
      \begin{fmffile}{Z0ee2}
      \begin{fmfgraph*}(30,30)
       \fmfpen{thick}
       \fmfleft{i1,i2,i3} \fmfright{o1,o2,o3} \fmftop{i1,v1,o1} \fmfbottom{i3,v3,o3}
       \fmf{phantom}{i2,v2,o2} \fmf{phantom}{v1,v2,v3} \fmffreeze
       \fmf{photon}{i2,v2} \fmf{fermion}{o3,v2} \fmf{fermion}{v2,o1}
       \fmflabel{$(Z_0)_\m (p+q)$}{i2} \fmflabel{$e^+ (p)$}{o3} \fmflabel{$e^-(q)$}{o1}
      \end{fmfgraph*}
      \end{fmffile}}~\Bigg)
  + $\MZO$~\Bigg(~~~~~~~~~~~
      \raisebox{-5.85ex}[0cm][0cm]{\unitlength=0.7mm
      \begin{fmffile}{G0ee2}
      \begin{fmfgraph*}(30,30)
       \fmfpen{thick}
       \fmfleft{i1,i2,i3} \fmfright{o1,o2,o3} \fmftop{i1,v1,o1} \fmfbottom{i3,v3,o3}
       \fmf{phantom}{i2,v2,o2} \fmf{phantom}{v1,v2,v3} \fmffreeze
       \fmf{dashes}{i2,v2} \fmf{fermion}{o3,v2} \fmf{fermion}{v2,o1}
       \fmflabel{$G_0 (p+q)$}{i2} \fmflabel{$e^+ (p)$}{o3} \fmflabel{$e^-(q)$}{o1}
      \end{fmfgraph*}
      \end{fmffile}}~\Bigg)
= 0
\vskip 1cm
\caption{Ward Identities for the decay $Z_0 \to e^+ e^-$.
         The first term is the usual one in WI while the second one is the correction coming from the Goldstone boson $G_0$.}\label{WIZ0ee}
\end{figure}
The Feynman amplitude of the process is given by (for the Feynman rules see Appendices~\ref{sec:Electroweak interactions} and \ref{sec:Yuk int})
\be
 \G_\m (Z_0 \to e^+ e^-) = -\frac{i}{2} \,  g_{Z_0} \bar u(q) \g_\m \( v_e^{Z_0} -a_e^{Z_0} \g_5 \) v(p) \label{Z0eeamp}
\ee
where $v_e=-1/2+2\sin^2\theta_W$, $a_e=-1/2$ and $g_{Z_0}=\sqrt{g_1^2+g_2^2}$. The WI for the process is given in Fig.~\ref{WIZ0ee}.
The first term is
\bea
 -i (p+q)^\m \G_\m (Z_0 \to e^+ e^-) &=& -\frac{1}{2} g_{Z_0} \, \bar u(q) (\slashed p + \slashed q) \( v_e^{Z_0} -a_e^{Z_0} \g_5 \) v(p) \nn\\
                                     &=& +\frac{1}{2} g_{Z_0} m_e \, \bar u(q) \( v_e^{Z_0} + a_e^{Z_0} \g_5 \) v(p)\nn\\
                                     &&  -\frac{1}{2} g_{Z_0} m_e \, \bar u(q) \( v_e^{Z_0} - a_e^{Z_0} \g_5 \) v(p)\nn\\
                                     &=& g_{Z_0} m_e a_e  \, \bar u(q) \g_5 v(p) \nn\\
                                     &=& -\frac{1}{2} m_e \sqrt{g_1^2+g_2^2}  \, \bar u(q) \g_5 v(p) \label{WIZ0eeamp}
\eea
where $m_e$ is the electron mass and we used the equation of motion of the spinors
\bea
 \bar u (\slashed q - m_e) &=& 0 \label{eombaru}\\
 (\slashed p + m_e) v &=& 0 \label{eomv}
\eea
The second term is
\bea
 \MZO \G (G_0 \to e^+ e^-) &=& \MZO \frac{y_e}{\sqrt2} \bar u(q) \g_5 v(p) \nn\\
                           &=& \frac{1}{2} v \, \sqrt{g_1^2+g_2^2} \, \frac{m_e}{v} \bar u(q) \g_5 v(p) \nn\\
                           &=& \frac{1}{2} m_e \, \sqrt{g_1^2+g_2^2} \, \bar u(q) \g_5 v(p) \label{WIG0eeamp}
\eea
where we used $y_e/\sqrt2 = m_e/v$ and eq.~(\ref{Z0mass}).
Summing the eq. (\ref{WIZ0eeamp}) and (\ref{WIG0eeamp}) we get
\be
 -i (p+q)^\m \G_\m (Z_0 \to e^+ e^-) + \MZO \G (G_0 \to e^+ e^-) = 0 \label{WIZ0eeTOT}
\ee
in agreement with the result showed in Fig.~\ref{GoldWI}.
\subsection{Ward Identities and Anomaly Cancellation} \label{subsect:Ward Identities and Anomaly Cancellation}
Let us go back to the anomaly cancellation procedure.
We focus on the $CBB$ anomaly. The WIs~(\ref{WIrhoCBBtrian}), (\ref{WImuCBBtrian}) and (\ref{WInuCBBtrian})
modified because of the presence of GS and GCS couplings. Using Fig.~\ref{GoldWI}, we get the WIs for the $CBB$ amplitude expressed in diagrammatic
form in Fig.~\ref{Fig:WIanoCBB}.
     \begin{figure}[tb]
  \vskip 1.5cm
%
$(p+q)^\r ~\Bigg($~~~~~~~~~~~~
      \raisebox{-4ex}[0cm][0cm]{\unitlength=0.4mm
      \begin{fmffile}{CBBloop}
      \begin{fmfgraph*}(60,40)
       \fmfpen{thick} \fmfleft{ii0,ii1,ii2} \fmfstraight \fmffreeze \fmftop{ii2,t1,t2,t3,oo2}
       \fmfbottom{ii0,b1,b2,b3,oo1} \fmf{phantom}{ii2,t1,t2} \fmf{phantom}{ii0,b1,b2} \fmf{phantom}{t1,v1,b1}
       \fmf{phantom}{t2,b2} \fmf{phantom}{t3,b3}
       \fmffreeze
       \fmf{photon}{ii1,v1} \fmf{fermion}{t3,v1} \fmf{fermion,label=$\psi$}{b3,t3} \fmf{fermion}{v1,b3}
       \fmf{photon}{b3,oo1} \fmf{photon}{t3,oo2}
       \fmflabel{$C_\r(p+q)$}{ii1}
       \fmflabel{$B_\m(p)$}{oo1}
       \fmflabel{$B_\n(q)$}{oo2}
      \end{fmfgraph*}
      \end{fmffile}}      ~~+~~~~~~
      \raisebox{-4ex}[0cm][0cm]{\unitlength=0.5mm
      \begin{fmffile}{GCS_CBB}
      \begin{fmfgraph*}(30,30)
       \fmfpen{thick}
       \fmfleft{i1} \fmfright{o1,o2}
       \fmf{boson}{i1,v1} \fmf{boson}{v1,o1} \fmf{boson}{v1,o2}
       \fmffreeze
       \fmflabel{$C$}{i1}\fmflabel{$B$}{o1}\fmflabel{$B$}{o2}
      \end{fmfgraph*}
      \end{fmffile}} $\Bigg)$
$+2ib_3 \Bigg($
      \raisebox{-4ex}[0cm][0cm]{ \unitlength=0.5mm
      \begin{fmffile}{axionBB}
      \begin{fmfgraph*}(30,30)
      \fmfpen{thick} \fmfleft{i1} \fmfright{o1,o2}
      \fmf{dashes}{i1,v1} \fmf{photon}{v1,o2}
      \fmf{photon}{v1,o1} \fmffreeze \fmflabel{$B$}{o1} \fmflabel{$B$}{o2} \fmflabel{$\f$}{i1}
      \end{fmfgraph*}
      \end{fmffile}} $\Bigg)=0$
        \vskip 2cm

~~~~~~$p^\m ~\Bigg($~~~~
      \raisebox{-4ex}[0cm][0cm]{\unitlength=0.4mm
      \begin{fmffile}{CBBloop2}
      \begin{fmfgraph*}(60,40)
       \fmfpen{thick} \fmfleft{ii0,ii1,ii2} \fmfstraight \fmffreeze \fmftop{ii2,t1,t2,t3,oo2}
       \fmfbottom{ii0,b1,b2,b3,oo1} \fmf{phantom}{ii2,t1,t2} \fmf{phantom}{ii0,b1,b2} \fmf{phantom}{t1,v1,b1}
       \fmf{phantom}{t2,b2} \fmf{phantom}{t3,b3}
       \fmffreeze
       \fmf{photon}{ii1,v1} \fmf{fermion}{t3,v1} \fmf{fermion,label=$\psi$}{b3,t3} \fmf{fermion}{v1,b3}
       \fmf{photon}{b3,oo1} \fmf{photon}{t3,oo2}
       \fmflabel{$C$}{ii1} \fmflabel{$B$}{oo1} \fmflabel{$B$}{oo2}
      \end{fmfgraph*}
      \end{fmffile}}      ~~~+~~~~~~
      \raisebox{-4ex}[0cm][0cm]{\unitlength=0.5mm
      \begin{fmffile}{GCS_CBB}
      \begin{fmfgraph*}(30,30)
       \fmfpen{thick}
       \fmfleft{i1} \fmfright{o1,o2}
       \fmf{boson}{i1,v1} \fmf{boson}{v1,o1} \fmf{boson}{v1,o2}
       \fmffreeze
       \fmflabel{$C$}{i1}\fmflabel{$B$}{o1}\fmflabel{$B$}{o2}
      \end{fmfgraph*}
      \end{fmffile}} $\Bigg)=0$
      \vskip 2cm
~~~~~~$q^\n ~\Bigg($~~~~
      \raisebox{-4ex}[0cm][0cm]{\unitlength=0.4mm
      \begin{fmffile}{CBBloop2}
      \begin{fmfgraph*}(60,40)
       \fmfpen{thick} \fmfleft{ii0,ii1,ii2} \fmfstraight \fmffreeze \fmftop{ii2,t1,t2,t3,oo2}
       \fmfbottom{ii0,b1,b2,b3,oo1} \fmf{phantom}{ii2,t1,t2} \fmf{phantom}{ii0,b1,b2} \fmf{phantom}{t1,v1,b1}
       \fmf{phantom}{t2,b2} \fmf{phantom}{t3,b3}
       \fmffreeze
       \fmf{photon}{ii1,v1} \fmf{fermion}{t3,v1} \fmf{fermion,label=$\psi$}{b3,t3} \fmf{fermion}{v1,b3}
       \fmf{photon}{b3,oo1} \fmf{photon}{t3,oo2}
       \fmflabel{$C$}{ii1} \fmflabel{$B$}{oo1} \fmflabel{$B$}{oo2}
      \end{fmfgraph*}
      \end{fmffile}}      ~~~+~~~~~~
      \raisebox{-4ex}[0cm][0cm]{\unitlength=0.5mm
      \begin{fmffile}{GCS_CBB}
      \begin{fmfgraph*}(30,30)
       \fmfpen{thick}
       \fmfleft{i1} \fmfright{o1,o2}
       \fmf{boson}{i1,v1} \fmf{boson}{v1,o1} \fmf{boson}{v1,o2}
       \fmffreeze
       \fmflabel{$C$}{i1}\fmflabel{$B$}{o1}\fmflabel{$B$}{o2}
      \end{fmfgraph*}
      \end{fmffile}} $\Bigg)=0$
      \vskip 1cm
\caption{The Ward Identities for the amplitude $C_\r(p+q)\to B_\m(p) \, B_\n(q)$
include the GCS as well as the axionic (GS) couplings.
The GS coupling arises from Lagrangian~(\ref{Laxionchiral}) while the GCS coupling comes from Lagrangian~(\ref{LGCSchiral}).
Each depicted diagram also contains the exchange $(\m,p)\leftrightarrow (\n, q)$.} \label{Fig:WIanoCBB}
\end{figure}
The Feynman rules for the GS mechanism are given in Appendix~\ref{sec:C-B model}.
We denote by $(\GS)^{BB}_{\m \n}$ the contribution coming from the GS coupling $\f F^B \tilde F^B$ in Lagrangian~(\ref{Laxionchiral})
\be
 (\GS)^{BB}_{\m \n}=-2 i b^{CBB}_2 \e_{\m \n \a \b} p^\a q^\b \label{GS_BB}
\ee
and we denote by $(\GCS)^{CBB}_{\r \m \n}$ the contribution from the GCS coupling $C B \tilde F^B$ in
Lagrangian~(\ref{LGCSchiral})
\be
(\GCS)^{CBB}_{\r \m \n}=2 d^{CBB} \e_{\r \n \m \a} (p-q)^\a   \label{GCS_CBB}
\ee
The WIs in Fig. \ref{Fig:WIanoCBB} correspond to
\bea
 (p+q)^\r \Big(\D_{\r \m \n}^{CBB}(p,q;0) +(\GCS)^{CBB}_{\r\m\n}\Big)+2i b_3 (\GS)^{BB}_{\m\n}&=&0\nn \\
 p^\m  \Big(\D_{\r \m \n}^{CBB}(p,q;0) +(\GCS)^{CBB}_{\r\m\n}\Big) &=&0\nn \\
 q^\n  \Big(\D_{\r \m \n}^{CBB}(p,q;0) +(\GCS)^{CBB}_{\r\m\n}\Big) &=&0
\label{WardIdanoCBB}
\eea
Using eq.~(\ref{WIrhoCBBtrian}), (\ref{WImuCBBtrian}) and (\ref{WInuCBBtrian}), the last two identities imply
\be
 \( 2+\b \) \, \frac{\cA^{CBB}}{16\pi^2}  - 2 d^{CBB} = 0 \qquad \Rightarrow
 \qquad d^{CBB} = { 2+\b \over 2} \,\frac{\cA^{CBB}}{16\pi^2} \label{dCBB}
\ee
and the first identity becomes
\be
 - \b \, \frac{ \cA^{CBB} }{8 \pi^2} + 4 \ { 2+\b \over 2}   \frac{ \cA^{CBB} }{16\pi^2} + 4 b^{CBB}_2 b_3 =0
\qquad \Rightarrow \qquad b^{CBB}_2 b_3 =- \frac{ \cA^{CBB}}{16 \pi^2}  \label{bCBB}
\ee
In the same way\footnote{The momenta of the $CCB$ triangle are so that on the incoming $(p+q)_\r$ line we have the $B$ vector.},
the cancellation of the remaining anomalies gives
\bea
&&b^{CCC}_2 b_3 =-\frac{\cA^{CCC}}{48\pi^2} \label{bCCC} \\
&&b^{CCB}_2 b_3 = -\frac{\cA^{CBB}}{16 \pi^2} \qquad d^{CCB} ={\b' \over 4} \frac{ \cA^{CCB} }{8 \pi^2}
\label{bdCCB}
\eea
where we used different shift parameters $\b'$ and $\b''$ respectively for the new triangles $CCB$ and $CCC$.
However as stated in the end of Section \ref{sect:Triangular Anomaly} and done in eq.~(\ref{bCCC}),
$\b''$ must be fixed to $\b''=-2/3$ because we have three equal external vectors.
It is worth noting that the GCS coefficients are determined in terms of the $\cA$'s by the WIs, while the $b_2$'s depend
only on the free parameter $b_3$, which is related to the mass of the anomalous $U(1)$.
Moreover the GCS couplings depends on the scheme of the anomaly distribution so they do not contain any
physical information (see Section \ref{sect:Computation of the amplitude}).
We can always cancel them with a particular choice of the shift parameters,
in fact they are also called GCS counterterms because they parametrize the arbitrariness of the scheme
of the anomaly distribution in the anomaly cancellation procedure. On the contrary, the GS coupling does not depend
to any shift parameters and so they have some physical relevance as we will show in the computation of the amplitudes
in Section \ref{sect:Computation of the amplitude}.

\subsection{Effective Lagrangian} \label{subsect:Effective Lagrangian}
Now we show the anomaly cancellation using the variation of the effective Lagrangian. As before we focus on the
$CBB$ amplitude. The WIs~(\ref{WIrhoCBBtrian}), (\ref{WImuCBBtrian}) and (\ref{WInuCBBtrian}) induce the following
variations on the 1 loop effective Lagrangian. For
\bea
 C_\m &\to& C_\m - \pd_\m \e^{C} \label{Ctrasf}\\
 B_\m &\to& B_\m - \pd_\m \e^{B} \label{Btrasf}
\eea
we have
\bea
 \d_C \L_{\chiral}^{CBB} &=& \e_C \frac{\b}{2} \frac{1}{32 \pi^2} \cA^{CBB} \Eps F_{\m \n}^B F_{\r \s}^B
 \label{L_CBBvarC}\\
 \d_B \L_{\chiral}^{CBB} &=& -2 \e_B \frac{2+\b}{4} \frac{1}{32 \pi^2} \cA^{CBB} \Eps F_{\m \n}^B F_{\r \s}^C
  \label{L_CBBvarB}
\eea
The requirement of gauge invariance implies
\bea
 \d_C \L_{1 loop}^{CBB} &=& \d_C \L_{\chiral}^{CBB}+ \d_C \L_{\axion}^{BB}+ \d_C \L_{\GCS}^{CBB} = 0 \label{gaugeinvC}\\
 \d_B \L_{1 loop}^{CBB} &=& \d_B \L_{\chiral}^{CBB}+ \d_B \L_{\axion}^{BB}+ \d_B \L_{\GCS}^{CBB} = 0 \label{gaugeinvB}
\eea
The gauge variations for the axion Lagrangian are
\bea
 \d_C \L_{\axion}^{BB} &=& - \e_C {1\over2} b^{CBB}_2 b_3 \, \Eps  F_{\m \n}^{B} F_{\r \s}^{B} \label{Laxion_CBBvarC}\\
 \d_B \L_{\axion}^{BB} &=& 0 \label{Laxion_CBBvarB}
\eea
and for the GCS terms are
\bea
 \d_C \L_{\GCS}^{CBB} &=& + d^{CBB} \Eps \pd_\m \e_C B_\n F_{\r \s}^{B}= - \e_C d^{CBB} \Eps \pd_\m \(B_\n F_{\r \s}^{B}\) \nn\\
               &=& - \frac{1}{2} d^{CBB} \e_C \Eps F_{\m \n}^{B} F_{\r \s}^{B} \label{LGCS_CBBvarC} \\
 \d_B \L_{\GCS}^{CBB} &=& + d^{CBB} \Eps C_\m \pd_\n \e_B F_{\r \s}^{B}= - \e_B d^{CBB} \Eps \pd_\n \(C_\m F_{\r \s}^{B}\)   \nn\\
               &=& + \frac{1}{2} d^{CBB} \e_B \Eps F_{\m \n}^{B} F_{\r \s}^{C} \label{LGCS_CBBvarB}
\eea
where we used $\pd_\m F_{\r \s}^{B,C}=0$.
Then eq.~(\ref{gaugeinvB}) implies
\be
 -\frac{2+\b}{2} \frac{1}{32 \pi^2} \cA^{CBB}+\frac{1}{2} d^{CBB}=0 \quad \Rightarrow \quad d^{CBB}=\frac{2+\b}{32 \pi^2} \cA^{CBB}
\ee
Substituting this result into eq.~(\ref{gaugeinvC}), we get
\bea
 &&\frac{\b}{2} \frac{1}{32 \pi^2} \cA^{CBB}- {1\over2} b^{CBB}_2 b_3-\frac{1}{2} \frac{2+\b}{32 \pi^2} \cA^{CBB}=
 -\frac{1}{2} \frac{2}{32 \pi^2} \cA^{CBB}- {1\over2} b^{CBB}_2 b_3 \nn\\
 &&\quad\Rightarrow b^{CBB}_2 b_3=-\frac{1}{16 \pi^2} \cA^{CBB}
\eea
obtaining the same relations as eq.~(\ref{dCBB}) and (\ref{bCBB}).

We showed the anomaly cancellation procedure in two different ways, the WI method or
the effective Lagrangian one. Dependently on what we need to compute we will choose the most convenient procedure.
\section{Supersymmetric Green-Schwarz mechanism} \label{sect:Supersymmetric Green-Schwarz mechanism}
In this section we illustrate how to supersymmetrize what we did in the previous section.
The supersymmetric version of eq.~(\ref{Laxionchiral}) and (\ref{LGCSchiral}) are
(see \cite{Klein:1999im} and \cite{Andrianopoli:2004sv})
\bea
 \L_{\axion} &=& {1\over4} \left. \( S + S^\dagger + 4 b_3 V^C \)^2 \right|_{\thth}
  - {1\over2} \Big\{ \[b^{CCC}_2 S ~\Tr\( W^{C} W^{C} \) \right. \nn\\
 && \left. +b^{CBB}_2 S ~\Tr\( W^{B} W^{B} \) + b^{CCB}_2 S ~W^{B} ~W^{C} \]_{\th^2} +h.c. \Big\}~
 \label{Laxionsusychiral}\\
 \L_{\GCS} &=&+ d^{CCB} \[ \( V^B D^\a V^C - V^C D^\a V^B\) W^C_\a + h.c. \]_{\thth} +\nn\\
          &&-  d^{CBB} \[ \( V^B D^\a V^C - V^C D^\a V^B\) W^B_\a + h.c. \]_{\thth}
 \label{LGCSsusychiral}
\eea
where the $V$'s and the $W$'s are the gauge superfields and the corresponding superfield strength, and
S is the \Stuckelberg multiplet
\be
 S =  s+ i\sqrt2 \th \psi_S + \th^2 F_S - i \th \s^\m \bar\th \pd_\m s +
               {\sqrt2\over2}  \th^2 \bar\th \bar\s^\m \pd_\m \psi_S - {1\over4} \thth \Box s \label{Smult}
\ee
which transforms under the $U(1)_C$ as
\bea
 V^C &\to& V^C + i \( \Lambda^C - \Lambda^{C\dag} \) \nn\\
  S  &\to& S - 4 i ~b_3 ~\Lambda^C \label{U1CTrans}
\eea
The component fields of $S$ are the complex scalar $s=\a+i\f$, where $\a$ is the \emph{saxion} and $\f$ is the axion,
and the spinorial field $\psi_S$ which is called \emph{axino}.
To fix the $b$'s and $d$'s parameter we act on the bosonic part as we did in
Section~\ref{sect:Anomaly cancellation. Green-Schwarz mechanism},
since the supersymmetric part will be fixed by supersymmetry, or we act directly on the superfields.
We show the last method for the $CBB$ case. Under the gauge transformations
\bea
 V^C &\to& V^C + i \( \Lambda^C - \Lambda^{C\dag} \) \nn\\
 V^B &\to& V^B + i \( \Lambda^B - \Lambda^{B\dag} \)
\eea
we have
\bea
 \d_C \L_{\chiral}^{CBB} &=&-i \frac{\b}{2} \frac{1}{32 \pi^2} \cA^{CBB} \[ \Lambda^C~\Tr\( W^{B} W^{B} \)\]_{\th^2} +h.c.
 \label{L_CBBvarCsusy}\\
 \d_B \L_{\chiral}^{CBB} &=& 2i \frac{2+\b}{4} \frac{1}{32 \pi^2} \cA^{CBB} \[ \Lambda^B~\Tr\( W^{B} W^{C} \)\]_{\th^2} +h.c
  \label{L_CBBvarBsusy}
\eea
\bea
 \d_C \L_{\axion}^{BB} &=& 2i  b^{CBB}_2 b_3 \[ \Lambda^C~\Tr\( W^{B} W^{B} \)\]_{\th^2} +h.c.
 \label{Laxion_CBBvarCsusy}\\
 \d_B \L_{\axion}^{BB} &=& 0  \label{Laxion_CBBvarBsusy}
\eea
\bea
 \d_C \L_{\GCS}^{CBB} &=& + \frac{i}{2} d^{CBB} \[ \Lambda^C~\Tr\( W^{B} W^{B} \)\]_{\th^2} +h.c.
 \label{LGCS_CBBvarCsusy} \\
 \d_B \L_{\GCS}^{CBB} &=&  - \frac{i}{2} d^{CBB} \[ \Lambda^B~\Tr\( W^{B} W^{C} \)\]_{\th^2} +h.c
 \label{LGCS_CBBvarBsusy}
\eea
The requirement of gauge invariance will impose the following conditions
\bea
 &&\!\!\!\!\!\!\!\!\!\!\!\!\!\!\d_C \L_{1 loop}^{CBB}=
    \d_C \L_{\chiral}^{CBB}+ \d_C \L_{\axion}^{BB}+ \d_C \L_{\GCS}^{CBB} \\
 &&=\[-i \frac{\b}{2} \frac{1}{32 \pi^2} \cA^{CBB}+2i  b^{CBB}_2 b_3+ \frac{i}{2} d^{CBB}\]
    \[ \Lambda^C~\Tr\( W^{B} W^{B} \)\]_{\th^2} +h.c.= 0 \nn\\
\nn\\
 &&\!\!\!\!\!\!\!\!\!\!\!\!\!\!\d_B \L_{1 loop}^{CBB} =
    \d_B \L_{\chiral}^{CBB}+ \d_B \L_{\axion}^{BB}+ \d_B \L_{\GCS}^{CBB} \\
 &&=\[2i \frac{2+\b}{4} \frac{1}{32 \pi^2} \cA^{CBB}- \frac{i}{2} d^{CBB}\]
    \[ \Lambda^C~\Tr\( W^{B} W^{C} \)\]_{\th^2} +h.c.=0  \nn
\eea
whose solutions are the eq.~(\ref{dCBB}) and (\ref{bCBB}). Analogously for the other anomalies.

In this section we do not give the component fields expansion for eq. ~(\ref{Laxionsusychiral}) and
(\ref{LGCSsusychiral}), leaving it for the details of our model in Section~\ref{subsec: Anomalies and GS mechanism}.
\section{Computation of the amplitudes} \label{sect:Computation of the amplitude}
In this section we show how we compute the 1 loop amplitudes\footnote{For a massive fermionic triangle see
Appendix~\ref{app:Fermionic loop diagram}.}
involving three external vectors.
As usual we focus on the $CBB$ case since the generalization to the other cases is straightforward.

Before going on we recall the Rosenberg parametrization\footnote{For a demonstration of the Rosenberg parametrization
see Appendix~\ref{subsec:Rosenberg}.} \cite{Rosenberg:1962pp}.
The amplitudes~(\ref{AVVtrian}), (\ref{VAVtrian}), (\ref{VVAtrian}) and (\ref{AAAtrian})
can be written as
\bea
&& \G_{\r \m \n}  (p,q;0)=\nn\\
&&\quad{1\over{ \pi^2}} \Big(
I_1(p,q;0) \,\e[p,\m,\n,\r] + I_2(p,q;) \, \e[q,\m,\n,\r]+  I_3(p,q;0)  \, \e[p,q,\m,\r]{p}_{\n} \nn\\
&&\quad+  I_4(p,q;0) \, \e[p,q,\m,\r]{q}_{\n} +  I_5(p,q;0) \, \e[p,q,\n,\r]p_\m +
I_6(p,q;0) \, \e[p,q,\n,\r]q_\m \Big)  \nn\\
\label{Ros}
\eea
with $\e[p,q,\r,\s]=\epsilon_{\m\n\r\s} p^\m q^\n$ and where
\bea
 I_3(p,q;0) &=& -\int_0^1 dx
    \int_0^{1-x} dy \frac{x y}{y (1-y) p^2 +  x(1-x) q^2 + 2 x y \,p\cdot q}  \nn \\
 I_4(p,q;0) &=& \int_0^1 dx
    \int_0^{1-x} dy \frac{x (x-1)}{y (1-y) p^2 + x(1-x) q^2 + 2 x y \,p\cdot q}  \nn \\
 I_5(p,q;0) &=& -I_4(q,p;0) \nn \\
 I_6(p,q;0) &=& -I_3(p,q;0)  \label{I's}
\eea
In terms of the Rosenberg parametrization  the $\b$ dependence of (\ref{AVVtrianbeta}) is
contained only in $I_1$ and $I_2$ (which are superficially divergent). However, using the
WIs~(\ref{WIrhoAVVtrianbetafinal}), (\ref{WImuAVVtrianbeta}) and (\ref{WInuAVVtrianbeta}) it is
possible to show that they can be expressed in terms of $I_3 \dots I_6$ as
\bea
 I_1^{AVV}(p,q,\b;0) &=& p \cdot q \, I_3(p,q;0) + q^2 \, I_4(p,q;0) +\frac{2+\b}{8}\nn\\
 I_2^{AVV}(p,q,\b;0) &=& -I_1^{AVV}(q,p,\b;0) \label{I12AVV}
\eea
Analogously for the VAV, VVA and AAA triangles.

We can act in the same way for the $CBB$ amplitude. The total fermionic triangle can be written as
\be
 \D_{\r \m \n}^{CBB} = -{\cA^{CBB}\over2}   \G_{\r \m \n}(p,q;0) \label{DeltaCBB}
\ee
where $\G_{\r \m \n}(p,q;0)$ is the parametrization~(\ref{Ros}).
Imposing the eq.~(\ref{WIrhoCBBtrian}), (\ref{WImuCBBtrian}) and (\ref{WInuCBBtrian}) we find that
\bea
 I_1^{CBB}(p,q,\b;0) &=& p \cdot q \, I_3(p,q;0) + q^2 \, I_4(p,q;0) +\frac{2+\b}{8}\nn\\
 I_2^{CBB}(p,q,\b;0) &=& -I_1^{CBB}(q,p,\b;0) \label{I12CBB}
\eea
and the amplitude, as we know, is not uniquely defined. The anomaly cancellation will solve this problem.
In the standard case we have $\cA^{CBB}=0$ so the amplitude is uniquely defined but identically zero.
Instead with the $\GS$ mechanism, the final amplitude will not be zero.
In this case we have
\be
 A^{CBB}_{\r \m \n} =\D^{CBB}_{\r \m \n} + (\GCS)^{CBB}_{\r \m \n}
              =\D^{CBB}_{\r \m \n} + 2 d^{CBB} \e_{\r \n \m \a} (p-q)^\a   \label{ACBB}
\ee
The GCS terms can be reabsorbed by the following redefinitions
\be
 A^{CBB}_{\r \m \n} =\D^{CBB}_{\r \m \n} + (\GCS)^{CBB}_{\r \m \n}
                    =-{\cA^{CBB} \over 2 } \tilde \G_{\r \m \n} \label{ACBBtilde}
\ee
where
\bea
 \tilde \G_{\r \m \n}  &=&
 {1\over{ \pi^2}} \Big( \tilde I_1 \e[p,\m,\n,\r] +\tilde I_2\e[q,\m,\n,\r]+ I_3 \e[p,q,\m,\r]{p}^{\n} \nn\\
   &&+  I_4 \e[p,q,\m,\r]{q}^{\n} +  I_5 \e[p,q,\n,\r]p^\m +
    I_6(\e[p,q,\n,\r]q^\m \Big) \label{Rostilde}
\eea
with
\bea
 \( {\cA^{CBB}\over{2 \pi^2}} \) \tilde I_1(p,q) &=& \( {\cA^{CBB}\over{2 \pi^2}}  \) I_1(p,q)-2 d^{CBB} \nn\\
 \( {\cA^{CBB}\over{2 \pi^2}} \) \tilde I_2(p,q) &=& \( {\cA^{CBB}\over{2 \pi^2}}  \) I_2(p,q)+2 d^{CBB}
\label{ItildeCBB}
\eea
Using eq. (\ref{dCBB}) and (\ref{I12CBB}) we get
\bea
 \tilde I_1(p,q) &=& p \cdot q I_3(p,q) + q^2 I_4(p,q) \nn\\
 \tilde I_2(p,q) &=& -\tilde I_1(q,p) \label{I12tilde}
\eea
that relate $\tilde I_1$ and $\tilde I_2$ to the other $I_i$'s.
So we have removed the $\b$ dependence of the fermionic triangle, thanks to
the GCS coupling, and the amplitude is uniquely defined and it obeys the following WIs
\bea
 (p+q)^\r A^{CBB}_{\r \m \n} &=& \frac{\cA^{CBB}}{32 \pi^2} \e_{\m \n \a \b} p^\a q^\b=-2 i b_3 (\GS)^{BB}_{\m \n}\nn \\
 p^\m A^{CBB}_{\r \m \n} &=&0\nn \\
 q^\n A^{CBB}_{\r \m \n} &=&0
\label{WI_ACBB}
\eea
Finally the theory is gauge invariant, but in this case the amplitude is different from zero.
This result does not depend on the scheme of the anomaly distribution.

\newpage\thispagestyle{empty}

\chapter{MiAUMSSM. Model building} \label{chap:MiAUMSSM. Model building}
One of the most attractive scenario for physics beyond the SM is the existence of additional massive neutral gauge
bosons (for a recent review see \cite{Langacker:2008yv} and references therein;
for some recent works we suggest \cite{Demir:2005ti}-\cite{Kors:2004dx}).  They could be one of the first discoveries at LHC if their mass is
in the range of a few TeV. Many different models have been developed in the past in order to investigate this
possibility. The mass could be acquired in a variety of ways: from Kaluza-Klein modes to a standard Higgs mechanism
or even the \Stuckelberg mechanism~\cite{StuckLag}. The latter is common to low energy effective
field theories which appear anomalous. The anomaly cancellation is achieved by the GS mechanism which
ensure the consistency of these models~\cite{GSmechanism,Sagnotti:1992qw}.

In string theory anomalous $U(1)$'s are very common. D-brane models contain several abelian factors,
living on each stack of branes, and they are typically anomalous \cite{Sagnotti:1992qw}-\cite{Buican:2006sn}. In
the presence of these anomalous $U(1)$'s, the \Stuckelberg mixing with the axions cancels mixed
anomalies, and renders the ``anomalous" gauge fields massive. The masses depend non-trivially on the internal volumes and on other
moduli, allowing the physical masses of the anomalous $U(1)$ gauge bosons to be much smaller than the string scale
(even at a few TeV range) \cite{Antoniadis:2002cs}.
For further works on these topics see \cite{Pradisi:1988xd}-\cite{DeRydt:2007vg}.

This chapter is entirely based on \cite{mypaper}, where we presented for the first time our model: the
Minimal Anomalous $U(1)'$ Extension of the MSSM (MiAUMSSM).
\newpage\section{Preliminaries} \label{sect:MiAUMSSM Intro}
We study an extension of the MSSM\footnote{Since we do not discuss phenomenological consequences,
the treatment of all this section is before the insertion of the coupling constant using $V \to 2 g V$.}
by the addition of an abelian vector multiplet $V^{(0)}$.
We assume that all the MSSM fields are charged under the additional vector multiplet $V^{(0)}$, with charges that are
given in Table~\ref{QTable}, where $Q_i, L_i$ are the left handed quarks and leptons respectively while $U^c_i,
D^c_i, E^c_i$ are the right handed up and down quarks and the electrically charged leptons. The superscript $c$
stands for charge conjugation. The index $i=1,2,3$ denotes the three different families. $H_{u,d}$ are the two Higgs
scalars.
  \begin{table}[h]
  \centering
  \begin{tabular}[h]{|c|c|c|c|c|}
   \hline & SU(3)$_c$ & SU(2)$_L$  & U(1)$_Y$ & ~U(1)$^{\prime}~$\\
   \hline $Q_i$   & $\bth$       &  $\btw$       &  $1/6$   & $Q_{Q}$ \\
   \hline $U^c_i$   & $\bar \bth$  &  $\bon$       &  $-2/3$  & $Q_{U^c}$ \\
   \hline $D^c_i$   & $\bar \bth$  &  $\bon$       &  $1/3$   & $Q_{D^c}$ \\
   \hline $L_i$   & $\bon$       &  $\btw$       &  $-1/2$  & $Q_{L}$ \\
   \hline $E^c_i$   & $\bon$       &  $\bon$       &  $1$     & $Q_{E^c}$\\
   \hline $H_u$ & $\bon$       &  $\btw$       &  $1/2$   & $Q_{H_u}$\\
   \hline $H_d$ & $\bon$       &  $\btw$       &  $-1/2$  & $Q_{H_d}$ \\
   \hline
  \end{tabular}
  \caption{Charge assignment.}\label{QTable}
  \end{table}
In order to gain in flexibility, our model is only string inspired: we do not commit to a specific
brane model and this is why the charges are not fixed, even if the effective cut-off is related to the mass of the
$Z'$.
Since our model is an extension of the MSSM, the gauge invariance of the superpotential~(\ref{MSSMsuperpot}), that contains the Yukawa
couplings and a $\m$-term, put constraints on the above charges
  \bea
   \QU &=& - \QQ - \QHu  \nn\\
   \QD &=& - \QQ + \QHu  \nn\\
   \QE &=& -\QL  + \QHu  \nn\\
   \QHd  &=& - \QHu \label{Qconstraints}
  \eea
Thus, $\QQ$, $\QL$ and $\QHu$ are free parameters of the model.

The extra vector multiplet generically is anomalous and consistency of the model requires an additional
\Stuckelberg multiplet $S$ with the proper couplings as well as GCS terms.
For the complete Langrangian of the model see Appendix~\ref{app:MiAUMSSM Lagrangian}.
\subsection{Anomalies and GS mechanism} \label{subsec: Anomalies and GS mechanism}
As showed in Section \ref{sect:Anomaly cancellation. SM and MSSM}, the MSSM is anomaly free.
All the anomalies that involve only the $SU(3)$, $SU(2)$ and $U(1)_Y$ factors vanish identically.
However, triangles with $U(1)'$ in the external legs in general are potentially anomalous.
These anomalies are\footnote{We are working in an effective field theory framework and we ignore
throughout the Thesis all the gravitational effects.  In particular, we do not consider the gravitational anomalies
which, however, could be cancelled by the GS mechanism.}
\bea
   U(1)'-U(1)'-U(1)'~~~:     &&\ \cA^{(0)} = \sum_f Q_f^3                        \label{Triangles1}\\
   U(1)'-U(1)_Y - U(1)_Y~~~: &&\ \cA^{(1)} = \sum_f Q_f Y_f^2                     \label{Triangles2}\\
   U(1)'-SU(2)-SU(2)~~~:     &&\ \cA^{(2)} = \sum_f Q_f \Tr[T_{k_2}^{(2)} T_{k_2}^{(2)}] \label{Triangles3}\\
   U(1)'-SU(3)-SU(3)~~~:     &&\ \cA^{(3)} = \sum_f Q_f \Tr[T_{k_3}^{(3)} T_{k_3}^{(3)}] \label{Triangles4}\\
   U(1)'-U(1)'-U(1)_Y~~~:    &&\ \cA^{(4)} = \sum_f Q_f^2  Y_f   \label{Triangles5}
\eea
where $f$ runs over the fermions in Table \ref{QTable}, $Q_f$ is the corresponding $U(1)'$ charge, $Y_f$ is the
hypercharge and $T_{k_a}^{(a)}$, $a=2,3;\,\, k_a=1,\ldots,{\rm dim G}^{(a)}$ are the generators of the
$G^{(2)}=SU(2)$ and $G^{(3)}=SU(3)$ algebras respectively. In our notation $\Tr[T_j^{(a)} T_k^{(a)}] = {1\over
2}\d_{jk}$. All the remaining anomalies that involve $U(1)'$s vanish identically due to group theoretical arguments
(see Chapter 22 of \cite{Weinberg2}). Using the charge constraints (\ref{Qconstraints}) we get
  \bea
   \cA^{(0)} &=& 3\ \Big\{ Q_{H_u}^3 + 3 \QHu Q_L^2 + Q_L^3 - 3 Q_{H_u}^2\ \( \QL + 6 \QQ \) \Big\} \label{A0}\\
   \cA^{(1)} &=& -{3\over2} \(3\QQ + \QL  \) \label{A1}\\
   \cA^{(2)} &=&  {3\over2} \(3\QQ  + \QL \) \label{A2}\\
   \cA^{(3)} &=& 0 \label{A3}\\
   \cA^{(4)} &=& -6 \QHu \(3\QQ + \QL  \) \label{A4}
\eea
Notice that the mixed anomaly between the anomalous $U(1)$ and the $SU(3)$ nonabelian factors $\cA^{(3)}$ vanishes
identically.


Many models have been developed in the past where all the anomalies (\ref{A0}-\ref{A4}) vanish by constraining the
charges $Q_f$ (see \cite{Langacker:2008yv} and references therein). On the contrary, in this Thesis we assume
that the $U(1)'$ is anomalous, i.e. (\ref{A0})-(\ref{A4}) do not vanish. Consistency of the model is achieved by the
GS mechanism showed in Sections~\ref{sect:Anomaly cancellation. Green-Schwarz mechanism} and
\ref{sect:Supersymmetric Green-Schwarz mechanism}. The axionic Lagrangian now reads
\bea
   \L_{\axion} &=& {1\over4} \left. \( S + S^\dagger + 4 b_3 V^{(0)} \)^2 \right|_{\thth} \nn\\
                &&- {1\over2} \left\{ \[\sum_{a=0}^2 b^{(a)}_2 S ~\Tr\( W^{(a)} W^{(a)} \) + b^{(4)}_2 S ~W^{(1)} ~W^{(0)} \]_{\th^2} +h.c. \right\}
   \qquad
   \label{Laxion}
\eea
where the index $a=0,\ldots,3$ runs over the $U(1)',\, U(1)_Y,\, SU(2)$ and $SU(3)$ gauge groups respectively.
The \Stuckelberg multiplet is given in (\ref{Smult}) and transforms under the $U(1)'$ as
\bea
   V^{(0)} &\to& V^{(0)} + i \( \Lambda - \Lambda^\dag \) \nn\\
   S  &\to& S - 4 i ~b_3 ~\Lambda \label{U1'}
  \label{U1Trans}\eea
where $b_3$ is a constant. The lowest component of $S$ is a complex scalar field $s=\a+i \f $. We assume that the
real part $\a$ gets an expectation value by an effective potential of stringy or different origin and contributes to
the coupling constants as
  \be
    \frac{1}{16 g_a^2 \t_a}=\frac{1}{16 \gt_a^2 \t_a} -{1\over2} b^{(a)}_2 \langle \a\rangle \label{couplingconst}
   \ee
where $g_a$ is the redefined coupling constant and the gauge factors $\t_a$ take the values $1,1,1/2,1/2$. The first
line in (\ref{Laxion}) is the \Stuckelberg Lagrangian \cite{StuckLag} of the $U(1)'$ which is gauge invariant  and
provides the kinetic terms and the axion-$U(1)'$ mixing.
The second line is not gauge invariant and provides the GS couplings that participate in the anomaly cancellation procedure. Notice
that in (\ref{Laxion}) the sum over $a$ omits the $a=3$ case since there is no mixed anomaly between the $U(1)'$ and
the $SU(3)$ factors as from eq.(\ref{A3}), i.e. $b_2^{(3)}=0$. The values of the other constants, $b_2^{(a)}$, are
fixed by the anomalies.

At first sight our Lagrangian (see Appendix~\ref{app:MiAUMSSM Lagrangian}) may not look like the most general possible one. In
particular, an explicit Fayet-Iliopoulos term $\xi V^{(0)}$ could be added. It is well known that in certain
string models (see, e.g. \cite{Poppitz:1998dj,Kiritsis:2008ry}), an one-loop FI term is absent, even if $Tr(Q) \neq 0$.
This  is in  apparent conflict with the observation \cite{Fischler:1981zk} that in field theory a quadratically
divergent FI term is always generated at one loop. The solution to this paradox is that in the low-energy Lagrangian
there should be a counterterm, which compensates precisely, i.e. both the divergent and the finite part of, the
one-loop contribution. We do not write explicitly this counterterm, since its exact expression is model and
regularization dependent, but we implicitly assume that such a cancellation occurs. As mentioned before, also the
terms responsible for the cancellation of gravitational anomalies are omitted.

Expanding $\L_{\axion}$ in component fields, using the Wess-Zumino gauge and substituting $\a$ by its VEV we get
   \bea
    \L_{\axion} &=& {1\over2} \( \pd_\m \f +2 b_3 V^{(0)}_\m \)^2
                   +{i\over4} \psi_S \s^\m \pd_\m \psib_S +{i\over4} \psib_S \sb^\m \pd_\m \psi_S \label{Laxioncomponent}\\
                 && +{1\over2} F_S \bar F_S + 2 b_3 \langle\a\rangle D^{(0)}-\sqrt2 b_3(\psi_S \l^{(0)}+h.c.)\nn\\
                 &&- {1\over4} \f \, \Eps \sum_{a=0}^2 b^{(a)}_2 \Tr \(  F_{\m \n}^{(a)} F_{\r \s}^{(a)} \)- {1\over4} b^{(4)}_2 \Eps \f F_{\m \n}^{(1)} F_{\r \s}^{(0)}\nn\\
                 &&+{1\over2} b^{(4)}_2 \langle\a\rangle F_{\m \n}^{(1)} F^{(0) \m \n}- b^{(4)}_2  \langle\a\rangle D^{(1)} D^{(0)} \nn\\
                 &&-{1\over2} \left\{\sum_{a=0}^2b^{(a)}_2 \[- 2\f \Tr \( \l^{(a)} \s^\m D_\m \lb^{(a)} \) +
                       {i\over\sqrt2} \Tr \( \l^{(a)} \s^\m \sb^\n F_{\m \n}^{(a)} \) \psi_S \right. \right. \nn\\
                 &&\left. - F_S \Tr \(\l^{(a)} \l^{(a)}\) - \sqrt2 \psi_S \Tr \(\l^{(a)} D^{(a)}\)\]\nn\\
                 &&+ b^{(4)}_2 \bigg[ \(-\f \l^{(1)} \s^\m \pd_\m \lb^{(0)}
                 +i\langle\a\rangle  \l^{(1)} \s^\m \pd_\m \lb^{(0)}  -{1\over2}F_S \l^{(1)} \l^{(0)}\right.\nn\\
                 && \left. \left.- {1\over\sqrt2} \psi_S  \l^{(1)} D^{(0)}
                 +{i\over2\sqrt2} \l^{(1)} \s^\m \sb^\n F_{\m \n}^{(0)} \psi_S\)
                 + (0 \leftrightarrow 1) \ \bigg] +h.c. \right\}\nn
   \eea
where we omit terms which are coming from $\la \a \ra W^{(a)} W^{(a)}$, since they are absorbed in the coupling
constant redefinition (\ref{couplingconst}). As stressed in the previous chapter,
this mechanism cancels some mixed anomalies and in addition provides a
mass term to the anomalous $U(1)$. Therefore, the anomalous $U(1)$ behaves $almost$ like the usual $Z'$ extensively
studied in the past.

For what concern the GCS terms we have
   \bea
    \L_{\GCS} &=&+ d_4     \[ \( V^{(1)} D^\a V^{(0)} - V^{(0)} D^\a V^{(1)}\) W^{(0)}_\a + h.c. \]_{\thth} +\nn\\
             &&-  d_5     \[ \( V^{(1)} D^\a V^{(0)} - V^{(0)} D^\a V^{(1)}\) W^{(1)}_\a + h.c. \]_{\thth} +\nn\\
             &&-  d_6 \Tr \bigg[ \( V^{(2)} D^\a V^{(0)} - V^{(0)} D^\a V^{(2)}\) W^{(2)}_\a +\nn\\
                          &&\qquad \quad+{1\over6} V^{(2)} D^\a V^{(0)} \bar D^2 \(  \[D_\a V^{(2)},V^{(2)}\] \) + h.c. \bigg]_{\thth}
   \label{GCS_1}
\eea
and their expression in component fields
   \bea
    \L_{\GCS}   &=&+  d_4 ~\Eps V^{(0)}_\m V^{(1)}_\n F_{\r \s}^{(0)} - d_5 ~\Eps V^{(0)}_\m V^{(1)}_\n F_{\r \s}^{(1)}\nn\\
                &&-  d_6 ~\Eps V^{(0)}_\m \, \Tr \[  V^{(2)}_\n F_{\r \s}^{(2)}
                  -{i\over3} V^{(2)}_\n \[V^{(2)}_\r, V^{(2)}_\s\] \] \nn\\
                &&+  d_4 \( \l^{(0)} \s^\m \lb^{(0)} V^{(1)}_\m -  \l^{(0)} \s^\m \lb^{(1)} V^{(0)}_\m   + h.c.  \) \nn\\
                &&-  d_5 \( \l^{(1)} \s^\m \lb^{(1)} V^{(0)}_\m  -  \l^{(1)} \s^\m \lb^{(0)} V^{(1)}_\m + h.c. \)\nn\\
                &&-  d_6 \Tr \[\l^{(2)} \s^\m \lb^{(2)} V^{(0)}_\m - \l^{(2)} \s^\m \lb^{(0)} V^{(2)}_\m +h.c.\]
   \eea
   %
The constants $d_4$, $d_5$ and $d_6$ are fixed by the cancellation of the mixed anomalies.

The cancellation of anomalies goes in the same way as in Sections~\ref{sect:Anomaly cancellation. Green-Schwarz mechanism} and
\ref{sect:Supersymmetric Green-Schwarz mechanism} so we give only the results
\bea
b^{(1)}_2 b_3 = -\frac{\cA^{(1)}}{128\pi^2}  \qquad\qquad\qquad  &&b^{(2)}_2 b_3 = -\frac{\cA^{(2)}}{64 \pi^2}\nn\\
b^{(0)}_2 b_3 = -\frac{\cA^{(0)}}{384\pi^2}  \qquad\qquad\qquad  &&b^{(4)}_2 b_3 = -\frac{\cA^{(4)}}{128 \pi^2}\nn\\
d_4 =- \frac{\cA^{(4)}}{384 \pi^2} \qquad\qquad\qquad d_5=\frac{\cA^{(1)}}{192\pi^2}&& \qquad\qquad d_6= \frac{\cA^{(2)}}{96 \pi^2}
\label{bsds}
\eea
For simplicity we chose a symmetric distribution for all the anomalies (see the discussion at the end of Section \ref{sect:Triangular Anomaly}).
For the anomaly cancellation in the broken phase we remand to a specific case in Section \ref{Sec: Anomaly cancellation broken}
and to our paper \cite{mypaper}.

Before going on we want to spend other few words on these anomalies.
In our model there are two extra states in the neutral fermionic sector, namely the axino and the primeino (see
Section \ref{Neutralinos}) which do not contribute to the fermionic loop. The remaining MSSM fermionic states are a
bino, a wino and the two higgsinos. Both $U(1)_Y$ and $SU(2)$ gauginos do not contribute to the fermionic loop due
to group theoretical arguments (see Section 28.1 of \cite{Weinberg3}). The higgsino eigenstates do not participate
because the $\tilde{h}_u$ contribution is cancelled by the $\tilde{h}_d$ one. This is due to the fact that each
diagram is proportional to an odd product of charges and the two higgsinos have opposite charges (see Table
\ref{QTable} and the constraints (\ref{Qconstraints})).
Therefore only SM fermions give the relevant contribution to the anomalies~(\ref{A0})-(\ref{A4}).

\subsection{Soft breaking terms} \label{subsec: softterms}
Since we added new matter content to the MSSM Lagrangian we expect that new soft breaking terms can be added to the theory.
The total soft breaking Lagrangian can be written as
   \be
    \L_{soft}=\L_{soft}^{MSSM}+\L_{soft}^{new} \label{Lsoft}
   \ee
where $\L_{soft}^{MSSM}$ is the MSSM soft breaking Lagrangian given in eq.~(\ref{LsoftMSSM}) and
   \bea
     \L_{soft}^{new}=- {1\over2}  \(M_0 \l^{(0)} \l^{(0)} + h.c. \) - {1\over2}  \(\frac{M_S}{2} \psi_S \psi_S  + h.c. \)
   \label{Lsoftnew}
   \eea
where $\l^{(0)}$ is the gaugino of the added $U(1)'$ and $\psi_S$ is the axino. We allow a soft mass term for the
axino since it couples only through GS interactions and not through Yukawa interactions \cite{Girardello:1981wz}.
Notice also that a mass term for the axion $\f$ is not allowed since it transforms non trivially under the anomalous
$U(1)'$ gauge transformation (\ref{U1Trans}).

\section{Model setup}\label{Consequences}

In this section we analyze the effects of the additional terms on the rest of the Lagrangian.
From now on we work with the Lagrangian (\ref{lagrangian}) after substitution $V^{(a)} \to g_a V^{(a)}$.

\subsection{Kinetic diagonalization of U(1)'s}

As we mentioned before, the \Stuckelberg multiplet contains a complex scalar field whose real part gets an
expectation value that modifies the coupling constant (\ref{couplingconst}). Therefore, the second line in
(\ref{Laxion}) contributes to the kinetic terms for the gauge fields and the term $\langle \a\rangle W^{(1)}W^{(0)}$
gives a kinetic mixing between the $V^{(1)}$ and $V^{(0)}$ gauge bosons. We have
\be
  \left. \( {1\over4}  W^{(0)} W^{(0)} + {1\over4} W^{(1)} W^{(1)}   +{\d\over2} W^{(1)} W^{(0)} \) \right|_{\th^2}
\label{kinmixlag} \ee
with $\d = - 4 b^{(4)}_2 g_0 g_1  \langle \a\rangle $. In order to diagonalize the kinetic terms, we use the matrix
\be
 \( \begin{array}{c} V^{(0)}\\
                     V^{(1)} \end{array} \) = \( \begin{array}{cc} C_\d &  0 \\
                    -S_\d  &  1 \end{array} \)
 \( \begin{array}{c} V_C\\
                     V_B \end{array} \)
\label{cdelta} \ee
where $C_\d = 1/\sqrt{1-\d^2}$ and $S_\d =  \d C_\d$.
Let us stress that in this case the mixing is a consequence of the anomaly cancellation procedure. Note that, since
$b_2^{(4)}\sim b_3^{-1}\sim M_{V^{(0)}}^{-1}$ (see eq. (\ref{bsds})), where $M_{V^{(0)}}$ is the mass of the
anomalous $U(1)$ that we assume to be in the TeV range, this mixing is tiny and can be ignored for our purposes.

\subsection{$D$ and $F$ terms}

The additional fields give rise also to $D$ and $F$ terms. More precisely, $D$ term contributions come from: (i) the
kinetic terms of chiral multiplets and (ii) the axionic Lagrangian (\ref{Laxion}), providing
\bea
 &&\!\!\!\!\!\!\!\!\!
   \L_D=\frac{1}{2}\sum_{a=0}^3 D^{(a)}_{k_a} D^{(a)}_{k_a} + \sum_{a=0}^3 g_a D^{(a)}_{k_a} z_i^\dag (T^{(a)}_{k_a})_j^i z^j
         + 4 g_0 b_3 \langle\a\rangle D^{(0)} + \d D^{(1)} D^{(0)}+\nn\\
 &&+2 \[\sum_{a=0}^2 g_a^2 \ b^{(a)}_2  \sqrt2 \psi_S \Tr \(  \l^{(a)} D^{(a)} \)
   + g_0 g_1 {b^{(4)}_2\over\sqrt2} \psi_S \( \l^{(1)} D^{(0)} + \l^{(0)} D^{(1)} \) +h.c. \]\nn\\
\eea
where $a=0,1,2,3$ denotes, as usual, the gauge group factors, $z_i$ are the lowest components of the $i$-th chiral
multiplet (except the multiplet which contains the axion) and $T^{(a)}_{k_a}$, $k_a=1,\ldots,{\rm dim G}^{(a)}$, are
the generators of the corresponding gauge groups, ${\rm G}^{(a)}$.
Solving the equations of motion for the D's and substituting back we obtain
\bea
 &&\!\!\!\!\!\!\!\!\!
   \L_{D_C}=- {1\over2} \left\{   \[C_\d g_0 \sum_f Q_f |z_f|^2 - S_\d g_1 \sum_f Y_f |z_f|^2 \] \right.
                  + C_\d 4 g_0 b_3 \langle\a\rangle \nn\\
                  && ~~~~~~ +2\sqrt2 b^{(0)}_2 g_0^2 \[ \psi_S \( C_\d^2 \l_C\)+h.c. \]
                  +2\sqrt2 b^{(1)}_2 g_1^2 \[ \psi_S \( S_\d^2\l_C -S_\d \l_B \)+h.c. \]\nn\\
                  &&~~~~~~+\sqrt2 b^{(4)}_2 g_0 g_1 \[ \psi_S \( C_\d \l_B - 2 C_\d S_\d \l_C\)+h.c. \]\Bigg\}^2\label{dcterm}
\eea
\bea
\L_{D_B} ~&=&- {1\over2} \Bigg\{ g_1 \sum_f Y_f |z_f|^2
                +2\sqrt2 b^{(1)}_2 g_1^2 \[ \psi_S \( \l_B -S_\d \l_C  \)+h.c. \]+\nn\\
               &&~~~~~~+\sqrt2 b^{(4)}_2 g_0 g_1 \[  \psi_S  C_\d  \l_C +h.c. \]\Bigg\}^2\label{dbterm}
\eea
\bea
\L_{D^{(2)}} &=& -\frac{1}{2} \sum_k\left\{g_2 z_i^\dag (T^{(2)}_k)_j^i z^j
          + b^{(2)}_2 g_2^2 \[  \sqrt2 \psi_S  \l^{(2)}_k  + h.c. \]\right\}^2\label{d2term}\\
      \L_{D^{(3)}} &=& -\frac{1}{2} \sum_k \left\{ g_3 z_i^\dag (T^{(3)}_k)_j^i z^j
          \right\}^2\label{d3term}
     \eea
Similarly, the $F$ term contributions are
  \bea
   &&\L_F = \sum_{f \in MSSM} \( F^f F_f^\dag-\frac{\pd W}{\pd z^f} F^f-\frac{\pd W^\dag}{\pd z^\dag_f} F^\dag_f\) \\
   &&\phantom{\L_F = }
     +\frac{1}{2} F_S F_S^\dag +
   \frac{1}{2} \left\{4 F_S \[ \sum_{a=0}^2 g_a^2 b^{(a)}_2 \Tr \(  \l^{(a)} \l^{(a)} \) + g_0 g_1 b^{(4)}_2 \l^{(1)} \l^{(0)} \] +h.c. \right\}
   \nn
  \eea
where the first line is the standard MSSM $F$ term contribution while the second line contains the new \Stuckelberg terms.
Solving the equation of motion
  \bea
   \L_{F_S} &=& -8 \[ \sum_a b^{(a)}_2  g_a^2 \Tr \(  \l^{(a)} \l^{(a)} \)+ g_1 g_0  b^{(4)}_2 \l^{(1)} \l^{(0)}\]\nn\\
      &&~~~ \times \[ \sum_a b^{(a)}_2 g_a^2 \Tr \(  \lb^{(a)} \lb^{(a)} \) + g_1 g_0 b^{(4)}_2 \lb^{(1)} \lb^{(0)}\]
\label{fsterm}
  \eea
Eq. (\ref{fsterm}) can also be written in the basis (\ref{cdelta}), but we will not need this term in the following.

We would like to mention that no $D$ and $F$ terms are coming from the GCS since they include only vector multiplets in
an antisymmetric form. Our results are in accordance with \cite{DeRydt:2007vg}.

 \subsection{Scalar potential}

As we have seen in the previous section, the additional $F$ terms (\ref{fsterm}) do not give any contribution to the
scalar potential. The $D_B$, $D^{(2)}$ and $D^{(3)}$ terms (see eq. (\ref{dbterm}), (\ref{d2term}) and
(\ref{d3term})) provide the usual contributions to the MSSM potential. The only new contribution comes from the
first line of (\ref{dcterm}). Thus the scalar potential can be written as
  \bea
   V&=&V_{MSSM}+V_{D_C} \label{V}\\
   V_{D_C} &=& {1\over2} \left\{   \[C_\d g_0 \sum_f Q_f |z_f|^2 - S_\d g_1 \sum_f Y_f |z_f|^2 \]
                  + C_\d 4 g_0 b_3 \langle\a\rangle \right\}^2
  \eea
Solving the equations for the minima of the potential
  \be
   \frac{\pd V}{\pd z_f} =0
  \ee
we get $\la z_f \ra =0$ for all the sfermions as in the
 MSSM case. Inserting back these VEVs into (\ref{V}) we get the following Higgs scalar potential
  \bea
   V_h &=& \Big\{ |\m|^2+m^2_{h_u} + 4 g_0^2 b_3 \langle\a\rangle C_\d X_\d \Big\} \Big(|h_u^0|^2 + |h_u^+|^2\Big) \nn\\
          &&+\Big\{ |\m|^2+m^2_{h_d} - 4 g_0^2 b_3 \langle\a\rangle C_\d  X_\d \Big\} \Big(|h_d^0|^2 + |h_d^-|^2\Big)\nn\\
       && + \Big\{\frac{1}{2} \( g_0 X_\d \)^2 +\frac{1}{8}(g_1^2+g_2^2)\Big\} \Big(|h_u^0|^2 +|h_u^+|^2 - |h_d^0|^2 - |h_d^-|^2 \Big)^2\nn\\
       &&+\Big\{ b\, (h_u^+ h_d^- - h_u^0 h_d^0) + h. c.\Big\}    + \half g_2^2 |h_u^+ h_d^{0\dag} + h_u^0 h_d^{-\dag}|^2
  \eea
which can be brought to the same form of the MSSM potential (see eq.~(\ref{VHiggscomplete})), after the following redefinitions
  \bea
    m^2_{h_u} + 4 g_0^2 b_3 \langle\a\rangle C_\d X_\d &\to& \tilde m^2_{h_u} \nn\\
    m^2_{h_d} - 4 g_0^2 b_3 \langle\a\rangle C_\d X_\d &\to& \tilde m^2_{h_d} \nn\\
   \(\( g_0 X_\d \)^2 +\frac{1}{4}(g_1^2+g_2^2) \)v^2&\to& \tilde M^2_{Z_0} \label{tildeMs}
  \eea
  where
  \be
   g_0 X_\d= C_\d g_0 \QHu - \half S_\d g_1\label{xdelta}
  \ee
At the minimum, we recover the MSSM result $\la h_u^+ \ra=\la h_d^- \ra=0$ for the Higgs charged components.
Defining as usual $\la h^0_i \ra= v_i/\sqrt2$ , $v_u^2+v_d^2 = v^2$ and $v_u/v_d=\tan\b$ we can still write the tree level
conditions for the electroweak symmetry breaking as
  \bea
   b^2 &>& \( |\m|^2+ \tilde m^2_{h_u} \) \( |\m|^2+ \tilde m^2_{h_d} \) \\
   2 b &<&  2 |\m|^2+ \tilde m^2_{h_u} + \tilde m^2_{h_d}
  \eea
in complete analogy with the MSSM case (using $\tilde{m}$'s).

\subsection{Higgs sector}

It is worth noting that in our model there is no axi-Higgs mixing. This is due to the fact that we do not consider
scalar potential terms for the axion  (on the contrary to \cite{Coriano':2005js}).

After the electroweak symmetry breaking we have four gauge generators that are broken, so we have four longitudinal
degrees of freedom. One of them is the axion, while the other three are the usual NG bosons coming from the Higgs
sector.

As it was mentioned above, the potential has the standard MSSM form, upon the redefinitions (\ref{tildeMs}). Since the
discussion goes in the same way as Section~\ref{subsect:Higgs Bosons}, here we give only the main ideas.
Higgs scalar fields consist of eight real scalar degrees of freedom: three of them are the NG bosons $G^0$, $G^\pm$.
The remaining five Higgs scalar mass eigenstates consist of two CP-even neutral scalars $h^0$ and $H^0$, one CP-odd neutral scalar $A^0$ and
a charge $+1$ scalar $H^+$ as well as its charge conjugate $H^-$ with charge $-1$. The gauge-eigenstate fields can be
expressed in terms of the mass eigenstate fields as
    \bea
     \( \begin{array}{c} h_u^0 \\
                         h_d^0 \end{array} \) &=& {1\over \sqrt{2}} \(\begin{array}{c} v_u \\
                                                                           v_d \end{array}\) +
                                          {1\over \sqrt{2}} R_\alpha \(\begin{array}{c} h^0 \\
                                                                                      H^0 \end{array}\) +
                                          {i\over \sqrt{2}} R_{\beta}\(\begin{array}{c} G^0 \\
                                                                                        A^0 \end{array}\)\label{HiggsMin}\\
     \( \begin{array}{c} h_u^+ \\
                         h_d^{-*} \end{array} \) &=&  R_{\beta}\(\begin{array}{c} G^+ \\
                                                                                        H^+ \end{array}\)
    \eea
where the orthogonal rotation matrices $R_\alpha, R_{\beta}$ are the same as in
(\ref{Ralpha and Rbeta}). Acting with these matrices on the gauge eigenstate fields we obtain the diagonal mass terms.
Replacing the tilde parameters (\ref{tildeMs}) we obtain the masses (compare with eq.~(\ref{eq:m2A0}), (\ref{eq:m2hH}) and (\ref{eq:m2Hpm}))
    \bea
     m_{A^0}^2 &=& 2|\m|^2 + m^2_{h_u} + m^2_{h_d} \label{eq:m2A0new}\\
     m^2_{h^0, H^0} &=& \frac{1}{2} \Bigg\{ m^2_{A^0} +
     \(\( g_0 X_\d \)^2 +\frac{1}{4}(g_1^2+g_2^2) \)v^2 \nn\\
              &&\mp \[\(m_{A^0}^2
              - \(\( g_0 X_\d \)^2 +\frac{1}{4}(g_1^2+g_2^2) \)v^2\)^2 \right. \nn\\
     &&\left. + 4 \(\( g_0 X_\d \)^2 +\frac{1}{4}(g_1^2+g_2^2) \)v^2
     m_{A^0}^2  \sin^2 (2\beta)\]^\half \Bigg\}\>\>\>\>\>{} \label{eq:m2hHnew}\\
     m^2_{H^\pm} &=& m^2_{A^0} + m_W^2 =  m^2_{A^0} +  g_2^2 \frac{v^2}{4} \label{eq:m2Hpmnew}
    \eea
and the mixing angles (compare with eq.~(\ref{alphaangle}))
    \bea
    {\sin 2\alpha\over \sin 2 \beta} &=& -{m_{H^0}^2 + m_{h^0}^2 \over m_{H^0}^2 - m_{h^0}^2}  \nn\\
     \quad {\tan 2\a\over \tan 2 \b} &=&
           {m_{A^0}^2+
           \(\( g_0 X_\d \)^2 +\frac{1}{4}(g_1^2+g_2^2) \)v^2 \over  m_{A^0}^2-
     \(\( g_0 X_\d \)^2 +\frac{1}{4}(g_1^2+g_2^2) \)v^2} \label{alphaanglenew}
    \eea
Notice that only the $h^0$ and $H^0$ masses modified with respect to the MSSM, due to the additional anomalous
$U(1)'$.

\subsection{Neutral Vectors}   \label{vectmasssol}

There are two mass-sources for the gauge bosons: (i) the \Stuckelberg mechanism and (ii) the Higgs mechanism. In
this extension of the MSSM, the mass terms for the gauge fields are given by
     \be
      \L_M = \frac{1}{2} \(C_\m \ B_\m \ V^{(2)}_{3\m} \) M^2
                          \( \begin{array}{c} C^\m\\ B^\m\\ V^{(2)\m}_{3} \end{array} \)
     \ee
$C_\mu ,\, B_\mu$ are the spin 1 components of the vector multiplets $V_C,\,V_B$. The gauge boson mass matrix is
     \be
      M^2= \( \begin{array}{ccc} M^2_C & ~~~g_0 g_1 \frac{v^2}{2} X_\d& ~~~-g_0 g_2 \frac{v^2}{2} X_\d  \\
                 ... & g_1^2 \frac{v^2}{4} & -g_1 g_2 \frac{v^2}{4}   \\
                 ... & ...  & g_2^2 \frac{v^2}{4}  \\\end{array} \)
  \label{BosonMasses} \ee
where $M^2_C =16 g_0^2 b_3^2 C_\d^2 + g_0^2 (v^2) X_\d^2$ and the lower dots denote the obvious terms under
symmetrization. After diagonalization, we obtain the eigenstates
 \bea
   A_\m &=&\frac{g_2 B_\m + g_1 V^{(2)}_{3\m}}{\sqrt{g_1^2+g_2^2}}  \label{photonnew}\\
    Z_{0\m} &=& \frac{g_2 V^{(2)}_{3\m} - g_1 B_\m}{\sqrt{g_1^2+g_2^2}}+g_0 \QHu\frac{\sqrt{g_1^2+g_2^2}  v^2}{2 M_{V^{(0)}}^2} C_\m+{\cal O}[g_0^3,M_{V^{(0)}}^{-3}]  \label{Z0new}\\
Z'_\m  &=& C_\m +\frac{g_0 \QHu v^2}{2 M_{V^{(0)}}^2}\(g_1 B_\m- g_2 V^{(2)}_{3\m}\) +{\cal
O}[g_0^3,M_{V^{(0)}}^{-3}] \label{Zprime}
    \eea
and the corresponding masses
   \bea
   M^2_{\g}&=&0\\
    M^2_{Z_0} &=&\frac{1}{4} \(g_1^2+g_2^2\) v^2
                 -(\QHu)^2\frac{\(g_1^2+g_2^2\) g_0^2  v^4}{4 M_{V^{(0)}}^2}+{\cal O}[g_0^3,M_{V^{(0)}}^{-3}]
\label{Z0massnew}\\
M^2_{Z'}  &=&M_{V^{(0)}}^2+g_0^2 \[(\QHu)^2 \(1+\frac{g_1^2 v^2+g_2^2 v^2}{4M_{V^{(0)}}^2}\)-\frac{\langle\a\rangle
g_1^3 \cA^{(4)}}{64 \pi ^2M_{V^{(0)}}}\] v^2+{\cal O}[g_0^3,M_{V^{(0)}}^{-3}]\nn\\
  \label{Zpmass}
\eea
where $M_{V^{(0)}}=4 b_3 g_0$ is the mass parameter for the anomalous $U(1)$ and it is assumed to be in the TeV
range. Due to their complicated form, the eigenstates and eigenvalues of $M^2$ (\ref{BosonMasses}) are expressed as
power expansions in  $g_0$ and $1/M_{V^{(0)}}$ keeping only the leading terms. Higher terms are denoted by ${\cal
O}[g_0^3,M_{V^{(0)}}^{-3}]$.

The first eigenstate (\ref{photonnew}) corresponds to the photon and it is exact to all orders. It slightly differs
from the usual MSSM expression (\ref{photon}) due to the kinetic mixing between $V^{(0)}$ and $V^{(1)}$.

For the rest of the Thesis, we neglect the kinetic mixing contribution since they are higher loop effects which go
beyond the scope of the Thesis. Then the rotation matrix from the hypercharge to the photon basis, up to
${\cal O}[g_0^3,M_{V^{(0)}}^{-3}]$  is
    \bea
     \( \begin{array}{c} Z'_\m\\
                         Z_{0 \m}\\
                         A_\m \end{array} \)
       &=&O_{ij}
     \( \begin{array}{c} V^{(0)}_\m\\
                         V^{(1)}_\m\\
                         V^{(2)}_{3\m} \end{array} \)\label{Oij}\\
      &=&\( \begin{array}{ccc}      1& g_1 \frac{g_0 \QHu v^2}{2 M_{V^{(0)}}^2}  ~~& -g_2 \frac{g_0 \QHu v^2}{2 M_{V^{(0)}}^2}  \\
                                g_0 \QHu \frac{\sqrt{g_1^2+g_2^2}  v^2}{2M_{V^{(0)}}^2} ~~~& - \frac{g_1}{{\sqrt{g_1^2+g_2^2}}}
& \frac{g_2}{{\sqrt{g_1^2+g_2^2}}}  \\
                                0 & \frac{g_2}{{\sqrt{g_1^2+g_2^2}}} & \frac{g_1}{{\sqrt{g_1^2+g_2^2}}}  \\ \end{array} \)
     \( \begin{array}{c} V^{(0)}_\m\\
                         V^{(1)}_\m\\
                         V^{(2)}_{3\m} \end{array} \) \nn
\eea
where $i,j=0,1,2$.

\subsection{Sfermions}

In general, the contributions to the sfermion masses are coming from (i) the $D$ and $F$ terms in the superpotential and
(ii) the soft-terms. However, in our case, the new contribution comes only from the $D_C$ terms
\bea
 V^{D_C}_\text{mass} &=& \bigg\{ \( C_\d g_0 \QHu + \half S_\d g_1 \) \( \frac{v_u^2-v_d^2}{2}\) + 4 C_\d g_0 b_3 \langle\a\rangle  \bigg\}
 \times\nn\\
   &&\bigg\{ \sum_f  \( C_\d g_0  Q_f - S_\d g_1 Y_f \) |y_f|^2 \bigg\}
\eea
where the $y_f$ stand for all possible sfermions.

\subsection{Neutralinos \label{Neutralinos}}

With respect to the MSSM, now we have two new fields: $\psi_S$ and $\l^{(0)}$. Thus, we have
    \be
    \L_{\mbox{neutralino mass}} = -\frac{1}{2} (\psi^{0})^T {\bf M}_{\tilde N} \psi^0 + h. c.
    \ee
where
   \be
    (\psi^{0})^T= (\psi_S, \ \l_C,\ \l_B,\ \l^{(2)},\ \tilde h_d^0,\ \tilde h_u^0) \label{neutrbase}
    \ee
The neutralino mass matrix $ {\bf M}_{\tilde N} $ gets contributions from (i) the MSSM terms, (ii) the $h-\tilde h-
\l^{(0)}$ couplings, (iii) the new soft-breaking terms $\L^{new}_{soft}$, (iv) the \Stuckelberg action and (v) the
 $D$ terms. Finally, we obtain the symmetric matrix
   \be
    {\bf M}_{\tilde N}
     =   \(\begin{array}{cccccc}
           \frac{M_S}{2}~~~ & m_{SC} & m_{SB} & {2g_2^3 b_2^{(2)}\over\sqrt2}  \, \Delta v^2 & 0 & 0 \\
           \dots & M_0 C_\d^2+M_1 S_\d^2~~~ &  -M_1 S_\d  & 0 & - g_0 v_d X_\d  ~~~& g_0 v_u X_\d   \\
           \dots & \dots & M_1 & 0 & -\frac{g_1 v_d}{2} & \frac{g_1 v_u}{2} \\
           \dots & \dots& \dots & M_2 & \frac{g_2 v_d}{2} & -\frac{g_2 v_u}{2} \\
           \dots & \dots & \dots & \dots & 0 & -\m  \\
          \dots & \dots & \dots & \dots & \dots & 0
         \end{array}\) \label{massmatrixNeutr} ~~~~
   \ee
where $M_0$, $M_1$, $M_2$ and $M_S$ are the masses coming from the soft breaking terms (\ref{Lsoft}), and
       \bea
    m_{SC}&=&\sqrt2\Bigg\{2\(C_\d^2 g_0^2 b_2^{(0)}+S_\d^2 g_1^2 b_2^{(1)}- C_\d S_\d g_0 g_1 b_2^{(4)} \)
                   \( g_0 X_\d \, \Delta v^2 +  C_\d M_{V^{(0)}} \la \a \ra\)\nn\\
         &&+\half \(-2S_\d g_1^2 b_2^{(1)}+C_\d g_0 g_1 b_2^{(4)} \)g_1 \, \Delta v^2+  \frac{C_\d}{2}M_{V^{(0)}}\Bigg\}\label{msc}\\
     m_{SB} &=& \sqrt2\left\{ \( C_\d g_0 g_1 b_2^{(4)}-2 S_\d g_1^2 b_2^{(1)} \)
                \(g_0 X_\d \, \Delta v^2  + C_\d M_{V^{(0)}} \la \a \ra\)
               + b_2^{(1)} g_1^3 \, \Delta v^2\right\}\nn
   \eea
with $\Delta v^2=v_u^2-v_d^2~$. It is worth noting that the $D$ terms and kinetic mixing terms are only higher order
corrections and they can be neglected in the computations of the eigenvalues and eigenstates.

\newpage\thispagestyle{empty}

\chapter{MiAUMSSM. Phenomenology} \label{chap:MiAUMSSM. Phenomenology}

In this chapter we discuss some phenomenology consequences of the model showed in Chapter~\ref{chap:MiAUMSSM. Model building}.
We focus on the simple but still interesting case $\QHu=0$.
In this case there is no mixing between the $V^{(0)}$ and the other SM gauge fields therefore $Z'=V^{(0)}$ (see
(\ref{Oij})).
Then the $Z'$ couples only to SM fermions and their corresponding superpartners. There is no tree level
interaction between the $Z'$ and the $W$'s or the neutralinos or the charginos since the corresponding
couplings are proportional to $\QHu$.
Moreover, as we will see in the following, this simple choice implies interesting properties of the $Z'$ involved in fermionic triangles.

First of all we will study the tree level decays of the $Z'$ into fermions and the anomalous decays into neutral gauge bosons.
We will check the anomaly cancellation in the broken phase and we will also give some prediction about these processes in LHC.

Then we will pass to the supersymmetric sector. We will consider the case in which the axino is the LSP and we will check
if the prediction for the relic density is in agreement or not with the current experimental bounds.

This chapter is entirely based on our papers \cite{mypaper} and \cite{mypaper2}.
\newpage\section{Tree level decays} \label{sect::Tree level decays}
First of all we discuss the tree level decays of the $Z'$. As suggested in the introduction, the tree level final states
can only be fermions or sfermions. Since this is a standard part we do not spend much time in details. We remember that
we are in the case $\QHu=0$.
\subsection{Fermions} \label{subsec: Fermions}
The Lagrangian describing the interaction of the $Z'$ with SM fermions is given by
\be
 \L^{int}_{Z'} = - {1\over2} \, g_{Z'} \, \bar \Psi_f \, \g^\m \( v_f^{Z'} - a_f^{Z'} \g_5 \) \Psi_f Z'_\m
 \label{ZpfermintLag}
\ee
where $g_{Z'}=g_0$, $\Psi_f$ is the usual Dirac spinor (see eq.~(\ref{chiralLDirac})) and the vectorial and axial couplings are defined as
\bea
   v_f^{Z'} &=& Q^{Z'}_{f_L}+Q^{Z'}_{f_R} \nn\\
   a_f^{Z'} &=& Q^{Z'}_{f_L}-Q^{Z'}_{f_R} \label{vfafZp}
\eea
and listed in Table~\ref{Table:ZpcouplingsQH0}.
\begin{table}[h]
\centering
\begin{tabular}[h]{|c|c|c|}
 \hline                          & $v_f^{Z'}$ & $a_f^{Z'}$ \\
 \hline $\n_e$, $\n_\m$, $\n_\t$ & $Q_L$      & $Q_L$\\
 \hline $e$, $\m$, $\t$          & $2 Q_L$    & $0$\\
 \hline $u$, $c$, $t$            & $2 Q_Q$    & $0$\\
 \hline $d$, $s$, $b$            & $2 Q_Q$    & $0$\\
 \hline
\end{tabular}
\caption{Couplings of the SM fermions with the $Z'$ gauge boson.}\label{Table:ZpcouplingsQH0}
\end{table}

The decay rate is given by
\be
 \G(Z' \to f \bar f) = C_f \frac{ g_{Z'}^2 }{48\pi} \MZp \sqrt{ 1 - 4\frac{m_f^2}{\MZp^2} }
         \[ \( v_f^{Z'} \)^2 R_v^f +
            \( a_f^{Z'} \)^2 R_a^f \]  \label{DecZpff}
\ee
where the colour factor $C_f=1(3)$ for leptons (quarks) and the kinematic factors
\bea
 R_v^f &=&  1 + 2\frac{m_f^2}{\MZp^2}  \nn\\
 R_a^f &=&  1 - 4\frac{m_f^2}{\MZp^2}  \label{RvRaf}
\eea
take into account the fermion mass $m_f$.
\subsection{Sfermions} \label{subsec: Sfermions}
The Lagrangian describing the interaction of the $Z'$ with sfermions is given by
\be
 \tilde \L^{int}_{Z'} = - \frac{i}{2} g_{Z'} \( v_f^{Z'} \pm a_f^{Z'} \) \tilde f^\dag_{L,R} \, \pd_\m \tilde f_{L,R} \, Z'_\m
 \label{ZpsfermintLag}
\ee
where the $+(-)$ is for left (right) handed sfermions respectively and with $v_f^{Z'}$ and $a_f^{Z'}$ given in eq.~(\ref{vfafZp}).
The decay rate for $Z'$ decay to sfermions is calculated to be
\be
 \G(Z' \to \tilde f_{L,R} \, \tilde f^\dag_{L,R} ) =
   C_f \frac{ g_{Z'}^2 }{48\pi} \frac{\( v_f^{Z'} \pm a_f^{Z'} \)^2}{4} \MZp
   \( 1 - 4\frac{m_{\tilde{f}_{L,R}}^2}{\MZp^2} \)^{3/2}
\label{DecZpsfsf}
\ee
where the colour factor $C_f$ is defined as in the fermion case and $m_{\tilde{f}_{L,R}}$ is the ${\tilde{f}_{L,R}}$ sfermion mass.

In the case of non-negligible sfermion mixing (such as the top squark) we must work in the mass eigenstate basis.
The decay width is similar to (\ref{DecZpsfsf}) with the appropriate couplings, except that the
phase space factor in the case of different mass decay products is $[1-2(m_1^2+m_2^2)/M_{Z'}^2+(m_1^2-m_2^2)^2/M_{Z'}^4]^{3/2}$.
For details see for instance \cite{Gherghetta:1996yr}.
\subsection{Numerical results} \label{subsec:Tree level decay width}
The tree level decay width for a $Z'$, is given by the sum of the decay rates given in the previous sections
\be
  \G_{Z'} = \sum_{f \in SM} \G \(Z' \to f \bar f\) + \sum_{\tilde f} \G \(Z' \to \tilde f \tilde f^\dag \)
\ee
We focus on the case $\MZp=1$ TeV. The reason will be clear in Section \ref{sec: LHC}.
Since we suppose that the LSP has a mass higher than 500 GeV (see Section \ref{sect:Axino Relic Density}
for scenarios in which this happens), the sfermion masses are heavier that $\MZp/2$, so they cannot be produced by an on-shell $Z'$ decay.
We have
\bea
 \G_{Z'} &=& \sum_{f \in SM} \G \(Z' \to f \bar f\) \nn\\
         &=& \sum_{f \in SM} C_f \frac{ g_{Z'}^2 }{48\pi} \MZp \sqrt{ 1 - 4\frac{m_f^2}{\MZp^2} }
         \[ \( v_f^{Z'} \)^2 R_v^f +
            \( a_f^{Z'} \)^2 R_a^f \] \nn\\
         &\simeq& \[ 119.37 \, (g_0 \QL)^2 + 477.02 \, (g_0 \QQ)^2 \] \GeV \label{ZpWidth}
\eea
where we used eq.~(\ref{DecZpff}).
Since the $Z'$ will be detected at LHC through is decays into electric charged leptons we give also the branching ratio (BR) for the decay
into an electron-positron pair
\bea
  \BR(Z' \to e^- e^+) &=& \G(Z' \to e^- e^+)/\G_{Z'} \nn\\
         &\simeq& \frac{26.52 \, (g_0 \QL)^2}{119.37 \, (g_0 \QL)^2 + 477.02 \, (g_0 \QQ)^2} \label{ZpeeBR}
\eea
The results are plotted in Fig.~\ref{Fig:Zpwidth} in the form of contour plots in the plane $g_0 Q_L, g_0 Q_Q$. Our
choices for $g_0$, $Q_Q$, $Q_L$ and $\MZp$ are in agreement with the current experimental bounds \cite{Abulencia:2006iv}.
\begin{figure}[t]
 \centering
 \includegraphics[scale=0.33]{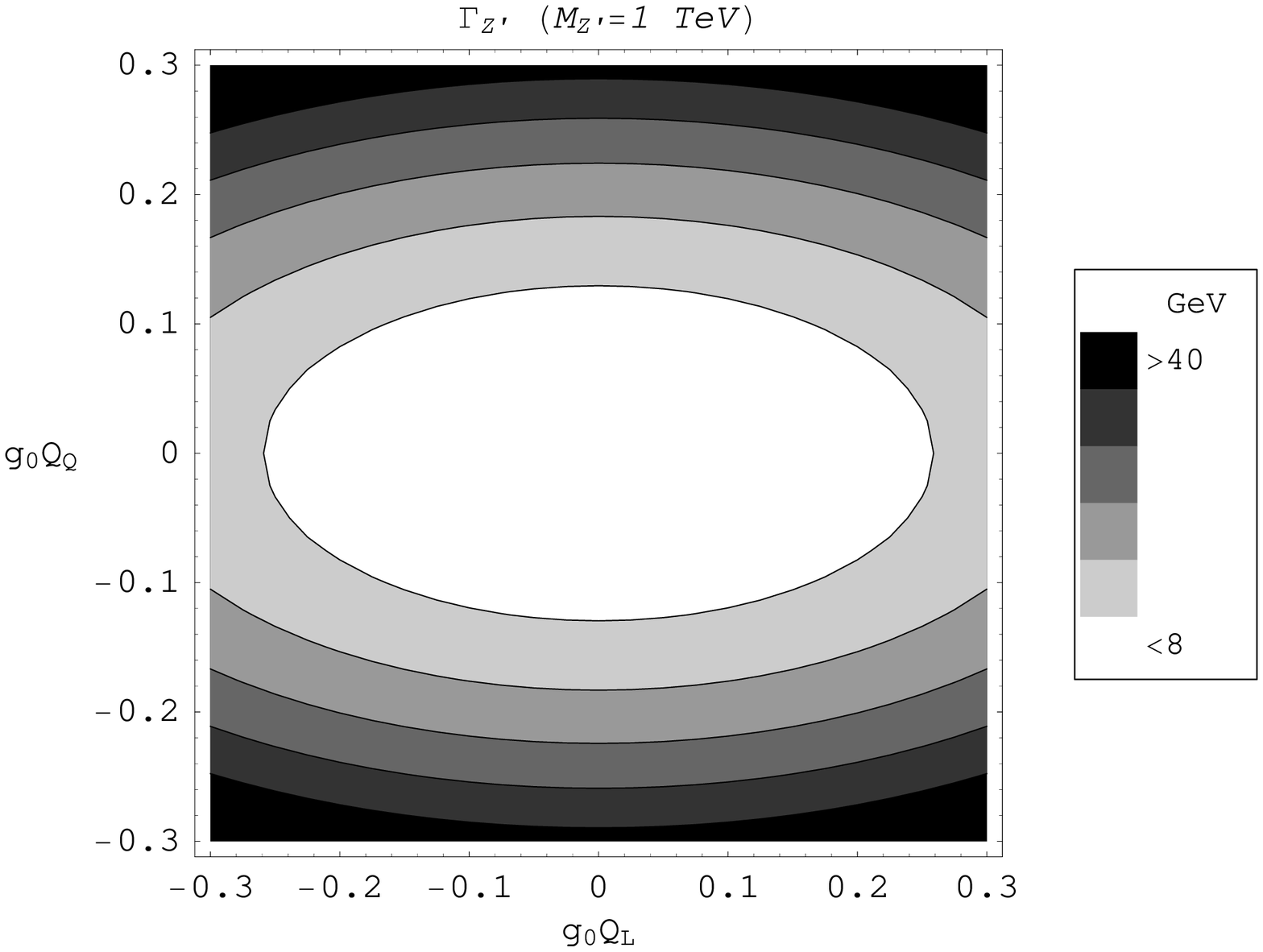}
~
 \includegraphics[scale=0.33]{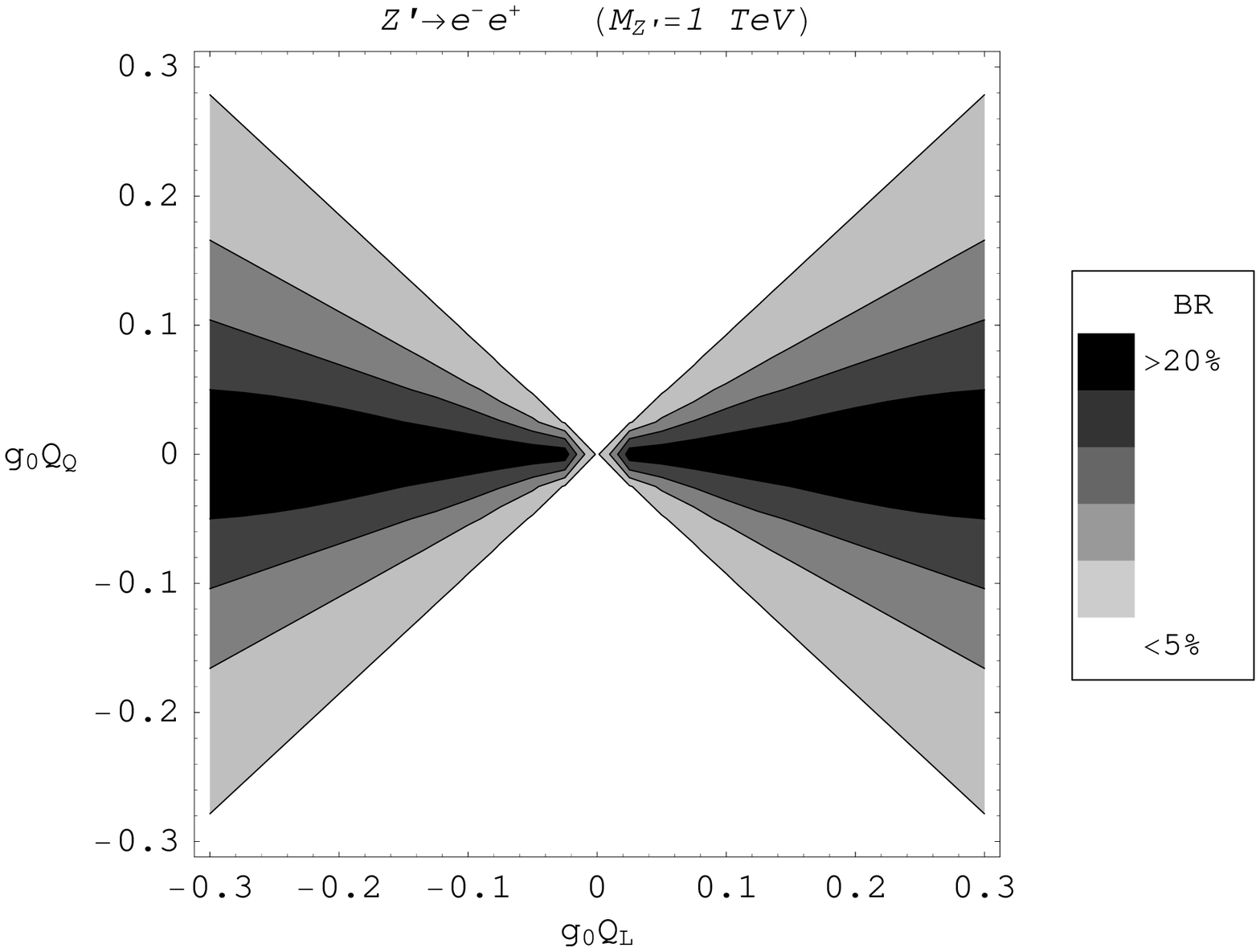}
(a)~~~~~~~~~~~~~~~~~~~~~~~~~~~~~~~~~~~~~~~~~~~~~~~~~~~~~~(b)~~
\caption{(a): Tree level decay width for a $Z'$ with mass equal to 1 TeV. The darker shaded regions correspond to larger decay widths.
         (b): Tree level branching ratio for $Z' \to e^- e^+$ with mass equal to 1 TeV.
            The darker shaded regions correspond to larger values for the branching ratio.} \label{Fig:Zpwidth}
\end{figure}
The tree level decay width has elliptic contours since eq.~(\ref{ZpWidth}) is the equation of an ellipse
in the plane $g_0 Q_L, g_0 Q_Q$. Concerning the $\BR(Z' \to e^- e^+)$ let us observe that if the quarks and the leptons have the same mass and
the same couplings, then the branching ratio will be simply the inverse of the total number of fermions
which is the number of leptons (6) plus the number of quarks (6) times the color factor (3)
\bea
 \BR(Z' \to e^- e^+) = \frac{1}{6 + 3 \times 6} = \frac{1}{24} \simeq 0.04
\eea
This is the reason why $\BR(Z' \to e^- e^+)$ has a lot of values less then 5\%.
In a region of parameters the most important quantity is the number of the possible final states.
\section{Anomalous Decays} \label{Sec: Anomalous decays}
In this section we compute the amplitudes for the on-shell decays of the $Z'$ into neutral gauge bosons.
The SM fermion interaction terms with SM neutral gauge bosons are
\bea
 \L^{int}_{Z'} &=& J^\m_{Z'} Z'_\m \nn = - {1\over2} \, g_{Z'} \, \bar \Psi_f \, \g^\m \( v_f^{Z'} - a_f^{Z'} \g_5 \) \Psi_f Z'_\m\\
 \L^{int}_{Z_0} &=& J^\m_{Z_0} Z_{0 \m} =- {1\over2} \, g_{Z_0} \, \bar \Psi_f \, \g^\m \( v_f^{Z_0} - a_f^{Z_0} \g_5 \) \Psi_f Z_{0 \m} \nn\\
 \L^{int}_{\g} &=& J^\m_{\g} A_\m =- e \, q_f \bar \Psi_f \, \g^\m \Psi_f A_\m  \label{neutralcurrents}
\eea
where $q_f$ denote the electric charges, $v_f^{Z_0}$ and $a_f^{Z_0}$ are the vectorial and axial couplings with
$Z_0$ and $v_f^{Z'}$ and $a_f^{Z'}$ are the vectorial and axial couplings with $Z'$, respectively.
They are listed in Table~\ref{Table:couplingsQH0}.
  \begin{table}[h]
  \centering
  \begin{tabular}[h]{|c|c|c|c|c|c|}
   \hline                          & $q_f$  &  $v_f^{Z_0}$                & $a_f^{Z_0}$ & $v_f^{Z'}$ & $a_f^{Z'}$ \\
   \hline $\n_e$, $\n_\m$, $\n_\t$ & $0$    &  $1/2$                      &   $1/2$  & $Q_L$ & $Q_L$\\
   \hline $e$, $\m$, $\t$          & $-1$   &  $-1/2+2 \sin^2 \theta_W$   &  $-1/2$  & $2 Q_L$ & $0$\\
   \hline $u$, $c$, $t$            & $2/3$  &  $1/2-4/3 \sin^2 \theta_W$  &   $1/2$  & $2 Q_Q$ & $0$\\
   \hline $d$, $s$, $b$            & $-1/3$ &  $-1/2+2/3 \sin^2 \theta_W$ &  $-1/2$  & $2 Q_Q$ & $0$\\
   \hline
  \end{tabular}
  \caption{Couplings of the SM fermions with the neutral gauge bosons.}\label{Table:couplingsQH0}
  \end{table}

We remind that, for $\QHu=0$, the $Z'$ is decoupled for the
SM neutral gauge sector, so $g_{Z'}=g_0$ and the SM coupling constants are given in eq.~(\ref{gZ0}) and (\ref{e}).
For the corresponding Feynman rules see Appendices~\ref{sec:Electroweak interactions} and \ref{sec:Zp int}.

A priori we could have three possibilities:
$Z' \to \g \, \g$, $Z' \to Z_0 \, \g$ and $Z' \to Z_0 \, Z_0$. But the on-shell decay $Z' \to \g \, \g$ is prohibited by the
Landau-Yang theorem \cite{Landau-Yang}, for which we give a brief demonstration in Appendix~\ref{app:Landau Yang}.
This decay is permitted only if at least one of the two photons is off-shell. Moreover with the choice $\QHu=0$, we have that the electric charged
fermions have only vectorial couplings with the $Z'$ (see Table \ref{Table:couplingsQH0}),
so the fermionic triangle loop for the decay $Z' \to \g \, \g$ has
three vectorial vertices, so the corresponding amplitude is zero because of the Furry theorem \cite{Furry}.
So in our case, the only possible decays in the neutral gauge sector (on and off-shell) are $Z' \to Z_0 \, \g$ and $Z' \to Z_0 \, Z_0$.
However we remind that we are interested only in on-shell decays.

Since there is no tree level interaction between the $Z'$ and neutralinos or charginos, no contribution to the fermionic triangles is given
by the SUSY sector so, we get the same results, for what the decays of interest are concerned, of non-SUSY models.

\subsection{Anomaly cancellation} \label{Sec: Anomaly cancellation broken}
Before going on in our computations we check that the anomaly cancellation still holds in the broken phase.
We focus on the process $Z' \to Z_0 \g$ since the extension to the other cases is straightforward.
In the broken phase, additional contributions coming from the NG boson ($G^0$) exchange must be added.
The corresponding WI's, given in diagrammatic form in Fig.~\ref{Fig: WIZpZ0gbroken}, are
\bea
 (p+q)^\r \( \D^{Z' Z_0 \g}_{\r\m\n} + (\GCS)^{Z' Z_0 \g}_{\r\m\n} \) + i \MZp (\GS)^{Z_0 \g}_{\m\n} &=& 0
          \label{WIrhoZpZ0gbroken}\\
 p^\m \( \D^{Z' Z_0 \g}_{\r\m\n} + (\GCS)^{Z' Z_0 \g}_{\r\m\n} \) + i \MZO (\NG)^{Z' \g}_{\r\n} &=& 0
          \label{WImuZpZ0gbroken}\\
 q^\n \( \D^{Z' Z_0 \g}_{\r\m\n} + (\GCS)^{Z' Z_0 \g}_{\r\m\n} \) &=& 0
          \label{WInuZpZ0gbroken}
\eea
where $\D^{Z' Z_0 \g}_{\r\m\n}$ is the fermionic loop with external $Z'$, $Z_0$ and $\g$, $(\GCS)^{Z' Z_0 \g}_{\r\m\n}$ is the
GCS coupling for the vectors cited above, $(\GS)^{Z_0 \g}_{\m\n}$ is the GS coupling between the axion, the $Z_0$ and the photon and
$(\NG)^{Z' \g}_{\r\n}$ is the fermionic loop with external $Z'$, $G^0$ and $\g$.
\begin{figure}[t]
 \vskip 1.5cm
%
$(p+q)^\r ~\Bigg($~~~~~~~~~~~~
 \raisebox{-4.3ex}[0cm][0cm]{\unitlength=0.4mm
 \begin{fmffile}{ZpZ0gloop3}
 \begin{fmfgraph*}(60,40)
  \fmfpen{thick} \fmfleft{ii0,ii1,ii2} \fmfstraight \fmffreeze \fmftop{ii2,t1,t2,t3,oo2}
  \fmfbottom{ii0,b1,b2,b3,oo1} \fmf{phantom}{ii2,t1,t2} \fmf{phantom}{ii0,b1,b2} \fmf{phantom}{t1,v1,b1}
  \fmf{phantom}{t2,b2} \fmf{phantom}{t3,b3}
  \fmffreeze
  \fmf{photon}{ii1,v1} \fmf{fermion}{t3,v1} \fmf{fermion,label=$\psi$}{b3,t3} \fmf{fermion}{v1,b3}
  \fmf{photon}{b3,oo1} \fmf{photon}{t3,oo2}
  \fmflabel{$Z'_\r(p+q)$}{ii1}
  \fmflabel{$Z_{0\m}(p)$}{oo1}
  \fmflabel{$\g_\n(q)$}{oo2}
 \end{fmfgraph*}
 \end{fmffile}}      ~+~~~
 \raisebox{-4ex}[0cm][0cm]{\unitlength=0.5mm
 \begin{fmffile}{GCS_ZZgnew}
 \begin{fmfgraph*}(30,30)
  \fmfpen{thick}
  \fmfleft{i1} \fmfright{o1,o2}
  \fmf{boson}{i1,v1} \fmf{boson}{v1,o1} \fmf{boson}{v1,o2}
  \fmffreeze
  \fmflabel{$Z'$}{i1}\fmflabel{$Z_0$}{o1}\fmflabel{$\g$}{o2}
 \end{fmfgraph*}
 \end{fmffile}} $\Bigg)$
$+i\MZp \Bigg($
 \raisebox{-4ex}[0cm][0cm]{ \unitlength=0.5mm
 \begin{fmffile}{axionZ0g}
 \begin{fmfgraph*}(30,30)
 \fmfpen{thick} \fmfleft{i1} \fmfright{o1,o2}
 \fmf{dashes}{i1,v1} \fmf{photon}{v1,o2}
 \fmf{photon}{v1,o1} \fmffreeze \fmflabel{$Z_0$}{o1} \fmflabel{$\g$}{o2} \fmflabel{$\f$}{i1}
 \end{fmfgraph*}
 \end{fmffile}} $\Bigg)=0$
  \vskip 2.2cm
~~~~~~$p^\m ~\Bigg($~~
 \raisebox{-4.3ex}[0cm][0cm]{\unitlength=0.4mm
 \begin{fmffile}{ZZg_131}
 \begin{fmfgraph*}(60,40)
  \fmfpen{thick} \fmfleft{ii0,ii1,ii2} \fmfstraight \fmffreeze \fmftop{ii2,t1,t2,t3,oo2}
  \fmfbottom{ii0,b1,b2,b3,oo1} \fmf{phantom}{ii2,t1,t2} \fmf{phantom}{ii0,b1,b2} \fmf{phantom}{t1,v1,b1}
  \fmf{phantom}{t2,b2} \fmf{phantom}{t3,b3}
  \fmffreeze
  \fmf{photon}{ii1,v1} \fmf{fermion}{t3,v1} \fmf{fermion,label=$\psi$}{b3,t3} \fmf{fermion}{v1,b3}
  \fmf{photon}{b3,oo1} \fmf{photon}{t3,oo2}
  \fmflabel{$Z'$}{ii1} \fmflabel{$Z_0$}{oo1} \fmflabel{$\gamma$}{oo2}
 \end{fmfgraph*}
 \end{fmffile}}      ~+~~~~
 \raisebox{-4.3ex}[0cm][0cm]{\unitlength=0.5mm
 \begin{fmffile}{GCS_ZZgnew}
 \begin{fmfgraph*}(30,30)
  \fmfpen{thick}
  \fmfleft{i1} \fmfright{o1,o2}
  \fmf{boson}{i1,v1} \fmf{boson}{v1,o1} \fmf{boson}{v1,o2}
  \fmffreeze
  \fmflabel{$Z'$}{i1}\fmflabel{$Z_0$}{o1}\fmflabel{$\g$}{o2}
 \end{fmfgraph*}
 \end{fmffile}} $\Bigg)$
 $+i \MZO
 \Bigg($~~
 \raisebox{-4.3ex}[0cm][0cm]{\unitlength=0.4mm
 \begin{fmffile}{ZpgGoldloop1}
 \begin{fmfgraph*}(60,40)
  \fmfpen{thick} \fmfleft{ii0,ii1,ii2} \fmfstraight \fmffreeze \fmftop{ii2,t1,t2,t3,oo2}
  \fmfbottom{ii0,b1,b2,b3,oo1} \fmf{phantom}{ii2,t1,t2} \fmf{phantom}{ii0,b1,b2} \fmf{phantom}{t1,v1,b1}
  \fmf{phantom}{t2,b2} \fmf{phantom}{t3,b3}
  \fmffreeze
  \fmf{photon}{ii1,v1} \fmf{fermion}{t3,v1} \fmf{fermion,label=$\psi$}{b3,t3} \fmf{fermion}{v1,b3}
  \fmf{dashes}{b3,oo1} \fmf{photon}{t3,oo2}
  \fmflabel{$Z'$}{ii1} \fmflabel{$G^0$}{oo1} \fmflabel{$\g$}{oo2}
 \end{fmfgraph*}
 \end{fmffile}}  $ \Bigg)  =0$
 \vskip 2.2cm
~~~~~~$q^\n ~\Bigg($~~
 \raisebox{-4.3ex}[0cm][0cm]{\unitlength=0.4mm
 \begin{fmffile}{ZZg_131}
 \begin{fmfgraph*}(60,40)
  \fmfpen{thick} \fmfleft{ii0,ii1,ii2} \fmfstraight \fmffreeze \fmftop{ii2,t1,t2,t3,oo2}
  \fmfbottom{ii0,b1,b2,b3,oo1} \fmf{phantom}{ii2,t1,t2} \fmf{phantom}{ii0,b1,b2} \fmf{phantom}{t1,v1,b1}
  \fmf{phantom}{t2,b2} \fmf{phantom}{t3,b3}
  \fmffreeze
  \fmf{photon}{ii1,v1} \fmf{fermion}{t3,v1} \fmf{fermion,label=$\psi$}{b3,t3} \fmf{fermion}{v1,b3}
  \fmf{photon}{b3,oo1} \fmf{photon}{t3,oo2}
  \fmflabel{$Z'$}{ii1} \fmflabel{$Z_0$}{oo1} \fmflabel{$\gamma$}{oo2}
 \end{fmfgraph*}
 \end{fmffile}}      ~+~~~~
 \raisebox{-4.3ex}[0cm][0cm]{\unitlength=0.5mm
 \begin{fmffile}{GCS_ZZgnew}
 \begin{fmfgraph*}(30,30)
  \fmfpen{thick}
  \fmfleft{i1} \fmfright{o1,o2}
  \fmf{boson}{i1,v1} \fmf{boson}{v1,o1} \fmf{boson}{v1,o2}
  \fmffreeze
  \fmflabel{$Z'$}{i1}\fmflabel{$Z_0$}{o1}\fmflabel{$\g$}{o2}
 \end{fmfgraph*}
 \end{fmffile}}      $\Bigg)=0$
 \vskip 1cm
\caption{The Ward identities for the amplitude $Z' \to Z_0 \, \g$ in the broken phase.} \label{Fig: WIZpZ0gbroken}
\end{figure}
The fermionic triangle $\D_{\r \m \n}^{Z' Z_0 \g}$ is given by
\be
 \D_{\r \m \n}^{Z' Z_0 \g}=-{1\over4} g_0 g_{Z_0} e \sum_f t_f^{Z' Z_0 \g} \, \G_{\r \m \n}^{VAV}(p,q;m_f) \label{DeltaZpZ0g}
\ee
where $\G_{\r \m \n}^{VAV}(p,q;m_f)$ is given by (\ref{VAVmasstrian}) and
\be
 t_f^{Z' Z_0 \g}=v_f^{Z'} a_f^{Z_0} q_f \label{tfZpZ0g}
\ee
So we have (compare with eq.~(\ref{WI_AVVmass}))
\bea
 (p+q)^\r \D_{\r \m \n}^{Z' Z_0 \g} &=&  {1\over4} g_0 g_{Z_0} e \sum_f t_f^{Z' Z_0 \g}  \[ {1\over{6\pi^2}} \] \e[p,q,\m,\n]
          \label{WIrhoDeltaZpZ0g}\\
 p^\m \D_{\r \m \n}^{Z' Z_0 \g}     &=&  {1\over4} g_0 g_{Z_0} e \sum_f t_f^{Z' Z_0 \g} {1\over{\pi^2}} \[ {1\over{6}} - m_f^2 I_0 (p,q;m_f)\] \e[q,p,\n,\r] ~~~~~~~~
          \label{WImuDeltaZpZ0g}\\
 q^\n \D_{\r \m \n}^{Z' Z_0 \g}     &=&  {1\over4} g_0 g_{Z_0} e \sum_f t_f^{Z' Z_0 \g}  \[ {1\over{6\pi^2}} \] \e[q,p,\r,\m]
          \label{WInuDeltaZpZ0g}
\eea
since we choose a symmetric distribution of the anomaly. The integral $I_0$ is given in eq.~(\ref{I_0integral}) and $\e[p,q,\r,\s]$
is defined after (\ref{Ros}).
Using eq.~(\ref{gZ0}) and (\ref{e}) and computing $\sum_f t_f^{Z' Z_0 \g}$ we get
\bea
 (p+q)^\r \D_{\r \m \n}^{Z' Z_0 \g} &=&  {1\over{8\pi^2}} g_0 g_1 g_2 \( 3 \QQ + \QL \) \e[p,q,\m,\n]
          \label{WIrhoDeltaZpZ0g1}\\
 p^\m \D_{\r \m \n}^{Z' Z_0 \g}     &=&  {1\over{4\pi^2}} g_0 g_1 g_2  \[ {3 \QQ + \QL \over{2}} - \sum_f t_f^{Z' Z_0 \g}  m_f^2 I_0\] \e[q,p,\n,\r] ~~~~~~~~
          \label{WImuDeltaZpZ0g1}\\
 q^\n \D_{\r \m \n}^{Z' Z_0 \g}     &=&  {1\over{8\pi^2}} g_0 g_1 g_2 \( 3 \QQ + \QL \) \e[q,p,\r,\m]
          \label{WInuDeltaZpZ0g1}
\eea
The corresponding GCS coupling is given by the proper combination of the GCS couplings in the hypercharge basis (see Fig. \ref{Fig:DecGCSGS}a)
\begin{figure}[t]
\centering
\vskip 1cm
~~~
 \raisebox{-4.3ex}[0cm][0cm]{\unitlength=0.5mm
 \begin{fmffile}{GCS_ZZgnew}
 \begin{fmfgraph*}(30,30)
  \fmfpen{thick}
  \fmfleft{i1} \fmfright{o1,o2}
  \fmf{boson}{i1,v1} \fmf{boson}{v1,o1} \fmf{boson}{v1,o2}
  \fmffreeze
  \fmflabel{$Z'$}{i1}\fmflabel{$Z_0$}{o1}\fmflabel{$\g$}{o2}
 \end{fmfgraph*}
 \end{fmffile}}
 ~$=~ R_{011}^{Z' Z_0 \g}\Bigg($~~~~~
 \raisebox{-4ex}[0cm][0cm]{\unitlength=0.5mm
 \begin{fmffile}{GCS_pYY1bb}
 \begin{fmfgraph*}(30,30)
  \fmfpen{thick}
  \fmfleft{i1} \fmfright{o1,o2}
  \fmf{boson}{i1,v1} \fmf{boson}{v1,o1} \fmf{boson}{v1,o2}
  \fmffreeze
  \fmflabel{$V^{(0)}$}{i1}\fmflabel{$V^{(1)}$}{o1}\fmflabel{$V^{(1)}$}{o2}
 \end{fmfgraph*}
 \end{fmffile}}
 ~~$\Bigg)+~ R_{022}^{Z' Z_0 \g}\Bigg($~~~~~
 \raisebox{-4ex}[0cm][0cm]{\unitlength=0.5mm
 \begin{fmffile}{GCS_p22}
 \begin{fmfgraph*}(30,30)
  \fmfpen{thick}
  \fmfleft{i1} \fmfright{o1,o2}
  \fmf{boson}{i1,v1} \fmf{boson}{v1,o1} \fmf{boson}{v1,o2}
  \fmffreeze
  \fmflabel{$V^{(0)}$}{i1}\fmflabel{$V^{(2)}$}{o1}\fmflabel{$V^{(2)}$}{o2}
 \end{fmfgraph*}
 \end{fmffile}}
$\Bigg)$ ~~~~~(a)
\vskip 2cm
~~~
 \raisebox{-4ex}[0cm][0cm]{ \unitlength=0.5mm
 \begin{fmffile}{axionZ0g}
 \begin{fmfgraph*}(30,30)
 \fmfpen{thick} \fmfleft{i1} \fmfright{o1,o2}
 \fmf{dashes}{i1,v1} \fmf{photon}{v1,o2}
 \fmf{photon}{v1,o1} \fmffreeze \fmflabel{$Z_0$}{o1} \fmflabel{$\g$}{o2} \fmflabel{$\f$}{i1}
 \end{fmfgraph*}
 \end{fmffile}}
 ~$=~ R_{011}^{Z' Z_0 \g}\Bigg($~~~
 \raisebox{-4ex}[0cm][0cm]{ \unitlength=0.5mm
 \begin{fmffile}{axionYY}
 \begin{fmfgraph*}(30,30)
 \fmfpen{thick} \fmfleft{i1} \fmfright{o1,o2}
 \fmf{dashes}{i1,v1} \fmf{photon}{v1,o2}
 \fmf{photon}{v1,o1} \fmffreeze \fmflabel{$V^{(1)}$}{o1} \fmflabel{$V^{(1)}$}{o2} \fmflabel{$\f$}{i1}
 \end{fmfgraph*}
 \end{fmffile}}
 ~~~~$\Bigg)+~ R_{022}^{Z' Z_0 \g}\Bigg($~~~
 \raisebox{-4ex}[0cm][0cm]{ \unitlength=0.5mm
 \begin{fmffile}{axion22}
 \begin{fmfgraph*}(30,30)
 \fmfpen{thick} \fmfleft{i1} \fmfright{o1,o2}
 \fmf{dashes}{i1,v1} \fmf{photon}{v1,o2}
 \fmf{photon}{v1,o1} \fmffreeze \fmflabel{$V^{(2)}$}{o1} \fmflabel{$V^{(2)}$}{o2} \fmflabel{$\f$}{i1}
 \end{fmfgraph*}
 \end{fmffile}}
~~$\Bigg)$ ~~~~~(b)
\vskip 1cm
\caption{Decomposition of the $(\GCS)^{Z' Z_0 \g}$ and $(\GS)^{Z_0 \g}$ diagrams in terms of the diagrams in the hypercharge basis.
As usual each depicted diagram also contains the exchange $(\m,p)\leftrightarrow (\n,q)$
The corresponding Feynman rules are given in Appendix~\ref{sec:Zp int}. The rotation factors are given in eq.~(\ref{ZpZ0grotfact}).}
\label{Fig:DecGCSGS}
\end{figure}
\bea
 (\GCS)^{Z' Z_0 \g}_{\r\m\n} &=& \( 16 g_0 g_1^2 R_{011}^{Z' Z_0 \g} d_5 + 8 g_0 g_2^2 R_{022}^{Z' Z_0 \g} d_6 \) \e[\r,\n,\m,p-q]
\eea
where the rotation factors are
\bea
 R_{011}^{Z' Z_0 \g} &=& - \frac{g_1 g_2}{g_1^2+g_2^2} \nn\\
 R_{022}^{Z' Z_0 \g} &=&   \frac{g_1 g_2}{g_1^2+g_2^2} \label{ZpZ0grotfact}
\eea
So we get
\bea
 (\GCS)^{Z' Z_0 \g}_{\r\m\n} &=& \frac{16 g_0 g_1 g_2}{g_1^2+g_2^2} \( -g_1^2 d_5 + g_2^2 {d_6\over2} \) \e[\r,\n,\m,p-q] \nn\\
                             &=& \frac{16 g_0 g_1 g_2}{192(g_1^2+g_2^2)\pi^2} \( - g_1^2 \cA^{(1)} + g_2^2 \cA^{(2)} \) \e[\r,\n,\m,p-q] \nn\\
                             &=&  \frac{g_0 g_1 g_2}{8\pi^2} \( 3 \QQ + \QL \) \e[\r,\n,\m,p-q] \label{GCSZpZ0g}
\eea
where we used eq.~(\ref{bsds}), (\ref{A1}) and (\ref{A2}). Then using eq.~(\ref{WInuDeltaZpZ0g1}) we have
\bea
 q^\n \( \D^{Z' Z_0 \g}_{\r\m\n} + (\GCS)^{Z' Z_0 \g}_{\r\m\n} \)
    &=& \frac{g_0 g_1 g_2}{8\pi^2} \( 3 \QQ + \QL \) \(\e[q,p,\r,\m] + \e[\r,q,\m,p-q] \) \nn\\
    &=& \frac{g_0 g_1 g_2}{8\pi^2} \( 3 \QQ + \QL \) \(\e[q,p,\r,\m] + \e[\r,q,\m,p] \) \nn\\
    &=& 0
\eea
and eq.~(\ref{WInuZpZ0gbroken}) is verified. From eq.~(\ref{WIrhoDeltaZpZ0g1}) and (\ref{GCSZpZ0g}) we easily obtain
\bea
 (p+q)^\r \( \D^{Z' Z_0 \g}_{\r\m\n} + (\GCS)^{Z' Z_0 \g}_{\r\m\n} \) = \frac{3 g_0 g_1 g_2}{8\pi^2} \( 3 \QQ + \QL \) \e[p,q,\m,\n]
 \label{WIrhoDeltaGCSZpZ0g1}
\eea
The corresponding GS coupling is given by the proper combination of the GS couplings in the gauge interaction basis (see Fig. \ref{Fig:DecGCSGS}b)
\bea
 (\GS)^{Z_0 \g}_{\m\n} &=& \( 8 i g_1^2 R_{011}^{Z' Z_0 \g} b_2^{(1)} - 4 i g_2^2 R_{022}^{Z' Z_0 \g} b_2^{(2)} \) \e[\m,\n,p,q] \nn\\
                       &=& \frac{i}{b_3} \frac{3 g_1 g_2}{32\pi^2} \( 3 \QQ + \QL \) \e[p,q,\m,\n]
\eea
where we used eq.~(\ref{bsds}), (\ref{A1}), (\ref{A2}) and (\ref{ZpZ0grotfact}).
We remember that since $\QHu=0$, $\MZp=4 g_0 b_3$ (see eq.~(\ref{Zpmass})). Then using eq.~(\ref{WIrhoDeltaGCSZpZ0g1}) we have
\bea
 &&(p+q)^\r \( \D^{Z' Z_0 \g}_{\r\m\n} + (\GCS)^{Z' Z_0 \g}_{\r\m\n} \) + i \MZp (\GS)^{Z_0 \g}_{\m\n}= \nn\\
    &&\qquad\qquad = \frac{3 g_1 g_2}{8\pi^2} \( 3 \QQ + \QL \) \(g_0 + i\MZp \frac{i}{4 b_3} \)\e[p,q,\m,\n] \nn\\
    &&\qquad\qquad = \frac{3 g_1 g_2}{8\pi^2} \( 3 \QQ + \QL \) \(g_0 - \frac{4 g_0 b_3}{4 b_3} \)\e[p,q,\m,\n] \nn\\
    &&\qquad\qquad = 0
\eea
and eq.~(\ref{WIrhoZpZ0gbroken}) is verified. From eq.~(\ref{WImuDeltaZpZ0g1}) and (\ref{GCSZpZ0g}) we easily obtain
\bea
 p^\m \( \D^{Z' Z_0 \g}_{\r\m\n} + (\GCS)^{Z' Z_0 \g}_{\r\m\n} \) = -{1\over{4\pi^2}} g_0 g_1 g_2  \sum_f \[ t_f^{Z' Z_0 \g}  m_f^2 I_0 \] \e[q,p,\n,\r]
 \label{WImuDeltaGCSZpZ0g1}
\eea
Computing the Feynman amplitude for $(\NG)^{Z' \g}_{\r\n}$ diagram we find
\bea
 (\NG)^{Z' \g}_{\r\n} &=& -{i\over{4\pi^2}} g_0 e \Bigg[ \sum_{e,\m,\t,d,s,b} \(\frac{y_{f_d}}{\sqrt2} \cos\b \, v_{f_d}^{Z'} q_{f_d} m_{f_d} I_0 \)+ \nn\\
                       &&\phantom{{i\over{4\pi^2}} g_0 e \Bigg[}
                                                       + \sum_{u,c,t} ~ \(\frac{y_{f_u}}{\sqrt2} \sin\b \, v_{f_u}^{Z'} q_{f_u} m_{f_u} I_0 \) \Bigg]\e[q,p,\n,\r] \nn\\
                        &=& - {i\over{4\pi^2}} g_0 \frac{g_1 g_2}{\frac{1}{2}\sqrt{g_1^2+g_2^2}v}
                            \Bigg[- \sum_{e,\m,\t,d,s,b} \(v_{f_d}^{Z'} \frac{1}{2} q_{f_d} m_{f_d}^2 I_0 \)+\nn\\
 &&\phantom{- {i\over{4\pi^2}} g_0 \frac{g_1 g_2}{\frac{1}{2}\sqrt{g_1^2+g_2^2}v}\Bigg[}
   + ~~\sum_{u,c,t}~~ \(v_{f_u}^{Z'} \frac{1}{2} q_{f_u} m_{f_u}^2 I_0 \) \Bigg]\e[q,p,\n,\r] \nn\\
 &=& - {i\over{4\pi^2}} \frac{g_0 g_1 g_2}{\MZO} \sum_f \[ v_f^{Z'} a_f^{Z_0} q_f  m_f^2 I_0 \] \e[q,p,\n,\r] \nn\\
 &=& - {i\over{4\pi^2}} \frac{g_0 g_1 g_2}{\MZO} \sum_f \[ t_f^{Z' Z_0 \g}  m_f^2 I_0 \] \e[q,p,\n,\r]
\label{G0Zpg1}
\eea
where we used $y_{f_u}\sin\b/\sqrt2=m_{f_u}/v$, $-y_{f_d}\cos\b/\sqrt2=m_{f_d}/v$, eq.~(\ref{e}) and (\ref{Z0massnew}) for $\QHu=0$, the values for $a_f^{Z_0}$ given in Table \ref{Table:couplingsQH0}
and the definition of $t_f^{Z' Z_0 \g}$ given in eq.~(\ref{tfZpZ0g}).
Then summing eq.~(\ref{WImuDeltaGCSZpZ0g1}) and (\ref{G0Zpg1}) we get
\bea
  &&p^\m \( \D^{Z' Z_0 \g}_{\r\m\n} + (\GCS)^{Z' Z_0 \g}_{\r\m\n} \) + i \MZO  (\NG)^{Z' \g}_{\r\n}= \nn\\
  &&\qquad\qquad = {1\over{4\pi^2}} g_0 g_1 g_2  \sum_f \[ t_f^{Z' Z_0 \g}  m_f^2 I_0 \] \(-1 + i \MZO \frac{-i}{\MZO} \) \e[q,p,\n,\r] \nn\\
  &&\qquad\qquad = 0
\eea
and also eq.~(\ref{WImuZpZ0gbroken}) is verified. So we checked the consistence of the anomaly cancellation in the broken phase.
\subsection{$Z' \to Z_0 \ \g$}
We calculate the decay rate for $Z' \to Z_0 \ \g$.
We compute all the relevant diagrams in the $R_\xi$ gauge, thus removing the interaction vertex $V^\m \partial_\m
G_V$ that involves the massive gauge bosons and the \Stuckelberg or NG boson. Therefore, the only diagrams that
remain are the fermionic loop, the GCS vertex and a not anomalous remnant contribution (Fig. \ref{diagzp}). It is
possible to show that the last blob-diagram, that involves several diagrams, is equal to zero. For the interested
reader we give further details in Appendix~\ref{notanomappdx}.
\begin{figure}[tb]
 \vskip 1.5cm
 \centering
 \raisebox{-5.2ex}[0cm][0cm]{\unitlength=0.7mm
 \begin{fmffile}{BBBMSSM_ZZgNEW}
 \begin{fmfgraph*}(30,30)
  \fmfpen{thick} \fmfleft{o1} \fmfright{i1,i2} \fmf{boson}{i1,v1,i2}\fmf{boson}{o1,v1}
  \fmfv{decor.shape=circle,decor.filled=hatched, decor.size=.30w}{v1}
  \fmflabel{$Z'^\r(p+q)$}{o1}\fmflabel{$Z_0^\m(p)$}{i1}\fmflabel{$\g^\n(q)$}{i2}
 \end{fmfgraph*}
 \end{fmffile}}
~~~=~~~~~~~~~~~~~~~~~~~~~~~~~~~~~~~~~~~~~~~~~~~~~~~~~~~~~~
 \vskip2.5cm
 \raisebox{-4.3ex}[0cm][0cm]{\unitlength=0.4mm
 \begin{fmffile}{ZZg_131}
 \begin{fmfgraph*}(60,40)
  \fmfpen{thick} \fmfleft{ii0,ii1,ii2} \fmfstraight \fmffreeze \fmftop{ii2,t1,t2,t3,oo2}
  \fmfbottom{ii0,b1,b2,b3,oo1} \fmf{phantom}{ii2,t1,t2} \fmf{phantom}{ii0,b1,b2} \fmf{phantom}{t1,v1,b1}
  \fmf{phantom}{t2,b2} \fmf{phantom}{t3,b3}
  \fmffreeze
  \fmf{photon}{ii1,v1} \fmf{fermion}{t3,v1} \fmf{fermion,label=$\psi$}{b3,t3} \fmf{fermion}{v1,b3}
  \fmf{photon}{b3,oo1} \fmf{photon}{t3,oo2}
  \fmflabel{$Z'$}{ii1} \fmflabel{$Z_0$}{oo1} \fmflabel{$\gamma$}{oo2}
 \end{fmfgraph*}
 \end{fmffile}}
~~~~~+~~~~
 \raisebox{-5.2ex}[0cm][0cm]{\unitlength=0.7mm
 \begin{fmffile}{GCS_ZZg}
 \begin{fmfgraph*}(30,30)
  \fmfpen{thick}
  \fmfleft{i1} \fmfright{o1,o2}
  \fmf{boson}{i1,v1} \fmf{boson}{v1,o1} \fmf{boson}{v1,o2}
  \fmffreeze
  \fmflabel{$Z'$}{i1}\fmflabel{$Z_0$}{o1}\fmflabel{$\g$}{o2}
 \end{fmfgraph*}
 \end{fmffile}}
~~~~~+~~~~
 \raisebox{-5.2ex}[0cm][0cm]{\unitlength=0.7mm
 \begin{fmffile}{BBBMSSM_NAZZg}
 \begin{fmfgraph*}(30,30)
  \fmfpen{thick} \fmfleft{o1} \fmfright{i1,i2} \fmf{boson}{i1,v1,i2}\fmf{boson}{o1,v1}
  \fmfv{decor.shape=circle,decor.filled= shaded, decor.size=.30w}{v1}
  \fmflabel{$Z'$}{o1}\fmflabel{$Z_0$}{i1}\fmflabel{$\g$}{i2}\fmflabel{~~Not An.}{v1}
 \end{fmfgraph*}
 \end{fmffile}}
 \vskip 1.5cm
 \caption{Diagrams for $Z' \to Z_0 \ \g$.} \label{diagzp}
\end{figure}
The decay rate for the process is given by
\be
 \G \(Z' \to Z_0 \g \) = \frac{p_F}{32 \pi^2 \MZp^2} \int |A^{Z' Z_0 \g}|^2 d\Omega
 \label{decay1}
\ee
where $A^{Z' Z_0 \g}$ is the total scalar amplitude and $p_F$ is the momentum of the outgoing vectors in the CM frame
\be
 p_F = \frac{\MZp}{2} \left( 1 - \frac{\MZO^2}{\MZp^2} \right)
\label{decay11}
\ee
The square of the total scalar amplitude is given by
\be
 |A^{Z' Z_0 \g}|^2 = \frac{1}{3} \sum_{\l'} \e^{\r_1}_{(\l')} \e^{* \r_2}_{(\l')} \
                   \sum_{\l^0} \e^{\m_1}_{(\l^0)} \e^{* \m_2}_{(\l^0)} \
                  \sum_{\l_\g} \e^{\n_1}_{(\l^\g)} \e^{* \n_2}_{(\l^\g)} \
                  A_{\r_1 \m_1 \n_1}^{Z' Z_0 \g} A_{\r_2 \m_2 \n_2}^{* \, Z' Z_0 \g}
\ee
where $\e$ are the polarizations of the gauge bosons, and $A_{\r \m \n}^{Z' Z_0 \g}$ is the Feynman amplitude of the process.
The factor $1/3$ comes from the average over the $Z'$ helicity states. The polarizations obey to the following
completeness relations
\bea
 \sum_{\l'} \e^{\r_1}_{(\l')} \e^{* \r_2}_{(\l')} &=& - \eta^{\r_1 \r_2} + \frac{k^{\r_1}_{(\l')} k^{\r_2}_{(\l')}}{\MZp^2} \\
 \sum_{\l^0} \e^{\m_1}_{(\l^0)} \e^{* \m_2}_{(\l^0)} &=& - \eta^{\m_1 \m_2} + \frac{k^{\m_1}_{(\l^0)} k^{\m_2}_{(\l^0)}}{\MZO^2} \\
 \sum_{\l_\g} \e^{\n_1}_{(\l^\g)} \e^{* \n_2}_{(\l^\g)} &\to& - \eta^{\n_1\n_2} \label{helicities}
\eea
where (\ref{helicities}) gives only the relevant part of the sum over helicities. Other terms are omitted since they
give vanishing contributions to the decay.
The amplitude is given by the sum of the fermionic triangle $\D_{\r \m \n}^{Z' Z_0 \g}$ plus the proper GCS vertex
\bea
 A_{\r \m \n}^{Z' Z_0 \g}&=& \D_{\r \m \n}^{Z' Z_0 \g} + (\GCS)_{\r \m \n}^{Z' Z_0 \g}
\eea
where $\D_{\r \m \n}^{Z' Z_0 \g}$ and $(\GCS)_{\r \m \n}^{Z' Z_0 \g}$ are given in eq.~(\ref{DeltaZpZ0g}) and (\ref{GCSZpZ0g}).
It is convenient to express the triangle amplitude by using the Rosenberg parametrization~\cite{Rosenberg:1962pp}
(see also Appendix~\ref{subsec:Rosenberg})
\bea
 \D_{\r \m \n}^{Z' Z_0 \g}&=& -{1\over4\pi^2} g_0 g_{Z_0} e \Big(A_1 \e[p,\m,\n,\r] + A_2\e[q,\m,\n,\r]+ A_3 \e[p,q,\m,\r]{p}_{\n} \nn\\
    &&+  A_4 \e[p,q,\m,\r]{q}_{\n} +  A_5 \e[p,q,\n,\r]p_\m +  A_6\e[p,q,\n,\r]q_\m \Big)
\eea
where
\be
 A_i=\sum_f t_f^{Z' Z_0 \g} I_i  \qquad \text{for } i=3, \dots , 6
\ee
$I_3,~I_4,~I_5$ and $I_6$ are finite integrals (their explicit forms are given in (\ref{I's_mass})), $\e[p,q,\r,\s]$
is defined after (\ref{Ros_mass}) and $t_f^{Z' Z_0 \g}$ is given in eq.~(\ref{tfZpZ0g}).
$A_1$ and $A_2$ are naively divergent by power counting and so they must be
regularized. We compute them by using the Ward identities. In this way it is possible to express $A_1$ and $A_2$ in
terms of the finite integrals $I_3,~I_4,~I_5$ and $I_6$. The GCS term has the following tensorial structure
\be d^{Z' Z_0 \g}\Big(\e[p,\m,\n,\r]-\e[q,\m,\n,\r] \Big) \ee
so it can be absorbed by shifting the first two coefficients of the Rosenberg parametrization for the triangle
(see Section \ref{sect:Computation of the amplitude}). The resulting amplitude can be written as
\bea
 \D_{\r \m \n}^{Z' Z_0 \g}&=&
    -{1\over4\pi^2} g_0 g_{Z_0} e \Big(\tilde A_1 \e[p,\m,\n,\r] +   \tilde A_2\e[q,\m,\n,\r]+  A_3 \e[p,q,\m,\r]{p}_{\n} \nn\\
    &&+  A_4 \e[p,q,\m,\r]{q}_{\n} +  A_5 \e[p,q,\n,\r]p_\m + A_6\e[p,q,\n,\r]q_\m \Big) \label{ampZ'Zg}
\eea
The WIs (\ref{WIrhoZpZ0gbroken}), (\ref{WImuZpZ0gbroken}) and (\ref{WInuZpZ0gbroken}) now read
\bea
 (p+q)^\r A_{\r \m \n}^{Z' Z_0 \g}+i M_{Z'} (\GS)^{Z_0 \g}_{\m \n}&=&0\\
  p^\m A_{\r \m \n}^{Z' Z_0 \g}+i M_{Z_0} (\NG)^{Z' \g}_{\r \n}&=&0\\
  q^\n  A_{\r \m \n}^{Z' Z_0 \g}&=& 0
\eea
where $M_{Z'}= 4b_3 g_0$ and $M_{Z_0}$ are the $Z'$ and $Z_0$ masses respectively.
Using the results of the previous section we obtain
\bea
 (p+q)^\r A_{\r \m \n}^{Z' Z_0 \g} &=& {1\over4\pi^2} g_0 g_{Z_0} e {1\over2}\sum_f t_f^{Z' Z_0 \g} ~\e[p,q,\m,\n]\\
 p^\m A_{\r \m \n}^{Z' Z_0 \g}     &=& -{1\over4\pi^2} g_0 g_{Z_0} e \sum_f t_f^{Z' Z_0 \g} m_f^2 ~I_0 ~\e[q,p,\n,\r]\\
 q^\n A_{\r \m \n}^{Z' Z_0 \g}     &=& 0
\eea
and inserting (\ref{ampZ'Zg}) into the above identities we get
\bea
  \tilde A_1 &=&  \(q^2 A_4 + p \cdot q A_3 \) \nn\\
  \tilde A_2 &=&  \(p^2 A_5 + p \cdot q A_6 + (\NG)^{Z' \g} \)\label{Z'ZgA1}
\eea
with
\be
 (\NG)^{Z' \g}=\sum_f t_f^{Z' Z_0 \g} \ m_f^2 I_0
\ee
where $I_0$ is the integral given in (\ref{I_0integral}). Substituting $\tilde A_1, ~ \tilde A_2$ from
(\ref{Z'ZgA1}) into the amplitude (\ref{ampZ'Zg}) and performing all the contractions we finally obtain
\bea
  |A^{Z' Z_0 \g}|^2 &=& g_0^2 g_{Z_0}^2 e^2 \frac{
  \left(M_{Z'}^2-M_{Z_0}^2\right)^2 \left(M_{Z'}^2+M_{Z_0}^2\right)}{96 M_{Z_0}^2 M_{Z'}^2 \pi^4} \times\nn\\
  &&\[\sum_f  t_f^{Z' Z_0 \g}  \Big(  (I_3+I_5) M_{Z_0}^2 + m_f^2 ~I_0  \Big)\]^2 \label{Z'Zg}
\eea
\subsection{$ Z' \to Z_0 \ Z_0 $}
The computations are similar to the previous case so we point out only the differences with the other decay. Mutatis
mutandis, the decay rate for the process is given in (\ref{decay1}) with the proper amplitude and
\be
 p_F = \frac{\MZp}{2} \sqrt{ 1 - \frac{4\MZO^2}{\MZp^2} }
\label{decay12}
\ee
The square of the total scalar amplitude is given by
\be
 |A^{Z' Z_0 Z_0}|^2 =  \frac{1}{3} \sum_{\l'} \e^{\r_1}_{(\l')} \e^{* \r_2}_{(\l')} \
                                    \sum_{\l^0} \e^{\n_1}_{(\l^0)} \e^{* \n_2}_{(\l^0)} \
                                    \sum_{\l^0} \e^{\m_1}_{(\l^0)} \e^{* \m_2}_{(\l^0)} \
                      A_{\r_1 \m_1 \n_1}^{Z' Z_0 Z_0} A_{\r_2 \m_2 \n_2}^{* \, Z' Z_0 Z_0}
\ee
where the amplitude $A_{\r \m \n}^{Z' Z_0 Z_0}$ is always the sum of the fermionic triangle and the GCS term. The contribution
to the fermionic triangle is
\bea
 \D_{\r \m \n}^{Z' Z_0 Z_0}&=&-{1\over8} g_0 g_{Z_0}^2 \Bigg[\sum_f \(
     v_f^{Z'} a_f^{Z_0} v_f^{Z_0} \, \G_{\r \m \n}^{VAV} +
     v_f^{Z'} v_f^{Z_0} a_f^{Z_0} \, \G_{\r \m \n}^{VVA} \)+ \nn\\
    &&\phantom{-{1\over8} g_0 g_{Z_0}^2 \Bigg[}
      \sum_n \( a_n^{Z'} v_n^{Z_0} v_n^{Z_0} \, \G_{\r \m \n}^{AVV}+
     a_n^{Z'} a_n^{Z_0} a_n^{Z_0} \, \G_{\r \m \n}^{AAA} \) \Bigg]
\eea
where $n$ runs over all the neutrinos while the $\G_{\r \m \n}$'s are given by (\ref{AVVmasstrian}), (\ref{VAVmasstrian}),
(\ref{VVAmasstrian}), (\ref{AAAmasstrian}). Using the fact that for the three neutrino families we have $v_n^{Z'}=a_n^{Z'}$
and $v_n^{Z_0}=a_n^{Z_0}$ we write the total amplitude (the sum of the triangles plus GCS terms) as
\bea
     A_{\r \m \n}^{Z' Z_0 Z_0}&=&
        -{1\over8\pi^2} g_0 g_{Z_0}^2  \Big(\tilde A_1 \e[p,\m,\n,\r] +
         \tilde A_2 \e[q,\m,\n,\r]+
          A_3 \e[p,q,\m,\r]{p}^{\n} \nn\\
        &&+  A_4 \e[p,q,\m,\r]{q}^{\n} +  A_5 \e[p,q,\n,\r]p^\m +
         A_6\e[p,q,\n,\r]q^\m \Big)
\eea
with
\be
 A_i=2 \sum_f t_f^{Z' Z_0 Z_0} I_i  \qquad \text{for } i=3, \dots , 6
\ee
where
\be
 t_f^{Z' Z_0 Z_0} = \tilde v_f^{Z'} a_f^{Z_0} v_f^{Z_0}
\ee
with $\tilde v_n^{Z'}=2 v_n^{Z'}$ for neutrinos and $\tilde v_f^{Z'}=v_f^{Z'}$ for the other fermions.
The WIs now read
\bea
 (p+q)^\r A_{\r \m \n}^{Z' Z_0 Z_0} +i M_{Z'} (\GS)^{Z_0 Z_0}_{\m \n}&=&0\\
 p^\m A_{\r \m \n}^{Z' Z_0 Z_0} +i M_{Z_0} (\NG)^{Z' Z_0}_{\r \n}&=&0\\
 q^\n A_{\r \m \n}^{Z' Z_0 Z_0} +i M_{Z_0} (\NG)^{Z_0 Z'}_{\m \r}&=&0
\eea
leading to
\bea
 (p+q)^\r A_{\r \m \n}^{Z' Z_0 Z_0}&=&  {1\over8\pi^2} g_0 g_{Z_0}^2 \sum_f t_f^{Z' Z_0 Z_0} \e[p,q,\m,\n]\\
 p^\m A_{\r \m \n}^{Z' Z_0 Z_0}&=&   -{1\over8\pi^2} g_0 g_{Z_0}^2 \sum_f t_f^{Z' Z_0 Z_0} m_f^2 I_0 \e[q,p,\n,\r]\\
 q^\n A_{\r \m \n}^{Z' Z_0 Z_0}&=&   -{1\over8\pi^2} g_0 g_{Z_0}^2 \sum_f t_f^{Z' Z_0 Z_0} m_f^2 I_0 \e[q,p,\r,\m]
\eea
From these equations we find the following values for $\tilde A_1$ and $\tilde A_2$
\bea
 \tilde A_1 &=& \(q^2 A_4 + p \cdot q A_3 - (\NG)^{Z' Z_0} \) \\
 \tilde A_2 &=& \(p^2 A_5 + p \cdot q A_6 + (\NG)^{Z' Z_0} \)
\eea
with
\be
 (\NG)^{Z' Z_0}=\sum_f t_f^{Z' Z_0 Z_0} \ m_f^2 I_0
\ee
where $I_0$ is the integral given in (\ref{I_0integral}). Substituting back into the amplitude and performing all
the contractions we finally obtain
\bea
 &&\!\!\!\!\!\!\!\!\!\!\!\!
|A^{Z' Z_0 Z_0}|^2 =
 g_0^2 g_{Z_0}^4 \frac{ \left(M_{Z'}^2-4 M_{Z_0}^2\right)^2}{192 M_{Z_0}^2\pi^4}
 \[ \sum_f t_f^{Z' Z_0 Z_0} \bigg( 2(I_3+I_5) M_{Z_0}^2+ m_f^2 I_0 \bigg)\]^2 \qquad \label{ampzprimozozo}
\eea
\subsection{Numerical Results} \label{subsec: Numerical Results}
     \begin{figure}[p]
      \centering
      \includegraphics[scale=0.33]{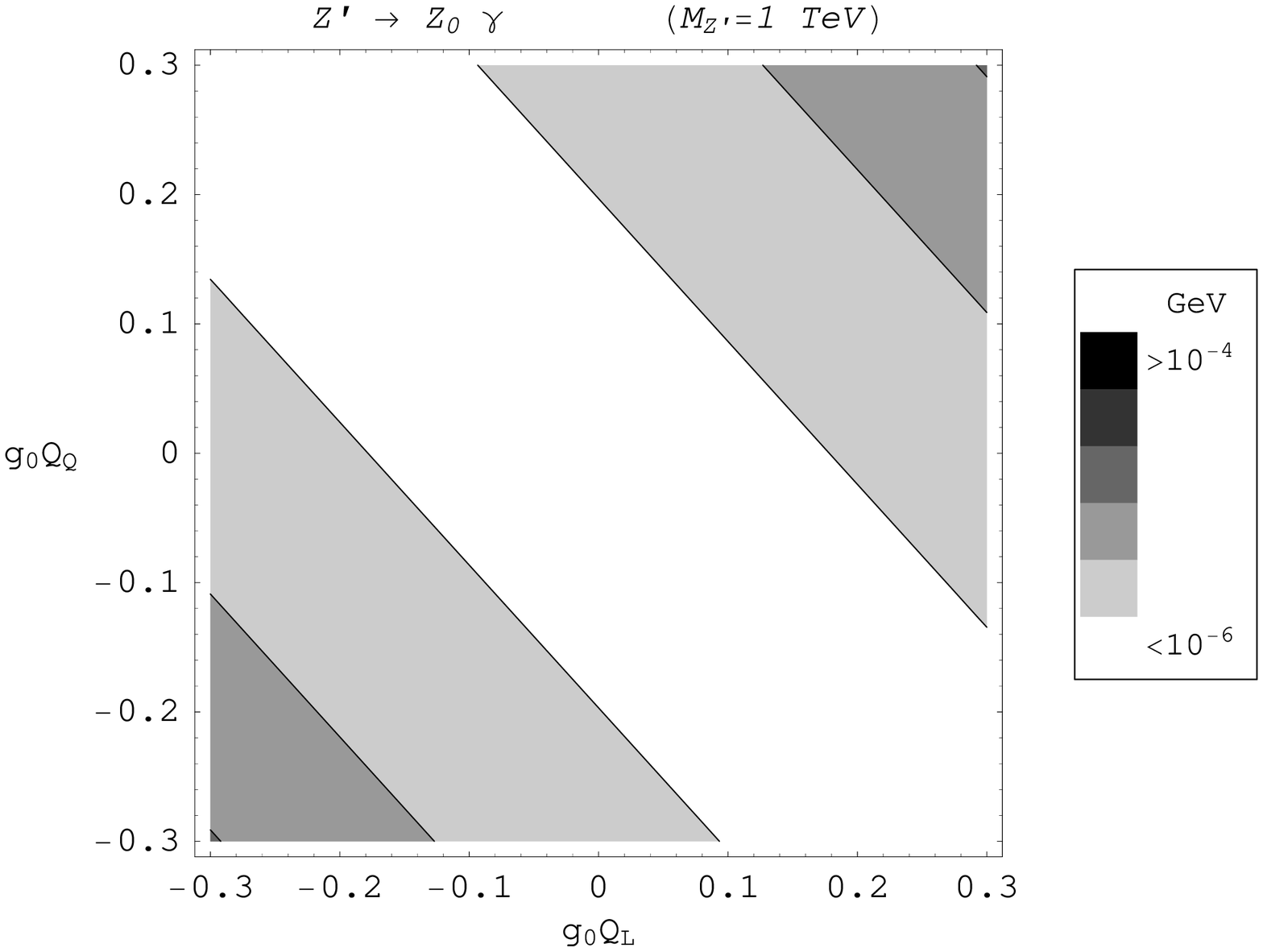}
      ~~
      \includegraphics[scale=0.33]{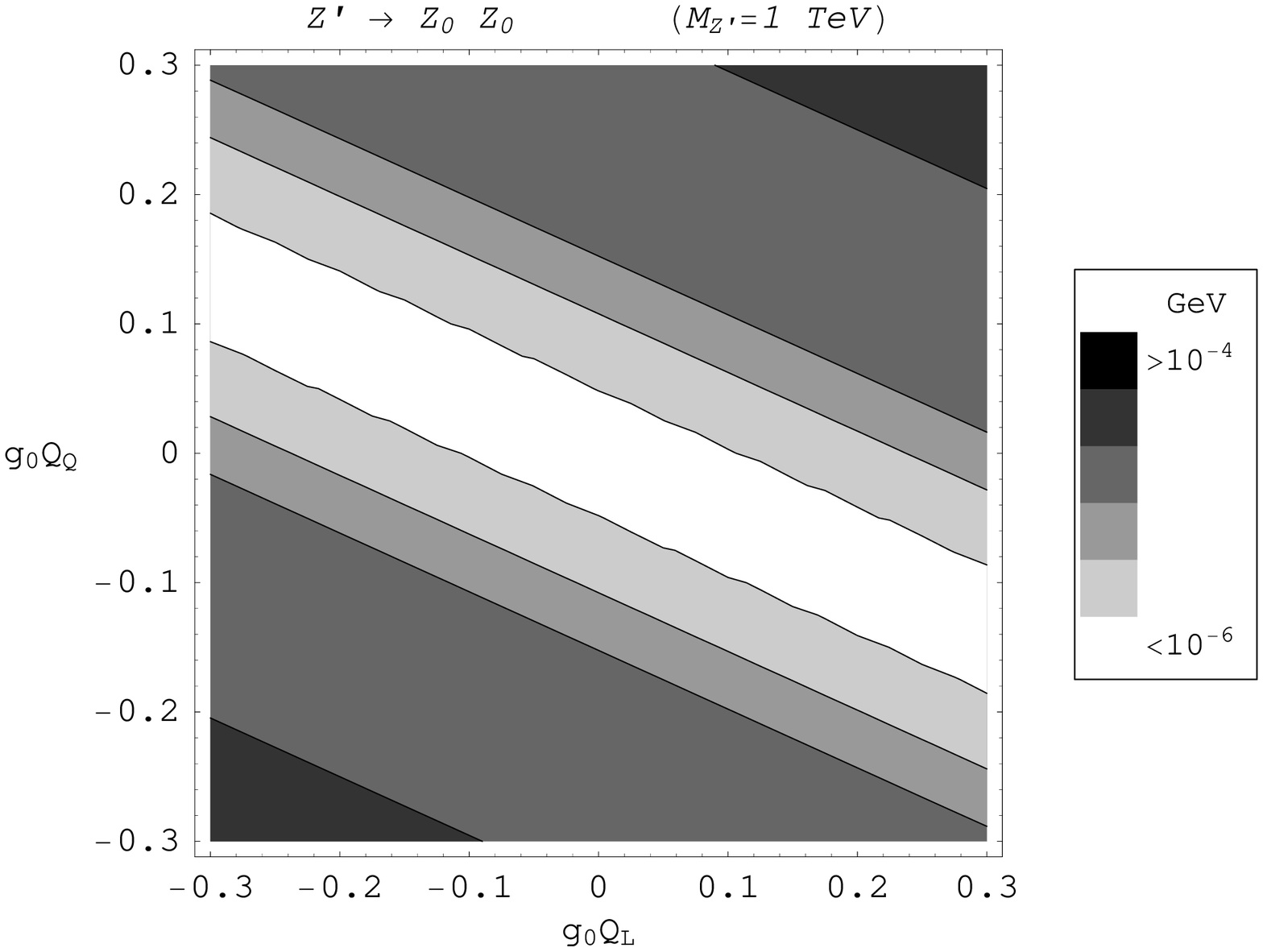}
      \caption{$\MZp=1$ TeV.} \label{Contourplot 1 TeV}
     \end{figure}
     \begin{figure}[hp]
      \centering
      \includegraphics[scale=0.33]{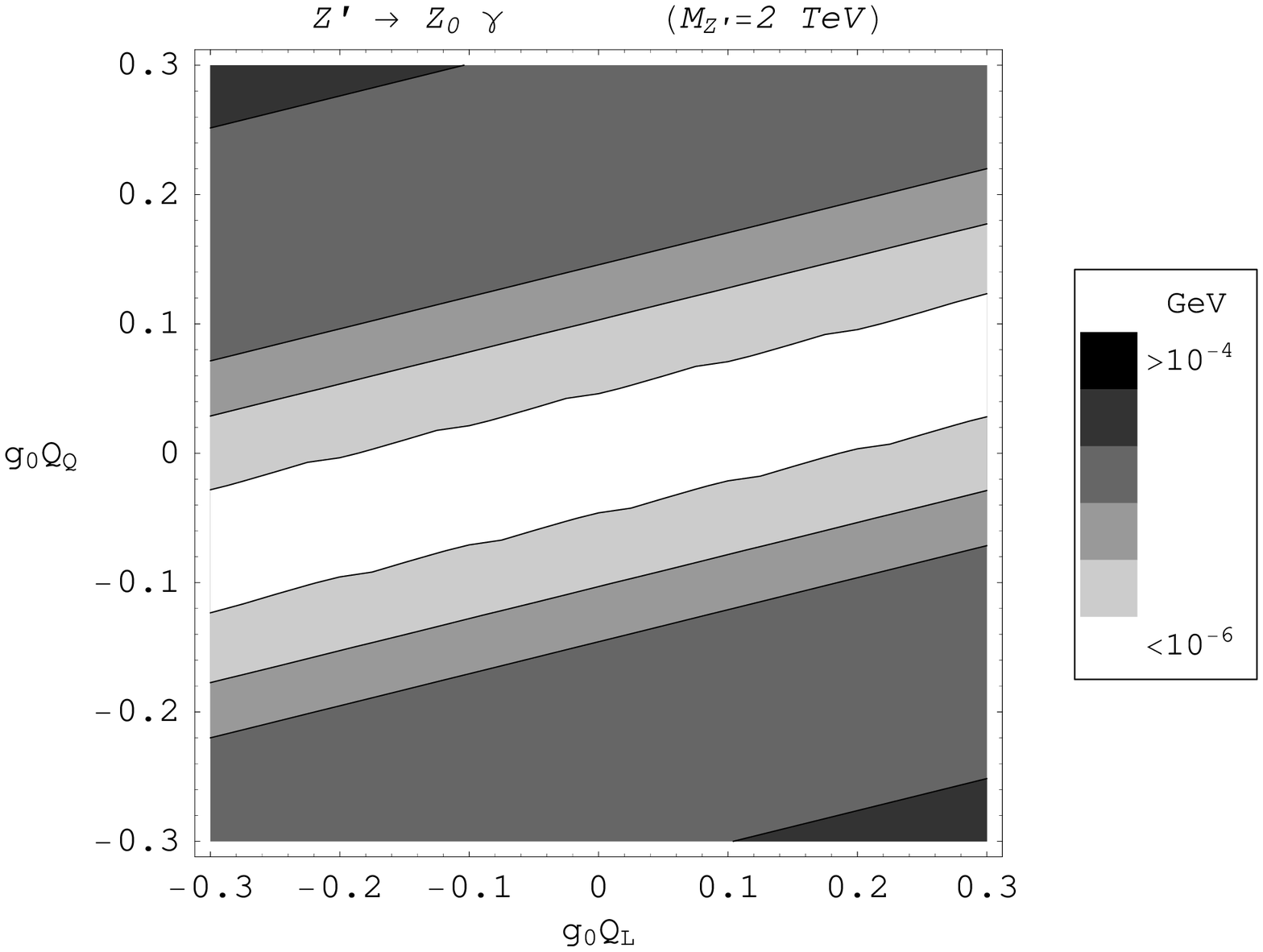}
      ~~
      \includegraphics[scale=0.33]{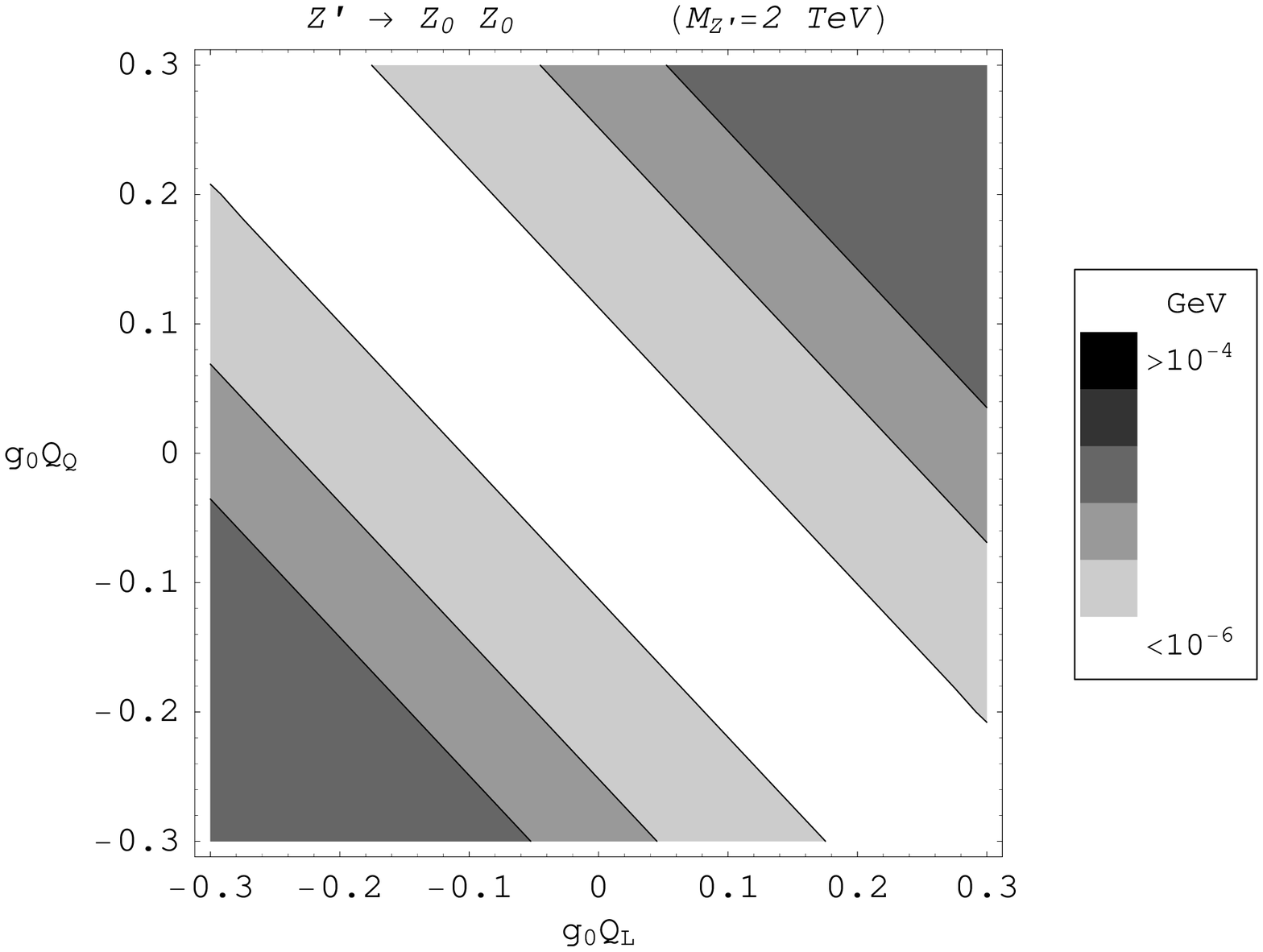}
      \caption{$\MZp=2$ TeV.} \label{Contourplot 2 TeV}
     \end{figure}
      \begin{figure}[hp]
      \centering
      \includegraphics[scale=0.33]{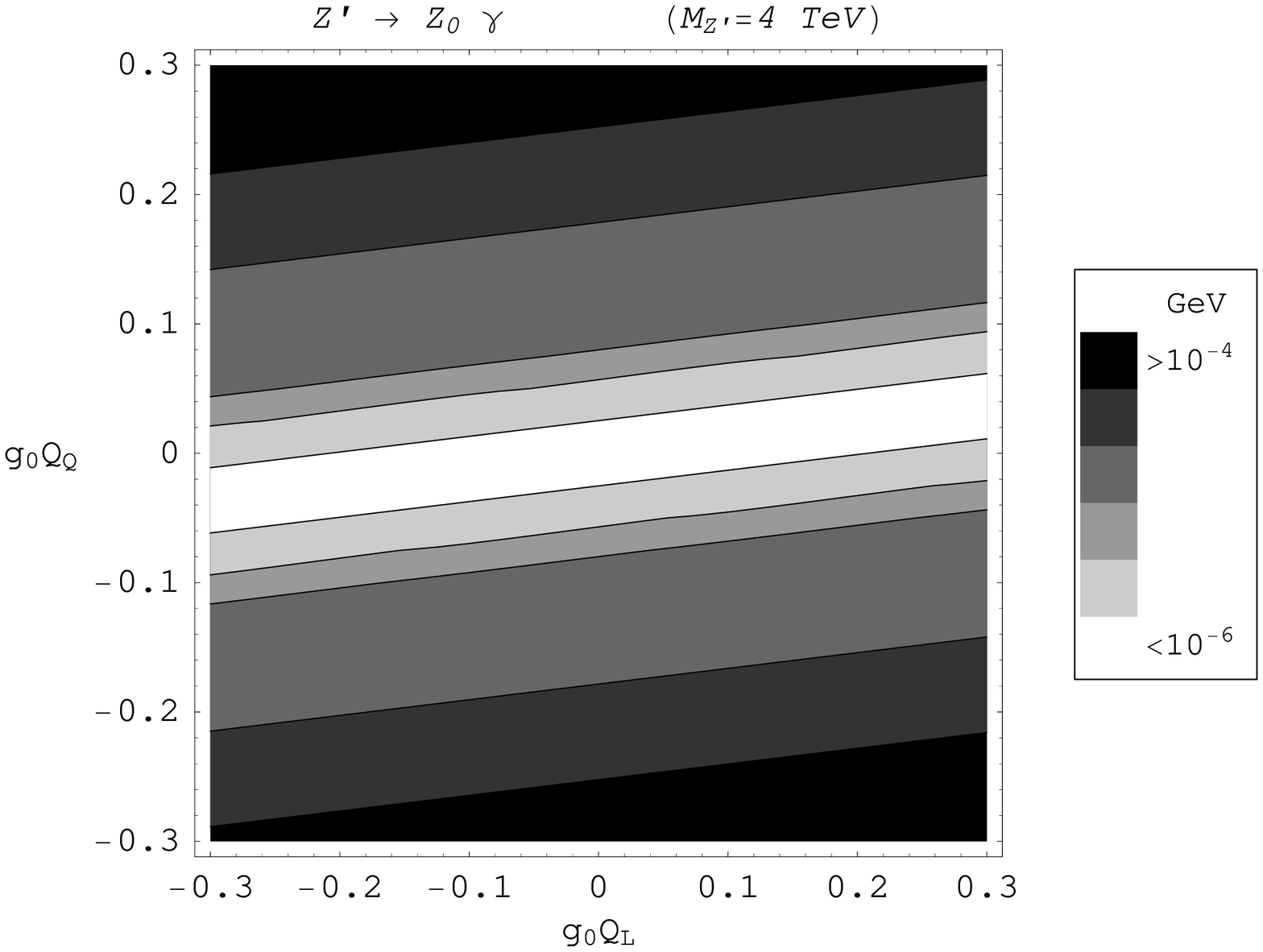}
      ~~
      \includegraphics[scale=0.33]{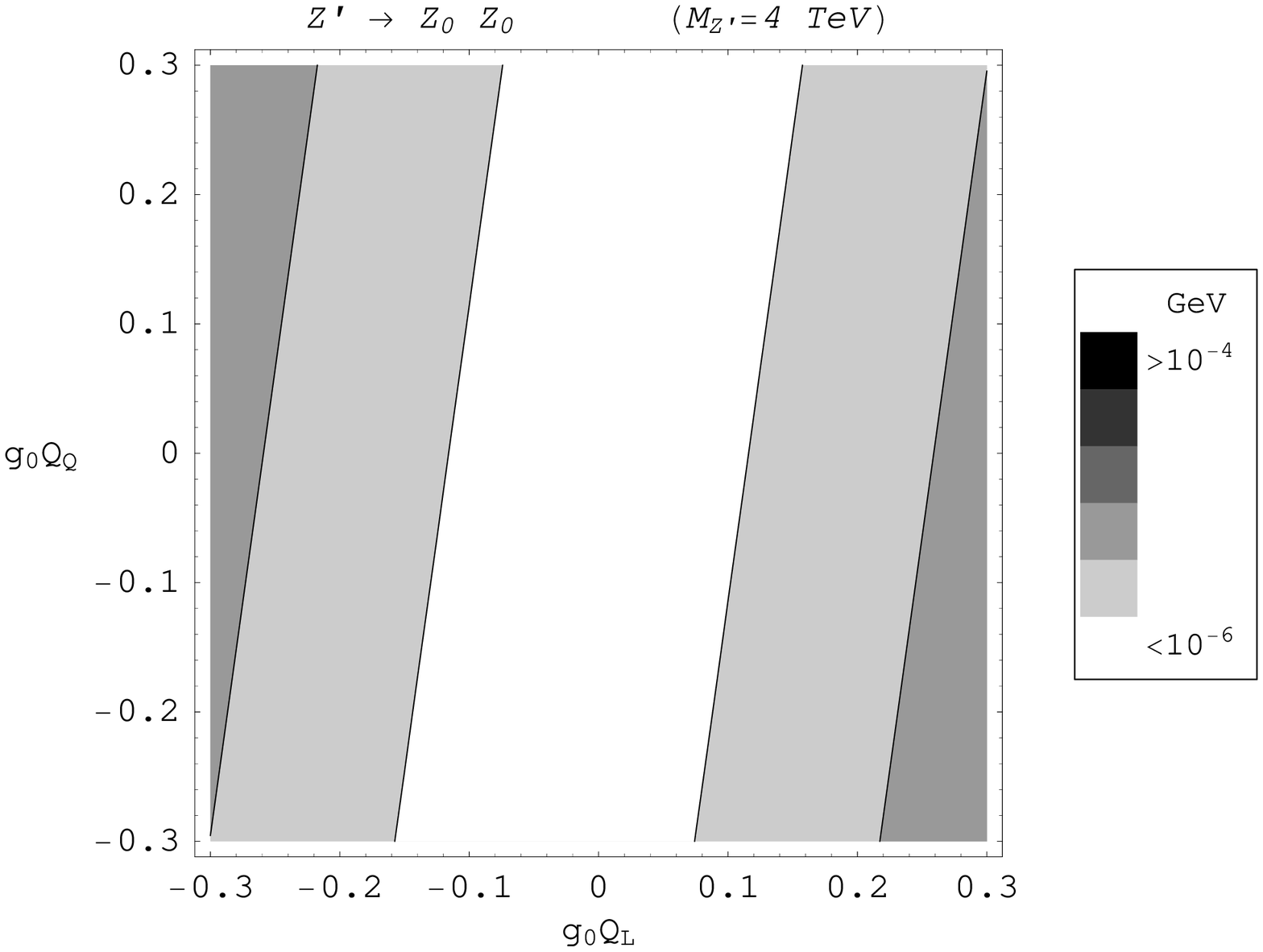}
      \caption{$\MZp=4$ TeV.} \label{Contourplot 4 TeV}
     \end{figure}
We show some numerical computations for the two decay rates $\Gamma (Z'\to Z_0 \gamma)$ and $\Gamma
(Z'\to Z_0 Z_0)$. They depend on the free parameters of the model, i.e. the couplings $g_0 Q_Q$, $g_0 Q_L$ and the mass of the
$Z'$, since $Q_{H_u}=0$. We show our results in Fig. \ref{Contourplot 1
TeV}-\ref{Contourplot 4 TeV} in the form of contour plots in the plane $g_0 Q_L, g_0 Q_Q$ for $M_{Z'}=1,2$ and $4$ TeV. Our
choices for $g_0$, $Q_Q$, $Q_L$ and $\MZp$ are in agreement with the current experimental bounds \cite{Abulencia:2006iv}.

The darker shaded regions correspond to larger decay rates. The white region corresponds to the value $10^{-6}$ GeV
that can be considered as a rough lower limit for the detection of the corresponding process. It is worth noting
that increasing $M_{Z'}$ the mean value of the decay rate of $Z'\to Z_0 \gamma$ grows while the one of $Z'\to Z_0
Z_0$ decreases. We would also like to mention that increasing $M_{Z'}$ the iso-decay rate contours in the plot
rotate clockwise getting  more and more parallel to the $g_0 Q_L$-axis. This effect is due to the fact that the
contribution of the triangle diagram with the top quark circulating inside the loop becomes the dominant
contribution for high $M_{Z'}$. In this case the decays strongly depend on the top quark coupling $g_0 Q_Q$ while the
lepton couplings $g_0 Q_L$ become irrelevant. Finally, we find that the region that gives the largest values (of order of
$10^{-4}$ GeV) of the decay $Z' \to Z_0 \, \gamma$ is for $M_{Z'}\sim 4$ TeV and for $g_0 Q_Q\sim 0.3$, $g_0 Q_L\sim -0.2$.
\section{LHC prediction} \label{sec: LHC}
      \begin{figure}[p]
      \centering
      \includegraphics[scale=0.38]{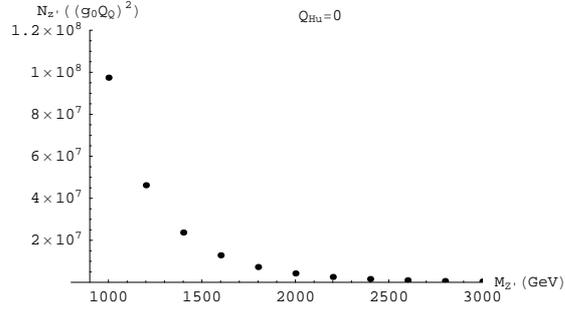}
      \caption{Number of $Z'$ produced at LHC in 1 year for $\L=10^{34} cm^{-2} s^{-1}$ and $\sqrt s = 14$ TeV,
               in units of $(g_0 Q_Q)^2$, in function of the mass of the $Z'$.} \label{NZp}
     \end{figure}
      \begin{figure}[hp]
      \centering
      \includegraphics[scale=0.33]{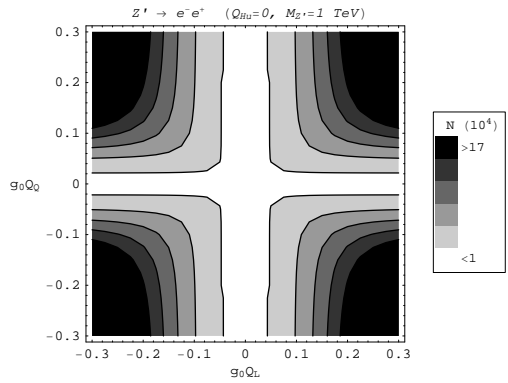}
      \caption{Number of $Z'\to e^- e^+$ at LHC in 1 year for $\L=10^{34} cm^{-2} s^{-1}$, $\sqrt s = 14$ TeV and
               $\MZp=1$ TeV.} \label{NZpee}
     \end{figure}
      \begin{figure}[hp]
      \centering
      \includegraphics[scale=0.33]{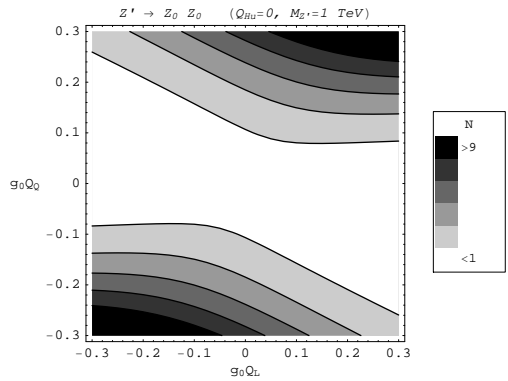}
      \caption{Number of $Z'\to Z_0 Z_0$ at LHC in 1 year for $\L=10^{34} cm^{-2} s^{-1}$, $\sqrt s = 14$ TeV and
               $\MZp=1$ TeV.} \label{NZpZZ}
     \end{figure}
To estimate the number of the decays that can be observed at LHC we shall use the narrow width
approximation,
\be
 N_{Z' \to \text{particles}}= N_{Z'} \ \BR (Z' \to \text{particles})
\ee
where $N_{Z'}$ is the total number of $Z'$ produced and $\BR (Z' \to \text{particles})$ is the branching
ratio of the decay of interest. We remind that we can use such approximation if $\(\G_{Z'}/\MZp\)^2<<1$.
For our choice of parameters, the maximum value will be for $g_0 \QL = g_0 \QL = 0.3$ and $\MZp = 1$ TeV, so
$\(\G_{Z'}/\MZp\)^2 \simeq 3 \times 10^{-3}$ and the latter inequality holds.

The total number of $Z'$ is $N_{Z'}=\s_{Z'} \, \L \, t \ $ where $\L=10^{34} {\rm \,\,cm^{-2} s^{-1}}$ the luminosity and $t=$1 year.
Finally $\s_{Z'}$ is the $Z'$ production cross section \cite{Langacker:2008yv}
\be
 \frac{d\s_{Z'}}{dy}= \frac{4\pi^2 x_1 x_2}{3M_{Z'}^3} \sum_{i}
 \big[f_{q_i}(x_1)f_{\bar q_i}(x_2) +f_{\bar q_i}(x_1)f_{q_i}(x_2)\big] \Gamma (Z' \to q_i \bar q_i)
\label{Zpproduction}
\ee
where $f_{q_i,\bar q_i}$ are the quark $q_i$ (or antiquark $\bar q_i$) structure functions
in the proton, and the momentum fractions are
\be
 x_{1,2}=(M_{Z'}/\sqrt{s}) e^{\pm y}
\label{xval}
\ee
The formula (\ref{Zpproduction}) does not take into account the gluon contribution inside the proton. The gluons can only contribute
with the fermionic loop triangle. There is no GCS contribution since there is no anomaly (see eq.~(\ref{A3})).
However the $q \, \bar q \, g$ vertex is vectorial like,
as the corresponding vertex with the $Z'$ for $\QHu=0$ (see Table \ref{Table:ZpcouplingsQH0}),
so we have a $VVV$ triangle which is zero because of the Furry theorem \cite{Furry}.
Then, in this case, we do not have any gluon contribution.

We integrate numerically the PDFs using a Mathematica package \cite{PDF}.
In Fig.~\ref{NZp} we show the result for $N_{Z'}$ at $\sqrt s = 14$ TeV. We can see that the number of the $Z'$ produced
falls off exponentially with $M_{Z'}$, so we shall focus on the case $M_{Z'} \sim 1$ TeV.
As said before, we assume that the LSP has a mass higher than 500 GeV (see Section~\ref{sect:Axino Relic Density}), so that we can ignore sparticles contribution
in the computation of the branching ratios.

In Fig.~\ref{NZpee}, we estimate the number of decays $Z' \to e^- e^+$ for 1 year of integrated luminosity.
We can see that in a wide region of parameter the number of events is larger then ten thousand.
This gives a lot of chance to discover a $Z'$ in LHC.

In the case $M_{Z'} = 1$ TeV the most favorite anomalous decay is $Z'\to Z_0 Z_0$.
In Fig.~\ref{NZpZZ}, we estimate the number of decays for 1 year of integrated luminosity which
turns out to be $N_{Z'\to Z_0 Z_0} \sim 10$ for large values of the couplings $g_0 \QL$ and $g_0 \QQ$.
\section{Axino Dark Matter} \label{Sec: Axino DM}
Assuming the conservation of R-parity the LSP is a good weak interacting massive particle (WIMP) dark matter candidate.
As in the MSSM the LSP is given by a linear combination of fields in the
neutralino sector.
The general form of the neutralino mass matrix is given
in~(\ref{massmatrixNeutr}). This is a six-by-six matrix.
From the point of view of the strength of the interactions the two extra
 states are
not on the same footing with respect
to the standard ones. The axino and the extra gaugino $\l^{(0)}$ dubbed primeino are in fact extremely weak interacting massive particle (XWIMP).
Thus we are interested in situations in which the extremely weak sector is
decoupled
from the standard one and the LSP
belongs to this sector.
This can be achieved at tree level with the choice $\QHu=0$.
Written
in the interaction eigenstate basis
$(\psi^{0})^T= (\psi_S, \ \l^{(0)},\ \l^{(1)} ,\ \l^{(2)}_3,\ \tilde h_d^0,\ \tilde
h_u^0)$, the neutralino mass matrix $ {\bf M}_{\tilde N} $ becomes
   \be
    {\bf M}_{\tilde N}
     =   \(\begin{array}{cccccc}
          \frac{M_S}{2} & \frac{M_{V^{(0)}}}{\sqrt2} &   0   &   0   &
     0
    & 0 \\
                  \dots &       M_0                  &   0   &   0   &
     0
    & 0   \\
                  \dots &      \dots                 &  M_1  &   0   &
-\frac{g_1
v_d}{2}& \frac{g_1 v_u}{2} \\
                  \dots &      \dots                 & \dots &  M_2  &
\frac{g_2
v_d}{2} & -\frac{g_2 v_u}{2} \\
                  \dots &      \dots                 & \dots & \dots &
     0
    & -\m  \\
                  \dots &      \dots                 & \dots & \dots &
   \dots
    & 0
         \end{array}\) \label{massmatrix} ~~~~
   \ee
where $M_S,~M_0,~M_1,~M_2$ are the soft masses coming from the soft
breaking terms
(\ref{Lsoft}) while $M_{V^{(0)}}$ is given in
eq.~(\ref{Zpmass}).
It is worth noting that the D terms and kinetic mixing terms
can be neglected in the tree-level computations of the eigenvalues and
eigenstates.

Moreover we make the assumption that $M_0\gg M_S,M_{V^{(0)}}$ so that the axino is the LSP.
This assumption is motivated by the interaction strengths of the two extra states.
Since the axino interacts via the vertex shown in Fig.~\ref{Upint}b (which is of the order of $\sim g_0 g_a^2/\MZp$) while the primeino interacts via the vertex in Fig.~\ref{Upint}a (which is of the order of $\sim g_a$),
if we do not assume the decoupling of the axino, the dominant contribution in the (co)annihilation processes
is that of the primeino, which is of the type of a standard gaugino interaction.
\begin{figure}[t]
\vskip 2.5cm
\centering
      \raisebox{-4.7ex}[0cm][0cm]{\unitlength=0.7mm
      \begin{fmffile}{lfsf1}
      \begin{fmfgraph*}(40,40)
       \fmfpen{thick}
       \fmfleft{i1} \fmfright{o1,o2}
       \fmf{fermion}{i1,v1} \fmf{fermion}{v1,o2}
       \fmf{dashes}{v1,o1}
       \fmffreeze
       \fmf{boson}{i1,v1}
       \fmflabel{$\l^{(a)}$}{i1} \fmflabel{$f$}{o2} \fmflabel{$\tilde f$}{o1}

      \end{fmfgraph*}
      \end{fmffile}}
~~~~~~~~~~~~~
      \raisebox{-4.7ex}[0cm][0cm]{\unitlength=0.7mm
      \begin{fmffile}{axilA}
      \begin{fmfgraph*}(40,40)
       \fmfpen{thick}
       \fmfleft{i1} \fmfright{o1,o2}
       \fmf{fermion}{i1,v1} \fmf{fermion}{v1,o2}
       \fmf{boson}{v1,o1}
       \fmffreeze
       \fmf{boson}{v1,o2}
       \fmflabel{$\psi_S$}{i1} \fmflabel{$\l^{(a)}$}{o2} \fmflabel{$V_\n^{(a)}$}{o1}
      \end{fmfgraph*}
      \end{fmffile}}
      \vskip 1.5cm
(a)~~~~~~~~~~~~~~~~~~~~~~~~~~~~~~~~~~~~~~~(b)
      \caption{(a) Gaugino-fermion-sfermion interaction vertex. (b) Axino-gaugino-vector interaction vertex.}
\label{Upint}
\end{figure}
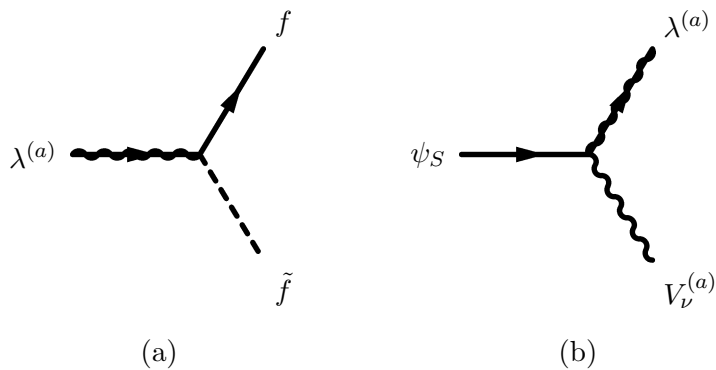

\subsection{Axino Interactions}
 The axino interactions can be read off from the interaction
lagrangian~(\ref{Laxioncomponent}). The relevant term, written in terms of four
components Majorana spinors\footnote{The gamma matrices $\gamma^\mu$ are
in the
Weyl representation.}, is given by:
\be
 \L =i \sqrt{2} g_1^2 b_2^{(1)}  \bar\Lambda^{(1)} \g_5 [\g^\m,\g^\n](\partial_\m V^{(1)}_\n)    \Psi_S +
        i {\frac{\sqrt{2}}{2}} g_2^2 b_2^{(2)}  \bar\Lambda^{(2)}_3 \g_5 [\g^\m,\g^\n](\partial_\m V^{(2)}_{3\n}) \Psi_S
\ee
where the $b_2^{(a)}$
coefficients are given in~(\ref{bsds}). The related interaction vertex Feynman rule is
\be
  C^{(a)} [\g^\m,\g^\n] i k_\m \label{lpg}
 \ee
where $k_\m$ is the momentum of the outgoing vector and the $C^{(a)}$'s are
\bea
 C^{(1)}=\sqrt{2}g_1^2 b_2^{(1)} \nn\\
 C^{(2)}={\frac{\sqrt{2}}{2}} g_2^2 b_2^{(2)}
 \label{vertice}
\eea
The factors~(\ref{vertice}) contains the parameters $b_2^{(a)}$ which are
related to
the anomalous $U(1)$ (see eq.~(\ref{bsds})). Therefore $C^{(a)}\ll g_a$ and
the axino
interactions will be extremely weak, being suppressed by an order of
magnitude
factor with respect to the weak interactions.
At tree level there is only one type of annihilation diagram, represented
in
Fig.~\ref{figura1}.
\begin{figure}[t]
\vskip 2.5cm
\centering
      \raisebox{-4.7ex}[0cm][0cm]{\unitlength=0.7mm
      \begin{fmffile}{axiaxi}
      \begin{fmfgraph*}(40,40)
       \fmfpen{thick} \fmfleft{i1,i2} \fmfright{o1,o2}
       \fmf{fermion}{i1,v1} \fmf{fermion}{v2,i2}
       \fmf{photon}{o1,v1} \fmf{photon}{o2,v2}
       \fmf{plain,label=$\l$}{v1,v2}  \fmf{photon}{v1,v2}
       \fmflabel{$\Psi_S(p_1)$}{i1} \fmflabel{$\Psi_S(p_2)$}{i2}
          \fmflabel{$A^\r (k_1)$}{o1} \fmflabel{$A^\n(k_2)$}{o2}
      \end{fmfgraph*}
      \end{fmffile}}
      \vskip 1.5cm
      \caption{Annihilation of two axinos into two gauge vectors via the exchange of a gaugino.}
      \label{figura1}
\end{figure}
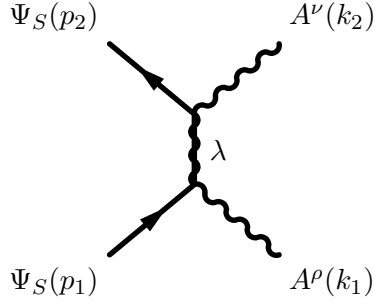
We denoted with $p_1$ and $p_2$ the incoming momenta of the axinos while
$k_1$ and
$k_2$ are the two outcoming momenta of the gauge bosons in the final
state.
We will concentrate on the case with two photons in the final state. In
this case
the result for the differential cross section is given by
\be
\ \frac{d\s}{d\Omega}=\frac{4M_S^2 \omega_1}{16 \pi^2
(\omega_1+\omega_2)^2(\sqrt{M_S^2-E_2^2})}\sum_{i,j=1}^2\mathcal{M}_i
\mathcal{M}^*_j
\label{axinoaxinodiffcs}
\ee
where $\omega_1$ and $\omega_2$ are the energies of the two outcoming
photons.
Each amplitude $\mathcal{M}_i$ is proportional to the related coefficient
$C^{(a)}$ whose generic form is given in~(\ref{vertice}).
The cross section~(\ref{axinoaxinodiffcs}), being extremely weak, cannot
give a
relic density in the WMAP preferred range. Thus we are forced to consider
a scenario
in which coannihilations between the axino and the NSLP became sizable.
Several scenarios can be considered for the NLSP. We can split them into two major classes:
one in which the NSLP is either a pure bino or a pure wino, and thus a coannihilation with a third MSSM particle is needed in order to recover the
WMAP result, and one in which the NSLP is a generic MSSM neutralino with a non negligible bino and/or wino
component.
In both classes in order to have effective coannihilations the NSLP (and eventually the other MSSM particle involved in the coannihilation process) must be almost degenerate in mass.
As a first example we consider a pure bino as the NLSP.
The allowed coannihilation processes with the axino are those which involve an exchange
of a photon
or a $Z_0$ in the intermediate state and with a SM fermion-antifermion
pair, Higgses and $W$'s in the final state. The diagram with the fermion-antifermion in the final state is sketched in Fig.~\ref{figuracoann}.
The differential cross section in the center of mass frame has the
following general form
\be
 \frac{d \s}{d \Omega} \propto \frac{1}{s}
\frac{p_f}{p_i} |\M|^2
\ee
where $s$ is the usual Mandelstam variable and $p_{f,i}$ is the spatial
momentum of the outgoing (incoming) particles.
On dimensional ground $ |\M|^2$ has at least a linear dependence on $p_f$
and this
implies that the dominant contribution comes from the diagram with the SM
fermion-antifermion pair $f$ and $\bar{f}$ in the final state:
\be
\Psi_S \l^{(a)}\rightarrow f \bar{f}
\ee
\begin{figure}[t]
\vskip 2.5cm
\centering
      \raisebox{-4.7ex}[0cm][0cm]{\unitlength=0.7mm
      \begin{fmffile}{axibino}
      \begin{fmfgraph*}(40,40)
       \fmfpen{thick} \fmfleft{i1,i2} \fmfright{o1,o2}
       \fmf{fermion}{i1,v1} \fmf{fermion}{v1,i2}
       \fmf{photon,label=$\g/Z_0$}{v1,v2}
       \fmf{fermion}{v2,o1} \fmf{fermion}{o2,v2}
       \fmffreeze
       \fmf{photon}{i1,v1}
       \fmflabel{$\lambda(p_2)$}{i1} \fmflabel{$\psi_S(p_1)$}{i2}
       \fmflabel{$f(p_4)$}{o1} \fmflabel{$f(p_3)$}{o2}
      \end{fmfgraph*}
      \end{fmffile}}
      \vskip 1.5cm
      \caption{Coannihilation of an axino and a bino into a $f \bar f$
pair via the exchange of a photon or a $Z_0$.}
      \label{figuracoann}
\end{figure}
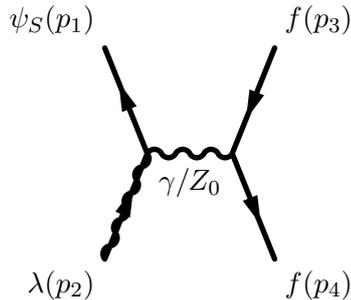
The resulting differential cross section, computed in the center of mass
frame, is
\be
\frac{d\s}{d \Omega}=\sum_{f} c_f
\frac{\sqrt{(E_3-m_{f})^2}}{64 \pi^2 (E_1+E_2)^2
\sqrt{(E_1^2-M_{S}^2)}}(\M_\g^2+\M_{Z_0}^2+\M^*_\g \M_{Z_0}+\M_\g
\M_{Z_0}^*)
\label{sezcoan}
\ee
where the sum is extended to all the SM fermions (with mass $m_f$) while
$c_f$ is a color factor. Details of the amplitude computation can be found in Appendix~\ref{nnAmp}.

\subsection{Axino Relic Density} \label{sect:Axino Relic Density}
In this section we compute the relic density of the axino.
The case of the axino as a cold dark matter candidate has been studied for the first time
in~\cite{axinoCDM}.
As we said in the previous section we study two scenarios: the first in which
 the axino coannihilates with only one NLSP degenerate in mass (a generic MSSM neutralino), the second in which there is an additional supersymmetric particle (either a chargino or a stau) involved in the coannihilation process with the axino and the NLSP.

Just to fix the notation we briefly review the relic density computation for $N$ interacting species~\cite{Griest:1990kh,Edsjo:1997bg,Edsjo:2003us}.
The Boltzmann equation for $N$ particle species is given by:
\be
\frac{dn}{dt}=-3Hn-\sum_{i,j=1}^N \langle \s_{ij} v_{ij} \rangle
(n_i n_j - n^{eq}_i n^{eq}_j) \label{boltz3}
\ee
where $n_i$ denotes the number density per unit of comoving volume of the
species $i=1,\ldots,N$ ($i=1$ refers to the LSP, $i=2$ refers to the NSLP, and so on), $n=\sum_i
n_i$, $H$ is
the Hubble constant, $\s_{ij}$ is the annihilation cross section between a
species
$i$ and a species $j$, $ v_{ij}$ is the modulus of the relative velocity
while
$n_i^{eq}$ is the equilibrium number density of the species $i$:
\begin{equation}
 n_i^{eq}=g_i\left(1+\Delta_i\right)^{3/2} e^{-\Delta_i x_f}
\end{equation}
where $g_i$ are the internal degrees of freedom, $\Delta_i=(m_i-m_1)/m_1$ while $x_f=m_1/T$ is the freeze-out temperature. Numerical computation gives $x_f\simeq 20$~\cite{Edsjo:1997bg}.

Eq.~(\ref{boltz3}) can be rewritten in a useful way by defining the thermal average of the effective
cross section

\be
\langle \s_{eff} v \rangle \equiv \sum_{i,j=1}^N\langle \s_{ij}
v_{ij} \rangle \frac{n_i^{eq}}{n^{eq}} \frac{n_j^{eq}}{n^{eq}}
\label{sigmaeff}
\ee
obtaining
\be
\frac{dn}{dt}=-3Hn-\langle \s_{eff} v \rangle (n^2 -
(n^{eq})^2) \label{Boltz}
\ee
where $n^{eq}=\sum_i n_i^{eq}$.
As a rule of thumb~\cite{Jungman:1995df} a first order estimate of the relic density is given by
\begin{equation}
 \Omega_\chi h^2\simeq \frac{10^{-27}\,{\rm cm^3\, s^{-1}}}{\langle \s_{eff} v \rangle}
\label{oh2vssigmaeff}
\end{equation}
We take into account the two cases $N=2$ and $N=3$:
\begin{itemize}
 \item $N=2$ case. Assuming that the relative velocities are all equal $v_{ij}\equiv v$ we get:
\begin{equation}
 \langle \s_{eff}^{(2)} v \rangle =\langle\sigma_{22}v\rangle\frac{\langle\sigma_{11}v\rangle/\langle\sigma_{22}v\rangle
+2\langle\sigma_{12}v\rangle/\langle\sigma_{22}v\rangle Q+Q^2}{(1+Q)^2}
\label{sigma2eff}
\end{equation}
where $Q=n_2^{eq}/n_1^{eq}$. The first term in the numerator can be neglected because the axino annihilation cross section is suppressed by a factor $(C^{(a)})^4$ with respect to the MSSM neutralino annihilations (see the previous section) and thus $\langle\sigma_{11}v\rangle\ll \langle\sigma_{22}v\rangle$. The second term involves the coannihilation cross section. Let us consider the case in which the NLSP is a generic MSSM neutralino (a linear combination of $\l^{(1)},\ \l^{(2)}_3,\ \tilde h_d^0,\ \tilde h_u^0$) with a non vanishing bino or wino components. As we saw in the previous section each amplitude is generically proportional to $C^{(a)}g_i$ with $i=1,2$. Without loss of generality we consider the diagram which involves the bino component $\Psi_S \l^{(1)}\rightarrow f \bar{f}$ and a photon exchange in the intermediate channel, i.e the $\M_\g^2$ amplitude in (\ref{sezcoan}).
We get
\be
C^2_\g= (C^{(1)} \cos\theta_W)^2= 2 (b_2^{(1)})^2 g_{1}^4 \cos^2\theta_W
\label{C}
\ee
From the expression of the mixed $U(1)'-U(1)_Y - U(1)_Y$ anomaly (see eq.~(\ref{A1})) and
from the eq.~(\ref{bsds}) we have the following relation
\be
b_2^{(1)}=\frac{3(3Q_Q+Q_L)}{256 \pi^2 b_3}
\ee
where $b_3=M_{Z'}/4g_0$. With the assumption $M_{Z'}=1$ TeV as in Section~\ref{sec: LHC} we finally get
\be
\frac{C^2_\g}{e^2}\simeq 5.76 \times 10^{-12} (3 g_0 Q_Q + g_0 Q_L)^2
\, {\rm GeV}^{-2}
\ee
where $e$ is the electric charge.
We get similar expressions for the other three terms in (\ref{sezcoan}).
This result has to be compared to the typical weak cross section $\langle\sigma_{22}v\rangle\simeq 10^{-9}\, {\rm GeV}^{-2}$. As long as the charges and the coupling constant of the extra $U(1)$ satisfy the perturbative requirement
\be g_0^2\cdot (3 Q_Q + Q_L)^2<16
\label{chargebound}
\ee
the following upper bound is satisfied:
\begin{equation}
 \frac{\langle\sigma_{12}v\rangle}{\langle\sigma_{22}v\rangle}\lesssim 10^{-6}
\label{binosigma12neg}
\end{equation}
in the case of a pure bino, while
\begin{equation}
 \frac{\langle\sigma_{12}v\rangle}{\langle\sigma_{22}v\rangle}\lesssim 10^{-5}
\label{winosigma12neg}
\end{equation}
in the case of a pure wino.
The previous ratios can be arbitrarly small, thus we can not derive a lower bound, but they cannot vanish since this would correspond to a complete decoupling of the axino. In this case the lightest MSSM state is the LSP.
Accordingly to eqs.~(\ref{oh2vssigmaeff}),~(\ref{sigma2eff}),~(\ref{binosigma12neg}) and~(\ref{winosigma12neg}) the relic density gets rescaled as~\cite{Feldman:2006wd}
\be
\( \Omega h^2\)^{(2)} \simeq \[ \frac{1+Q}{Q} \]^2 \( \Omega h^2\)^{(1)}
\label{2rescaling}
\ee
We performed a random sampling of MSSM models in which the NLSP is a pure bino or a mixed bino-higgsino (the case of a pure wino falls back into the $N=3$ case due to the wino-chargino mass degeneracy) and we computed the relic density in presence of coannihilations using the DarkSUSY package~\cite{Gondolo:2004sc}.
These two situations are easily realized in some corners of the mSUGRA parameter space. Thus in our scan we assumed this scenario in order to fix the pattern of the supersymmetry breaking parameters at weak scale. We emphasize here that this choice is completely arbitrary, and it is assumed only for simplicity, since in our model the supersymmetry breaking mechanism is not specified.
In the former case there is no model which satisfies the WMAP constraints~\cite{wmapdata}:
\be
0.0913\le\Omega h^2\le 0.1285
\ee
since the annihilation cross section of a pure bino is too low and the rescaling~(\ref{2rescaling}) is not enough to get the right relic density.
In the latter case the higgsino component tends to increase the annihilation cross section and thus we find models which satisfy the WMAP constraints. The results are summarized in the Fig.~\ref{DMbino-higgsino} for $\Delta_2=1\%$ and $\Delta_2=5\%$.
\begin{figure}[t]
      \begin{center}
      \includegraphics[scale=0.65]{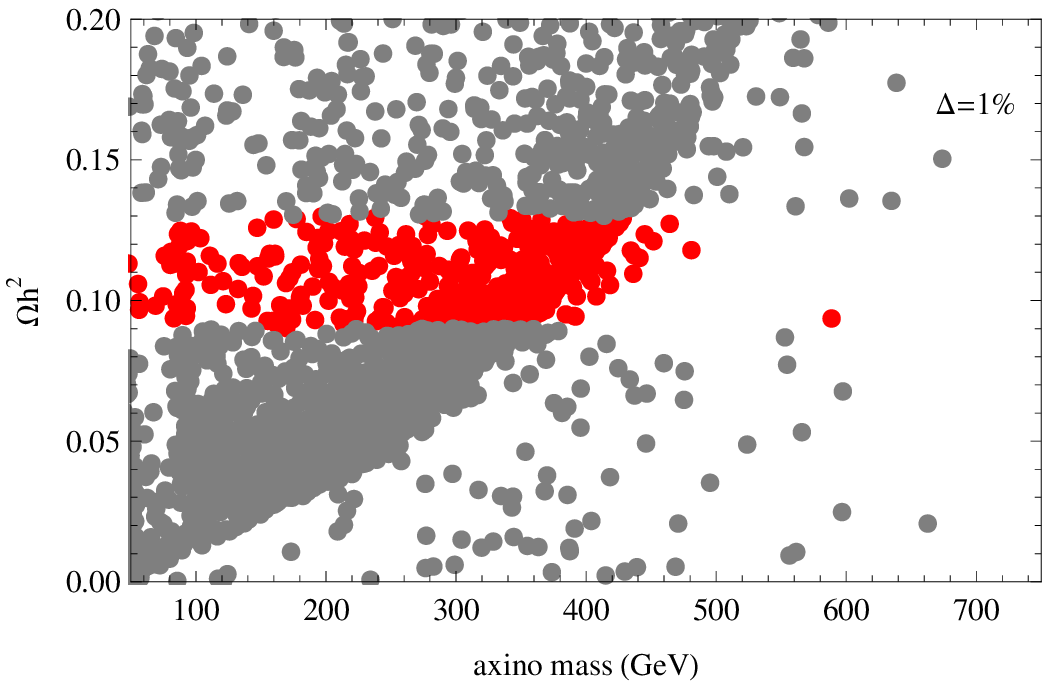}
      \includegraphics[scale=0.65]{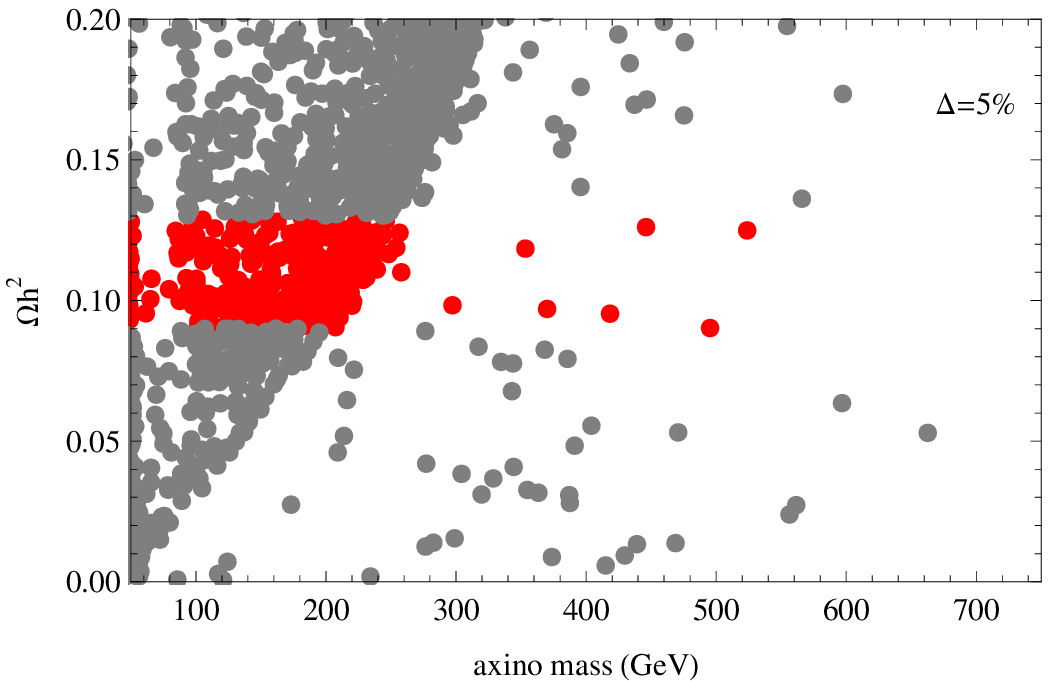}
      \caption{Axino relic density in the case in which the NSLP is a linear combination bino-higgsino. Red (darker) points denote models which satisfy WMAP data. Left panel: $\Delta_2=1\%$. Right panel:  $\Delta_2=5\%$.}
\label{DMbino-higgsino}
\end{center}
\end{figure}
In order to fulfill the WMAP data (red (darker) points in the plots in Fig.~\ref{DMbino-higgsino}) the axino mass must be in the range $50\; \text{GeV}\lesssim M_S\lesssim 700\; \text{GeV}$ in the limit $\Delta_2\to 0$, where the lowest bound is given by the current experimental constraints~\cite{Amsler:2008zzb}.
The range $500\; \text{GeV}\lesssim M_S\lesssim 700\; \text{GeV}$ is compatible with the LSP mass range
assumed in Sections~\ref{subsec:Tree level decay width} and \ref{sec: LHC}.
\item $N=3$ case. This is the case in which there is a third MSSM particle almost degenerate in mass with the LSP and the NLSP. Typical situations of this kind arise when the NLSP and the next to next to lightest supersymmetric particle (NNLSP) are respectively the bino and the stau or the wino and the lighest chargino.
Expanding in an explicit way all the terms in the sum~(\ref{sigmaeff}) we get:
\bea
 \la \s_\text{eff}^{(3)} v \ra &=& \la \s_{11} v \ra \g_1^2 + \la \s_{12} v \ra  \g_1 \g_2 + \la \s_{13} v \ra  \g_1 \g_3 +\nn\\
                 &&\la \s_{21} v \ra  \g_2 \g_1 + \la \s_{22} v \ra  \g_2^2 + \la \s_{23} v \ra  \g_2 \g_3 +\nn\\
                 &&\la \s_{31} v \ra  \g_3 \g_1 + \la \s_{32} v \ra  \g_3 \g_2 + \la \s_{33} v \ra  \g_3^2 \nn\\
               &=& \Big[ \la \s_{11} v \ra  (\numeq_1)^2 + \la \s_{12} v \ra  \numeq_1 \numeq_2 + \la \s_{13} v \ra  \numeq_1 \numeq_3 + \nn\\
                 &&\ \la \s_{21} v \ra  \numeq_2 \numeq_1 + \la \s_{22} v \ra  (\numeq_2)^2 + \la \s_{23} v \ra  \numeq_2 \numeq_3 +\nn\\
                 &&\phantom{\Big[}
                  \la \s_{31} v \ra  \numeq_3 \numeq_1 + \la \s_{32} v \ra  \numeq_3 \numeq_2 + \la \s_{33} v \ra  (\numeq_3)^2 \Big]
                 \frac{1}{(\numeq)^2} \nn\\
               &\simeq& 
\frac{ \big[ \la \s_{22} v \ra  (\numeq_2)^2 + 2 \la \s_{23} v \ra  \numeq_2 \numeq_3 +
                 \la \s_{33} v \ra  (\numeq_3)^2 \big] }{(\numeq)^2}
\eea
where in the last line we have neglected the terms $\la \s_{11} v \ra $, $\la \s_{12} v \ra $ and $\la \s_{13} v \ra $ since these are the thermal averaged cross sections which involve the axino.
By introducing a new set of variables defined by
\be
 Q_i=\frac{\numeq_i}{\numeq_1}= \frac{g_i}{g_1} (1 + \D_i)^{3/2} e^{-x_f \D_i} \qquad \text{for} \ i=2,3
\ee
where $g_i$ are the internal degrees of freedom of the particle species, $x_f=m_1/T$ and $\D_i=(m_i-m_1)/m_1$, we obtain
\bea
 \la \s_\text{eff}^{(3)} v \ra
               &\simeq& \frac{\la \s_{22} v \ra  Q_2^2 + 2 \la \s_{23} v \ra  Q_2 Q_3 + \la \s_{33} v \ra  Q_3^2}{\(1+Q_2+Q_3\)^2}
\eea
Under the assumption $(m_3-m_2)/m_1 \ll 1 /x_f$, $Q_3/Q_2 \simeq g_3/g_2$ we finally get
\be
 \la \s_\text{eff}^{(3)} v \ra \simeq \frac{Q_2^2}{\[1+ \(1 + \frac{g_3}{g_2}\) Q_2\]^2} \la \s_\text{MSSM} v \ra
\label{sigma3eff}
\ee
where
\be
 \la \s_\text{MSSM} v \ra  = \la \s_{22} v \ra  + 2 \frac{g_3}{g_2} \la \s_{23} v \ra  + \(\frac{g_3}{g_2}\)^2 \la \s_{33} v \ra
\ee
In order to compute the rescaling factor between the relic density of our model and the MSSM relic density we have to express $\s_\text{MSSM}$ in terms of a two coannihilating species effective cross section.
This is given by
\bea
 \la \s_\text{eff}^{(2)} v \ra &=& \frac{ \la \s_{22} v \ra  (\numeq_2)^2 + 2 \la \s_{23} v \ra  \numeq_2 \numeq_3 + \la \s_{33} v \ra  (\numeq_3)^2}{(\numeq)^2} \nn\\
               &=& \frac{ \la \s_{22} v \ra  (\numeq_2)^2 + 2 \la \s_{23} v \ra  \numeq_2 \numeq_3 + \la \s_{33} v \ra  (\numeq_3)^2}{(\numeq_2 + \numeq_3)^2} \nn\\
               &=& \frac{ \la \s_{22} v \ra  (\numeq_2)^2 + 2 \la \s_{23} v \ra  \numeq_2 \numeq_3 + \la \s_{33} v \ra  (\numeq_3)^2}{(\numeq_2)^2 (1 + \numeq_3/\numeq_2)^2}\nn\\
               &=& \frac{ \la \s_{22} v \ra   + 2 \la \s_{23} v \ra  Q_{23} + \la \s_{33} v \ra  Q_{23}^2}{ (1 + Q_{23})^2}
\eea
where
\bea
 Q_{23}&=&\numeq_3/\numeq_2= \frac{g_3}{g_2} \(1 + \frac{m_3-m_2}{m_2}\)^{3/2} e^{-x_f \frac{m_3-m_2}{m_2}}\nn\\
       &\simeq& \frac{g_3}{g_2}
\eea
since $(m_3-m_2)/m_1 \ll 1 /x_f$ and $m_2>m_1$ then $(m_3-m_2)/m_2 \ll 1 /x_f$.
We remind the reader that the values of $\numeq_2$, $\numeq_3$ and $\numeq$ are different with respect to those in the former case since now there are only two species in the thermal bath.
\begin{figure}[ht!]
      \begin{center}
      \includegraphics[scale=0.65]{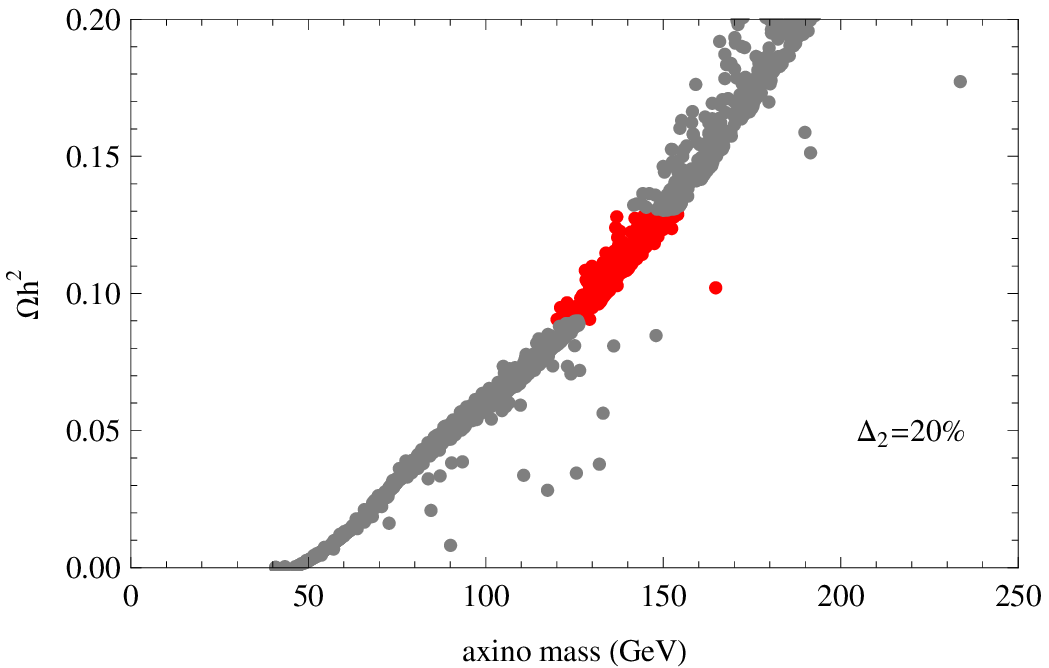}
      \includegraphics[scale=0.65]{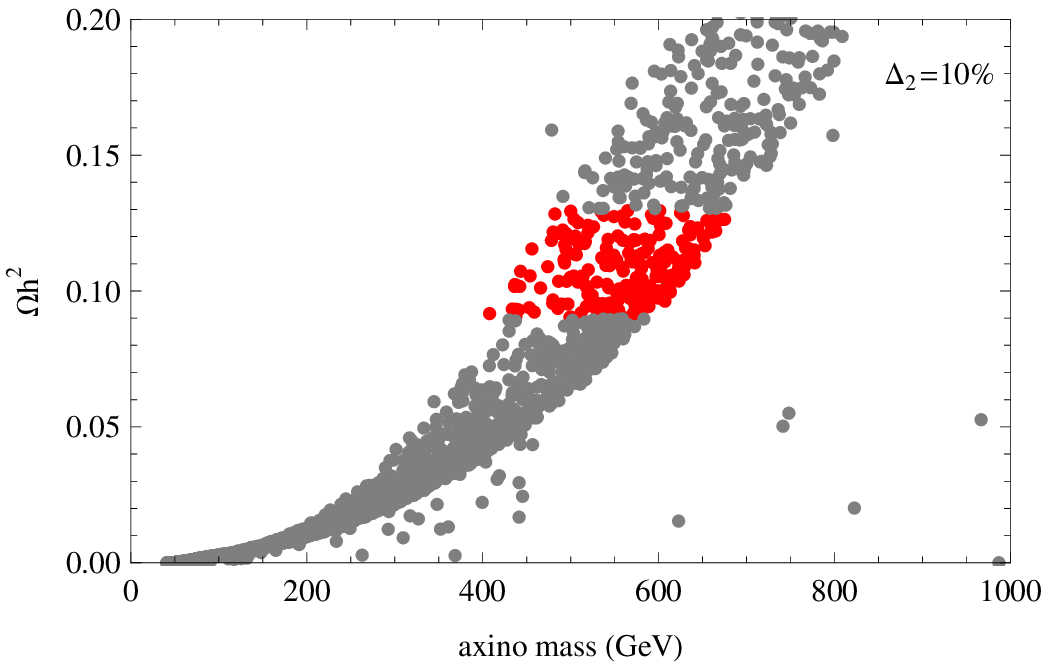}
      \includegraphics[scale=0.65]{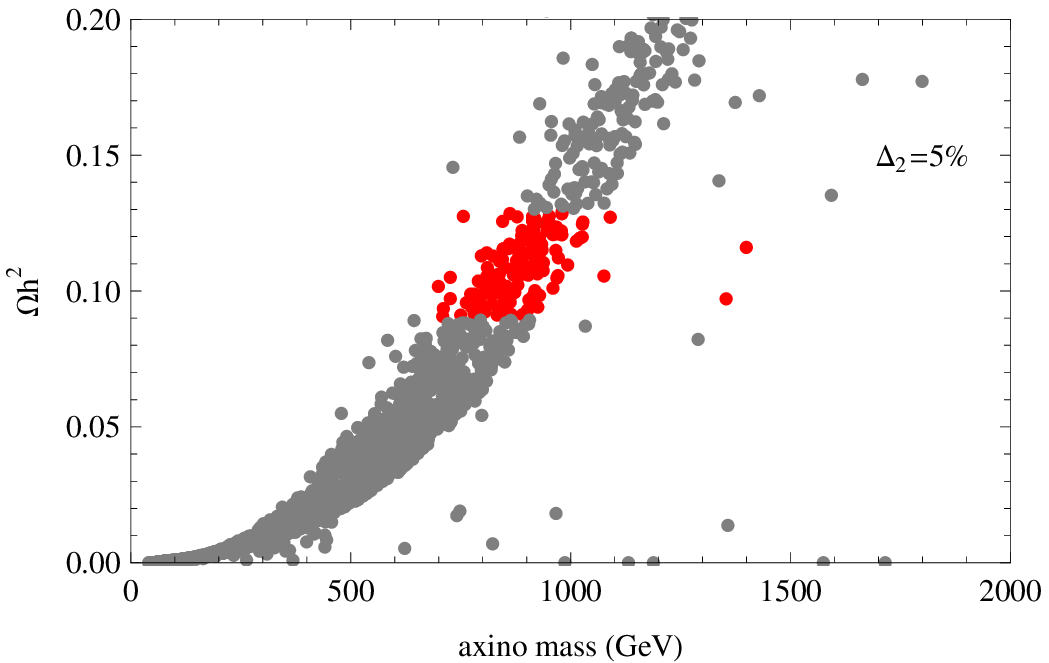}
      \includegraphics[scale=0.65]{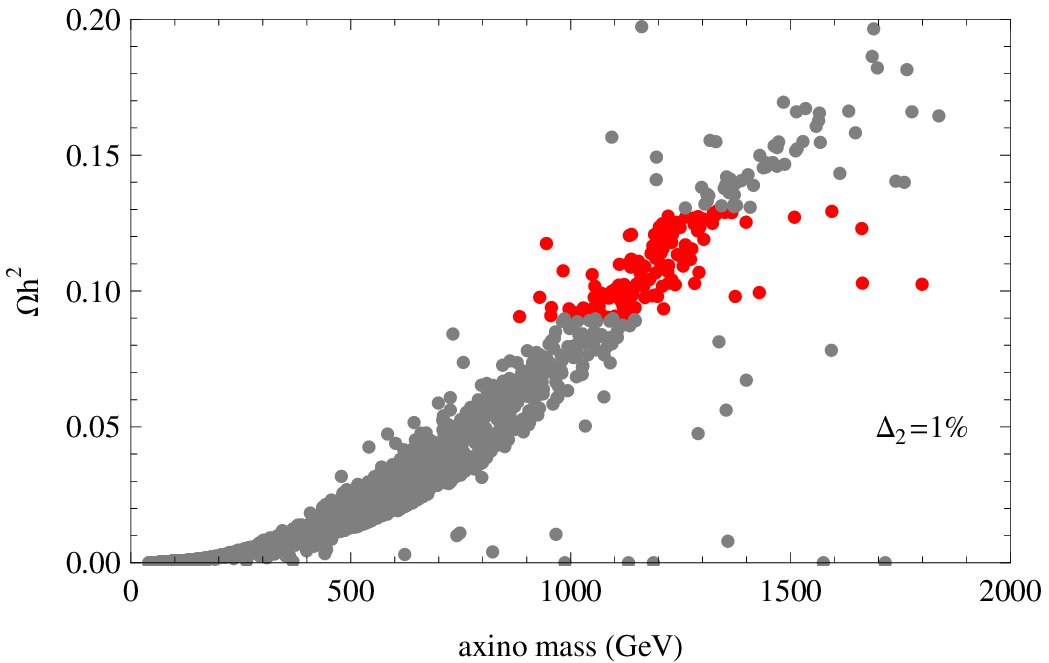}
     \caption{Axino relic density in the case in which the NSLP is a wino while the NNLSP is the lightest chargino. Red (darker) points denote models which satisfy WMAP data. Upper left panel: $\Delta_2=20\%$. Upper right panel:  $\Delta_2=10\%$. Lower left panel: $\Delta_2=5\%$. Lower right panel: $\Delta_2=1\%$.}
\label{DMwino-chargino}
\end{center}
\end{figure}
We then find
\bea
 \la \s_\text{eff}^{(2)} v \ra &\simeq& \frac{ \la \s_{22} v \ra  + 2 \frac{g_3}{g_2} \la \s_{23} v \ra  + \(\frac{g_3}{g_2}\)^2 \la \s_{33} v \ra  }{ \(1 + \frac{g_3}{g_2}\)^2} \nn\\
                     &\simeq& \frac{\la \s_\text{MSSM} v \ra  }{\(1 + \frac{g_3}{g_2}\)^2}
\eea
and inserting back this relation into~(\ref{sigma3eff}) we obtain
\be
 \la \s_\text{eff}^{(3)} v \ra \simeq \[ \frac{\(1 + \frac{g_3}{g_2}\) Q_2}{1+ \(1 + \frac{g_3}{g_2}\)  Q_2} \]^2 \la \s_\text{eff}^{(2)} v \ra
\ee
The rescaling factor between the three and two particle species relic density is given by the following relation
\be
 \( \Omega h^2\)^{(3)} \simeq \[ \frac{1+ \(1 + \frac{g_3}{g_2}\)  Q_2}{\(1 + \frac{g_3}{g_2}\) Q_2} \]^2 \( \Omega h^2\)^{(2)}\label{rescrelic2to3}
\ee
We performed a random sampling of MSSM models with bino-stau and wino-chargino coannihilations. The first situation is realized in some corners of the mSUGRA parameter space\footnote{or in the so called Constrained MSSM (CMSSM).} while the second situation is naturally realized in anomaly mediated supersymmetry breaking scenarios. For each model we computed the relic density $\( \Omega h^2\)^{(2)}$ for the two coannihilating species with the DarkSUSY package~\cite{Gondolo:2004sc}. We finally computed $\( \Omega h^2\)^{(3)}$ using~(\ref{rescrelic2to3}).
The bino-stau models which satisfy the WMAP constraints
have an axino mass in the range $100\; \text{GeV}\lesssim M_S\lesssim 350\; \text{GeV}$ in the limit $\Delta_2\to 0$. As the mass gap increases the number of allowed models drastically decreases and eventually vanishes for $\Delta\simeq 5\%$.
In the wino-chargino case, models which satisfy the WMAP constraints are shown in Fig.~\ref{DMwino-chargino} for four reference values of $\Delta_2$.
The space of parameters with $\Delta_2\lesssim 5\%$ and an axino mass $M_S\gtrsim 700$ GeV is favored while as the mass gap increases lower axino masses become favored, e.g. $100\; \text{GeV} \lesssim M_S<200$ GeV ($\Delta_2\simeq 20\%$).
The range $M_S\gtrsim 700$ GeV is compatible with the LSP mass range assumed in Sections~\ref{subsec:Tree level decay width} and  \ref{sec: LHC}.
\end{itemize}

Finally let us comment on the differences between our scenario and that studied in the work~\cite{Feldman:2006wd}. In our framework the $U(1)'$ does not arise from a hidden sector and thus all the MSSM fields can be charged under this extra abelian gauge group.
This is the most relevant feature which could also be detected experimentally (see for example~\cite{Langacker:2008yv}).
Moreover in our scenario the axino interactions are suppressed with respect to the weak interactions due to the GS couplings while in~\cite{Feldman:2006wd} the mechanism to suppress the couplings and give an XWIMP is provided by the kinetic mixing between the $U(1)'$ and $U(1)_Y$.

\newpage\thispagestyle{empty}

\chapter{Conclusions} \label{chap:Conclusions}
In this Thesis, we have reviewed an extension of the MSSM by the addition of an anomalous abelian vector multiplet and showed
some original results concerning the phenomenology of an anomalous $Z'$.

In Chapter~\ref{chap:OverviewMSSM} we introduced the MSSM and its main properties such as the gauge coupling unification.

In Chapter~\ref{chap:Anomalies} we presented the chiral anomalies in quantum field theory and their cancellation by the standard
procedure or by a string inspired mechanism firstly discovered by Green and Schwarz.
Then we discussed the Green-Schwarz mechanism in a standard and in a supersymmetric framework showing which are its main features and
physical consequences.

In Chapter~\ref{chap:MiAUMSSM. Model building} we introduced our model, the MiAUMSSM.
We showed how the MSSM modified by the addition of an anomalous $U(1)$ gauge field, a \Stuckelberg multiplet and Chern-Simons terms.
We checked the correct electroweak symmetry breaking and we found new contributions to the mass matrices for almost all
the particle in the spectrum with tree level or 1-loop intensity.

In Chapter~\ref{chap:MiAUMSSM. Phenomenology} we presented the first phenomenological results of our model for the choice $\QHu=0$.
We computed the $Z'$ production at LHC for variable $\MZp$, noticing that the number of the $Z'$ produced
falls off exponentially with its mass, so the most favoured case is $M_{Z'} \sim 1$ TeV.
Then we estimated the number of leptonic decays and anomalous decays for a $Z'$ with $M_{Z'} \sim 1$ TeV
for 1 year of integrated luminosity in LHC.
We saw that in a wide region of parameter we have more than $10^4$ $Z' \to e^- e^+$ decays,
while the maximum number of $Z' \to Z_0 Z_0$ is of the order of 10.
Then we studied a possible dark matter candidate in the framework of our model.
In the decoupling limit $\QHu=0$ and under the assumption $M_0\gg M_S,M_{V^{(0)}}$ the axino turns out to be the LSP.
Being an XWIMP, in order to satisfy the WMAP constraints on the relic density we must have at least a NLSP
almost degenerate in mass with the axino. We considered the case with two and three coannihilating particles
and we found some configuration which satisfies the WMAP constraints.
In the exact degeneracy limit $\Delta_2\to 0$ the allowed models have an axino mass
in the range $50\; \text{GeV}\lesssim M_S\lesssim 700\; \text{GeV}$ for the bino-higgsino
coannihilation case while $900\; \text{GeV}\lesssim M_S\lesssim 2\; \text{TeV}$ for the wino-chargino coannihilation case.

The eventual discover of extra gauge bosons and dark matter particles are some of the main issues in today high energy physics.
In the next months, the new colliders and satellite experiments, such as LHC and PAMELA will give us clues
on the fundamental theory underlying the Standard Model.

\newpage\thispagestyle{empty}

\appendix


\chapter{Gammology} \label{app:Gammology}
\section{Notations and conventions} \label{sec:Notations and conventions}
We use the space-time metric $\eta_{\m \n} = \text{diag}(+,-,-,-)$ and the spinorial conventions
  \be
   \e_{21}=\e^{12}=1 \qquad \e_{12}=\e^{21}=-1 \qquad \e_{11}=\e^{11}=\e_{22}=\e^{22}=0 \label{epsmetric}
  \ee
  \be
   \psi^\a = \e^{\a \b} \psi_\b \qquad \psi_\a = \e_{\a \b} \psi^\b \qquad
   \bar\psi^\ad = \e^{\ad \bd} \bar\psi_\bd \qquad \bar\psi_\ad = \e_{\ad \bd} \bar\psi^\bd \label{psi1}
  \ee
  \be
   \psi \chi = \psi^\a \chi_\a   \qquad \bar\psi \bar\chi = \bar\psi_\ad \bar\chi^\ad \label{psi2}
  \ee
The Dirac matrices are
\be
\g^\m=\( \begin{array}{cc}   0   &  \s^\m \\
                         \sb^\m  &  0 \end{array} \) \qquad \text{where} \quad
\left\{\begin{array}{rcl} \s^\m&=&(1,-\vec\s)\\
                         \sb^\m&=&(1,\vec\s) \ \end{array} \right. \label{gmatrices}
\ee
and we define
\be
\g_5=\( \begin{array}{cc} -1 &  0 \\
                           ~0 &  1 \end{array} \) \label{g5matrix}
\ee
\section{Properties of $\g$ matrices and trace theorems} \label{sec:Properties of g matrices and trace theorems}
The basic properties of $\g$ matrices are
\be
 \{ \g^\m , \g^\n\} = 2 \eta^{\m\n} \qquad \text{and} \qquad  \{ \g^\m , \g_5\} = 0 \label{ganticomm}
\ee
The trace theorems are
\bea
 &&\Tr 1 = 4 \nn\\
 &&\Tr \( \text{odd number of $\g$'s matrices} \)=0 \nn\\
 &&\Tr (\g^\m \g^\n) = 4 \eta^{\m\n} \\
 &&\Tr (\g^\m \g^\n \g^\r \g^\s) = 4 \[ \eta^{\m\n} \eta^{\r\s} - \eta^{\m\r} \eta^{\n\s} + \eta^{\m\s} \eta^{\n\r} \] \nn\\
 &&\Tr \g_5 = 0 \\
 &&\Tr \[\g_5 \(\text{odd number of $\g$'s matrices} \) \]=0 \nn\\
 &&\Tr (\g_5 \g^\m \g^\n) = 0 \nn\\
 &&\Tr (\g_5 \g^\m \g^\n \g^\r \g^\s) = 4 i \e^{\m \n \r \s} \nn\\
 &&\Tr (\g^\m \g^\n \g^\r \g^\s \dots) = \Tr (\dots \g^\s \g^\r \g^\m \g^\n) \label{tracetheorems}
\eea
where $\e^{\m \n \r \s}= 1 (- 1)$ for $\m$, $\n$, $\r$, $\s$ an even (odd) permutation of 0,1,2,3; and 0 if two indices are the same.

Other useful results for simplifying trace calculations are ($\as=a_\m \g ^\m$)
\bea
 \g_\m \g^\m &=& 4 \nn\\
 \g_\m \as \g^\m &=& - 2 \as \nn\\
 \g_\m \as \bs \g^\m &=& 2 a \cdot b \nn\\
 \g_\m \as \bs \cs \g^\m &=& -2 \cs \bs \as \label{gotherresults}
\eea

\newpage\thispagestyle{empty}
\chapter{Lagrangians} \label{app:Lagrangians}
\section{MSSM Lagrangian} \label{app:MSSM Lagrangian}
The Lagrangian of the MSSM contains several terms
  \be \label{lagrangian_MSSM}
   \L^{MSSM} = \L_Q^{MSSM} + \L_L^{MSSM} + \L_{gauge}^{MSSM} + \L_H^{MSSM} + \L_{W}^{MSSM} + \L_{Soft}^{MSSM}
  \ee
where
\bea
 \L_Q^{MSSM} &=& \( Q_i^\dag e^{V^{(3)}} e^{V^{(2)}} e^{{1\over6}V^{(1)}} Q_i \right. \\
             &&~+\left.(U^c_i)^\dag e^{-V^{(3)}} e^{-{2\over3}V^{(1)}} U^c_i +
               (D^c_i)^\dag e^{-V^{(3)}} e^{{1\over3}V^{(1)}} D^c_i\)_{\thth} \nn\\
\nn\\
 \L_L^{MSSM} &=& \( L_i^\dag e^{V^{(2)}} e^{-{1\over2}V^{(1)}} L_i + (E^c_i)^\dag e^{V^{(1)}} E^c_i \)_{\thth} \\
\nn\\
 \L_H^{MSSM} &=& \( H_u^\dag e^{V^{(2)}} e^{{1\over2}V^{(1)}} H_u +
               H_d^\dag e^{V^{(2)}} e^{-{1\over2}V^{(1)}} H_d \)_{\thth} \\
\nn\\
 \L_{gauge}^{MSSM} &=& \( {1\over8 g_3^2} \Tr\( W^{(3)} W^{(3)} \) + {1\over8 g_2^2} \Tr\( W^{(2)} W^{(2)} \)  \right.\nn\\
         &&~+ \left.{1\over16 g_1^2} \Tr\( W^{(1)} W^{(1)} \) \)_{\th^2} +h.c.\\
\nn\\
 \L_{W}^{MSSM} &=&\(  y_u^{i j} Q_i U^c_j H_u - y_d^{i j} Q_i D^c_j H_d - y_e^{i j} L_i E^c_j H_d + \m H_u H_d\)_{\th^2} + h.c. \quad
\eea
\bea
\L_{soft}^{MSSM} &=& - {1\over2}  \sum_{a=1}^3 \(M_a \l^{(a)} \l^{(a)} + \conj \)
              -\( m^2_{q_{ij}} \qt_i \qt_j^\dag + m^2_{u_{ij}} \ut^c_i \utcs_j + m^2_{d_{ij}} \dt^c_i \dtcs_j  \right. \nn\\
                 &&\left.+m^2_{l_{ij}} \lt_i \lt_j^\dag + m^2_{e_{ij}} \et^c_i \etcs_j + m^2_{h_u} |h_u|^2 + m^2_{h_d} |h_d|^2 \) \nn\\
                 &&-\( a_u^{ij} \qt_i \qt_j h_u - a_d^{ij} \qt_i \dt_j h_d - a_e^{ij} \lt_i \et_j h_d + b h_u h_d + \conj \)
   \eea
Notice that in order to include the coupling constants in the gauge interactions we need to substitute $V \to 2 g V$.

\section{MiAUMSSM Lagrangian} \label{app:MiAUMSSM Lagrangian}
The Lagrangian of the model contains several terms
  \be \label{lagrangian}
   \L = \L_Q + \L_L + \L_{gauge} + \L_H + \L_{W} + \L_{\axion} + \L_{\GCS} + \L_{Soft}
  \ee
where
\bea
 \L_Q &=& \( Q_i^\dag e^{V^{(3)}} e^{V^{(2)}} e^{{1\over6}V^{(1)}} e^{Q_Q V^{(0)}}Q_i \right. \\
             &&+\left.(U^c_i)^\dag e^{-V^{(3)}} e^{-{2\over3}V^{(1)}} e^{Q_{U^c} V^{(0)}}U^c_i +
               (D^c_i)^\dag e^{-V^{(3)}} e^{{1\over3}V^{(1)}} e^{Q_{D^c} V^{(0)}}D^c_i\)_{\thth} \nn\\
\nn\\
 \L_L &=& \( L_i^\dag e^{V^{(2)}} e^{-{1\over2}V^{(1)}} e^{Q_L V^{(0)}}L_i + (E^c_i)^\dag e^{V^{(1)}} e^{Q_{E^c} V^{(0)}}E^c_i \)_{\thth} \\
\nn\\
 \L_H &=& \( H_u^\dag e^{V^{(2)}} e^{{1\over2}V^{(1)}} e^{Q_{H_u} V^{(0)}}H_u +
               H_d^\dag e^{V^{(2)}} e^{-{1\over2}V^{(1)}} e^{Q_{H_d} V^{(0)}}H_d \)_{\thth} \\
\nn\\
 \L_{gauge} &=& \( {1\over8 g_3^2} \Tr\( W^{(3)} W^{(3)} \) + {1\over8 \tilde g_2^2} \Tr\( W^{(2)} W^{(2)} \)  \right. \nn\\
         &&+ \left.{1\over16 \tilde g_1^2} \Tr\( W^{(1)} W^{(1)} \) + {1\over16 \tilde g_0^2} \Tr\( W^{(0)} W^{(0)} \)\)_{\th^2} +h.c.\\
\nn\\
 \L_{W} &=&\(  y_u^{i j} Q_i U^c_j H_u - y_d^{i j} Q_i D^c_j H_d - y_e^{i j} L_i E^c_j H_d + \m H_u H_d\)_{\th^2} + h.c.  \\
\nn\\
 \L_{\axion} &=& {1\over4} \left. \( S + \bar S + 4 b_3 V^{(0)} \)^2 \right|_{\thth} \nn\\
                &&- {1\over2} \left\{ \[\sum_{a=0}^2 b^{(a)}_2 S ~\Tr\( W^{(a)} W^{(a)} \) + b^{(4)}_2 S ~W^{(1)} ~W^{(0)} \]_{\th^2} +h.c. \right\}
\eea
  \bea
   \L_{\GCS} &=&+ d_4     \[ \( V^{(1)} D^\a V^{(0)} - V^{(0)} D^\a V^{(1)}\) W^{(0)}_\a + h.c. \]_{\thth} +\nn\\
             &&-  d_5     \[ \( V^{(1)} D^\a V^{(0)} - V^{(0)} D^\a V^{(1)}\) W^{(1)}_\a + h.c. \]_{\thth} +\nn\\
             &&-  d_6 \Tr \bigg[ \( V^{(2)} D^\a V^{(0)} - V^{(0)} D^\a V^{(2)}\) W^{(2)}_\a +\nn\\
                          &&\qquad \quad+{1\over6} V^{(2)} D^\a V^{(0)} \bar D^2 \(  \[D_\a V^{(2)},V^{(2)}\] \) + h.c. \bigg]_{\thth}
    \eea
   \bea
 \L_{soft} &=& - {1\over2}  \sum_{a=0}^3 \(M_a \l^{(a)} \l^{(a)} + \conj \)  - {1\over2}  \(\frac{M_S}{2} \psi_S \psi_S  + h.c. \)\nn\\
              &&-\( m^2_{q_{ij}} \qt_i \qt_j^\dag + m^2_{u_{ij}} \ut^c_i \utcs_j + m^2_{d_{ij}} \dt^c_i \dtcs_j  \right. \nn\\
                 &&\left.+m^2_{l_{ij}} \lt_i \lt_j^\dag + m^2_{e_{ij}} \et^c_i \etcs_j + m^2_{h_u} |h_u|^2 + m^2_{h_d} |h_d|^2 \) \nn\\
                 &&-\( a_u^{ij} \qt_i \qt_j h_u - a_d^{ij} \qt_i \dt_j h_d - a_e^{ij} \lt_i \et_j h_d + b h_u h_d + \conj \)
   \eea
where $\L_Q$, $\L_L$ and $\L_H$ provide the kinetic terms and the gauge interactions of the matter particles such as (s)quarks,
(s)leptons, Higgs(ino)s; $\L_{gauge}$ contains the kinetic terms for the gauge supermultiplet; $\L_W$ is the usual MSSM
superpotential; $\L_{\axion}$ provides the kinetic term of the St\"uckelberg multiplet and its Green-Schwarz interactions used
in the anomaly cancellation procedure; $\L_{\GCS}$ contains the Generalized Chern Simons interactions giving trilinear gauge boson
couplings needed to complete the anomaly cancellation procedure; finally, $\L_{Soft}$ contains the usual soft breaking
terms of the MSSM as well as the new terms for the primeino and the axino.

Notice that in order to include the coupling constants in the gauge interactions we need to redefine them as shown in equation
(\ref{couplingconst}) and to substitute $V \to 2 g V$.

\newpage\thispagestyle{empty}
\chapter{Feynman rules} \label{app:Feynman rules}

In this appendix we collect the Feynman rules for the interaction vertices that are involved in our computations.
We give the rules for the $C-B$ model described in Section \ref{sect:Triangular Anomaly} and some rules for the
SM and the MSSM.
Moreover the terms involved in the GS mechanism shown in Section \ref{sect:Anomaly cancellation. Green-Schwarz mechanism}
and \ref{subsec: Anomalies and GS mechanism} are included.
The generalization to the other anomalous diagrams is straightforward, so it is omitted.
The solid lines represent fermions and the wiggle lines are gauge fields. Dashed lines are scalars.
%
%
%
%
%
%
%
%
%
%
%
%
%
%
%
%
%


\section{$C-B$ model} \label{sec:C-B model}
\subsection*{Tree level}
\bea
 \text{fermionic vertex (B):}
 \raisebox{-3.8ex}[0cm][0cm]{\unitlength=0.4mm
 \begin{fmffile}{Bvertex}
 \begin{fmfgraph*}(60,40)
 \fmfpen{thick} \fmftop{o1} \fmfbottom{i1,i2} \fmf{boson}{c1,o1} \fmf{fermion}{i1,c1}  \fmf{fermion}{c1,i2}
 \fmflabel{$B_\m$}{o1} \fmflabel{$f$}{i1} \fmflabel{$f$}{i2}
 \end{fmfgraph*}
 \end{fmffile}}
 &=& - \frac{i}{2} \g_\m \( v_f^{B} -a_f^{B} \g_5 \) \label{Bvertex} \\
\nn\\
\nn\\
\nn\\
 \text{fermionic vertex (C):}
 \raisebox{-3.8ex}[0cm][0cm]{\unitlength=0.4mm
 \begin{fmffile}{Cvertex}
 \begin{fmfgraph*}(60,40)
 \fmfpen{thick} \fmftop{o1} \fmfbottom{i1,i2} \fmf{boson}{c1,o1} \fmf{fermion}{i1,c1}  \fmf{fermion}{c1,i2}
 \fmflabel{$C_\m$}{o1} \fmflabel{$f$}{i1} \fmflabel{$f$}{i2}
 \end{fmfgraph*}
 \end{fmffile}}
 &=& - \frac{i}{2} \g_\m \( v_f^{C} -a_f^{C} \g_5 \) \label{Cvertex}\\\nn
 \eea
\vskip 0.25cm
\begin{flushleft}
The vectorial couplings $v_f^{B,C}$ and the axial couplings $a_f^{B,C}$ are given in eq.~(\ref{chiralvec}) and (\ref{chiralaxi}).
\end{flushleft}
\subsection*{GS mechanism}
%
%
\bea
 \text{GS vertex:}
 \raisebox{-3.8ex}[0cm][0cm]{\unitlength=0.4mm
 \begin{fmffile}{GSvertex}
 \begin{fmfgraph*}(60,40)
 \fmfpen{thick} \fmftop{o1} \fmfbottom{i1,i2}
 \fmf{dashes,label=$\downarrow$}{c1,o1} \fmf{photon,label=\begin{rotate}{22}$\!\!\!\!\underleftarrow{}$\end{rotate}}{i1,c1}  \fmf{photon,label.side=right,label=\begin{rotate}{-22}$\!\!\underrightarrow{}$\end{rotate}}{c1,i2}
 \fmflabel{$\f$}{o1} \fmflabel{$B_\m(p)$}{i1} \fmflabel{$B_\n(q)$}{i2}
 \end{fmfgraph*}
 \end{fmffile}}
  &=& -i \, b^{CBB} \, \e_{\m\n\r\s} p^\r q^\s \label{GSvertex} \\
\nn\\
\nn\\
\nn\\
\nn\\
 \text{GCS vertex:}
 \raisebox{-3.8ex}[0cm][0cm]{\unitlength=0.4mm
 \begin{fmffile}{GCSvertex3}
 \begin{fmfgraph*}(60,40)
 \fmfpen{thick} \fmftop{o1} \fmfbottom{i1,i2}
 \fmf{boson,label=$\downarrow$}{c1,o1}
 \fmf{boson,label=\begin{rotate}{22}$\!\!\!\!\underleftarrow{}$\end{rotate}}{i1,c1}
 \fmf{boson,label.side=right,label=\begin{rotate}{-22}$\!\!\underrightarrow{}$\end{rotate}}{c1,i2}
 \fmflabel{$C_\r \ (p+q)$}{o1} \fmflabel{$B_\m (p)$}{i1} \fmflabel{$B_\n (q)$}{i2}
 \end{fmfgraph*}
 \end{fmffile}}  &=& 2 \, d^{CBB} \, \e_{\r\n\m\a} p^\a \label{GCSvertex}\\\nn
 \\\nn
 \eea
The GS coupling $b^{CBB}$ and the GCS coupling $d^{CBB}$ are given respectively in eq.~(\ref{bCBB}) and eq.~(\ref{dCBB}).
\section{Electroweak interactions} \label{sec:Electroweak interactions}
\subsection*{QED}
\bea
 \text{fermionic vertex:}
 \raisebox{-3.8ex}[0cm][0cm]{\unitlength=0.4mm
 \begin{fmffile}{QEDvertex2}
 \begin{fmfgraph*}(60,40)
 \fmfpen{thick} \fmftop{o1} \fmfbottom{i1,i2} \fmf{boson}{c1,o1} \fmf{fermion}{i1,c1}  \fmf{fermion}{c1,i2}
 \fmflabel{$A_\m$}{o1} \fmflabel{$f$}{i1} \fmflabel{$f$}{i2}
 \end{fmfgraph*}
 \end{fmffile}}
  &=& -i \, e \, q_f \, \g_\m \label{QEDvertex} \\
\nn\\
\nn\\
\nn\\
\nn\\
 \text{scalar vertex:}
 \raisebox{-3.8ex}[0cm][0cm]{\unitlength=0.4mm
 \begin{fmffile}{scalarQEDvertex4}
 \begin{fmfgraph*}(60,40)
 \fmfpen{thick} \fmftop{o1} \fmfbottom{i1,i2}
 \fmf{boson,label=$\downarrow$}{c1,o1}
 \fmf{dashes,label=\begin{rotate}{22}$\!\!\!\!\underleftarrow{}$\end{rotate}}{i1,c1}
 \fmf{dashes,label.side=right,label=\begin{rotate}{-22}$\!\!\underrightarrow{}$\end{rotate}}{c1,i2}
 \fmflabel{$A_\m(p+q)$}{o1} \fmflabel{$\tilde f(p)$}{i1} \fmflabel{$\tilde f(q)$}{i2}

 \end{fmfgraph*}
 \end{fmffile}}
  &=& i \, e \, q_f (p-q)_\m \label{scalarQEDvertex}\\\nn
  \\\nn
 \eea
The coupling constant $e$ and the charges $q_f$ are given respectively in eq.~(\ref{e}) and in Table \ref{Table:couplingsQH0}.
\newpage\subsection*{$Z_0$ interactions}
%
%
\bea
 \text{fermionic vertex:}
 \raisebox{-3.8ex}[0cm][0cm]{\unitlength=0.4mm
 \begin{fmffile}{Z0vertex}
 \begin{fmfgraph*}(60,40)
 \fmfpen{thick} \fmftop{o1} \fmfbottom{i1,i2}
 \fmf{boson}{c1,o1} \fmf{fermion}{i1,c1}  \fmf{fermion}{c1,i2}
 \fmflabel{${Z_0}_\m$}{o1} \fmflabel{$f$}{i1} \fmflabel{$f$}{i2}
 \end{fmfgraph*}
 \end{fmffile}}
  &=& - \frac{i}{2} g_{Z_0} \g_\m \( v_f^{Z_0} -a_f^{Z_0} \g_5 \) \qquad \label{Z0vertex}\\\nn
 \eea
%
The coupling constant $g_{Z_0}$ and the vectorial and axial couplings $v_f^{Z_0}$ and $a_f^{Z_0}$ are given respectively
in eq.~(\ref{gZ0}) and in Table \ref{Table:couplingsQH0}.
\subsection*{$W^\pm$ interactions}
%
%
\bea
 \text{fermionic vertex:}
 \raisebox{-3.8ex}[0cm][0cm]{\unitlength=0.4mm
 \begin{fmffile}{Wvertex}
 \begin{fmfgraph*}(60,40)
 \fmfpen{thick} \fmftop{o1} \fmfbottom{i1,i2} \fmf{boson}{c1,o1} \fmf{fermion}{i1,c1}  \fmf{fermion}{c1,i2}
 \fmflabel{$W_\m$}{o1} \fmflabel{$l (q_u)$}{i1} \fmflabel{$\n_l (q_d)$}{i2}
 \end{fmfgraph*}
 \end{fmffile}}
  &=& - \frac{i}{2} g_W  \, \g_\m (1-\g_5) \label{Wvertex} \\
\nn\\
\nn\\
\nn\\
\nn\\
 \text{$Z_0$ vertex:}
 \raisebox{-3.8ex}[0cm][0cm]{\unitlength=0.4mm
 \begin{fmffile}{trivecvertex}
 \begin{fmfgraph*}(60,40)
 \fmfpen{thick} \fmftop{o1} \fmfbottom{i1,i2}
 \fmf{boson,label=$\downarrow$}{c1,o1} \fmf{photon,label=\begin{rotate}{22}$\!\!\!\!\underleftarrow{}$\end{rotate}}{i1,c1}  \fmf{photon,label.side=right,label=\begin{rotate}{-22}$\!\!\underrightarrow{}$\end{rotate}}{c1,i2}
 \fmflabel{${Z_0}_\r(p+q)$}{o1} \fmflabel{$W_\m(p)$}{i1} \fmflabel{$W_\n(q)$}{i2}
 \end{fmfgraph*}
 \end{fmffile}}
  &=& i \, e \, \cot \theta_W \big[ \eta_{\r\m} (2p+q)_\n + \eta_{\m\n} (q-p)_\r +\nn\\
  &\phantom{=}&\phantom{i \, e \, \cot \theta_W \big[}
  - \eta_{\r\n} (p+2q)_\m \big] \label{Z0Wvertex}\\\nn
 \eea
where $g_W$ and $\theta_W$ are given in Section~\ref{subsect:Vector Bosons}.
\section{Yukawa's interactions} \label{sec:Yuk int}
%
%
\bea
 \text{NG-Yukawa vertex:}
 \raisebox{-3.8ex}[0cm][0cm]{\unitlength=0.4mm
 \begin{fmffile}{Yukawavertex1}
 \begin{fmfgraph*}(60,40)
 \fmfpen{thick} \fmftop{o1} \fmfbottom{i1,i2} \fmf{dashes}{c1,o1} \fmf{fermion}{i1,c1}  \fmf{fermion}{c1,i2}
 \fmflabel{$G^0$}{o1} \fmflabel{$f$}{i1} \fmflabel{$f$}{i2}
 \end{fmfgraph*}
 \end{fmffile}}
  &=&
  \left\{ \begin{array}{cc}
                      SM: & \frac{y_f}{\sqrt2} \, \g_5\\
                      MSSM \ f_u: & \frac{y_f}{\sqrt2} \sin\b \, \g_5  \quad\\
                      MSSM \ f_d: & -\frac{y_f}{\sqrt2} \cos\b \, \g_5  \qquad\\
                     \end{array} \right. \label{NG-Yukawavertex} \\\nn
\eea
%
%
where $y_f$ is the Yukawa coupling of the fermion, and $\b$ is the angle in eq.~(\ref{deftanbeta}).
\section{$Z'$ interactions} \label{sec:Zp int}
\subsection*{Tree level}
%
%
\bea
 \text{fermionic vertex:}
 \raisebox{-3.8ex}[0cm][0cm]{\unitlength=0.4mm
 \begin{fmffile}{Zpvertex1}
 \begin{fmfgraph*}(60,40)
 \fmfpen{thick} \fmftop{o1} \fmfbottom{i1,i2}
 \fmf{boson}{c1,o1} \fmf{fermion}{i1,c1}  \fmf{fermion}{c1,i2}
 \fmflabel{$Z'_\m$}{o1} \fmflabel{$f$}{i1} \fmflabel{$f$}{i2}
 \end{fmfgraph*}
 \end{fmffile}}
  &=& - \frac{i}{2} g_{Z'} \g_\m \( v_f^{Z'} -a_f^{Z'} \g_5 \) \label{Zpvertex} \\\nn
\\\nn
 \eea
The coupling constant $g_{Z'}=g_0$ for $\QHu=0$ and the vectorial and axial couplings $v_f^{Z'}$ and $a_f^{Z'}$
are given in Table \ref{Table:couplingsQH0}.
\subsection*{GS mechanism}
%
%
\bea
 \text{GS vertex:}
 \raisebox{-3.8ex}[0cm][0cm]{\unitlength=0.4mm
 \begin{fmffile}{GSvertexZp}
 \begin{fmfgraph*}(60,40)
 \fmfpen{thick} \fmftop{o1} \fmfbottom{i1,i2}
 \fmf{dashes,label=$\downarrow$}{c1,o1} \fmf{photon,label=\begin{rotate}{22}$\!\!\!\!\underleftarrow{}$\end{rotate}}{i1,c1}  \fmf{photon,label.side=right,label=\begin{rotate}{-22}$\!\!\underrightarrow{}$\end{rotate}}{c1,i2}
 \fmflabel{$\f$}{o1} \fmflabel{$V^{(1)}_\m(p)$}{i1} \fmflabel{$V^{(1)}_\n(q)$}{i2}
 \end{fmfgraph*}
 \end{fmffile}}
  &=& -4 i  g_1^2 \, b_2^{(1)} \, \e_{\m\n\r\s} p^\r q^\s \label{GSvertexZp} \\
\nn\\
\nn\\
\nn\\
\nn\\
 \text{GCS vertex:}
 \raisebox{-3.8ex}[0cm][0cm]{\unitlength=0.4mm
 \begin{fmffile}{GCSvertexZp}
 \begin{fmfgraph*}(60,40)
 \fmfpen{thick} \fmftop{o1} \fmfbottom{i1,i2}
 \fmf{boson,label=$\downarrow$}{c1,o1}
 \fmf{boson,label=\begin{rotate}{22}$\!\!\!\!\underleftarrow{}$\end{rotate}}{i1,c1}
 \fmf{boson,label.side=right,label=\begin{rotate}{-22}$\!\!\underrightarrow{}$\end{rotate}}{c1,i2}
 \fmflabel{$V^{(0)}_\r \ (p+q)$}{o1} \fmflabel{$V^{(1)}_\m (p)$}{i1} \fmflabel{$V^{(1)}_\n (q)$}{i2}
 \end{fmfgraph*}
 \end{fmffile}}  &=& 16 g_0 g_1^2 \, d_5 \, \e_{\r\n\m\a} p^\a \label{GCSvertexZp}\\\nn
 \nn\\
 \eea
The GS coupling $b_2^{(1)}$ and the GCS coupling $d_5$ are given in eq.~(\ref{bsds}).

\newpage\thispagestyle{empty}
\chapter{More on amplitudes} \label{app:More on amplitudes}
\section{Linearly divergent integrals} \label{app:Linearly divergent integrals}
Consider a shift of integration variable in a one dimensional integral
\be
 \D(a) = \int_{-\infty}^\infty dx \[ f(x+a) -f(x) \]
\ee
It can be easily shown that the shift may be illegitimate if the integral is divergent. Too see this, we expand
\bea
 \D(a) &=& \int_{-\infty}^\infty dx \[ a f'(x) + \frac{a^2}{2} f''(x) + \dots \] \nn\\
       &=& a [ f(\infty) - f(-\infty) ] + \frac{a^2}{2} [ f'(\infty) - f'(-\infty) ] + \dots
\eea
where the primes mean differentiation. When the integral $\int_{-\infty}^\infty dx f(x)$ converges
(or at most diverges logarithmically) we have $0=f(\pm \infty)=f'(\pm \infty)=\dots$ so $\D(a)=0$.
Instead, for a linearly divergent integral, $0 \neq f(\pm \infty)$ and $0=f'(\pm \infty)=\dots$ so we get
\be
 \D(a) = a [ f(\infty) - f(-\infty) ] \neq 0
\ee
which corresponds to a ``surface term ''. The generalization to $n$ dimensions is straightforward
\bea
 \D(a) &=& \int d^n r  \[ f(r+a) -f(r) \] \nn\\
       &=& \int d^n r  \[ a^\t \frac{\pd}{\pd r_\t} f(r)
                        + \frac{1}{2} a^\t a^\s \frac{\pd^2}{\pd r_\t \pd r_\s} f(r)+ \dots \]
\eea
Applying Gauss's theorem, all terms but the first vanish upon integrating over the surface at $r \to \infty$
\be
 \D(a) = a^\t \frac{r_\t}{r} f(r) S_n(r)
\ee
where $S_n(r)$ is the surface area of the hypersphere with radius $r$.
For the case of a four-dimensional Minkowski space-time, we have
\be
 \D(a) = a^\t \int d^4 r \pd_\t f(r) = 2i \pi^2 a^\t \lim_{r \to \infty} r^2 r_\t f(r) \label{MinkIntdiv}
\ee
\section{Ward Identities and Goldstones} \label{app:Ward Identities}
In this section we give an intuitive demonstration of the WI used in Fig.~\ref{WIZ0ee}. For a more rigorous and general
demonstration we remand to \cite{Chanowitz:1985hj}.

An external $Z_0$ is created by the Lagrangian term
\be
  \L^{int}_{Z_0} = J^\m_{Z_0} Z_{0 \m} =- {1\over2} \, g_{Z_0} \, \bar \Psi_e \, \g^\m \( v_f^{Z_0} - a_f^{Z_0} \g_5 \) \Psi_e Z_{0 \m}
\ee
Therefore we expect the amplitude $\G_\m$ (involving an external incoming $Z_0$) to be given by the matrix element of $J^\m_{Z_0}$
\be
 \G^\m (Z_0) = \int d^4 x \, e^{-i k \cdot x} \la f | J^\m_{Z_0} | i \ra \label{GammaZ0demWI}
\ee
where $f \, (i)$ represents the final (initial) state.
From the equation of motion we know that
\bea
 \pd_\m J^\m_{Z_0} &=& {1\over2} \, g_{Z_0}\, a_e^{Z_0} \pd_\m  \( \bar \Psi_e  \g^\m  \g_5 \Psi_e \) \nn\\
                   &=&  i g_{Z_0} \, a_e^{Z_0} m_e \bar \Psi_e \g_5 \Psi_e \label{dmJmZ0}
\eea
since, as it is well known, the vectorial current is conserved while the axial one has the divergence proportional to the fermion mass.
We can show that the result is proportional to the axial Goldstone current coming from the Yukawa
coupling between the Higgs scalar and the electron.
In the SM (the extension to MSSM is straightforward) we give mass to the electron with the Yukawa Lagrangian
\be
 \L_{y_e}=- y_e \bar L_e H e_R + h.c.
\ee
where $y_e$ is the Yukawa coupling (taken to be real), $e_R$ is the right handed electron, and the leptonic and Higgs doublets are respectively
\be
 L_e = \( \begin{array}{c}
                      \n_e \\
                       e_L \end{array} \) \qquad
 H = \( \begin{array}{c}
                      i G^+ \\
                      \frac{(h^0 + v) + i G^0}{\sqrt2} \end{array} \)
\ee
with $e_L$ the left handed electron, $\n_e$ the corresponding neutrino, $h_0$ the real scalar Higgs field,
$G^0$ the real Goldstone scalar of the $Z_0$, $G^+$ the complex Goldstone scalar of $W^+$ and $v$ the Higgs VEV
which we take to be real. So we have
\bea
  \L_{y_e} &=& - \frac{y_e}{\sqrt2} \[ \bar e_L \( h^0 + v + i G^0 \) e_R + \bar e_R \( h^0 + v - i G^0 \) e_L \] + \dots \nn\\
           &=& - \frac{y_e}{\sqrt2} \[ \( h^0 + v \) \( \bar e_R e_L + \bar e_L e_R \)
                                           + i G^0  \(- \bar e_R e_L + \bar e_L e_R \) \] + \dots \nn\\
           &=& - \frac{y_e}{\sqrt2} \[ \( h^0 + v \) \bar \Psi_e \Psi_e
                                           + i G^0  \bar \Psi_e \g_5 \Psi_e \] + \dots \nn\\
           &=& - \frac{y_e}{\sqrt2} v \bar \Psi_e \Psi_e - \frac{y_e}{\sqrt2} h^0 \bar \Psi_e \Psi_e
              + J^5_{G^0} G^0  + \dots
\eea
where we omitted the terms in $G^+$, since we are not interested on them.
The Dirac spinor $ \Psi_e = \( \begin{array}{c} e_L\\ e_R \end{array} \)$ and the Goldstone current is
\be
 J^5_{G^0} = - i \frac{y_e}{\sqrt2}  \bar \Psi_e \g_5 \Psi_e
\ee
We can identify the electron mass $m_e = \frac{y_e}{\sqrt2} v$. Inserting these results into eq.~(\ref{dmJmZ0}), we find
\bea
 \pd_\m J^\m_{Z_0} &=& i g_{Z_0} \, a_e^{Z_0} m_e \bar \Psi_e \g_5 \Psi_e = i \MZO \frac{y_e}{\sqrt2} \bar \Psi_e \g_5 \Psi_e \nn\\
                   &=& -\MZO J^5_{G^0} \label{dmJmZ0G0}
\eea
where we used the definitions of $g_{Z_0}$, $a_e^{Z_0}$, $\MZO$ given after eq.~(\ref{Z0eeamp}) and
after eq.~(\ref{WIG0eeamp}).
Applying the result~(\ref{dmJmZ0G0}), we get
\bea
 0 &=& \int d^4 x \, \pd_\m \( e^{i k \cdot x} \la f | J^\m_{Z_0} | i \ra \)  \nn\\
 &=& i k_\m \int d^4 x \( e^{i k \cdot x} \la f | J^\m_{Z_0} | i \ra \) + \int d^4 x \, e^{i k \cdot x} \pd_\m \la f | J^\m_{Z_0} | i \ra \nn\\
 &=&i k_\m \int d^4 x \( e^{i k \cdot x} \la f | J^\m_{Z_0} | i \ra \)-\MZO \int d^4 x \, e^{i k \cdot x} \la f | J^5_{G^0} | i \ra
\eea
which implies the WI
\be
 - i k_\m \G^\m \( Z_0 \) + \MZO \G \(G^0\) = 0
\ee
where we used eq.~(\ref{GammaZ0demWI}) and
\be
 \G \(G^0\) = \int d^4 x \, e^{-i k \cdot x} \la f | J^5_{G^0} | i \ra
\ee
is the amplitude for an external incoming $G^0$ Goldstone boson. This result also includes the massless vector case which is recovered
putting the vector mass equal to zero.
\section{Fermionic loop diagram} \label{app:Fermionic loop diagram}  \label{subsec:Rosenberg}
In this section we give some general properties of the massive fermionic triangle diagram of Fig.~\ref{TriangleDiagram}.
To make the discussion self-consistent we will repeat some equations seen in the previous chapters.

Consider a case in which only a single fermion circulates in the loop and each coupling is either axial (A) or
vectorial (V) with charge equal to minus one. The fermionic triangles containing  an odd number of axial couplings,
denoted by AVV , VAV, VVA and AAA are
\bea
 \G_{\r \m \n}^{AVV}(p,q;m_f)&=&\int \frac{d^{4} \ell}{(2 \pi)^{4}} \, \Tr\(  \g_5  \g_\r  \frac{1}{\slashed \ell
   - \slashed q - m_f} \g_\n \frac{1}{\slashed \ell -m_f} \g_\m
  \frac{1}{\slashed \ell + \slashed p - m_f } \)   \nn\\
   &&+ (p \leftrightarrow q, \m \leftrightarrow \n)
   \label{AVVmasstrian}\\
 \G_{\r \m \n}^{VAV}(p,q;m_f)&=&\int \frac{d^{4} \ell}{(2 \pi)^{4}}
   \, \Tr\(  \g_\r  \frac{1}{\slashed \ell - \slashed q - m_f}
   \g_\n \frac{1}{\slashed \ell -m_f} \g_5 \g_\m
   \frac{1}{\slashed \ell + \slashed p - m_f } \)   \nn\\
   &&+ (q \leftrightarrow -(p+q), \n \leftrightarrow \r) \label{VAVmasstrian}\\
 \G_{\r \m \n}^{VVA}(p,q;m_f)&=&\int \frac{d^{4} \ell}{(2 \pi)^{4}}
   \, \Tr\(   \g_\r  \frac{1}{\slashed \ell - \slashed q - m_f}
   \g_5 \g_\n  \frac{1}{\slashed \ell -m_f}  \g_\m
   \frac{1}{\slashed \ell + \slashed p - m_f } \)  \nn\\
   &&+ (p \leftrightarrow -(p+q), \m \leftrightarrow \r)  \label{VVAmasstrian}\\
\G_{\r \m \n}^{AAA}(p,q;m_f)&=&\int \frac{d^{4} \ell}{(2 \pi)^{4}} \, \Tr\(  \g_5  \g_\r  \frac{1}{\slashed \ell -
   \slashed q - m_f} \g_5  \g_\n \frac{1}{\slashed \ell -m_f} \g_5 \g_\m
   \frac{1}{\slashed \ell + \slashed p - m_f } \)  \nn\\
   &&+ (p \leftrightarrow q, \m \leftrightarrow \n)  \label{AAAmasstrian} \eea
These integrals are superficially divergent (by power counting) and thus there is an ambiguity in their definition.
The internal momentum $\ell$ can, in fact, be arbitrarily shifted~(see Section 6.2 of \cite{Cheng:1985bj})
\be
 \ell_\s \to \ell_\s+\a\, p_\s+(\a-\b)q_\s
\ee
leading to
\be
 \G_{\r \m \n}^{AVV}(p,q,\b;m_f)=\G_{\r \m \n}^{AVV}(p,q;m_f)-\frac{\b}{8\pi^2}\epsilon_{\r\m\n\s}(p-q)^\s
 \label{Gshift}
\ee
     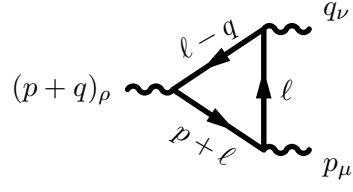
\begin{figure}[t] \vskip 1.5cm
      \centering
      \raisebox{-4.3ex}[0cm][0cm]{\unitlength=0.4mm
      \begin{fmffile}{loop}
      \begin{fmfgraph*}(60,40)
       \fmfpen{thick} \fmfleft{ii0,ii1,ii2} \fmfstraight \fmffreeze \fmftop{ii2,t1,t2,t3,oo2}
       \fmfbottom{ii0,b1,b2,b3,oo1} \fmf{phantom}{ii2,t1,t2} \fmf{phantom}{ii0,b1,b2} \fmf{phantom}{t1,v1,b1}
       \fmf{phantom}{t2,b2} \fmf{phantom}{t3,b3}
       \fmffreeze
       \fmf{photon}{ii1,v1}
       \fmf{fermion,label.side=right,label=\begin{rotate}{30}$\! \! \! \! \! \! \ell-q$\end{rotate}}{t3,v1}
       \fmf{fermion,label=$\ell$}{b3,t3}
       \fmf{fermion,label.dist=0.5cm,label=\begin{rotate}{-30} $\! \! \! \! \! \! p+\ell$\end{rotate}}{v1,b3}
       \fmf{photon}{b3,oo1} \fmf{photon}{t3,oo2}
       \fmflabel{$(p+q)_\r$}{ii1} \fmflabel{$p_\m$}{oo1} \fmflabel{$q_\n$}{oo2}
      \end{fmfgraph*}
      \end{fmffile}}
    \vskip 1.5cm
    \caption{The anomalous triangle diagram.}\label{TriangleDiagram}\end{figure}
The amplitudes (\ref{AVVmasstrian}),(\ref{VAVmasstrian}),(\ref{VVAmasstrian}) and (\ref{AAAmasstrian})
can be written using the Rosenberg parametrization \cite{Rosenberg:1962pp} as
\bea
 &&\!\!\!\!\!\!\!\!\!\!\G_{\r \m \n}  (p,q;m_f)=~~~\nn\\
 &&\!\!\!\!\!\!\!\!\!\!{1\over{ \pi^2}} \Big(
 I_1(p,q;m_f) \,\e[p,\m,\n,\r] + I_2(p,q;m_f) \, \e[q,\m,\n,\r]+  I_3(p,q;m_f)  \, \e[p,q,\m,\r]{p}_{\n} \nn\\
 &&\!\!\!\!\!\!\!\!\!\!+  I_4(p,q;m_f) \, \e[p,q,\m,\r]{q}_{\n} +  I_5(p,q;m_f) \, \e[p,q,\n,\r]p_\m +
 I_6(p,q;m_f) \, \e[p,q,\n,\r]q_\m \Big)  \nn\\
\label{Ros_mass}
\eea
with $\e[p,q,\r,\s]=\epsilon_{\m\n\r\s} p^\m q^\n$ and where
\bea
 I_3(p,q;m_f) &=& -\int_0^1 dx
 \int_0^{1-x} dy \frac{x y}{y (1-y) p^2 + x(1-x) q^2 + 2 x y \,p\cdot q - m_f^2}  \nonumber \\
 I_4(p,q;m_f) &=& \int_0^1 dx \int_0^{1-x} dy \frac{x (x-1)}{y (1-y) p^2 +  x(1-x) q^2 + 2 x y \,p\cdot q - m_f^2}  \nonumber \\
 I_5(p,q;m_f) &=& -I_4(q,p;m_f) \nonumber \\
 I_6(p,q;m_f) &=& -I_3(p,q;m_f)  \label{I's_mass}
\eea
are finite integrals in the Feynman parameters $x,y$ while $I_1$ and $I_2$ contain all the divergent contribution of the 1 loop integration.
We give a demonstration of the Rosenberg parametrization for the AVV triangle.
The extension to the other triangles is straightforward.
Our starting point is eq.~(\ref{AVVmasstrian}).
Lorentz invariance and momentum conservation imply the following expansion
\bea
 &&\!\!\!\!\!\!\!\!\!\!\!\!\!\!\!\!\!\!
   \G_{\r \m \n}(p,q;m_f)=
   B_1(p,q;m_f) \, \e[p,\m,\n,\r]     + B_2(p,q;m_f) \, \e[q,\m,\n,\r]+ \nn\\
 &&\!\!\!\!\!\!\!\!\!\!\!\!\!\!\!\!\!\!\phantom{\G_{\r \m \n}(p,q;m_f)=}
   B_3(p,q;m_f) \, \e[p,q,\m,\r] p_\n + B_4(p,q;m_f) \, \e[p,q,\m,\r] q_\n +\nn\\
 &&\!\!\!\!\!\!\!\!\!\!\!\!\!\!\!\!\!\!\phantom{\G_{\r \m \n}(p,q;m_f)=}
   B_5(p,q;m_f) \, \e[p,q,\n,\r] p_\m + B_6(p,q;m_f) \, \e[p,q,\n,\r] q_\m +\nn\\
 &&\!\!\!\!\!\!\!\!\!\!\!\!\!\!\!\!\!\!\phantom{\G_{\r \m \n}(p,q;m_f)=}
   B_7(p,q;m_f) \, \e[p,q,\m,\n] p_\r + B_8(p,q;m_f) \, \e[p,q,\m,\n] q_\r
\eea
where the $B$'s are functions of the vector momenta $p$, $q$ and of the fermion mass $m_f$.
Applying the Schouten identity
\bea
 &&k^\r \e[\a,\b,\m,\n]+ k^\a \e[\b,\m,\n,\r]+ k^\b \e[\m,\n,\r,\a]+\nn\\
 &&k^\m \e[\n,\r,\a,\b]+ k^\n \e[\r,\a,\b,\m]=0
\eea
on $p$ and $q$, we get
\bea
 &&\!\!\!\!\!\!\!\!\!\!\!\!\!\!\!\!\!\!
   \G_{\r \m \n}(p,q;m_f)=
   \[ B_1 + p \cdot q B_7 + q^2 B_8 \]\, \e[p,\m,\n,\r] + \nn\\
 &&\!\!\!\!\!\!\!\!\!\!\!\!\!\!\!\!\!\!\phantom{\G_{\r \m \n}(p,q;m_f)=}
   \[ B_2 - p^2 B_7 - p \cdot q B_8 \] \, \e[q,\m,\n,\r]+ \nn\\
 &&\!\!\!\!\!\!\!\!\!\!\!\!\!\!\!\!\!\!\phantom{\G_{\r \m \n}(p,q;m_f)=}
   \[ B_3 + B_7 \] \, \e[p,q,\m,\r] p_\n + \[ B_4 + B_8 \] \, \e[p,q,\m,\r] q_\n +\nn\\
 &&\!\!\!\!\!\!\!\!\!\!\!\!\!\!\!\!\!\!\phantom{\G_{\r \m \n}(p,q;m_f)=}
   \[ B_5 - B_7 \] \, \e[p,q,\n,\r] p_\m + \[ B_6 - B_8 \] \, \e[p,q,\n,\r] q_\m
\eea
If we identify
\bea
 I_1 &=& \pi^2 \[ B_1 + p \cdot q B_7 + q^2 B_8 \] \nn\\
 I_2 &=& \pi^2 \[ B_2 - p^2 B_7 - p \cdot q B_8 \] \nn\\
 I_3 &=& \pi^2 \[ B_3 + B_7 \] \nn\\
 I_4 &=& \pi^2 \[ B_4 + B_8 \] \nn\\
 I_5 &=& \pi^2 \[ B_5 - B_7 \] \nn\\
 I_6 &=& \pi^2 \[ B_6 - B_8 \]
\eea
we get the Rosenberg parametrization (\ref{Ros_mass}).

Now we want to compute the finite part of the parametrization i.e.
integrals $I_3 \dots I_6$. Let us go back to eq.~(\ref{AVVmasstrian}). It is easy to check that the exchanged contribution has the only the effect of
doubling the amplitude. So we get
\be
 \G_{\r \m \n}=2 \int \frac{d^{4} \ell}{(2 \pi)^{4}} \,
  \frac{\Tr \[ \g_5 \g_\r \( \slashed \ell - \slashed q + m_f \)
                    \g_\n \( \slashed \ell +m_f \)
                    \g_\m \( \slashed \ell + \slashed p + m_f \)
            \]}{\[ (\ell-q)^2 - m_f^2 \] \[ \ell^2 - m_f^2 \] \[ (\ell+p)^2 - m_f^2 \]}
\ee
Introducing the Feynman parameters $x,y,z$  
we obtain
\bea
 \G_{\r \m \n}&=&
   \int \frac{d^{4} \ell}{(2 \pi)^{4}}
  \int  \frac{4 N_{\r \m \n} \, \d(1-x-y-z) \, dx\,dy\,dz }{\left\{ \[ (\ell-q)^2 - m_f^2 \]x + \[ (\ell+p)^2 - m_f^2 \]y + \[ \ell^2 - m_f^2 \]z \right\}^3} \nn\\
 &=& 4 \int \frac{d^{4} \ell}{(2 \pi)^{4}}
  \int_0^1 dx \int_0^{1-x} dy \frac{N_{\r \m \n}}{\[ \ell^2 + 2y \ell \cdot p - 2x \ell \cdot q + y p^2 + q^2 - m_f^2\]^3}\nn\\
\eea
with
\bea
N_{\r \m \n} &=&  \Tr \[ \g_5 \g_\r \( \slashed \ell - \slashed q + m_f \)
                    \g_\n \( \slashed \ell +m_f \)
                    \g_\m \( \slashed \ell + \slashed p + m_f \) \] \\
             &=&  \Tr \[ \g_5 \g_\r \( \slashed \ell - \slashed q\)
                    \g_\n \( \slashed \ell\)
                    \g_\m \( \slashed \ell + \slashed p\) \]+
                 m_f^2 \Tr \[ \g_5 \g_\r \g_\n \g_\m \( \slashed \ell + \slashed p - \slashed q \) \] \nn
\eea
The numerator $N_{\r \m \n}$ has terms with three momenta and terms with one momenta.
The latter contribute to $I_1$ and $I_2$ in eq.~(\ref{Ros_mass}).
At this point we are not interested in an explicit computation of such integrals, so we leave them indicated as dots in the
following computations.
Using the change of variable\footnote{The change is allowed since we are interested only on the finite part of the integral.}
$\ell =\tilde \ell + k$ with $k=-yp + xq$ we get
\bea
 \G_{\r \m \n}&=& 4 \int \frac{d^{4} \tilde\ell}{(2 \pi)^{4}}
  \int_0^1 dx \int_0^{1-x} dy \frac{\tilde N_{\r \m \n}}{\[ \tilde\ell^2 -\Delta \]^3}
\eea
where
\be
 \Delta=- \[ y(1-y) p^2 + x (1-x) q^2 + 2 xy p \cdot q -m_f^2 \]
\ee
and
\bea
 \tilde N_{\r \m \n} &=& \Tr \[ \g_5 \g_\r \( \tilde{\slashed\ell} + \slashed k - \slashed q \)
                                      \g_\n \( \tilde{\slashed\ell} + \slashed k \)
                                      \g_\m \( \tilde{\slashed\ell} + \slashed k + \slashed p \) \] + \dots\nn\\
                     &=& \Tr \[ \g_5 \g_\r \( \slashed k - \slashed q \) \g_\n \slashed k \g_\m \( \slashed k + \slashed p \) \] + \dots\nn\\
                     &=& -2 \[ x (x-1) q_\n - xy p_\n \] q^\a p^\b \Tr \[ \g_5 \g_\r \g_\m \g_\a \g_\b \]+\nn\\
                      && -2 \[ xy q_\m - y (y-1) q_\m \] q^\a p^\b \Tr \[ \g_5 \g_\r \g_\n \g_\a \g_\b \] +\dots
\eea
where we omitted terms with odd powers of $\tilde \ell$ because their contribution is identically zero since we integrate
over all the space a function which is odd in $\tilde \ell$.
Then we find
\bea
\G_{\r \m \n}&=& - 8   \int_0^1 dx \int_0^{1-x} dy \int \frac{d^{4} \tilde\ell}{(2 \pi)^{4}}
                \frac{1}{\[ \tilde\ell^2 -\Delta \]^3} \times \nn\\
  &&\times \Big\{ \[ x (x-1) q_\n - xy p_\n \] q^\a p^\b \Tr \[ \g_5 \g_\r \g_\m \g_\a \g_\b \]  +\nn\\
  &&\phantom{\times \Big\{}
                  \[ xy q_\m - y (y-1) q_\m \] q^\a p^\b \Tr \[ \g_5 \g_\r \g_\n \g_\a \g_\b \] \Big\} + \dots
\eea
Having isolated the finite part, we can compute the integral over the momentum and take the trace over the Dirac matrices.
Using the result
\be
 \int \frac{d^{4} \tilde\ell}{(2 \pi)^{4}} \frac{1}{\[ \tilde\ell^2 -\Delta \]^3} =
 -\frac{i}{2} \frac{1}{(4 \pi)^2} \frac{1}{\Delta}
\ee
we obtain
\bea
&&\G_{\r \m \n}= \frac{1}{\pi^2}   \int_0^1 dx \int_0^{1-x} dy \Bigg\{
                 \frac{\[ x (x-1) q_\n - xy p_\n \] \e[p,q,\m,\r]}{y(1-y) p^2 + x (1-x) q^2 + 2 xy p \cdot q -m_f^2 } +\nn\\
&&\phantom{\G_{\r \m \n}=}
                 \frac{\[ xy q_\m - y (y-1) q_\m \] \e[p,q,\n,\r]}{y(1-y) p^2 + x (1-x) q^2 + 2 xy p \cdot q -m_f^2 } \Bigg\} + \dots
\eea
Comparing with eq.~(\ref{Ros_mass}) we identify
\bea
 &&I_3(p,q;m_f) = -\int_0^1 dx \int_0^{1-x} dy \frac{x y}{y (1-y) p^2 + x(1-x) q^2 + 2 x y \,p\cdot q - m_f^2}  \nonumber \\
 &&I_4(p,q;m_f) = \int_0^1 dx \int_0^{1-x} dy \frac{x (x-1)}{y (1-y) p^2 +  x(1-x) q^2 + 2 x y \,p\cdot q - m_f^2}  \nonumber \\
 &&I_5(p,q;m_f) = -\int_0^1 dx \int_0^{1-x} dy \frac{y (y-1)}{y (1-y) p^2 +  x(1-x) q^2 + 2 x y \,p\cdot q - m_f^2} \nonumber \\
 &&I_6(p,q;m_f) = \int_0^1 dx \int_0^{1-x} dy \frac{x y}{y (1-y) p^2 + x(1-x) q^2 + 2 x y \,p\cdot q - m_f^2} \qquad \quad
\eea
Is it easy to check that $I_5(p,q;m_f) = -I_4(q,p;m_f)$ and $I_6(p,q;m_f) = -I_3(p,q;m_f)$, reproducing exactly the definitions in eq.~(\ref{I's_mass}).

In terms of the Rosenberg parametrization  the $\b$ dependence of (\ref{Gshift}) is contained only in $I_1$ and
$I_2$ (which are superficially divergent). However, using the WIs,
\bea
(p+q)^\r \G_{\r \m \n}^{AVV}(p,q,\b;m_f) &=& {1\over{\pi^2}} \[ {\b\over4} + m_f^2 I_0 (p,q;m_f)\] \e[p,q,\m,\n] \nn \\
p^\m \G_{\r \m \n}^{AVV}(p,q,\b;m_f)     &=& -{2+\b\over8\pi^2}\,  \e[q,p,\n,\r] \nn \\
q^\n \G_{\r \m \n}^{AVV}(p,q,\b;m_f)     &=& -{2+\b\over8\pi^2}\,  \e[q,p,\r,\m] \label{WI_AVVmass}
\eea
it is possible to show that they can be expressed in terms of $I_3 \dots I_6$ as
\bea
I_1^{AVV}(p,q,\b;m_f) &=& p \cdot q \, I_3(p,q) + q^2 \, I_4(p,q) +\frac{2+\b}{8}\nn\\
I_2^{AVV}(p,q,\b;m_f) &=& -I_1^{AVV}(q,p,\b;m_f) \label{I12massAVV}
\eea
where $I_0$ is defined as
\be
 I_0 (p,q;m_f) = -\int_0^1 dx \int_0^{1-x} dy \frac{1}{y (1-y) p^2 +
    x(1-x) q^2 + 2 x y \,p\cdot q - m_f^2} \label{I_0integral}
\ee
The $I_0$ contribution is only present in a massive fermionic triangle. We encounter such contribution
in theories with chiral massive fermions such as SM, MSSM and MiAUMSSM.
Finally we remark that the $I_0$ contribution in eq. (\ref{WI_AVVmass}) is not $\b$ dependent so it is
fixed on the axial vertex. The same will happen for the VAV, VVA and AAA cases.
\section{Landau-Yang theorem} \label{app:Landau Yang}
The Landau-Yang theorem states that two massless spin-one objects cannot combine to form a spin-one object.
From a phenomenological point of view this means that a $Z'$ cannot decay into two on-shell photons.
Now we give a partial demonstration of this fact using the Rosenberg parametrization (\ref{Ros_mass}).
We put the $Z'$ on the $\r$ leg and the photons on the $\m,\n$ legs.
The amplitude of the process is given by
\bea
 A^{Z' \g \g}_{\r\m\n} &=& -{1\over2\pi^2} g_{Z'} e^2 \Big(\tilde A_1 \e[p,\m,\n,\r] + \tilde A_2\e[q,\m,\n,\r]+ A_3 \e[p,q,\m,\r]{p}_{\n} \nn\\
               &&+  A_4 \e[p,q,\m,\r]{q}_{\n} +  A_5 \e[p,q,\n,\r]p_\m +  A_6\e[p,q,\n,\r]q_\m \Big)
\eea
where $\tilde A_1$ and $\tilde A_2$ considered the GCS absorption (see Section \ref{sect:Computation of the amplitude}),
and $A_3$, $A_4$, $A_5$ and $A_6$ give the finite contribution of the amplitude.
The WIs on the photon lines are
\bea
 p^\m A^{Z' \g \g}_{\r\m\n}  (p,q;m_f) &=& 0 \\
 q^\n A^{Z' \g \g}_{\r\m\n}  (p,q;m_f) &=& 0
\eea
which imply that
\bea
 \tilde A_1 &=& p \cdot q A_3 +  q^2 A_4 = p \cdot q A_3\\
 \tilde A_2 &=& p \cdot q A_6 +  p^2 A_5 = p \cdot q A_6
\eea
since the photons are massless. Using this result we get
\bea
 \left| A^{Z' \g \g} \right|^2&\propto& \sum_{\l'} \e^{\r_1}_{(\l')} \e^{* \r_2}_{(\l')} \
        \sum_{\l_\g} \e^{\n_1}_{(\l^\g)} \e^{* \n_2}_{(\l^\g)} \
        \sum_{\l_\g} \e^{\m_1}_{(\l^\g)} \e^{* \m_2}_{(\l^\g)} \
        A_{\r_1 \m_1 \n_1}^{Z' \g \g} A_{\r_2 \m_2 \n_2}^{* \, Z' \g \g}= \nn\\
      &\propto& \MZp^6 \( A_3 + A_6\)^2
\eea
where we used the completeness relations
\bea
 \sum_{\l'} \e^{\r_1}_{(\l')} \e^{* \r_2}_{(\l')} &=& - \eta^{\r_1 \r_2} + \frac{k^{\r_1}_{(\l')} k^{\r_2}_{(\l')}}{\MZp^2} \\
 \sum_{\l_\g} \e^{\n_1}_{(\l^\g)} \e^{* \n_2}_{(\l^\g)} &\to& - \eta^{\n_1\n_2} \\
 \sum_{\l_\g} \e^{\m_1}_{(\l^\g)} \e^{* \m_2}_{(\l^\g)} &\to& - \eta^{\m_1\m_2}
\eea
From eq.~(\ref{I's_mass}) we have $A_6=-A_3$, so
\be
 \G(Z' \to \g \g) \propto \left| A^{Z' \g \g} \right|^2 = 0
\ee
in agreement with the Landau-Yang theorem.
\section{Treatment of non anomalous diagrams\label{notanomappdx}}

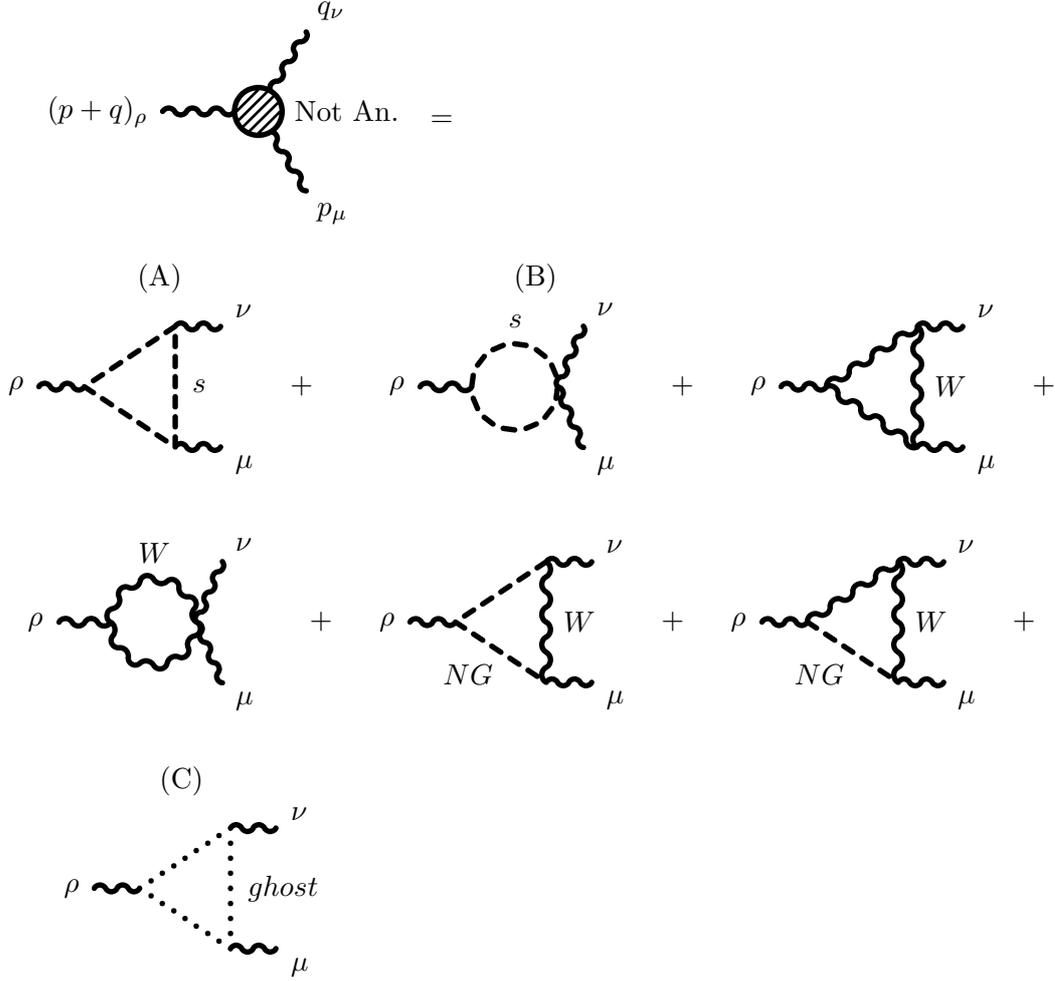
\begin{figure}[tb]
      \centering
        \vskip1.5cm
      \raisebox{-5.2ex}[0cm][0cm]{\unitlength=0.7mm
      \begin{fmffile}{BBBMSSM_NAnew}
      \begin{fmfgraph*}(30,30)
       \fmfpen{thick} \fmfleft{o1} \fmfright{i1,i2} \fmf{boson}{i1,v1,i2}\fmf{boson}{o1,v1}
       \fmfv{decor.shape=circle,decor.filled=shaded, decor.size=.30w}{v1}
       \fmflabel{$(p+q)_\r$}{o1}\fmflabel{$p_\m$}{i1}\fmflabel{$q_\n$}{i2}\fmflabel{~~Not An.}{v1}
      \end{fmfgraph*}
      \end{fmffile}}
       ~~~~~~~~~=~~~~~~~~~~~~~~~~~~~~~~~~~~~~~~~~~~~~~~~~~~~~~~~~~~~~~~

      \vskip 1.5cm
      (A)~~~~~~~~~~~~~~~~~~~~~~~~~~~~~~~~~~(B)~~~~~~~~~~~~~~~~~~~~~~~~~~~~~~~~~~~~~~~~~~~~
      \vskip 0.8cm
      \raisebox{-4.3ex}[0cm][0cm]{\unitlength=0.4mm
      \begin{fmffile}{ZZg_s_111}
      \begin{fmfgraph*}(60,40)
       \fmfpen{thick} \fmfleft{ii0,ii1,ii2} \fmfstraight \fmffreeze \fmftop{ii2,t1,t2,t3,oo2}
       \fmfbottom{ii0,b1,b2,b3,oo1} \fmf{phantom}{ii2,t1,t2} \fmf{phantom}{ii0,b1,b2} \fmf{phantom}{t1,v1,b1}
       \fmf{phantom}{t2,b2} \fmf{phantom}{t3,b3}
       \fmffreeze
       \fmf{photon}{ii1,v1} \fmf{dashes}{t3,v1} \fmf{dashes,label=$s$}{b3,t3} \fmf{dashes}{v1,b3} \fmf{photon}{b3,oo1}
       \fmf{photon}{t3,oo2}
       \fmflabel{$\r$}{ii1} \fmflabel{$\m$}{oo1} \fmflabel{$\n$}{oo2}
      \end{fmfgraph*}
      \end{fmffile}}
      ~~~~~+~~~~~
      ~~
      \raisebox{-4.3ex}[0cm][0cm]{\unitlength=0.4mm
       \begin{fmffile}{ZZg_2_17}
       \begin{fmfgraph*}(60,40)
        \fmfpen{thick} \fmfleft{ii1} \fmfright{oo1,oo2} \fmf{photon}{ii1,v1} \fmf{phantom,left,tension=0.3}{v1,v2,v1}
        \fmf{photon}{oo1,v2,oo2}\fmffreeze
        \fmf{dashes,left,tension=0.3,label=$s$}{v1,v2} \fmf{dashes,left,tension=0.3}{v2,v1}
        \fmflabel{$\r$}{ii1} \fmflabel{$\m$}{oo1} \fmflabel{$\n$}{oo2}
       \end{fmfgraph*}
       \end{fmffile}}
       ~~~~~+~~~~~~
       \raisebox{-4.3ex}[0cm][0cm]{\unitlength=0.4mm
       \begin{fmffile}{ZZg_W_111}
       \begin{fmfgraph*}(60,40)
        \fmfpen{thick} \fmfleft{ii0,ii1,ii2} \fmfstraight \fmffreeze \fmftop{ii2,t1,t2,t3,oo2}
        \fmfbottom{ii0,b1,b2,b3,oo1} \fmf{phantom}{ii2,t1,t2} \fmf{phantom}{ii0,b1,b2} \fmf{phantom}{t1,v1,b1}
        \fmf{phantom}{t2,b2} \fmf{phantom}{t3,b3}
        \fmffreeze
        \fmf{photon}{ii1,v1} \fmf{photon}{t3,v1} \fmf{photon,label=$W$}{b3,t3} \fmf{photon}{v1,b3} \fmf{photon}{b3,oo1}
        \fmf{photon}{t3,oo2}
        \fmflabel{$\r$}{ii1} \fmflabel{$\m$}{oo1} \fmflabel{$\n$}{oo2}
       \end{fmfgraph*}
       \end{fmffile}}
       ~~~~~+~~~~~
       \vskip 2.5cm
       \raisebox{-4.3ex}[0cm][0cm]{\unitlength=0.4mm
       \begin{fmffile}{ZZg_2W_17}
       \begin{fmfgraph*}(60,40)
        \fmfpen{thick} \fmfleft{ii1} \fmfright{oo1,oo2} \fmf{photon}{ii1,v1} \fmf{phantom,left,tension=0.3}{v1,v2,v1}
        \fmf{photon}{oo1,v2,oo2}\fmffreeze
        \fmf{photon,left,tension=0.3,label=$W$}{v1,v2} \fmf{photon,left,tension=0.3}{v2,v1}
        \fmflabel{$\r$}{ii1} \fmflabel{$\m$}{oo1} \fmflabel{$\n$}{oo2}
       \end{fmfgraph*}
       \end{fmffile}}
       ~~~~~+~~~~~
       \raisebox{-4.3ex}[0cm][0cm]{\unitlength=0.4mm
       \begin{fmffile}{ZZg_WGG_121}
       \begin{fmfgraph*}(60,40)
        \fmfpen{thick} \fmfleft{ii0,ii1,ii2} \fmfstraight \fmffreeze \fmftop{ii2,t1,t2,t3,oo2}
        \fmfbottom{ii0,b1,b2,b3,oo1} \fmf{phantom}{ii2,t1,t2} \fmf{phantom}{ii0,b1,b2} \fmf{phantom}{t1,v1,b1}
        \fmf{phantom}{t2,b2} \fmf{phantom}{t3,b3}
        \fmffreeze
        \fmf{photon}{ii1,v1} \fmf{dashes}{t3,v1} \fmf{photon,label=$W$}{b3,t3} \fmf{dashes,label=$NG$}{v1,b3}
        \fmf{photon}{b3,oo1} \fmf{photon}{t3,oo2}
        \fmflabel{$\r$}{ii1} \fmflabel{$\m$}{oo1} \fmflabel{$\n$}{oo2}
       \end{fmfgraph*}
       \end{fmffile}}
       ~~~~~+~~~~~
       \raisebox{-4.3ex}[0cm][0cm]{\unitlength=0.4mm
       \begin{fmffile}{ZZg_WWG_121}
       \begin{fmfgraph*}(60,40)
        \fmfpen{thick} \fmfleft{ii0,ii1,ii2} \fmfstraight \fmffreeze \fmftop{ii2,t1,t2,t3,oo2}
        \fmfbottom{ii0,b1,b2,b3,oo1} \fmf{phantom}{ii2,t1,t2} \fmf{phantom}{ii0,b1,b2} \fmf{phantom}{t1,v1,b1}
        \fmf{phantom}{t2,b2} \fmf{phantom}{t3,b3}
        \fmffreeze
        \fmf{photon}{ii1,v1} \fmf{photon}{t3,v1} \fmf{photon,label=$W$}{b3,t3} \fmf{dashes,label=$NG$}{v1,b3}
        \fmf{photon}{b3,oo1} \fmf{photon}{t3,oo2}
        \fmflabel{$\r$}{ii1} \fmflabel{$\m$}{oo1} \fmflabel{$\n$}{oo2}
       \end{fmfgraph*}
       \end{fmffile}}
       ~~~~~+~~~~~
      \vskip 1.5cm
      (C)~~~~~~~~~~~~~~~~~~~~~~~~~~~~~~~~~~~~~~~~~~~~~~~~~~~~~~~~~~~~~~~~~~~~~~~~~~~~~~
      \vskip 0.8cm
       \raisebox{-4.3ex}[0cm][0cm]{\unitlength=0.4mm
       \begin{fmffile}{ZZg_ghost_111}
       \begin{fmfgraph*}(60,40)
        \fmfpen{thick} \fmfleft{ii0,ii1,ii2} \fmfstraight \fmffreeze \fmftop{ii2,t1,t2,t3,oo2}
        \fmfbottom{ii0,b1,b2,b3,oo1} \fmf{phantom}{ii2,t1,t2} \fmf{phantom}{ii0,b1,b2} \fmf{phantom}{t1,v1,b1}
        \fmf{phantom}{t2,b2} \fmf{phantom}{t3,b3}
        \fmffreeze
        \fmf{photon}{ii1,v1} \fmf{dots}{t3,v1} \fmf{dots,label=$ghost$}{b3,t3} \fmf{dots}{v1,b3} \fmf{photon}{b3,oo1}
        \fmf{photon}{t3,oo2}
        \fmflabel{$\r$}{ii1} \fmflabel{$\m$}{oo1} \fmflabel{$\n$}{oo2}
       \end{fmfgraph*}
       \end{fmffile}}~~~~~~~~~~~~~~~~~~~~~~~~~~~~~~~~~~~~~~~~~~~~~~~~~~~~~~~~~~~~~~~~~~~~~~~~~~~~~~
       \vskip 1cm
      \caption{Non Anomalous diagrams for trilinear neutral gauge boson amplitudes.}\label{OtherDiagrams}
     \end{figure}
In this section we show that the non anomalous diagrams in Fig. \ref{diagzp} vanish. The
diagrams we consider, reported in Fig. \ref{OtherDiagrams}, have no specific assignment for the external legs, to keep
the discussion as general as possible.
All the factors which are not relevant for our aim are omitted and all the possible leg exchanges are understood.
Finally, we use dimensional regularization and the $R_\xi$ gauge with $\xi=1$, in such a way that each diagram vanishes separately.

A) The Scalar triangle loop is given by
       \bea
        D^A_{\m \n \r} (p,q) &=&  \int \frac{d^{2 \w} l}{(2 \pi)^{2 \w}}
        \frac{(2l+p-q)_\r (2l-q)_\n (2l+p)_\m}{\[(l-q)^2-m^2\]\[l^2-m^2\]\[(l+p)^2-m^2\]}+\nn\\
        &&+ \, (p \leftrightarrow q, \m \leftrightarrow \n) \nn\\
        &=& \int \frac{d^{2 \w} l}{(2 \pi)^{2 \w}}
        \frac{(2l+p-q)_\r (2l-q)_\n (2l+p)_\m}{\[(l-q)^2-m^2\]\[l^2-m^2\]\[(l+p)^2-m^2\]} \nn\\
        &&+\int \frac{d^{2 \w} l}{(2 \pi)^{2 \w}}
        \frac{(2l+q-p)_\r (2l-p)_\m (2l+q)_\n}{\[(l-p)^2-m^2\]\[l^2-m^2\]\[(l+q)^2-m^2\]}
       \eea
 Performing the change of variable $l_\m \to -l_\m$ in the second integral, one gets
       \bea
        \!\!\!\!\!\!\!\!
        D^A_{\m \n \r} (p,q)
        &=& \int \frac{d^{2 \w} l}{(2 \pi)^{2 \w}}
        \frac{(2l+p-q)_\r (2l-q)_\n (2l+p)_\m}{\[(l-q)^2-m^2\]\[l^2-m^2\]\[(l+p)^2-m^2\]} \nn\\
        &&\!\!\!\!\!\!\!
        -\int \frac{d^{2 \w} l}{(2 \pi)^{2 \w}}
        \frac{(2l+p-q)_\r (2l+p)_\m (2l-q)_\n}{\[(l-q)^2-m^2\]\[l^2-m^2\]\[(l+p)^2-m^2\]} = 0
       \label{diaga}\eea

 B) The ``Scalar bubble loop'' is given by
       \bea
        D^B_{\m \n \r} (p,q)
        &=& -2 \int \frac{d^{2 \w} l}{(2 \pi)^{2 \w}}
        \frac{(2l+p+q)_\r \emn}{\[l^2-m^2\]\[(l+p+q)^2-m^2\]} \nn\\
        &=& -2 \int \frac{d^{2 \w} l}{(2 \pi)^{2 \w}}
        \frac{(l+p+q)_\r \emn}{\[l^2-m^2\]\[(l+p+q)^2-m^2\]}\nn\\
        &&-2 \int \frac{d^{2 \w} l}{(2 \pi)^{2 \w}}
        \frac{(l)_\r \emn}{\[l^2-m^2\]\[(l+p+q)^2-m^2\]}
       \eea
 Performing the change of variable $l \to -l-p-q$ in the second integral one gets
       \bea
        D^B_{\m \n \r} (p,q)
        &=& -2 \int \frac{d^{2 \w} l}{(2 \pi)^{2 \w}}
        \frac{(l+p+q)_\r \emn}{\[l^2-m^2\]\[(l+p+q)^2-m^2\]}\nn\\
        &&+2 \int \frac{d^{2 \w} l}{(2 \pi)^{2 \w}}
        \frac{(l+p+q)_\r \emn}{\[(l+p+q)^2-m^2\]\[l^2-m^2\]}= 0
       \label{diagb}\eea

C) Since the ghost interact with neutral vectors only through the third component of $SU(2)$,
the Ghost triangle loop is proportional to
       \be
        \e_{3bc} \e_{3cd} \e_{3db} = -\d_{bd} \e_{3db} = 0
        \label{diagc}
       \ee
The other diagrams in Fig. \ref{OtherDiagrams} can also be shown to vanish after manipulations similar to the ones
used in (\ref{diaga}), (\ref{diagb}), (\ref{diagc}).
\section{Decay rates. General case}
In this section we compute the amplitudes for the decays $Z'\to Z_0 \, \g$ and $Z' \to Z_0 \, Z_0$
in the general case $Q_{H_u}\neq 0$, still neglecting the effects coming from the kinetic mixing.
We work in the limit
\be g_a
  v_{u,d}<<\m,M_0,M_1,M_2,M_S, M_{V^{(0)}}  \label{nosusylimit}
\ee
in which $m_{SC}\approx M_{V^{(0)}},\, m_{SB}\approx 0$ (see (\ref{msc}), (\ref{cdelta}),
(\ref{xdelta})). Hence,  (\ref{massmatrixNeutr}) takes the same form as in the symmetric phase in which
neutralinos and charginos do not contribute to the anomaly (see Section \ref{subsec: Anomalies and GS mechanism}).
In the limit (\ref{nosusylimit}) an extension of the SM by an extra $U(1)$ and our SUSY model
give the same results for what the decays of interest are concerned.

The SM fermion interaction terms with the neutral gauge bosons are given in eq.~(\ref{neutralcurrents})
where now
\bea
 &&v_f^{Z'} = Q^{Z'}_{f_L}+Q^{Z'}_{f_R}  \quad\qquad
   a_f^{Z'} = Q^{Z'}_{f_L}-Q^{Z'}_{f_R}  \nn\\
 &&v_f^{Z_0}= Q^{Z_0}_{f_L}+Q^{Z_0}_{f_R}  \quad\qquad
   a_f^{Z_0}= Q^{Z_0}_{f_L}-Q^{Z_0}_{f_R} \nn\\
 &&\ q_f      = Q_{f_L}=Q_{f_R}
\eea
The left and right fermions and the Dirac spinors are related using the usual definition
$ \Psi_f = \( \begin{array}{c} f_L\\ f_R \end{array} \)$.
The left and right charges are defined in the following way
\bea
 g_{Z'} Q^{Z'}_{f_L} &=& g_2 T_3 O_{02} + g_1 Y_{f_L} O_{01} + g_0 Q_{f_L} \\
 g_{Z'} Q^{Z'}_{f_R} &=& g_1 Y_{f_R} O_{01} + g_0 Q_{f_R} \\
 g_{Z_0} Q^{Z_0}_{f_L} &=& g_2 T_3 O_{12} + g_1 Y_{f_L} O_{11} + g_0 Q_{f_L} O_{10}\\
 g_{Z_0} Q^{Z_0}_{f_R} &=& g_1 Y_{f_R} O_{11} + g_0 Q_{f_R} O_{10} \\
 e Q_{f_L} &=& g_2 T_3 O_{22} + g_1 Y_{f_L} O_{21} = g_1 Y_{f_R} O_{21} = e Q_{f_R}
\eea
where $O_{ij}$ is given in (\ref{Oij}) and $T_3$ is the eigenvalue of $T^{(2)}_3$.
\subsection{$Z' \to Z_0 \ \g$}
The amplitude is given by the sum of the fermionic triangle $\D_{\r \m \n}^{Z' Z_0 \g}$ plus the proper GCS vertex
\bea
 A_{\r \m \n}^{Z' Z_0 \g}&=& \D_{\r \m \n}^{Z' Z_0 \g} + (\GCS)_{\r \m \n}^{Z' Z_0 \g} \nn\\
         \D_{\r \m \n}^{Z' Z_0 \g}&=& -{1\over4} g_{Z'} g_{Z_0} e \sum_f \( v_f^{Z'} a_f^{Z_0} q_f \G^{VAV}_{\r \m \n} +
                                                                           a_f^{Z'} v_f^{Z_0} q_f \G^{AVV}_{\r \m \n} \)
\eea
The resulting amplitude can be written as
\bea
 A_{\r \m \n}^{Z' Z_0 \g}&=&
   -{1\over4\pi^2} g_{Z'} g_{Z_0} e  \Big(\tilde A_1 \e[p,\m,\n,\r] + \tilde A_2\e[q,\m,\n,\r]+ A_3 \e[p,q,\m,\r]{p}^{\n}\nn\\
 &&+  A_4 \e[p,q,\m,\r]{q}^{\n} +  A_5 \e[p,q,\n,\r]p^\m +   A_6\e[p,q,\n,\r]q^\m \Big) \label{ZpZgapp}
\eea
with
\be
 A_i=\sum_f t_f^{Z' Z_0 \g} I_i  \qquad \text{for } i=3, \dots , 6
\ee
where
\be
 t_f^{Z' Z_0 \g} = \( v_f^{Z'} a_f^{Z_0}+a_f^{Z'} v_f^{Z_0} \) q_f
\ee
and the integrals $I_i$ given in (\ref{I's_mass}).
$\tilde A_1$ and $\tilde A_2$ are the new coefficients with the GCS absorbed similarly to (\ref{ItildeCBB}).

The Ward identities in Fig.~\ref{GoldWI} for the amplitude now read
\bea
 (p+q)^\r A_{\r \m \n}^{Z' Z_0 \g}+i M_{Z'} \[(\GS)^{Z_0 \g}_{\m \n}+(\NG)^{Z_0 \g}_{\m \n}\]&=&0 \label{WI-ZZg-general1}\\
  p^\m A_{\r \m \n}^{Z' Z_0 \g}+i M_{Z_0} \[ (\GS)^{Z' \g}_{\r \n}+(\NG)^{Z' \g}_{\r \n}\]&=&0 \label{WI-ZZg-general2}\\
  q^\n  A_{\r \m \n}^{Z' Z_0 \g}&=& 0 \label{WI-ZZg-general3}
\eea
where $M_{Z'}$ and $M_{Z_0}$ are the $Z'$ and $Z_0$ masses respectively.
In both (\ref{WI-ZZg-general1}) and (\ref{WI-ZZg-general2}) we have a $(\GS)$ and a
$(\NG)$ contribution due to the two Goldstone bosons which are a linear combination of the axion and $G^0$.
We use (\ref{WI-ZZg-general2}) and (\ref{WI-ZZg-general3}) to fix $\tilde A_1$ and $\tilde A_2$ while
(\ref{WI-ZZg-general1}) is automatically satisfied. Contracting with $p^\mu$ we get
\bea
 &&\!\!\!\!\!\!\!\!\!\!
   p^\m A_{\r \m \n}^{Z' Z_0 \g}=-\Bigg\{8 \[ 4 g_0 g_1^2 \ R_{101}^{Z' Z_0 \g} \ b_2^{(1)} b_3 +
                                             2 g_0 g_2^2 \ R_{202}^{Z' Z_0 \g} \ b_2^{(2)} b_3 +
                                             2 g_0^2 g_1 \ R_{001}^{Z' Z_0 \g} \ b_2^{(4)} b_3\] \nn\\
 &&\!\!\!\!\!\!\!\!\!\!\phantom{p^\m A_{\r \m \n}^{Z' Z_0 \g}=-\Bigg\{}
   + {1\over4\pi^2} g_{Z'} g_{Z_0} e \sum_f v_f^{Z'} a_f^{Z_0} q_f \ m_f^2 I_0 \Bigg\} \ \e[q,p,\n,\r]
\eea
where $I_0$ is the integral given in (\ref{I_0integral}).
The solution for $\tilde A_1$ and $\tilde A_2$ is
\bea
 \tilde A_1  &=& \(q^2 A_4 + p \cdot q A_3 \) \\
 \tilde A_2  &=& \(p^2 A_5 + p \cdot q A_6\) +(\GS)^{Z' \g} +(\NG)^{Z' \g}
\eea
with
\bea
 (\NG)^{Z' \g}&=&\sum_f v_f^{Z'} a_f^{Z_0} q_f \ m_f^2 I_0 \\
 (\GS)^{Z' \g}&=&\frac{64 \pi^2 g_0}{g_{Z'} g_{Z_0} e} \[ 2 g_1^2 \ R_{101}^{Z' Z_0 \g} \ b_2^{(1)} b_3 +
                                                       g_2^2 \ R_{202}^{Z' Z_0 \g} \ b_2^{(2)} b_3 +
                                                  g_0 g_1 \ R_{001}^{Z' Z_0 \g} \ b_2^{(4)} b_3\]\nn\\
\eea
The rotation factors are
\bea
 R_{101}^{Z' Z_0 \g}       &=& O_{01} O_{10} O_{21} \nn\\
 R_{202}^{Z' Z_0 \g} &=& O_{02} O_{10} O_{22} \nn\\
 R_{001}^{Z' Z_0 \g}      &=& O_{10} O_{21}
\eea
with $O_{ij}$ given by (\ref{Oij}). Substituting $\tilde A_1, ~ \tilde A_2$ into the amplitude (\ref{ZpZgapp}) and performing all the contractions
we finally obtain
\bea
 |A^{Z' Z_0 \g}|^2 &=& g_{Z'}^2 g_{Z_0}^2 e^2 \frac{
     \left(M_{Z'}^2-M_{Z_0}^2\right)^2 \left(M_{Z'}^2+M_{Z_0}^2\right)}{96 M_{Z_0}^2 M_{Z'}^2 \pi ^4} \times
    ~~~~~~~~~~~~~~~~~~~~\nn\\
    && \[\sum_f t_f^{Z' Z_0 \g} (I_3+I_5) M_{Z_0}^2   +(\GS)^{Z' \g} +(\NG)^{Z' \g}  \]^2
\eea
\subsection{$Z' \to Z_0 \ Z_0$}
The contribution to the fermionic triangle is
\bea
 \D_{\r \m \n}^{Z' Z_0 Z_0}&=&
   -{1\over8} g_{Z'} g_{Z_0}^2 \Bigg[\sum_f \Big( v_f^{Z'} a_f^{Z_0} v_f^{Z_0} \, \G_{\r \m \n}^{VAV} +
                                                  v_f^{Z'} v_f^{Z_0} a_f^{Z_0} \, \G_{\r \m \n}^{VVA} +\nn\\
  &&\phantom{-{1\over8} g_{Z'} g_{Z_0}^2 \Bigg[\sum_f \Big(}
     a_f^{Z'} v_f^{Z_0} v_f^{Z_0} \, \G_{\r \m \n}^{AVV} +
     a_f^{Z'} a_f^{Z_0} a_f^{Z_0} \, \G_{\r \m \n}^{AAA} \Big) \Bigg]
\eea
We write the total amplitude (the sum of the triangles plus GCS terms) as
\bea
     A_{\r \m \n}^{Z' Z_0 Z_0}&=&
        -{1\over8\pi^2} g_{Z'} g_{Z_0}^2 \Big[\tilde A_1 \e[p,\m,\n,\r] +
         \tilde A_2\e[q,\m,\n,\r]+
          A_3 \e[p,q,\m,\r]{p}_{\n} \nn\\
        &&+  A_4 \e[p,q,\m,\r]{q}_{\n} +  A_5 \e[p,q,\n,\r]p_\m +
         A_6\e[p,q,\n,\r]q_\m \Big]
\eea
with
\be
 A_i=\sum_f t_f^{Z' Z_0 Z_0} I_i  \qquad \text{for } i=3, \dots , 6
\ee
where
\be
 t_f^{Z' Z_0 Z_0}  =   \( a_f^{Z'} v_f^{Z_0} v_f^{Z_0}+ 2 v_f^{Z'} a_f^{Z_0} v_f^{Z_0}  + a_f^{Z'} a_f^{Z_0} a_f^{Z_0}\)
\ee
and the integrals $I_i$ are given in (\ref{I's_mass}). The Ward identities now read
\bea
 (p+q)^\r A_{\r \m \n}^{Z' Z_0 Z_0} +i M_{Z'} \[ (\GS)^{Z_0 Z_0}_{\m \n}+(\NG)^{Z_0 Z_0}_{\m \n} \]&=&0\label{WI-ZZZ-general1}\\
 p^\m A_{\r \m \n}^{Z' Z_0 Z_0} +i M_{Z_0} \[ (\GS)^{Z' Z_0}_{\r \n}+(\NG)^{Z' Z_0}_{\r \n} \]&=&0\label{WI-ZZZ-general2}\\
 q^\n A_{\r \m \n}^{Z' Z_0 Z_0} +i M_{Z_0} \[ (\GS)^{Z_0 Z'}_{\m \r}+(\NG)^{Z_0 Z'}_{\m \r} \]&=&0
 \label{WI-ZZZ-general3} \eea
where $M_{Z'}$ and $M_{Z_0}$ are the $Z'$ and $Z_0$ masses respectively.
In (\ref{WI-ZZZ-general1})-(\ref{WI-ZZZ-general3}) the $(\GS)$ and $(\NG)$ terms are present for the same reason
as in the preceding subsection.
We use (\ref{WI-ZZZ-general2}) and (\ref{WI-ZZZ-general3}) to fix $\tilde A_1$ and $\tilde A_2$ while
(\ref{WI-ZZZ-general1}) is automatically satisfied. Contracting with $p^\mu$ and $q^\nu$ we get
\bea
 &&\!\!\!\!\!\!\!\!\!\!\!\!\!\!\!\!\!\!\!
   p^\m A_{\r \m \n}^{Z' Z_0 Z_0}=
   -\Bigg\{8 \[4 g_0^3 \ R_{000}^{Z' Z_0 Z_0} \ b_2^{(0)} b_3 + 4 g_0 g_1^2 \ R_{101}^{Z' Z_0 Z_0} \ b_2^{(1)} b_3 +\right.\\
 &&\!\!\!\!\!\!\!\!\!\!\!\!\!\!\!\!\!\!\!\phantom{ p^\m A_{\r \m \n}^{Z' Z_0 Z_0}=-\Bigg\{8 }~
    \left. 2 g_0 g_2^2 \ R_{202}^{Z' Z_0 Z_0} \ b_2^{(2)} b_3 + 2 g_0^2 g_1 \ R_{001}^{Z' Z_0 Z_0} \ b_2^{(4)} b_3\] +\nn\\
 &&\!\!\!\!\!\!\!\!\!\!\!\!\!\!\!\!\!\!\!\phantom{ p^\m A_{\r \m \n}^{Z' Z_0 Z_0}=-\Bigg\{}
  + {1\over8\pi^2} g_{Z'} g_{Z_0}^2 \sum_f \(v_f^{Z'} a_f^{Z_0} v_f^{Z_0} +{1\over3} a_f^{Z'} a_f^{Z_0} a_f^{Z_0} \)\ m_f^2 I_0 \Bigg\} \
   \e[q,p,\n,\r] \nn\\
 &&\!\!\!\!\!\!\!\!\!\!\!\!\!\!\!\!\!\!\!
 q^\n A_{\r \m \n}^{Z' Z_0 Z_0}=
   -\Bigg\{8 \[4 g_0^3 \ R_{000}^{Z' Z_0 Z_0} \ b_2^{(0)} b_3 + 4 g_0 g_1^2 \ R_{101}^{Z' Z_0 Z_0} \ b_2^{(1)} b_3 +\right.\\
 &&\!\!\!\!\!\!\!\!\!\!\!\!\!\!\!\!\!\!\!\phantom{ p^\m A_{\r \m \n}^{Z' Z_0 Z_0}=-\Bigg\{8 }~
    \left. 2 g_0 g_2^2 \ R_{202}^{Z' Z_0 Z_0} \ b_2^{(2)} b_3 + 2 g_0^2 g_1 \ R_{001}^{Z' Z_0 Z_0} \ b_2^{(4)} b_3\] +\nn\\
 &&\!\!\!\!\!\!\!\!\!\!\!\!\!\!\!\!\!\!\!\phantom{ p^\m A_{\r \m \n}^{Z' Z_0 Z_0}=-\Bigg\{}
  + {1\over8\pi^2} g_{Z'} g_{Z_0}^2 \sum_f \(v_f^{Z'} a_f^{Z_0} v_f^{Z_0} +{1\over3} a_f^{Z'} a_f^{Z_0} a_f^{Z_0} \)\ m_f^2 I_0 \Bigg\} \
  \e[q,p,\r,\m]\nn
\eea
where $I_0$ is the integral given in (\ref{I_0integral}).
The solution for $\tilde A_1$ and $\tilde A_2$ is
\bea
 \tilde A_1 &=& \(q^2 A_4 + p \cdot q A_3 \) -\[ (\GS)^{Z' Z_0}+(\NG)^{Z' Z_0} \]\\
 \tilde A_2 &=& \(p^2 A_5 + p \cdot q A_6\) +~ (\GS)^{Z' Z_0}+(\NG)^{Z' Z_0}
\eea
 with
\bea
 (\NG)^{Z' Z_0}&=&\sum_f \(v_f^{Z'} a_f^{Z_0} v_f^{Z_0} +{1\over3} a_f^{Z'} a_f^{Z_0} a_f^{Z_0} \)\ m_f^2 I_0 \\
 (\GS)^{Z' Z_0}&=&\frac{64 \pi^2}{g_{Z'} g_{Z_0}^2} \Bigg[ 4 g_0^3 \ R_{000}^{Z' Z_0 Z_0}     \ b_2^{(0)} b_3 +
                                                       4 g_0 g_1^2 \ R_{101}^{Z' Z_0 Z_0}       \ b_2^{(1)} b_3
                                                       +\nn\\
                                     && \qquad \quad  +2 g_0 g_2^2 \ R_{202}^{Z' Z_0 Z_0} \ b_2^{(2)} b_3 +
                                                       2 g_0^2 g_1 \ R_{001}^{Z' Z_0 Z_0}      \ b_2^{(4)} b_3
                                                       \Bigg]
\eea
The rotation factors are
\bea
 R_{000}^{Z' Z_0 Z_0}     &=& O_{10} O_{10} \nn\\
 R_{101}^{Z' Z_0 Z_0}       &=& O_{01} O_{10} O_{11} \nn\\
 R_{202}^{Z' Z_0 Z_0} &=& O_{02} O_{10} O_{12} \nn\\
 R_{001}^{Z' Z_0 Z_0}      &=& O_{10} O_{11} + O_{01} O_{10} O_{10}
\eea
with $O_{ij}$ given by (\ref{Oij}). Substituting back into the
amplitude and performing all the contractions we finally obtain
\bea
 |A^{Z' Z_0 Z_0}|^2 &=&
 g_{Z'}^2 g_{Z_0}^4 \frac{ \left(M_{Z'}^2-4 M_{Z_0}^2\right)^2}{192 M_{Z_0}^2\pi^4}
 \times \nn\\
 && \[ \sum_f t_f^{Z' Z_0 Z_0} (I_3+I_5) M_{Z_0}^2 + (\GS)^{Z' Z_0}+(\NG)^{Z' Z_0}\]^2  \nn\\
\eea

%
%
%
%
%
%
%
%
%
%
%
%
%
%
%
%
%
%
%
%
\section{Amplitude for $\psi_S \, \l^{(1)} \to f \bar f$} \label{nnAmp}
In this Appendix we give some detail about the amplitude computation for the process $\l^{(1)} \, \psi_S \to f \bar f$,
\be
 \M = - i k^\m \bar v_S \g_5 \[ \g_\m,\g_\n \] u_1 \eta^{\n \r} \[e q_f \frac{C_\g}{k^2} \bar u_f \g_\r v_f +
   \frac{g_{Z_0}}{2}  \frac{C_{Z_0}}{k^2-\MZO^2} \bar u_f \g_\r (v_f^{Z_0} - a_f^{Z_0} \g_5) v_f \]
\ee
where $C_{\g}=C^{(1)} \cos\theta_W$, $C_{Z_0}=-C^{(1)} \sin\theta_W$ while $k^2=s$ is the momentum of the intermediate gauge boson.
The corresponding square modulus is
\bea
 |\M|^2 &=& - 64 \[
 T_a \(\frac{ a_f C_{Z_0} g_{Z_0} }{k^2-\MZO^2} \)^2
 +T_v \( \frac{ 2 C_\g e q_f}{k^2 }+ \frac{C_{Z_0} g_{Z_0} v_f}{k^2-\MZO^2} \)^2 \]
\label{nunuampl}
\eea
with
\bea
 T_v &=&m_f^4 (p_{\l_1} p_S)
     +M_1 M_S \Big[2 m_f^4+3 (p_f p_{\bar f}) m_f^2+(p_f p_{\bar f})^2\Big]+\nn\\
&&-(p_f p_{\bar f})
   \Big[(p_{\l_1} p_f) (p_f p_S)+(p_{\l_1} p_{\bar f}) (p_{\bar f} p_S)\Big] +m_f^2 \Big[(p_{\l_1} p_S) (p_f p_{\bar f})+\nn\\
&&-2 (p_{\l_1} p_f) (p_f p_S)-(p_{\l_1} p_{\bar f})
   (p_f p_S)-(p_{\l_1} p_f) (p_{\bar f} p_S)-2 (p_{\l_1} p_{\bar f}) (p_{\bar f} p_S)\Big] \nn\\
 T_a &=& \Big[(p_{\l_1} p_{\bar f}) (p_f p_S)+(p_{\l_1} p_f) (p_{\bar f} p_S)\Big] m_f^2-M_1 M_S
   \[m_f^4-(p_f p_{\bar f})^2\]+\nn\\
&&-(p_f p_{\bar f}) \Big[(p_{\l_1} p_f) (p_f p_S)+(p_{\l_1} p_{\bar f})(p_{\bar f} p_S) \Big]
\eea
where
$p_{\l_1}$, $p_{S}$, $p_f$ and $p_{\bar f}$ are the bino, axino and
SM fermions 4-momenta respectively.
Writing all the momenta in function of $s$ and integrating over the solid angle we get
\bea
\sigma&=&c_f \(g_1^2 b_2^{(1)}\)^2 \sqrt{s-4 m_f^2} \times \nn\\
&&
\times\frac{ \Big[-2 M_1^4+\(4
   M_S^2+s\) M_1^2-6 M_S s M_1-2 M_S^4+s^2+M_S^2 s \Big]}{12 \pi  \(\MZO^2-s\)^2 s^{5/2} \sqrt{M_1^4-2
   \(M_S^2+s\) M_1^2+\(M_S^2-s\)^2}} \times\nn\\
&&
\times\Bigg[\(s+2 m_f^2\) \Big(2 \cos\theta_W e q_f \(\MZO^2-s\) + \sin\theta_W g_{Z_0} v_f s \Big)^2 + \nn\\
&&\phantom{\times\Bigg[}
  \(s-4 m_f^2\) \(\sin\theta_W g_{Z_0} a_f s\)^2 \Bigg]
\eea

\newpage\thispagestyle{empty}

\newpage\thispagestyle{empty}
\newpage\thispagestyle{empty}

\end{document}